\documentclass[11pt]{book}

\usepackage{chhead}
\usepackage{graphicx}
\usepackage{subeqn}
\usepackage[small,nooneline,bf]{caption2}
\usepackage[latin1]{inputenc}
\usepackage{german}

\newcommand{\NPA}[3]{Nucl.\ Phys.\ {\bf A#1},\ #2 (#3)}
\newcommand{\NPB}[3]{Nucl.\ Phys.\ {\bf B#1},\ #2 (#3)}

\newcommand{\PLB}[3]{Phys.\ Lett.\ B\ {\bf #1},\ #2 (#3)}

\newcommand{\PRL}[3]{Phys.\ Rev.\ Lett.\ {\bf #1},\ #2 (#3)}
\newcommand{\PRB}[3]{Phys.\ Rev.\ B\ {\bf #1},\ #2 (#3)}

\newcommand{\PRD}[3]{Phys.\ Rev.\ D\ {\bf #1},\ #2 (#3)}

\newcommand{\ibid}[3]{{\bf #1},\ #2 (#3)}
\renewcommand\a{\alpha}
\renewcommand\b{\beta}
\newcommand\g{\gamma}
\renewcommand\d{\delta}
\newcommand\e{\epsilon}
\newcommand\z{\zeta}

\renewcommand\k{\kappa}
\renewcommand\l{\lambda}
\newcommand\m{\mu}
\newcommand\n{\nu}
\newcommand\x{\xi}
\newcommand\p{\pi}

\newcommand\s{\sigma}
\renewcommand\t{\tau}

\newcommand\D{\Delta}
\newcommand\G{\Gamma}
\renewcommand\L{\Lambda}

\newcommand{\non}{\nonumber\\}

\newcommand{\be}{\begin{equation}}
\newcommand{\ee}{\end{equation}}
\newcommand{\bea}{\begin{eqnarray}}
\newcommand{\eea}{\end{eqnarray}}
\newcommand{\ba}[1]{\begin{array}{#1}}
\newcommand{\ea}{\end{array}}

\newcommand{\bm}[1]{\mbox{\boldmath${#1}$}}
\newcommand{\uq}{\hat{\mathbf{q}}} 
\newcommand{\uk}{\hat{\mathbf{k}}}
\newcommand{\up}{\hat{\mathbf{p}}}
\newcommand{\vg}{\bm{\gamma}}
\newcommand{\gperp}{\bm{\gamma}_{\perp}}

\newcommand{\Tr}{{\rm Tr}}
\newcommand{\vv}{\bm{v}}
\newcommand{\vJ}{\bm{J}}

\begin{document}

\originalTeX

\pagestyle{ch}

%%%%%%%%%%%%%%%%%%%%%%%%%%%%%%%%%%%%%%%%%%%%%%%%%%%%%%%%%%%%%%%%%%%%%%%
%
%                   TITELBLATT
% 
%%%%%%%%%%%%%%%%%%%%%%%%%%%%%%%%%%%%%%%%%%%%%%%%%%%%%%%%%%%%%%%%%%%%%%%
%
\thispagestyle{empty}
\vspace{14cm}
\begin{center}
{\Huge\bf{Spin-one Color Superconductivity in}} 

\bigskip
{\Huge\bf{Cold and Dense Quark Matter}}\\

\vspace{2.5cm}
{\Large
Dissertation \\
zur Erlangung des Doktorgrades \\
der Naturwissenschaften
\\[.8cm]
vorgelegt beim Fachbereich Physik \\
der Johann-Wolfgang-Goethe-Universit\"at \\
in Frankfurt am Main
\\[1.2cm]
von \\
Andreas Schmitt \\
aus Frankfurt am Main
\\[1.2cm]
Frankfurt am Main, Mai 2004
\\[0.3cm]
%(D F 1)
}\end{center}
%\clearpage
%\thispagestyle{empty}
%\normalsize
%\vspace*{8cm}
%\noindent
%vom Fachbereich Physik der Johann Wolfgang Goethe--Universit\"at\\
%als Dissertation angenommen.\\[3cm]
%Dekan: Prof.\ Dr.\ \\[1cm]
%Gutachter: Prof.\ Dr.\ D.\ H.\ Rischke,\\[1cm]
%Datum der Disputation: 
%\clearpage

%
\newpage

\chapter*{Abstract}

In this thesis, several color-superconducting phases where 
quarks of the same flavor form Cooper pairs are investigated. 
In these phases, a Cooper pair carries total spin one. A systematic 
classification of theoretically possible phases, discriminated by the 
color-spin structure of the order parameter and the respective symmetry 
breaking pattern, is presented. In the weak-coupling 
limit, i.e., for asymptotically high densities, a universal form of the 
QCD gap equation is derived, applicable to arbitrary color-superconducting 
phases. 
It is applied to several spin-one and spin-zero phases in order to determine 
their energy gaps and critical temperatures. In some of the spin-one phases
the resulting gap function is anisotropic and has point or line nodes. 
It is shown that the phases with two different gaps violate the well-known
BCS relation between the critical temperature and the zero-temperature gap.
Moreover, the screening properties of color superconductors 
regarding gluons and photons are discussed. In particular, it turns out that, 
contrary to spin-zero color superconductors, spin-one color superconductors 
exhibit an electromagnetic Meissner effect. This property is proven by 
symmetry 
arguments as well as by an explicit calculation of the gluon and photon
Meissner masses.  

 %Abstract (only for arXive version) 

\chapter*{Acknowledgments} 

My special thanks go to two people who substantially contributed to this
thesis. 

First of all, I would like to thank my advisor Dirk Rischke for
introducing me to the research field of QCD and guiding me through the
conceptual and technical problems of color superconductivity. 
I thank him for innumerable fruitful discussions, hints, and suggestions, 
and his patience to discuss even the slightest technical details with me. 
I have
enjoyed the three years in his group last but not least because of his 
friendly and helpful way to cooperate with his students and colleagues. 
{\it Dank' Dir, Dirk!}

Second, I thank Qun Wang, without whom this thesis would not have been 
possible. He closely collaborated with me from the beginning of the work, and
I have largely benefitted from his experience, his knowledge, and his 
patience. Through many inspiring discussions we have approached the 
final results of this thesis.
Most important, these discussions always have been in a nice and friendly
atmosphere. It was a great pleasure to work with you, Qun! {\it Xiexie!}

Furthermore, thanks to all members of Dirk's group: Thanks to Igor Shovkovy,
who helped me to understand a lot of the problems that have been essential for 
this thesis. Thanks to Amruta Mishra and Philipp Reuter for sharing the 
office with me throughout the three years. We have always had a great 
atmosphere in 608, and I will miss this time. Special thanks to Philipp for 
discussing physics with me since our first semester. Thanks to Mei Huang, 
Defu Hou, Tomoi Koide, and Stefan R\"uster for inspiring comments and 
discussions in our group meetings (and in the {\it mensa}).   

Thanks to the ``trouble team'' Kerstin Paech, Alexander Achenbach, 
Manuel Reiter, and Gebhardt Zeeb, for helpfully caring about any kind of 
computer trouble. 

Thanks for valuable comments to M.\ Alford, W.\ Greiner, M.\ Hanauske,
P.\ Jaikumar, M.\ Lang, C.\ Manuel, J.\ Ruppert, K.\ Rajagopal, H.-c.\ Ren, 
T.\ Sch\"afer, J.\ Schaffner-Bielich, D.T.\ Son, and H.\ St\"ocker.   

This work was supported by GSI Darmstadt.     

 %Acknowledgments

\tableofcontents

%\listoffigures
%\listoftables

\chapter{Introduction} \label{intro}
The first three sections of this thesis serve as a  
motivation and as an introduction into the basics of the 
underlying theories that describe the physics. In these introductory sections,
we essentially address the following questions.
The first question is related to the second term of the thesis' title, 
``cold and dense quark matter''. It is, of course, ``Why do we study cold and
dense quark matter?''. This question naturally is connected with the questions
``Where and when (= under which conditions) does cold and dense quark matter 
exist?'' and, simply, but important, ``What {\em is} cold and dense 
quark matter?''. The first part of this introduction, 
Sec.\ \ref{colddense}, is dedicated to these questions, which will be
studied without discussing technical details in order to allow an easy 
understanding of the thesis' main goals. 

The second question addressed in the introduction, namely in 
Sec.\ \ref{BCS_He3}, is related to the first term of the thesis' title,
more precisely to its third word, ``superconductivity''. This
question simply is ``What is superconductivity?''. Since this is a theoretical
thesis, this question basically will turn out to be ``What are the  
theoretical means to explain superconductivity?''.
This section is a pure ``condensed-matter section'', or more precisely,
a ``low-energy condensed-matter section'', i.e., we focus
on many-particle (= many-fermion) systems adequately described solely by 
electromagnetic interactions. 
It introduces the basics of the famous BCS theory, developed in the fifties 
in order to explain the phenomena and the mechanism of superconductivity in 
metals and alloys. A special emphasis is put on the theoretical concept
of ``spontaneous symmetry breaking'', a concept widely and successfully 
applied to a variety of different fields in physics. 
%Since
%are closely connected with groups, an elementary group theoretical 
%discussion also enters this section. 
Although phenomenologically 
quite different from superconductors, a second physical system is
introduced in Sec.\ \ref{BCS_He3}, namely superfluid Helium 3 ($^3$He). From
the theoretical point of view, superfluidity is very similar to 
superconductivity. Furthermore, the theory of superfluid $^3$He contains 
important aspects (not relevant in ordinary superconductivity) 
which are 
applicable to {\em color} superconductivity, and especially to {\em spin-one}
color superconductivity. 

Therefore, in the last part of the introduction, Sec.\ \ref{CSC}, we approach
the question ``What is color superconductivity?'' by asking 
``How are color superconductors related to  
ordinary superconductors and superfluid $^3$He?''. It turns out that the 
methods on which
the theory of color superconductors is built on, do not differ essentially 
from the well-established methods introduced in Sec.\ \ref{BCS_He3}. 
However, we consider quark matter! Therefore, this ``condensed-matter physics 
of QCD'' is more than a transfer of well-known theories to a different 
physical system. Besides the physically completely different implications,
which extensively touch the field of astrophysics, also the theory gains
complexity due to the complicated nature of quarks and the involved 
properties of QCD. Of course, Sec.\ \ref{CSC} also discusses the question
``What is special about {\em spin-one} color superconductors?'', and, since 
particularly {\em spin-zero} color superconductors have been a matter of 
study in a large number of publications in recent years, ``Why should 
not only spin-zero,
but also spin-one color superconductors be considered?''. 
%Actually, 
%here we are at the point where most of the new results of this thesis
%belong to. While only a few of the new results presented in 
%Sec.\ \ref{mainpart}, mainly remarks or useful generalizations, concern 
%spin-{\em zero} superconductivity, the main part of the new results concern 
%spin-{\em one} color superconductors, which
%so far has been discussed in a much lower number of publications.

   \section{Cold and dense quark matter} \label{colddense}

\subsection{The phase diagram of QCD} \label{QCDphasediagram}

\begin{figure}[ht] 
\begin{center}
\includegraphics[width=14cm]{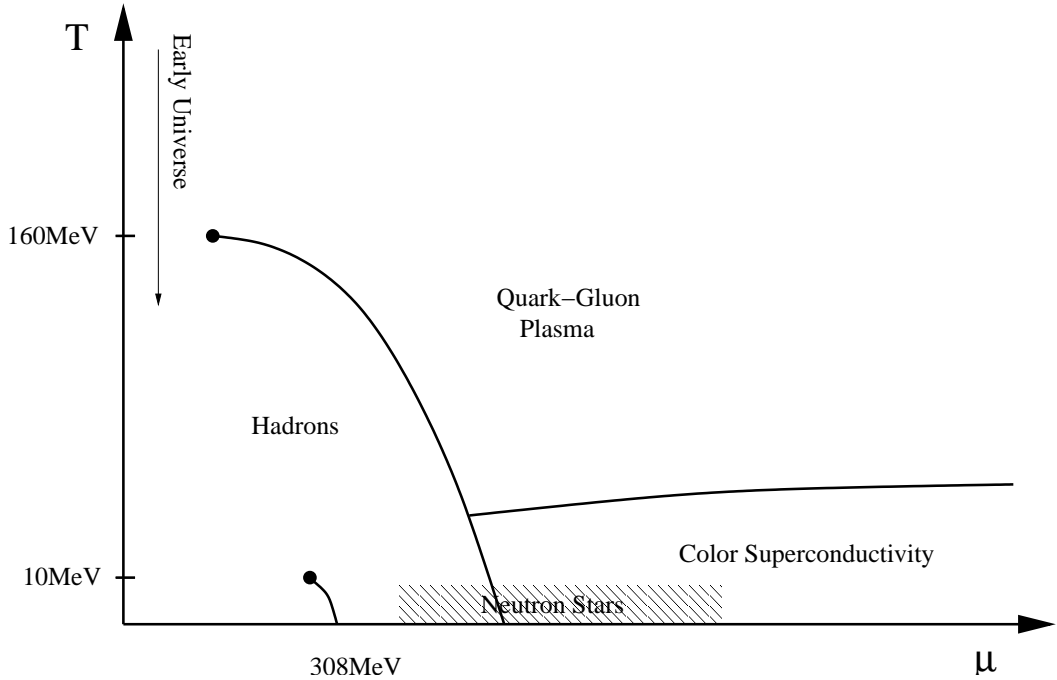}
\vspace{0.5cm}
\caption[QCD phase diagram]{Schematic view of the QCD phase diagram. 
The curves indicate 
the phase boundaries of: the liquid/gas phase transition of nuclear matter,
starting at $(T,\m)=(0,308\,{\rm MeV})$, ending in a critical point at
$T=10\,{\rm MeV}$); the hadron-quark phase transition, separating hadronic
matter from the quark-gluon plasma and ending in a critical point 
at $T=160\,{\rm MeV}$; the normal-conducting/superconducting phase
transition in the quark-gluon plasma.}
\label{QCDpd}
\end{center}
\end{figure}

In the phase diagram of quantum chromodynamics (QCD), Fig.\ \ref{QCDpd}, 
every point represents an infinitely large system of quarks and 
gluons in thermal equilibrium with a certain temperature $T$ and 
a quark chemical potential $\m$. Since the particle number density $\rho$ 
is a monotonously increasing function with $\m$, $\rho\sim\m^3$, we can,
for the following qualitative and introductory discussion,  
equivalently use ``system with high chemical potential'' and 
``dense system''. We expect that QCD is the 
suitable theory even for the extreme regions of this phase diagram, 
namely very hot or very dense systems. Due to a special property of QCD,
called asymptotic freedom \cite{asymp}, the coupling between quarks and
gluons becomes weaker in the case of a large momentum exchange or in the
case of a small mutual distance. Therefore, applying QCD to systems with
high temperature and/or large densities, we expect the quarks to be
in a deconfined phase \cite{collins}, contrary to the low 
temperature/density phase, where they are confined into hadrons. 

In recent years, these ``extreme''
systems have gained more and more attention in experimental as well as in 
theoretical physics. There are several reasons for this interest. First of all,
it is a general experience in physics that the study of systems in extreme
regions of the phase diagram (or in extreme energy regions, or with extreme 
velocities, masses, etc.) often are followed by genuinely new developments in 
theory and experiment, leading to a deeper understanding of the existing
theories (and of nature) or to a completely new theory. Therefore, 
it is an outstanding goal of research to go beyond temperatures and 
densities at which quarks are in the ordinary hadronic phase. Simply speaking,
take a system of quark matter and heat it up and/or squeeze it to a sufficient
amount and you will definitely learn a lot of new physics. Second, the
investigation of systems under extreme conditions might help to 
understand special regions (in space and in time) of the universe. In other
words, most likely there exists or existed quark matter in the deconfined
phase, also called quark-gluon plasma. For instance, in the early 
stages of its evolution the universe was very hot. While cooling down, the 
universe passed through a cross-over into the hadronic phase 
(note that the quark-hadron phase boundary ends in a critical point; 
thus, for small densities, as present in the early universe, there is no phase
transition in the strict sense from the quark-gluon plasma to the hadronic 
phase). 

Through collisions 
of heavy nuclei at ultrarelativistic energies one tries to imitate the 
situation of the early universe in the laboratory \cite{heavyion}. In these 
experiments, one expects to create a quark-gluon plasma at least for a short 
time, after which
the temporarily deconfined quarks again form hadrons. Since these hadrons 
are the particles that can be observed by the detectors (and not the
individual quarks), it is a subtle task to deduce the properties 
(or even the existence) of the quark-gluon plasma from the experimental data. 
Nevertheless,
in recent years, a fruitful, though often controversial, interplay between
different experimental groups as well as between experiment and theory 
has led to an established opinion that the quark-gluon plasma can be created 
in 
heavy-ion collisions (with bombarding energies experimentally accessible 
nowadays or in the near future) or that it already has been created. 
Still, there are a lot of open questions in this field (note 
for instance, that, in order to describe heavy-ion collisions properly,
one needs nonequilibrium methods). 

Theoretically, the region of high 
temperature and low density (strictly speaking, $\m=0$) can be described 
by ``lattice QCD'' \cite{lattice}. In lattice QCD, one calculates the
partition function of thermal QCD numerically on a lattice in the 
four-dimensional space spanned by the three spatial directions and the 
inverse temperature axis. Making use of the so-called Polyakov loop 
(or Wilson line)
as an order parameter, lattice QCD is able to make predictions for the nature
and the critical temperature of the quark-hadron phase transition 
\cite{wilson,polyakov}. Recently, also the technically involved problem of 
extending lattice QCD calculations to nonzero chemical potentials has been 
approached; for a review, see Ref.\ \cite{laermann}.

Let us now turn to the region of the phase diagram that is of special 
interest in this thesis, the region of cold and dense matter, 
named ``color superconductivity'' in Fig.\ \ref{QCDpd}. In 
this plot of the QCD phase diagram, no value of the chemical potential 
has been assigned to the intersection point of the quark-hadron phase 
boundary and the $T=0$ axis; unlike the gas-liquid nuclear matter phase 
transition, which, for $T=0$, occurs at the well-known value 
$\m=308\,{\rm MeV}$. The reason for
the missing number is simple: One does not know it. This region of the phase 
diagram, namely cold quark matter at densities ranging from a few times nuclear
matter ground state density, $\rho_0\sim 0.15\,{\rm fm}^{-3}$, up to 
infinite density, is poorly understood. Therefore, one (initially theoretical)
motivation of this thesis is to contribute to a further understanding of a 
special region in the QCD phase diagram. 
The rich phase structure of ordinary condensed-matter physics (one might 
call it ``condensed-matter physics of quantum electrodynamics (QED)'')
suggests that also in ``condensed-matter physics of QCD'' a variety of 
different phases appears. This connection with ordinary condensed-matter (and
solid-state) physics can be considered as another motivation for the study 
of cold
and dense quark matter. On the one hand, both research fields share the 
interest for similar physical systems, on the other hand, these physical 
systems differ in essential properties, wherefore it is promising
to learn from each other and induce new developments in both fields. This
interplay will be illustrated for instance in the next two sections,
where color superconductivity (Sec.\ \ref{CSC}) will be based on the 
theory of ordinary superconductivity and superfluidity (Sec.\ \ref{BCS_He3}).

Another motivation for the investigation of cold and dense matter is its 
relevance
for astrophysics. The densest matter systems have not been produced
in the laboratory, they rather exist in nature, namely in compact
stellar objects. Therefore, let us insert a short introduction about 
neutron stars. Elementary introductions about properties and the evolution 
of neutron stars can be found in text books such as 
Refs.\ \cite{phillips,weber}. For more specialized reviews and articles, 
treating neutron star properties of special interest for this thesis,
see Refs.\ \cite{ruderman,pons,prakash,svidzinsky}.

\subsection{Neutron stars} \label{neutronstars}

Neutron stars are compact stellar objects that originate from supernova 
explosions. First observations of neutron stars were done in the late 
sixties \cite{astro}. Neutron stars have a radius of about 10 km and a mass 
which is of the order of the sun's mass, $M\sim 0.1M_{S} - 2M_{S}$ 
\cite{weber}.
Consequently, the matter that a neutron star is composed of is extremely 
dense;
note for a comparison that the sun's radius is $6.7\cdot 10^5\,{\rm km}$.
The matter density increases from the surface to the center of the star. 
Therefore, the matter of a neutron star exists in different phases, each
phase forming a layer of the spherical star. At the surface of the neutron 
star, there is a thin crust of iron, followed by a layer that consists of 
neutron-rich nuclei in an electron gas. At still higher densities there is a
phenomenon called 
``neutron drip'', i.e., neutrons start to coexist individually
in equilibrium with nuclei and electrons. Even closer to the center of the 
star, approaching nuclear matter density $\rho_0$, nuclei cease to exist and a 
phase of neutrons, protons, and electrons
is the preferred state of matter. Neutron stars are called neutron stars 
because this phase (and the nuclear phase at lower densities) 
is very neutron-rich. Approaching even higher densities, a phase containing
pions, muons, and hyperons is predicted to be the favorite state, i.e., simply
speaking, the fundamental elements of neutrons and protons, $u$ and $d$ 
quarks, start to form different hadrons and also the heavier $s$ quarks
might be involved. At the core of the neutron star, matter density might very
well reach values an order of magnitude larger than $\rho_0$ \cite{weber}.
Therefore, pure quark matter is likely to be found in the interior of the
star. 

Properties of neutron stars are explored by theoretical considerations as well
as astrophysical observations. Theoretical models make use of two essential 
ingredients. First, the equation of state, which connects pressure with 
temperature and energy density. Second, general relativity, i.e., 
Einstein's field equations. Experimental data is essentially based on 
a certain  property of a neutron star, namely its rotation. Due to this 
rotation, neutron stars are pulsars, i.e., they emit electromagnetic 
radiation in periodic pulses. The rotating periods are in 
the range of milliseconds \cite{weber}. Observations have shown that the 
spinning-down of the star (the rotation frequency decreases due to 
radiative energy loss) is interrupted by sudden spin-ups, called glitches.

From the spectra and frequency of the radiation pulses, properties of the 
neutron star can be deduced. The most important properties,
more or less well known, are mass, radius, temperature, and magnetic
field of the star. The typical values for mass and radius have been quoted
above; let us now briefly discuss temperature and magnetic field. The 
temperature of neutron stars right after their creation is in the range 
of $10^{11}\,{\rm K}$, or 10 MeV. During its evolution,
the star cools down. This decrease of temperature is dominated by 
neutrino emission. After a time of about a million years, the star has 
cooled down to temperatures $\sim 10^5\,{\rm K}$, or 10 eV 
\cite{weber,prakash}.
Consequently, matter in the interior of neutron stars indeed is 
a realization of ``cold and dense matter'', where ``cold'', of course,
refers to the scale given in the QCD phase diagram, Fig.\ \ref{QCDpd}. 
As will be shown in Sec.\ \ref{mainpart},
temperatures present in old neutron stars are certainly lower than 
the critical temperatures of color superconductors.

Next, let us discuss the magnetic field. Indirect measurements suggest 
that the magnetic field 
at the surface of a neutron star is of the order of $10^{12}\,{\rm G}$
\cite{weber}. 
This is thirteen orders of magnitude larger than the magnetic field 
at the surface of the earth. Many questions concerning the magnetic fields
of neutron stars, especially its origin, are not well understood. On the
other hand, this physical quantity perhaps is the most important one in 
order to investigate possible superconducting phases in the interior 
of the star. Since color superconductivity and its magnetic properties 
will be a matter of investigation in the main part of this thesis, 
especially in Sec.\ \ref{mixingscreening}, let us now summarize the 
conventional picture of a neutron star (= without quark matter) regarding 
superconductivity and superfluidity. Recent developments in this interesting
field can be found in Refs.\ \cite{ruderman,svidzinsky,sedrakian,link,zhit}.

As a simplified picture, assume that the neutron star consists of 
two different phases. One of them containing neutron-rich
nuclei and forming the crust of approximately one kilometer. The other one
forming the core of the star and consisting of neutrons and protons (and 
electrons). For sufficiently low temperatures, $T\sim 10^9\,{\rm K}$, 
the neutrons are in a superfluid state while the protons
form a superconductor. Therefore, the core of the star is governed by
an interplay between a superfluid and a superconductor, both present 
in the same spatial region. There are at least two observed properties of the
star that are closely related to these exotic states in the interior. First, 
the above mentioned glitches, and second, the precession periods of the star.
The explanation of the glitches is closely related to the superfluidity of the 
neutrons. Due to the rotation of the star, an array of vortex lines 
is formed. This vortex array expands when the star spins down. Sudden
jumps of the rotation frequency are explained by the fact that the vortex
lines are pinned to the crust of the star \cite{ruderman}. The picture 
becomes more complicated if one includes the superconducting protons. If this
superconductor is of type-II, there are magnetic flux tubes
through which the magnetic field may penetrate the core of the star. Taking
into account an interaction between these flux tubes and the superfluid 
vortices, it has been shown that this picture of a neutron 
star contradicts the observed precession periods of about one year 
\cite{link}.
In Refs.\ \cite{sedrakian,zhit}, however,
the possibility is pointed out that protons form a type-I superconductor.

At the end of this astrophysical intermezzo, let us also mention that
not only conventional neutron stars and neutron stars with a quark core 
have been studied, but also the possibility of pure quark stars. 
Recent interpretations of observations regarding this 
question can be found in Refs.\ \cite{quarkstars}.

      %QCD phase diagram
   \section{Superconductivity and superfluidity} \label{BCS_He3}

Superconductivity as well as superfluidity are states of interacting 
many-fermion systems that are distinguished from the normal state by 
an order parameter. The transition from the normal state to the 
superconducting or superfluid state therefore is a phase transition. 
The order parameter characterizes the different phases and, as a function 
of temperature,   
changes its value at a certain temperature, 
the critical temperature $T_c$. While
this function is zero in the normal phase, it assumes a nonzero value in the
superconducting and superfluid phases. The concept of an order parameter 
and a critical temperature is common to all phase transitions. For instance, 
in the liquid/gas phase transition of water
the order parameter is the particle density, which discontinuously changes
from one phase to the other. Another example is ferromagnetism, where
the order parameter is given by the magnetization. 

In both superconductivity and superfluidity the order parameter is given
by a less trivial quantity. Although the mathematical structure of the 
order parameter can be quite different for superconductors and 
superfluids (and also for different kinds of these systems), the underlying
physical mechanism is the same in each case. This fact allows us to identify 
the order parameter as the quantity accounting
for existence (superconductor/superfluid) or non-existence (normal state)
of Cooper pairs. In the following we will elaborate on the properties and  
theories of superconductors and superfluids. In the case of the latter, we will
focus on superfluid $^3$He. One can find reviews of these theories in 
many textbooks such as Refs.\ \cite{gala,fetter,tinkham,vollhardt}.

\subsection{Superconductivity} \label{superconductivity}

The history of superconductivity started in 1911, when Kamerlingh Onnes 
discovered that the electric resistance of mercury became unmeasurably
small below temperatures of $4.2\,{\rm K}$ \cite{onnes}. Almost fifty
years later, in 1957, a theoretical model for the phenomenon of 
superconductivity, based on microscopic theories, was published \cite{bcs}.
This theory by Bardeen, Cooper, and Schrieffer (BCS theory)
has been very successful in explaining and predicting the 
properties of conventional superconductors until this day. 
However, in 1986, with 
the first discovery of high-$T_c$ superconductors \cite{bednorz}, a class
of superconductors was started to be studied, for which still no
satisfactory theoretical explanation has been found. These high-$T_c$ 
superconductors can have transition temperatures up to $125\,{\rm K}$.      
       
\begin{figure}[ht] 
\begin{center}
\vspace{0.5cm}
\includegraphics[width=10cm]{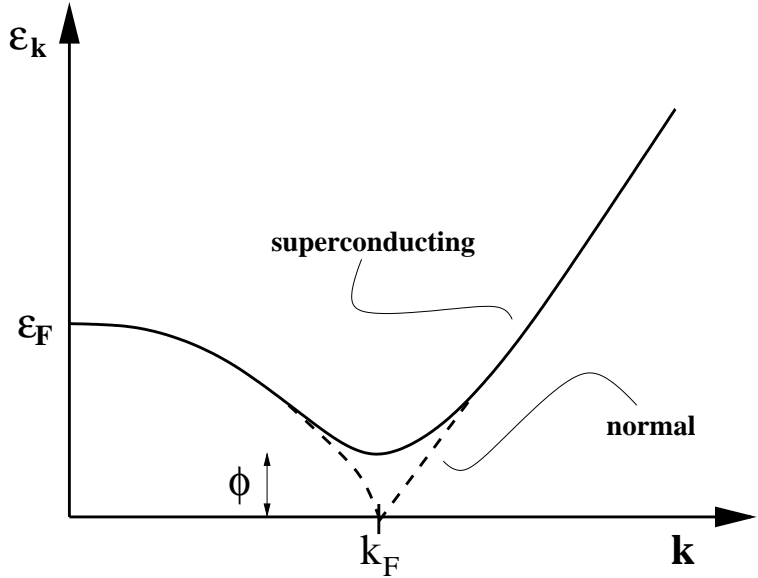}
\vspace{0.5cm}
\caption[Dispersion relation in a superconductor]{Fermionic 
(quasi)particle dispersion relations for the normal
and superconducting phases (non\-relativistic). }
\label{figexcite}
\end{center}
\end{figure}

As mentioned above, the nonzero value of the order parameter in the 
superconducting phase is equivalent to the existence of Cooper pairs. 
Let us explain this in some more detail. The physical system we are dealing
with is a metal or alloy. Theoretically, it can be 
described by an interacting many-electron system in the presence of 
phonons, i.e., 
lattice oscillations. Since electrons are fermions, they obey the Pauli 
exclusion principle and thus, at $T=0$, all quantum states
up to a certain energy, the Fermi energy $\e_F=\mu$ (where $\m$ is the 
electron chemical potential), are occupied, each by a 
single electron, while all energy states above $\e_F$ are empty. In 
momentum space, due to $\e_k^0=p^2/(2m)$, where $\e_k^0$ is the 
energy, $p\equiv \hbar k$ the momentum, and $m$ the mass of the electron, 
the boundary between occupied and empty states
is the surface of a sphere, the Fermi sphere, whose radius is given by
the Fermi momentum $p_F\equiv \hbar k_F$. Cooper showed that if there is an 
arbitrarily
weak attractive force between the fermions, a new ground state of the system
will be preferred, in which electrons at the vicinity of the Fermi surface
form pairs (comparable to a bound state). Then, the total energy of the 
system is reduced by the amount of the sum of the 
binding energies of the electron pairs. Moreover,
the single particle excitation energies are modified. They acquire a gap 
$\phi_k$, accounting for the fact that in the superconducting state one needs 
a finite amount of energy to excite an electron (more precisely, a 
quasiparticle) at the Fermi surface, which is not the case in the normal 
state, where an infinitesimally small energy is needed to excite an electron 
at the Fermi surface (cf.\ Fig.\ \ref{figexcite}). 

Coopers theorem applies to a superconductor since, indeed, and in spite 
of the repulsive Coulomb interaction, there is an attractive force between 
the electrons. This is provided by exchange of virtual phonons and was first 
pointed out by Fr\"ohlich \cite{froehlich}. In a crystal, the phonon 
energy is bounded by the Debye energy $\hbar\omega_D$. Thus, the exchanged 
momentum is much smaller than $\hbar k_F$. Therefore, due to Pauli blocking, 
only 
electrons at the vicinity of the Fermi surface, say, in an interval 
$[k_F-q,k_F+q]$, where $q\ll k_F$ is determined by the Debye energy, 
can interact via this
phonon exchange. This is the reason why superconductivity is a pure
Fermi surface phenomenon. 

\begin{figure}[ht] 
\vspace{0.5cm}
\begin{center}
\includegraphics[width=9cm]{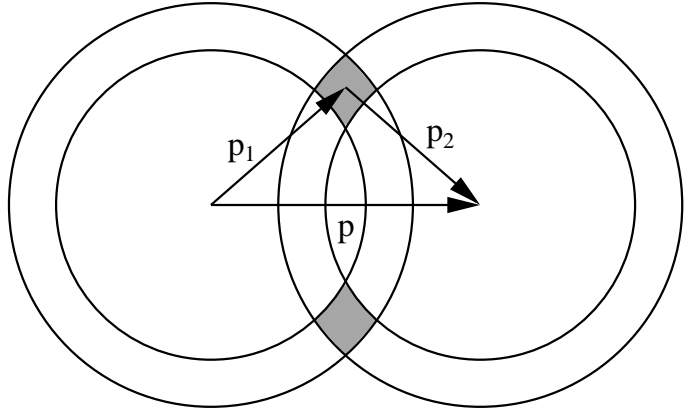}
\vspace{0.5cm}
\caption[Total momentum of a Cooper pair]{Schematic picture of the 
total momentum ${\bf p}$ of an electron 
Cooper pair, composed of the two electron momenta ${\bf p}_1$ and ${\bf p}_2$.
Only electrons with momenta in a small vicinity around the Fermi surface can
form Cooper pairs. Therefore, only electrons in the shaded areas (= volumes
in three-dimensional momentum space) can form
Cooper pairs with a given momentum ${\bf p}$. The shaded region and thus the
total binding energy is maximized for ${\bf p}=0$, i.e., 
${\bf p}_1=-{\bf p}_2$.}
\label{pairmomentum}
\end{center}
\end{figure}

In conventional superconductors, each Cooper pair has a vanishing
total spin, $S=0$,  as well as a vanishing angular momentum, $L=0$ ($s$-wave).
In high-$T_c$ superconductors, however, the situation seems to be 
more complicated. Superconductors with $d$-wave states as well as spin-triplet
superconductors, $S=1$, have been found experimentally and studied 
in theoretical models \cite{maeno,mackenzie}. 
But let us continue with the simplest 
situation of conventional superconductors. 
In this case, both electrons of a Cooper pair have momenta of equal absolute 
value but  
opposite direction. Consequently, the total momentum of a Cooper pair 
vanishes, which is illustrated in Fig.\ \ref{pairmomentum}. The fact
that all Cooper pairs must have the same total momentum can, roughly 
speaking, be explained by a Bose-Einstein condensation of the pairs. 
In a Bose-Einstein condensate, a macroscopic number (= proportional to 
the volume of the system) of bosons are in one quantum state, the ground
state. But note that in conventional (weak coupling)
superconductors the Cooper pairs are extensively overlapping in space.
Therefore, they are far from being considered as separated, well-defined
bosons and the picture of Bose-condensed Cooper pairs is questionable.
Nevertheless, one speaks of a ``condensation of Cooper pairs'' to 
describe the superconducting state.
However, in strong-coupling superconductors, where the spatial 
extension of a Cooper pair might be smaller than the average distance 
between the electrons, a real Bose-Einstein condensation might set in.
The difference between BCS Cooper pairing and a Bose-Einstein condensation
of Cooper pairs has been studied for instance in 
Refs.\ \cite{nozieres,babaev} and, for color superconductivity, in 
Ref.\ \cite{itakura}.

Let us now quote the essential equations for a quantitative treatment
of superconductivity according to BCS theory. The BCS gap equation for a 
temperature $T$ and 
a chemical potential $\m$ reads \cite{fetter},
\be \label{BCSgapeq}
1=g\int\frac{d^3{\bf k}}{(2\pi)^3} \frac{1}{2\e_k}\tanh\left(
\frac{\e_k}{2k_B T}\right) \,\, ,
\ee
Here, $k_B$ is the Boltzmann constant, $g$ the coupling constant of the 
attractive electron-electron interaction and $\e_k$ the quasiparticle 
excitation energy (see Fig.\ \ref{figexcite}),
\be \label{excitenonrel}
\e_k=\sqrt{(\e_k^0-\mu)^2 + \phi^2} \,\, .
\ee
Note that, in order to derive Eq.\ (\ref{BCSgapeq}), one assumes the 
interaction to be constant for electrons at the vicinity of the Fermi surface
and zero elsewhere. In this case, the gap function $\phi_k$ assumes a constant 
value $\phi$ at the Fermi surface and thus can be cancelled from the original 
integral equation which leads to Eq.\ (\ref{BCSgapeq}), where $\phi$
is only present in the excitation energies. Introducing a
cut-off for the momentum integration, naturally given by the limit of  
the phonon energy, $\hbar\omega_D$, i.e., integrating solely over the above
mentioned momentum interval around the Fermi surface,  the
gap equation can easily be solved for $T=0$. One obtains
\be \label{BCSgap}
\phi = 2\hbar\,\omega_D \,\exp\left(-\frac{1}{N(0)\,g}\right) \,\, ,
\ee
where 
\be
N(0) = \frac{m\,k_F}{2\pi^2\hbar^2}
\ee
is the density of states at the Fermi surface. In the limit $T\to T_c$,
the gap equation yields a relation between the zero-temperature gap $\phi$
and the critical temperature,
\be \label{BCSrelation}
\frac{k_B T_c}{\phi} = \frac{e^\g}{\pi} \simeq 0.57 \,\, ,
\ee
where $\g\simeq 0.577$ is the Euler-Mascheroni constant.

The relations given in Eqs.\ (\ref{BCSgap}) and (\ref{BCSrelation}), 
regarding the gap parameter and the critical temperature, respectively, 
are two fundamental results of BCS theory. In the following sections, 
especially in Sec.\ \ref{gapeqsolution}, it will be discussed
if and how these results have to be modified for {\it color} superconductors.

As discussed in Sec.\ \ref{neutronstars}, it is of great physical importance
if a superconductor is of type I or type II. This classification
has direct consequences for the behaviour of the superconductor in an
external magnetic field. Superconductors of type I are characterized
by one critical magnetic field $H_c$ (depending on temperature).  
Therefore, for a fixed temperature and an external magnetic field $H>H_c$ the 
system is in the normal-conducting state, whereas for a magnetic field $H<H_c$,
the system is in the superconducting state. Moreover, in this case, the
superconductor exhibits the Meissner effect, i.e., the magnetic 
field is expelled from the interior of the 
superconductor. More precisely, there is a finite penetration length $\l$ for
the magnetic field, or in other words, the magnetic photon acquires 
a mass (``Meissner mass''), $m_M\sim 1/\l$. Superconductors of type II,
however, are characterized by two different critical magnetic fields, 
$H_{c1}<H_{c2}$. For magnetic fields $H>H_{c2}$ the system is in the normal 
phase, and for $H<H_{c1}$ the system is superconducting with complete
expulsion of magnetic fields. For magnetic fields
with a strength between the two critical fields, $H_{c1}<H<H_{c2}$,
the system is superconducting but the magnetic field penetrates partially
into the superconductor through so-called flux tubes. These objects
are one-dimensional ``topological defects'' (= vortices), i.e., tubes
with a width of the order of the coherence length $\xi$ where the 
order parameter or, equivalently, the density of Cooper pairs, vanishes.
The type of a superconductor is determined by the Ginzburg-Landau
parameter 
\be
\k \equiv \frac{\l}{\xi} \,\, .
\ee
A superconductor is of type I if $\k<1/\sqrt{2}$ and of type II if
$\k>1/\sqrt{2}$. This difference between superconductors with certain
values for $\k$ and the existence of vortices was theoretically predicted by
Abrikosov in 1957 \cite{abrikosov}, based on the phenomenological 
Ginzburg-Landau Theory \cite{ginzburg}. This prediction turned out to be 
relevant especially for high-$T_c$ superconductors which are of type II
and thus show a pattern of flux tubes for suitable external magnetic fields.

\subsection{Superfluid $^3$He} \label{he3} 

\begin{figure}[ht] 
\begin{center}
\includegraphics[width=12cm]{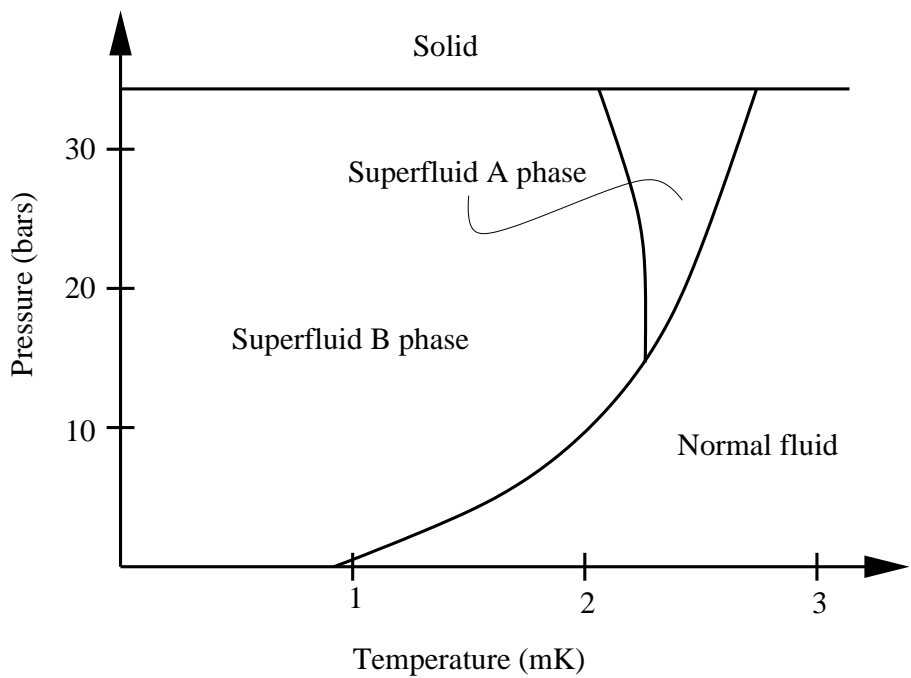}
\vspace{0.5cm}
\caption[$^3$He phase diagram, I]{Phase diagram for $^3$He. Two 
different phases can be found
in the superfluid region, namely the A and B phases.}
\label{Hephase1}
\end{center}
\end{figure}

Superfluidity in helium was first observed for the isotope $^4$He 
\cite{kapitza}. Its transition temperature was found to be $T_c=2\,{\rm K}$.
Since $^4$He atoms are bosons, their superfluid phase
is theoretically explained by a Bose-Einstein condensation of the atoms. 
The lighter 
isotope $^3$He, however, is fermionic since it is composed of three nucleons 
and two electrons, adding up to a non-integer total spin. Motivated
by the success of BCS theory, in the sixties, theoreticians 
applied the mechanism of Cooper pairing to systems with fermionic atoms.
Experimentally, superfluid $^3$He was first observed in 1971 at temperatures
around $T_c=2\,{\rm mK}$ \cite{osheroff}, cf.\ Fig.\ \ref{Hephase1}, i.e., 
three orders of magnitude smaller than in the case of $^4$He. The 
theoretical breakthrough concerning the explanation of the rich phase 
structure of superfluid $^3$He was done by Leggett in 1975 \cite{leggett}. 

Let us in the following elaborate on some aspects of the theory of 
superfluid $^3$He; it will turn out that, in order
to understand (spin-one) color superconductivity, it is very useful to be 
familiar not only with the BCS theory of conventional superconductors but 
also with the theory of superfluid $^3$He. 

Roughly speaking, superfluidity is the same as superconductivity, with the only
difference that the Cooper pairs are not charged. This simplified statement
has to be treated with great care (especially in the case of {\it color}
superconductivity/superfluidity!) and even can be misleading. 
But nevertheless, let us start with this statement for a brief theoretical 
introduction into superfluid $^3$He. Here, as for the electron liquid 
in metals, there is an attractive interaction between the fermions. This 
interaction is provided by the van-der-Waals force. Consequently, 
Cooper's theorem applies and Cooper pairs of atoms are formed below
the transition temperature. These pairs have total momentum ${\bf p}=0$, 
and there is a gap equation which has a nonzero solution for the gap 
that occurs 
in the quasiparticle excitation energies. In this sense, superfluidity
is very similar to superconductivity. But since $^3$He atoms (and thus 
also the Cooper pairs) are neutral, while electrons carry electric charge,
the physical implications are very different for the two systems. The 
question of electric (and color) neutrality in the case of a
color superconductor is much more complicated and will be discussed in 
Secs.\ \ref{CSC} and \ref{mixingscreening}.  

\begin{figure}[ht] 
\begin{center}
\includegraphics[width=9cm]{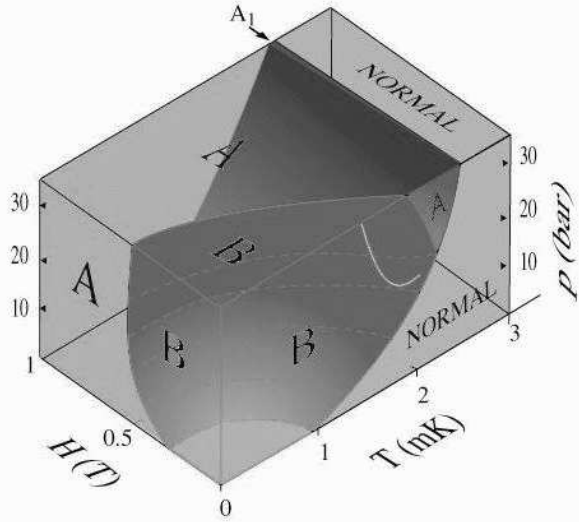}
\vspace{0.5cm}
\caption[$^3$He phase diagram, II]{Phase diagram for $^3$He including an 
external magnetic field $H$. 
For an increasing magnetic field, the B phase becomes more and more disfavored
while the A phase covers most of the superfluid region of the phase diagram, 
and an additional phase, the ${\rm A}_1$ phase, that is absent in the situation
of a vanishing magnetic field, occurs.}
\label{Hephase2}
\end{center}
\end{figure}

But note another difference between superfluid $^3$He and conventional 
superconductors. The atomic interaction potential 
becomes repulsive for short mutual distances. Therefore, a Cooper pair in 
$^3$He has total angular momentum $L=1$ ($p$-wave state), 
since, in this case, the 
pair wave function vanishes for zero distance. This has direct implications
for the total spin of the pair, as can be seen from the following 
simple symmetry argument.
Since the total wave function of the Cooper pair, consisting of two
fermions, has to be antisymmetric and the $p$-wave function is antisymmetric
with respect to exchange of the coordinates of the two fermions,
it must be symmetric with respect to exchange of the two spin states of the
fermions. Thus, in superfluid $^3$He, the Cooper pairs are in an $S=1$
state. This special feature of a nonzero angular momentum as well as a nonzero
spin is the origin for a rich phase structure, i.e., there are several 
different superfluid phases as can be seen in the phase diagrams shown in
Figs.\ \ref{Hephase1} and \ref{Hephase2}.
 
A simple way to understand the occurrence of several different superfluid 
phases is provided by the concept of spontaneous symmetry breaking.
This concept is used in many different fields of physics since it is
closely connected with the theory of phase transitions.
In order to make use of it in the discussion of color superconductivity, we 
explain some basic aspects in this introduction and apply them to 
superfluid $^3$He. 

\begin{figure}[ht] 
\begin{center}
\includegraphics[width=15cm]{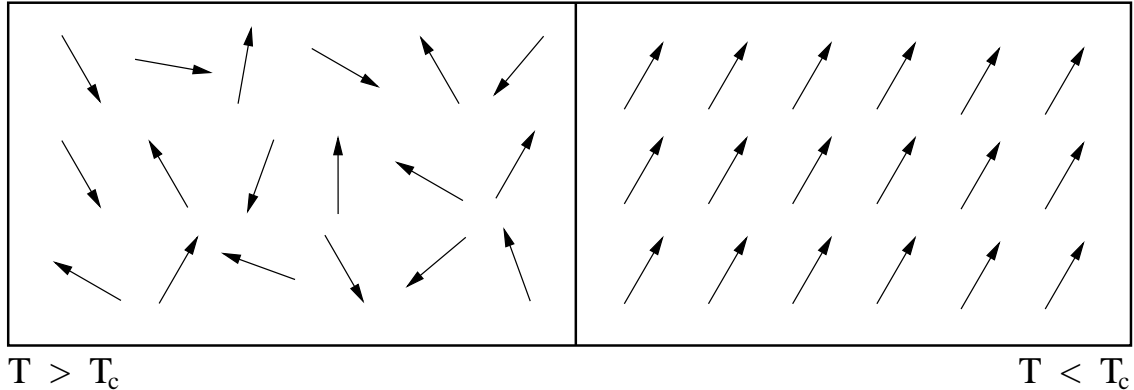}
\vspace{0.5cm}
\caption[Spontaneous symmetry breaking in a ferromagnet]{Illustration of the 
ferromagnetic phase transition. The symmetry
is spontaneously broken down from $SO(3)$ (left panel) to $U(1)$ 
(right panel).}
\label{ferro}
\end{center}
\end{figure}

First, let us discuss the simple example of ferromagnetism. Consider
a lattice of localized spins where each spin vector points in a random 
direction of
three-dimensional space, cf.\ left panel of Fig.\ \ref{ferro}.
Then, macroscopically, the system is invariant under arbitrary rotations in
real space, i.e., its symmetry group is $G=SO(3)$. This is the
non-magnetic phase, where the magnetization, which is the order parameter
for this phase transition, vanishes. For temperatures below the critical
temperature $T_c$, all microscopic spins align in one direction, causing
a finite magnetization, cf.\ right panel of Fig.\ \ref{ferro}. 
Obviously, the system is no longer invariant 
under arbitrary rotations. The symmetry is broken. But still, rotations around
the axis parallel to the direction of the magnetization do not change
the system macroscopically. Therefore, the symmetry is not 
completely broken, but there is a residual symmetry given by the 
subgroup $H=U(1)\subseteq G$. The term {\it spontaneous} symmetry breaking 
accounts for the fact 
that even an infinitesimally small external magnetic field causes the phase 
transition into the ferromagnetic phase.     

Now we extend our discussion, first on a purely mathematical level, 
to a symmetry group that is a direct product
of two groups $G=U(1)_a\times U(1)_b$. At this point, let us recall 
some related group theoretical facts, relevant for various sections 
of this thesis. All symmetries we are dealing with in this thesis
are described by Lie groups or direct products of several Lie groups. 
Remember that a 
Lie group is defined as a group that is a differentiable manifold and for 
which the group operations are continuous. The local structure of the 
Lie group is determined by the tangent space at the unit element. This 
vector space is a Lie algebra, i.e., the multiplication in this space
is given by the Lie product $[-,-]$ with the usual properties. 
Since all Lie groups 
considered in this thesis are matrix groups, $[-,-]$ always is the commutator
of two matrices. The basis elements of the Lie algebra
are the generators of the Lie group, which means that, via the
exponential map, the generators are mapped onto Lie group elements. 
A direct product of (Lie) groups $G_1\times G_2$, as in the current example, 
is again a (Lie) group with group elements $(g_1,g_2)$, $g_1\in G_1$, 
$g_2\in G_2$, and the group operation 
$(g_1,g_2)(h_1,h_2)=(g_1h_1,g_2h_2)$.
Denoting the Lie algebra of $G_{1,2}$ as ${\cal G}_{1,2}$, the Lie algebra
of the product group $G_1\times G_2$ is given by  
${\cal G}_1 \oplus{\cal G}_2$. This expression has to be understood as
the direct sum of the two vector spaces ${\cal G}_1$ and ${\cal G}_2$ with 
the additional property $[A_1,A_2]=0$ for all $A_1\in {\cal G}_1$, 
$A_2\in {\cal G}_2$. Remember also that a subgroup of $G_1\times G_2$ 
corresponds to a subalgebra of ${\cal G}_1 \oplus{\cal G}_2$. This
correspondence between Lie groups and Lie algebras is used in those sections
of this thesis where a detailed discussion of symmetry breaking patterns
is presented, see especially Sec.\ \ref{grouptheory}. In this introduction,
let us return to the elementary discussion of symmetry breaking patterns
of the group $G=U(1)_a\times U(1)_b$.

\begin{figure}[ht] 
\begin{center}
\vspace{0.5cm}
\includegraphics[width=12cm]{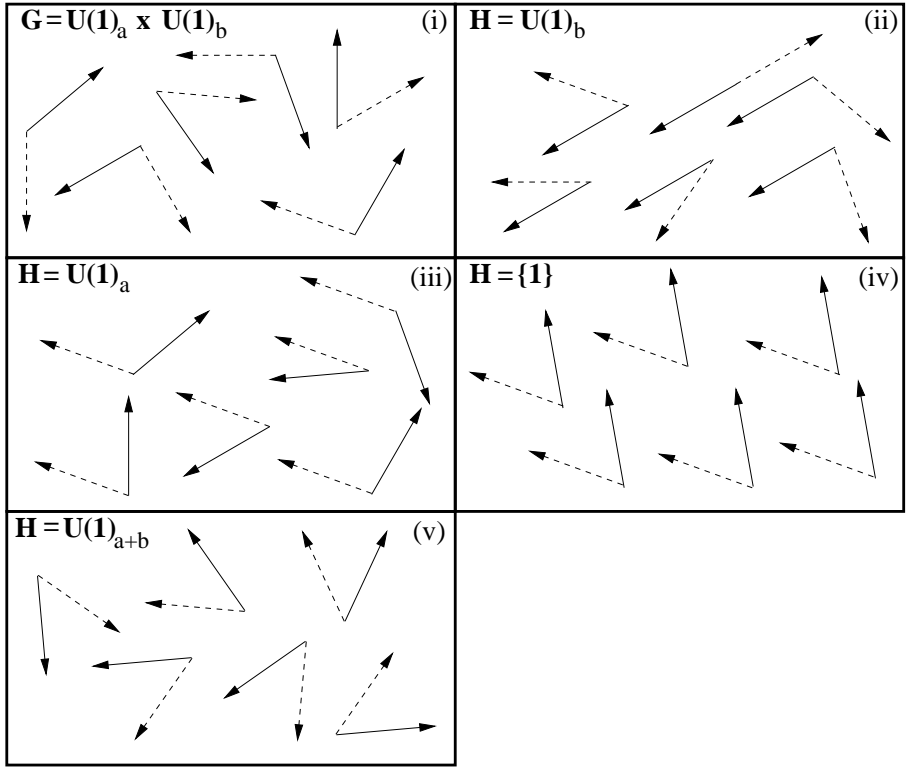}
\vspace{0.5cm}
\caption[Illustration of symmetry breaking patterns]
{Patterns of symmetry breaking for the case of a direct product group
$G=U(1)_a\times U(1)_b$. Solid arrows correspond to $U(1)_a$ while dashed
arrows correspond to $U(1)_b$.}
\label{figpatterns}
\end{center}
\end{figure}

In Fig.\ \ref{figpatterns}, panel $(i)$ shows a system with a symmetry given
by the group $G=U(1)_a\times U(1)_b$. The solid arrows correspond to 
$U(1)_a$ while 
the dashed ones correspond to $U(1)_b$. More precisely, the system
is described by a representation $a\otimes b$ of the group $G$, where
$U(1)_a$ and $U(1)_b$ act on vectors in $a$ (= solid arrows) and vectors in 
$b$ (= dashed arrows), respectively. In panel $(i)$, arbitrary rotations
of both kinds of arrows do not change the macroscopic properties of the system.
(Note that, unlike Fig.\ \ref{ferro}, Fig.\ \ref{figpatterns} has to be 
understood as a two-dimensional system; therefore, the rotation group is
the one-dimensional $U(1)\cong SO(2)$.) In panel $(ii)$, the symmetry is 
broken down to the residual subgroup $H=U(1)_b$,
i.e., the system is still invariant under rotations of the dashed arrows but
no longer invariant under rotations of the solid arrows. The corresponding 
situation with $H=U(1)_a$ is represented in panel $(iii)$.  
In panel $(iv)$, there is no nontrivial subgroup. Any rotation of either of 
the two classes of arrows changes the macroscopic properties of the system. 
Thus, the residual group only
consists of the unit element, $H=\{{\bf 1}\}$, or, more precisely, 
$H=\{({\bf 1},{\bf 1})\}$. In this case, the original symmetry is completely 
broken. The most interesting case is shown in panel $(v)$. A new 
symmetry arises through the relative orientation of the vectors, the angle 
between the solid and dashed arrows. Assuming that this microscopic angle
corresponds to a macroscopic observable, {\it separate} rotation of 
either solid or dashed arrows changes the system whereas any {\it joint}
rotation leaves the system invariant. Therefore, we denote the residual 
symmetry by $H=U(1)_{a+b}$. Mathematically speaking, this subgroup of $G$ is
generated by a linear combination of the generators of the original 
groups $U(1)_a$ and $U(1)_b$. In the following brief summary of different 
phases in superfluid $^3$He it is shown that this 
symmetry breaking pattern indeed is realized in nature. Another realization 
can be found in the Weinberg-Salam model of electroweak interactions. 
And, last but not least, for the study of color superconductivity, cf.\ for 
instance Sec.\ \ref{CSC}, it is an essential theoretical ingredient.     
  
In the case of superfluid $^3$He, the symmetry group 
\be
G=SO(3)_L\times SO(3)_S\times U(1)_N
\ee
is spontaneously broken. Here, $SO(3)_L$ and $SO(3)_S$ describe rotations
in angular momentum and spin space, respectively, and $U(1)_N$ accounts
for particle number conservation. The order parameter $\D$ 
is a complex $3\times 3$ 
matrix since it is an element of the nine-dimensional representation space 
$[{\bf 3}]_L\otimes [{\bf 3}]_S$, where $[{\bf 3}]_L$ and $[{\bf 3}]_S$
are the triplet representations of $SO(3)_L$ and $SO(3)_S$, respectively. 
The group $U(1)_N$ simply acts via multiplication of a phase factor on this
representation. Let us list the order parameters and corresponding 
residual symmetry groups of the three phases occurring in the phase 
diagrams in Figs.\ \ref{Hephase1} and \ref{Hephase2}. 
\begin{subequations} \label{BAA1}
\bea 
\mbox{B phase:} &\qquad& \D_B=\frac{1}{\sqrt{3}}
\left(\begin{array}{ccc} 1 & 0 & 0 \\ 0 & 1 & 0 \\
0 & 0 & 1 \end{array}\right) \,\, , \qquad H_B=SO(3)_{L+S} \,\, , 
\label{heB}\\ && \nonumber \\
\mbox{A phase:} &\qquad& \D_A=\frac{1}{\sqrt{2}}
\left(\begin{array}{ccc} 0 & 0 & 0 \\ 0 & 0 & 0 \\
1 & i & 0 \end{array}\right) \,\, , \qquad H_A=U(1)_S\times U(1)_{L+N} \,\, , 
\label{heA}\\ && \nonumber \\ 
\mbox{${\rm A}_1$ phase:} &\qquad& \D_{A_1}=\frac{1}{2}
\left(\begin{array}{ccc} 1 & i & 0 \\ -i & 1 & 0 \\
0 & 0 & 0 \end{array}\right) \,\, , \qquad H_{A_1}=U(1)_{S+N}\times U(1)_{L+N}
 \,\, . \label{heA1}
\eea
\end{subequations}
Here and in the following, we use the term ``order parameter'' somewhat 
sloppily for the pure matrix (or vector) structure. More rigorously,
the order parameter is this matrix multiplied by a gap function, 
since the order parameter has to be a function that vanishes in the normal 
phase. 

In the B phase, there is a residual group consisting of joint rotations in 
angular momentum and spin space. Furthermore, the particle number conservation
symmetry is completely broken. Note that this phase covers the largest 
superfluid region of the phase diagram for zero external magnetic field.
For a sufficiently large external magnetic field, however, the B phase 
disappears from the phase diagram. This can be understood from the following
symmetry argument. Since an external homogeneous magnetic field, 
pointing in a constant
direction, reduces the original spatial symmetries of the system from 
arbitrary rotations, $SO(3)$, to rotations around a fixed axis, $U(1)$, 
there is ``no space'' for a residual $SO(3)$ as it is present in the B phase. 
More precisely, in the presence of a magnetic field that reduces the 
original symmetry, the B phase (and also the A phase) is modified to 
the so-called ${\rm B}_2$ (${\rm A}_2$) phase whose residual symmetry 
is given by $H=U(1)_{L+S}$ ($H=U(1)_{L+N}$). 

In the A phase as well as in the ${\rm A}_1$ phase, $G$ is broken to a 
two-dimensional residual subgroup.
In both cases, the particle number conservation symmetry couples with 
the symmetries corresponding to spin and angular momentum.    
Here, an interesting question arises regarding superfluidity 
of these phases. Naively speaking, one expects a phase to be superfluid only 
if $U(1)_N$ is broken. In this sense, the B phase in $^3$He is a 
superfluid. But how about the A and ${\rm A}_1$ 
phases? Indeed, the superfluidity of these 
phases in the usual sense is questionable. For instance, in superfluid $^4$He 
and the B phase in $^3$He there is a (topologically) stable superflow. This 
is equivalent to the existence of line defects in the rotating bulk liquid
(as already mentioned for superfluid neutron matter in neutron stars, 
cf.\ Sec.\ \ref{neutronstars}). This superflow is unstable in 
the A phase  and only becomes stable in the presence of
a magnetic field. We do not elaborate further 
on this point (for more details regarding this interesting problem,
see Refs.\ \cite{vollhardt,volovik}) but, since in color superconductivity 
one encounters similar problems, we emphasize that the coupling
of the $U(1)_N$ symmetry with other symmetries is the origin of  
nontrivial properties regarding superfluidity.

\begin{figure}[ht] 
\begin{center}
\vspace{0.5cm}
\includegraphics[width=12cm]{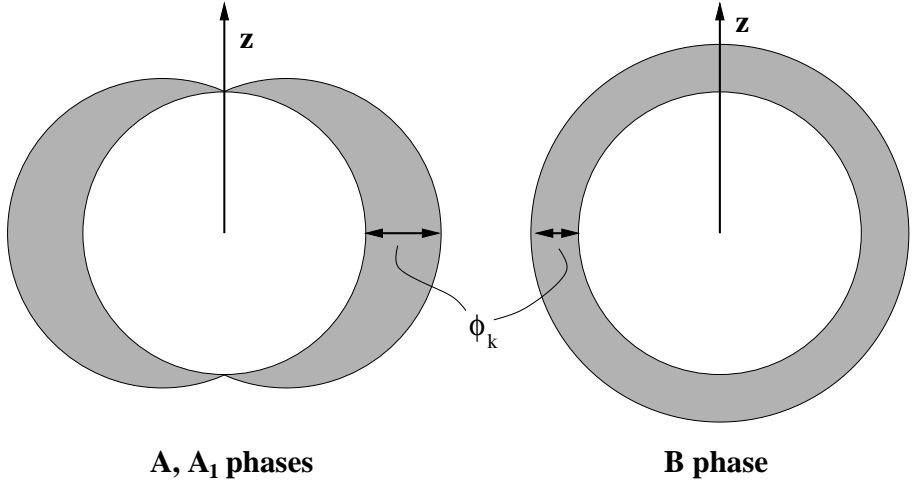}
\vspace{0.5cm}
\caption[Gap functions in $^3$He]{Angular dependence (schematic) of the 
gap functions $\phi_k$ in the 
A, ${\rm A}_1$, and B phases of superfluid $^3$He. Both plotted functions are 
symmetric with respect to rotations around the $z$-axis.}
\label{figgaps}
\end{center}
\end{figure}

Finally, we mention one more interesting feature of superfluid $^3$He, 
namely the anisotropy of the gap function. As can be seen from 
Eq.\ (\ref{BCSgap}), for 
conventional superconductors the gap function is constant on the Fermi sphere.
Of course, this is not necessarily valid in cases where the order parameter
breaks the rotational symmetry of the system. Therefore, in some phases 
of superfluid $^3$He, the gap function $\phi_k$ depends on the direction 
of the fermion momentum $\hbar{\bf k}$. In Fig.\ \ref{figgaps} we schematically
show this dependence for the three phases mentioned above. 

The gap
in the B phase is isotropic, while in both other phases there is an anisotropy.
Moreover, in the A phase as well as in the ${\rm A}_1$ phase, the gap 
function has nodal points at the north and south pole of the Fermi surface.   
Consequently, quasiparticles at the Fermi surface whose momentum points into 
a certain direction (parallel or antiparallel to the $z$-axis) can be excited 
by an infinitesimally small energy. Therefore, the nodal structure 
of the gap function is of physical relevance. For instance, the 
temperature dependence of the specific heat is different 
in the A phase compared to the B phase (power-law dependence versus 
exponential dependence). Again, for more details we refer 
the reader to the special condensed-matter literature 
\cite{vollhardt,wheatley}. 

Let us now, 
being well-prepared by low-energy condensed-matter physics, 
turn to high-energy condensed-matter physics, i.e., color superconductivity.

          %Superconductivity/fluidity
   \section{Color superconductivity} \label{CSC}

In order to provide a plain introduction into the theory of color 
superconductors (reviews about color superconductivity can be found in 
Refs.\ \cite{rajarev,alfordrev,schaeferrev1,schaeferrev2,rischkerev,renrev}), 
we connect the physics presented in the previous two
sections of this introduction: The theoretical models describing 
ordinary superconductivity and superfluidity, Sec.\ \ref{BCS_He3}, are 
applied to cold and dense (= deconfined) quark matter, 
Sec.\ \ref{colddense}. Since this transfer implies the inclusion 
of the theory of strong interactions, QCD, and the transition from a 
nonrelativistic to a relativistic treatment, it is not surprising that
a lot of new questions arise in the case of dense quark systems. 
Nevertheless, the basic mechanism, i.e., 
Cooper's theorem, explained in Sec.\ \ref{superconductivity}, can be directly 
applied. All one has to do is to replace the electron liquid, interacting
via exchange of virtual phonons, by a quark system interacting via the strong
interaction (thus, no lattice is required for color superconductivity). 
Due to asymptotic freedom, the strong interaction at (asymptotically) 
high densities is dominated by 
the exchange of a single gluon. And indeed, there is an attractive channel
of this interaction, providing the condition for the application of Cooper's
theorem. Consequently, for sufficiently low
temperatures, the quarks at their Fermi surface rearrange in order to form a 
ground state characterized by the existence of quark Cooper pairs 
\cite{barrois,bailin,alford1,rapp}. Note that,
unlike the case of electrons, it is not a priori clear that two quarks,
possibly distinguished by their flavor, their color, and their electric charges
have identical Fermi momenta. We comment on this question below and 
first assume that in this respect they behave like electrons in a conventional
superconductor. 

Then, analogous to a conventional superconductor, in a color superconductor the
quarks in the vicinity of the Fermi surface form Cooper pairs with zero 
total momentum and acquire an energy gap in 
their (quasiparticle) excitation spectrum. But, considering the different 
intrinsic properties of quarks compared to electrons,  
such as flavor, color, and electric charge, the following natural questions 
arise. Are quark Cooper pairs charged?
What is the total spin of a Cooper pair? Is there a Meissner effect in a 
color superconductor? Are {\it color} superconductors
also {\it electromagnetic} superconductors? How does the quasiparticle 
dispersion relation look like? How about superfluidity in a color 
superconductor? Are there vortices/flux tubes? Is a color superconductor 
of type I or II?

It turns out that color superconductors provide a multitude of theoretically
possible phases, and the answers to almost all of the above questions 
depend on the
specific phase and especially on the number of quark flavors $N_f$ 
involved in the system. Therefore, in the following, we briefly discuss 
several different phases in color superconductors. Common to all phases is 
the attractive color channel $[\bar{\bf 3}]_c^a$,
\be \label{colorchannels}
[{\bf 3}]_c \otimes [{\bf 3}]_c = [\bar{\bf 3}]_c^a \oplus [{\bf 6}]_c^s \,\, ,
\ee   
where the lower index ``$c$'' indicates ``color'', and the upper indices
``$a$'' and ``$s$'' stand for ``antisymmetric'' and ``symmetric'', 
respectively. In this equation, the coupling of two quarks forming a Cooper 
pair is described with the help of representation theory. The
complex vector space on the right-hand side of the equation is 
nine-dimensional, since each single-quark space $[{\bf 3}]_c$ accounts for 
three fundamental colors, say red, green, and blue. The tensor product
can be decomposed into a direct sum of two representations of the color gauge 
group $SU(3)_c$.
This is shown on the right hand side of the equation, where the antisymmetric
antitriplet $[\bar{\bf 3}]_c^a$ and the symmetric sextet $[{\bf 6}]_c^s$ are
three- and six-dimensional representations of $SU(3)_c$, respectively.
(Remember the analogous ansatz for mesons, composed of $u$, $d$, and $s$ 
quarks and the corresponding 
antiquarks, $[{\bf 3}] \otimes [\bar{\bf 3}]$, or for baryons, composed of 
three $u$, $d$, and $s$ quarks 
$[{\bf 3}] \otimes [{\bf 3}] \otimes [{\bf 3}]$, resulting
in the well-known multiplets.) The antisymmetry of the attractive color channel
implies that a quark Cooper pair always is composed of two quarks 
with different colors.
     
In order to proceed, one has to specify the number of flavors.

\subsection{Two- and three-flavor color superconductors} \label{2SCCFL}

Let us first consider the simplest case, a system of massless $u$ and $d$ 
quarks, i.e., 
$N_f=2$ \cite{bailin}. According to the number of flavors, the superconducting
phase in this system is commonly called the ``2SC phase''. In this phase,
besides the color antitriplet, Cooper pairs are formed in the flavor-singlet
$[{\bf 1}]_f^a$ and spin-singlet $[{\bf 1}]_J^a$ channels. While the former
is a representation of the flavor group $SU(2)_f$ (more precisely,
of both the left-handed and right-handed chirality groups $SU(2)_\ell$ and
$SU(2)_r$), the latter is a representation of the spin group $SU(2)_J$. 
Note that in relativistic theories one has to consider the total spin
$J=L+S$ rather than separately treating angular momentum $L$ and spin $S$.
This is in contrast to the nonrelativistic treatment of superfluid $^3$He,
presented in the previous section. Since both $[{\bf 1}]_f^a$, $[{\bf 1}]_J^a$,
as well as the color representation $[\bar{\bf 3}]_c^a$, are antisymmetric 
with respect to exchange of the corresponding quantum numbers of the single 
quarks, the total wave function of the quark Cooper pair is antisymmetric, as
required by the Pauli principle. Mathematically speaking, a quark Cooper 
pair in the 2SC phase is an element of the space 
$[\bar{\bf 3}]_c^a \otimes [{\bf 1}]_f^a \otimes [{\bf 1}]_J^a \cong
[\bar{\bf 3}]_c^a$. Therefore, the order parameter $\D$, which, as explained 
in Sec.\ \ref{BCS_He3}, directly connects the existence of Cooper pairs with
a spontaneous breakdown of symmetries, is a complex 3-vector. It is 
easy to show that, for $N_f=2$,  any choice of $\D$ leads to an equivalent 
symmetry
breaking pattern. For simplicity, one chooses $\D_i=\d_{i3}$.     
The original symmetry of the system,
\be \label{G2SC}
G=SU(3)_c\times SU(2)_f \times SU(2)_J \times U(1)_{em} \times U(1)_B \,\, ,
\ee
is broken to 
\be \label{H2SC}
H_{2SC}=SU(2)_c\times SU(2)_f \times SU(2)_J \times U(1)_{c+em} 
\times U(1)_{em+B} \,\, ,
\ee 
where $U(1)_{em}$ is the electromagnetic gauge group and $U(1)_B$ the 
baryon number conservation symmetry. Consequently, in the 2SC phase, 
the flavor and spin groups remain unbroken. However, the color gauge group
is broken down to $SU(2)_c$. Furthermore, there are two residual $U(1)$'s.
The notation for these $U(1)$'s in Eq.\ (\ref{H2SC}) is explained in 
Sec.\ \ref{he3}, see Fig.\ \ref{figpatterns} and Eqs.\ (\ref{BAA1}), 
i.e., $U(1)_{c+em}$ is generated
by a linear combination of the generators of $SU(3)_c$ and $U(1)_{em}$. 
Physically, in the 2SC phase, a Cooper pair carries spin zero, $J=0$, it is
composed of a $u$ and a $d$ quark, and it carries anti-blue (= anti-3)
color charge, since it is composed of a red and a green quark (using the
above convention for the order parameter). As can be seen from the residual
symmetry group $H_{2SC}$, the question of the electric charge of a Cooper pair
is more subtle. It will be discussed in the following section, 
Sec.\ \ref{higgs}. Nevertheless, let us already mention the fundamental 
differences between the symmetry groups in Eqs.\ (\ref{G2SC}) and (\ref{H2SC})
compared to the corresponding symmetry breaking patterns in $^3$He, 
Eqs.\ (\ref{BAA1}). While in the latter case, all groups correspond 
to global symmetries, here two local gauge groups are involved, 
accounting for the strong and the electromagnetic interaction. Note that 
in ordinary superconductors, the local group $U(1)_{em}$ is completely 
broken, whereas the occurrence of the $U(1)_{em}$ generator in the residual
group of the 2SC phase demands a careful interpretation 
(cf.\ Sec.\ \ref{higgs}). Similarly, due to the residual subgroup 
$U(1)_{em+B}$, the 2SC phase is not superfluid in the usual sense;   
cf.\ discussion below Eqs.\ (\ref{BAA1}) about the A and ${\rm A}_1$ phases of
superfluid $^3$He. 

Next, we consider a system of three massless quark flavors, $N_f=3$. 
As in the 2SC phase, the condensation of Cooper pairs occurs in the 
antisymmetric flavor and spin channels, which, together with the 
antisymmetric color channel, ensures the antisymmetry of the pair wave 
function. The only difference, caused by 
the different number of flavors, is the fact that the antisymmetric flavor
representation of $SU(3)_f$ is an antitriplet (instead of a singlet in the case
of $SU(2)_f$). Therefore, the order parameter in three-flavor color 
superconductors is an element of 
$[\bar{\bf 3}]_c^a\otimes[\bar{\bf 3}]_f^a\otimes[{\bf 1}]_J^a \cong 
[\bar{\bf 3}]_c^a\otimes[\bar{\bf 3}]_f^a$. Consequently, unlike the two-flavor
case, the order parameter $\D$ is a complex $3\times 3$ matrix. As discussed
for the case of superfluid $^3$He, this structure of the order parameter
allows for several different phases. However, in the case of three-flavor
color superconductors, it is generally believed that the only important
phase is the so-called color-flavor-locked (CFL) phase with an order
parameter $\D_{ij}=\d_{ij}$ \cite{alford2,pisarski}. In this phase,
the symmetry given by the group $G$ in Eq.\ (\ref{G2SC}) is spontaneously 
broken to
\be \label{HCFL}
H_{CFL}=SU(3)_{c+f}\times U(1)_{c+em} \,\, .
\ee
Since not obvious in our simplified notation, it should be mentioned that 
the order parameter in the CFL phase breaks
chiral symmetry in the form $SU(3)_c\times SU(3)_\ell\times SU(3)_r
\to SU(3)_{c+\ell+r}$ (in contrast to the 2SC phase, where 
$SU(2)_\ell\times SU(2)_r$ is unbroken). In the CFL phase, the order parameter
is invariant under joint rotations in color and flavor space. This, of course,
is the reason for the term ``color-flavor locking''. Furthermore,
as in the 2SC phase, there is a local residual $U(1)$ originating from both
the color and electromagnetic gauge groups. And, unlike the 2SC phase,
$U(1)_B$ is completely broken, which renders this phase superfluid. 

Let us now elaborate on the physical meaning of the residual gauge 
group $U(1)_{c+em}$, occurring in both the 2SC and CFL phases.

\subsection{Anderson-Higgs mechanism in color superconductors} 
\label{higgs}  

In this section, we point out a certain aspect of spontaneous symmetry 
breaking, which is relevant for broken gauge symmetries. In 
conventional superconductors, the electromagnetic gauge symmetry is broken
by the order parameter. As mentioned in Sec.\ \ref{superconductivity}, this
leads to the Meissner effect, or, equivalently, to a massive photon. 
This photon mass is generated by the so-called Anderson-Higgs mechanism 
\cite{anderson,higgs}, which is a general mechanism occurring in theories
with spontaneously broken gauge symmetries. In color superconductivity, 
not only the electromagnetic, but also the color gauge group is
involved. Let us first briefly introduce the mechanism in a general way.
Then, applying it to cold and dense quark matter,
we make use of the analogy between the Weinberg-Salam
model of electroweak interactions \cite{weinberg} and the two- and 
three-flavor phases of color superconductivity. 

Remember the following basics of spontaneous symmetry breaking 
in field theories \cite{khuang}. Consider a Lagrangian for a complex field 
$\phi\in V$, where $V$ is an $r$-dimensional representation of 
a Lie group $G$. 
Suppose that this Lagrangian is invariant under global transformations
in $G$. Furthermore, suppose that there is a nonzero ground 
state of the system (= lowest energy solution of the corresponding equations 
of motion) $\phi_0$ which is invariant under transformations of a subgroup 
$H$ of $G$, but not under all transformations in $G$.
In this case, the symmetry of the system is said to be spontaneously broken 
and there are ${\rm dim}\,G/H={\rm dim}\,G - {\rm dim}\,H$ massless 
``Goldstone bosons'', i.e., the originally 
$r$ degrees of freedom of the matter field generate ${\rm dim}\,G/H$
massless and $r-{\rm dim}\,G/H$ massive modes (present in the vicinity of the
vacuum state, i.e., for low energies). Obviously, $r-{\rm dim}\,G/H>0$
is a restriction for possible residual groups $H$. Here, the dimension of 
a Lie group is defined as the dimension of its Lie algebra.  

Now let us gauge the Lagrangian. This is done by introducing a covariant 
derivative containing gauge fields $A_a$, $a=1,\ldots,{\rm dim}\,G$.
Then, the Lagrangian is invariant under {\it local} transformations of the group
$G$; and besides the $r$ degrees of freedom of the field $\phi$, there
are $2\,{\rm dim}\,G$ (2 for each massless gauge field) additional degrees of 
freedom. In this case, spontaneous symmetry breaking caused by a nonzero
value of the ground state expectation value creates $r-{\rm dim}\,G/H$ 
massive modes, as in the
non-gauged case. But, instead of the Goldstone bosons there are now
${\rm dim}\,G/H$ {\it massive} gauge fields, leaving only the remaining 
${\rm dim}\,H$
gauge fields massless. In other words, the degrees of freedom of the 
``would-be'' Goldstone bosons are absorbed (``eaten up'')
by the gauge fields which acquire a third degree of freedom and thus 
become massive. 
%Let us remark that the representation space $V$ can be 
%decomposed into a direct sum of a ${\rm dim}\,G/H$-dimensional space 
%(``Goldstone space'') and its $r-{\rm dim}\,G/H$-dimensional complement
%(``Higgs space''). Then there exists a gauge transformation $g(x)\in G$ 
%such that for any field $\phi$, $g(x)\phi(x)$ is orthogonal to the
%Goldstone space. This gauge is called unitary gauge.  
This generation of masses for the gauge fields by 
spontaneuos symmetry breaking is called Anderson-Higgs mechanism. It was 
first introduced by Anderson in solid-state physics \cite{anderson}
and then applied by Higgs \cite{higgs}, and later by Weinberg and Salam in 
particle physics, especially in the theory of electroweak interactions 
\cite{weinberg}. 

In the Weinberg-Salam model, the Higgs field is introduced
in order to generate the masses of the electron, as well as for the 
$W^\pm$ and $Z$ gauge bosons, since explicit mass terms in the Lagrangian 
would violate gauge invariance. Let us focus on the gauge fields in this 
model. The gauge group is $G=SU(2)_I\times U(1)_Y$ (corresponding to isospin 
and hypercharge); since ${\rm dim}\,G=3+1=4$, there are originally four 
massless gauge fields, say $W_1$, $W_2$, $W_3$ for $SU(2)_I$, and $W_0$ 
for $U(1)_Y$. The vacuum
expectation value of the Higgs field, however, is only invariant under
$H=U(1)_{I+Y}$, which is generated by a linear combination of one generator
of $SU(2)_I$ and one of $U(1)_Y$. Therefore, there are 
${\rm dim}\,G/H=3$ massive
gauge fields, called $W^+$, $W^-$, and $Z$, and ${\rm dim}\,H=1$ massless
field $A$. Since the residual group $U(1)_{I+Y}$ contains joint rotations 
of the original two groups $SU(2)_I$ and $U(1)_Y$, the pair of fields $(Z,A)$  
is generated by an orthogonal rotation of the original pair $(W_3,W_0)$. 
The angle $\theta_W$ of the rotation is called Weinberg angle.
The resulting fields are the gauge bosons of the weak interaction
and the photon field $A$; the residual group is the gauge group of
electromagnetism, $U(1)_{I+Y}\equiv U(1)_{em}$. Furthermore,
the electromagnetic coupling constant is determined by the coupling constants
of the original theory and the Weinberg angle. These facts are summarized in 
the first column of Table \ref{analogy}.

\begin{table}
\begin{center}
\begin{tabular}{|c||c|c|}
\hline
 & {\bf Weinberg-Salam} & {\bf CFL phase} \\ 
\hline\hline 
{\bf gauge} & $SU(2)_I\times U(1)_Y$ & $SU(3)_c\times U(1)_{em}$  
\\
{\bf group}& isospin, hypercharge & color, electromagnetism \\  
\hline 
{\bf gauge fields} & $W_1,W_2,W_3,W_0$ & $A_1,\ldots,A_8,A$ \\
\hline
{\bf coupling constants} & $G$, $G'$ & $g$, $e$ \\
\hline
{\bf symmetry} & $SU(2)_I\times U(1)_Y$ & 
$SU(3)_c\times U(1)_{em}$ \\
{\bf breaking} &$\to U(1)_{em}$& $\to U(1)_{c+em}$\\ 

\hline &$W^+,W^-$ & $A_1,\ldots, A_7$  \\
{\bf new fields} & $Z=\cos\theta_W W_3 + \sin\theta_W W_0$ & 
$\tilde{A}_8=\cos\theta A_8 + \sin\theta A$ \\
 & $A=-\sin\theta_W W_3 + \cos\theta_W W_0$ &
$\tilde{A}=-\sin\theta A_8 + \cos\theta A$ \\ 
\hline
{\bf new coupling constant} & $e=G'\cos\theta_W$ & $\tilde{e}=e\cos\theta$ \\
\hline 
{\bf massive fields} & $W^+$, $W^-$, $Z$ & $A_1,\ldots,A_7,\tilde{A}_8$ \\
\hline
{\bf massless fields} & $A$ & $\tilde{A}$ \\
\hline
\end{tabular}
\caption[Analogy to Weinberg-Salam model]{Analogy between the Weinberg-Salam 
model of electroweak interaction 
and the CFL phase in a three-flavor color superconductor.
In both cases there is a mixing between the original gauge fields. 
In the case of the CFL phases, all gluons acquire a mass via the 
Anderson-Higgs mechanism (= color Meissner effect), while the (rotated) photon 
remains massless (= no electromagnetic Meissner effect). 
}
\label{analogy}
\end{center}
\end{table}
 
The second column of Table \ref{analogy} shows the analogous situation in a 
three-flavor color superconductor. Since the mechanism is exactly the same 
(the role of the nonvanishing vacuum expectation value of the Higgs field
is played by the order parameter of the superconducting state), no further
explanation is needed. Physically, one finds that the eighth gluon $A_8$ 
couples to the photon $A$, giving rise to a new (rotated) eighth gluon 
$\tilde{A}_8$ and a new (rotated) photon $\tilde{A}$ 
\cite{alford3,gorbar,litim,manuel}.
While all original gluons $A_1,\ldots,A_7$ and the new gluon $\tilde{A}_8$ 
acquire a mass via the Anderson-Higgs mechanism, the new photon $\tilde{A}$
remains massless. Consequently, there is a {\it color} Meissner effect 
for all gluons, i.e., color magnetic fields are expelled from the
CFL color superconductor. But there is no {\it electromagnetic} Meissner effect in 
the CFL phase, which means that this color superconductor is
no electromagnetic superconductor. One should emphasize that the original 
photon field $A$ has no physical meaning in the interior of the superconductor.
The real fields are the rotated fields. This, although at first sight 
peculiar, can again be understood with the analogous statement in the 
standard model. In this case, below the ``electroweak phase transition'', only
the gauge bosons $W^\pm$, $Z$, and $A$ are relevant, while the gauge 
fields of the original symmetry are not existent. Due to the huge difference
of the strong and electromagnetic coupling constants, the mixing angle
in a color superconductor is small. Therefore, one can interpret the absence 
of the electromagnetic Meissner effect as follows. An incoming photon is 
not absorbed at the surface of the superconductor (as in the case of a 
Meissner effect) but slightly rotated into a new photon that can 
penetrate the superconductor and thus create a magnetic field in the interior. 

The situation in the 2SC phase is very similar. While, for exactly the
same reason as in the CFL phase, there is no 
electromagnetic Meissner effect either, the only difference is the 
residual color group $SU(2)_c$, cf.\ Eq.\ (\ref{H2SC}). Therefore, 
only five gluons become massive, whereas the three remaining ones
can penetrate the superconductor. These three massless gluons are those 
that do not see the third color (remember that all Cooper pairs 
in the 2SC phase carry (anti-)blue color charge). 

A more detailed and quantitative discussion
of the Meissner effect (and the Meissner masses) as well as the 
extension to one-flavor color superconductors
are presented in Sec.\ \ref{mixingscreening}.

\subsection{One-flavor color superconductors}\label{oneflavor}

The apparently most trivial situation in cold and dense quark matter 
is a system with only one quark flavor, $N_f=1$. But in this case,
a complication enters the color-superconducting phase for the following 
reason. Due to the antisymmetry of the attractive color channel, cf.\ 
Eq.\ (\ref{colorchannels}), the Cooper pair condensation in a 
one-flavor system has to occur in a symmetric spin channel. Note 
that in a two- or three-flavor system, the antisymmetric spin channel where 
the Cooper pairs carry spin zero can be chosen because an 
antisymmetric flavor channel is available to make the total pair wave 
function antisymmetric. This is not the case for only
one quark flavor. Therefore, in the simplest case, a Cooper pair here
carries total spin one, $J=1$, and the order parameter $\D$ is an element of 
$[\bar{\bf 3}]_c^a\otimes [{\bf 3}]_J^s$, which is a representation 
of $SU(3)_c\times SU(2)_J$. Since integer spin representations are 
not only representations of $SU(2)_J$ but also of $SO(3)_J$, we could
equivalently choose the symmetry group $SU(3)_c\times SO(3)_J$. 
(The groups $SU(2)$ and $SO(3)$ have the same Lie algebra, i.e., they are 
locally isomorphic, however, globally, they are not isomorphic.) 
As in the three-flavor case and the case of
superfluid $^3$He, $\D$ is a complex $3\times 3$ matrix. 
Unlike the color-superconducting phases discussed in the previous sections,
here the order parameter potentially breaks rotational symmetry in real space.
Thus, as in superfluid $^3$He, anisotropy effects are expected.  

Spin-one (or one-flavor) color superconductivity was
first studied in Ref.\ \cite{bailin}. More recent works discussing spin-one
phases can be found in Refs.\ 
\cite{iwasaki,pisarski2,pisarski3,schaefer,alford4,buballa}.  
In the main part of this thesis, Sec.\ \ref{mainpart},
we study the properties of several possible spin-one color
superconducting phases, i.e., we discuss their gap functions, their 
gap parameters, and their critical temperatures. Moreover, we discuss their
response to electric and magnetic fields; in particular, we answer the question
if spin-one color superconductors exhibit an electromagnetic Meissner effect.
Finally, we study the effective potential in order to determine the 
preferred phase in a spin-one color superconductor.

\subsection{More on (color) superconductors} \label{more}

In the previous three introductory subsections about color superconductivity, 
we 
focussed on the symmetry aspects of the superconducting phases with $N_f=1,2,3$
and pointed out the similarities and differences to ordinary 
superconductivity and superfluid $^3$He. 
In this section, a brief introduction into color-superconducting phases 
beyond the (idealized) 2SC and CFL phases is given. Most 
of these phases are not treated in the main part of this thesis, wherefore
we cite some useful references.  
For the investigation of these phases, properties of realistic 
physical systems, especially of neutron stars, 
are implemented into the theory of color superconductors. 
This leads to superconducting phases which are not or only partially based 
on the traditional BCS theory. Some of these phases have their analogues 
in ordinary condensed-matter physics. 
Finally, the following discussion of realistic systems leads to an 
argument why spin-one color superconductors might be relevant in nature. 

It is essential to notice that the above introduced 2SC and CFL phases are 
based on one fundamental assumption. This assumption is the identity of the
Fermi momenta of all $N_f$ quark flavors. In conventional 
superconductors this assumption is valid, since both constituents of a Cooper
pair are electrons with identical properties, especially with
identical mass and chemical potential, and hence with identical
Fermi momenta. Also in superfluid $^3$He, one expects one single
spherical Fermi surface for all helium atoms. Nevertheless, even in 
conventional condensed-matter physics, this ideal scenario does not 
always seem to be an appropriate description of the system. 
There might be two (or more) species of particles having different Fermi 
surfaces. For instance, in an external magnetic field, due to Zeeman 
splitting, 
the electron energy depends on the spin projection and consequently 
there are two different Fermi momenta for spin-up and spin-down 
electrons. In this case, a superconducting state may be formed in which
the Cooper pairs carry nonzero total momentum, leading to a spatially
varying order parameter. This state (FFLO state) was already theoretically
predicted in 1964 \cite{FFLO}; experimental evidence for the existence of
the FFLO state has been discussed recently \cite{radovan}. The FFLO 
state can be understood as a displacement of the Fermi spheres, breaking
translational invariance. 

Note that a splitting of the Fermi momenta 
does not necessarily have to originate from an external magnetic field, 
it might as well be caused by different densities or masses of two 
fermion species. 
Besides the formation of Cooper 
pairs with nonzero momentum, theoretical studies have discussed the
possibility of a so-called ``interior gap'' (or ``breached pair'') for
systems with two different Fermi momenta \cite{liu}. In this case,
the fermions with smaller Fermi momentum have to be lifted to 
a higher energy in order to form Cooper pairs with the 
fermions with larger Fermi momentum. If this energy cost is lower 
than the energy gain of the formation of Cooper pairs, the 
superconducting state is preferred.   

Not only for electrons in a metal, but also in atomic systems, the situation 
of a splitting of Fermi surfaces is investigated \cite{mishra}. 
Experiments with $^6$Li and $^{40}$K (here, the splitting of Fermi 
surfaces also originates from the spin projections of the atoms) seem to 
suggest the observation of a superfluid state in a magnetic trap
\cite{ohara,regal}. (Note that these experiments are similar to the celebrated 
experiments with bosonic atoms where Bose-Einstein condensation of atoms in a 
magnetic trap has been observed.)           

In cold and dense quark matter, it is not surprising that, considering
realistic systems, there might be a separation of Fermi surfaces for different
quarks. Since quarks carry flavor, color, and electric charge, (and 
a mass), there are
a priori enough intrinsic properties to account for more than one single
Fermi momentum. These intrinsic properties have to be combined with 
the macroscopic properties of realistic systems, such as neutron stars.  
One of these properties is the overall electric charge neutrality.
A second one is the condition of $\b$-equilibrium (including electrons
into the system). Both requirements are implemented in color 
superconducting quark systems for instance in 
Refs.\ \cite{iida1,rajagopal,buballa2,kim,alford5,steiner,huang,neumann,ruester}.
Another complication arises when one relaxes the assumption of ultrarelativistic
quarks. In realistic systems, one at least has to take into account the
mass difference between the two light $u$ and $d$ quarks and the 
heavier $s$ quark. (Assuming that $0\simeq m_u =m_d \ll m_s$, the 
three quark Fermi momenta are given by $k_{F,u}=\m_u$, $k_{F,d} = \m_d$,
and $k_{F,s}=\sqrt{\m_s^2-m_s^2}$, where $\m_u$, $\m_d$, and $\m_s$ are the 
corresponding chemical potentials.)
In principle, also the condition of overall color neutrality has 
to be included. This is done in model calculations via introducing
color chemical potentials $\m_3$ and $\m_8$ (cf.\ for instance 
Refs.\ \cite{huang,neumann,mishra2}). 
However, for QCD calculations, it has been 
shown that color neutrality in a (two-flavor) color superconductor 
is automatically ensured via a nonvanishing color background field
\cite{rebhan,dietrich}.
 
Anyhow, the question arises if, in spite of the difference between the 
Fermi momenta of different quark flavors, the superconducting
phase is still favored. For color superconductors, there have been 
suggestions for possible scenarios similar to the ones 
for electrons or atoms mentioned above.   
For instance, the FFLO phase is discussed in Refs.\ \cite{alford6,bowers} 
(therein called LOFF phase). Besides the deplacement of the 
Fermi surface in the LOFF phase, also a deformation of the Fermi 
surfaces, breaking rotational invariance, has been proposed \cite{muether}.
An interior gap structure is studied in 
Ref.\ \cite{gubankova}. The mechanism of breached pairing in a 
two-flavor color superconductor leads to an interesting phase, called
the ``gapless 2SC'' phase \cite{shovkovy1} which also 
has been applied to neutron stars \cite{shovkovy2}. 
(Recently, also the possibility of a gapless CFL phase has been
investigated \cite{gCFL,matsuura}.) In this 
phase, a difference (if not too large) between the Fermi momenta 
of $u$ and $d$ quarks, leads to quasiparticle spectra, which, although 
exhibiting a nonzero gap parameter $\phi$, for certain 
quasiparticle momenta look like the normal spectrum, i.e., 
the spectrum is ``gapless''. Note the topological difference between 
this situation and 
the nodal structure of the gap in the A phase of superfluid $^3$He,
cf.\ Fig.\ \ref{figgaps}. In the latter, the vanishing gap leads to 
a gapless excitation energy for quasiparticles with two certain directions
of the momentum ${\bf k}$ (but with $|{\bf k}|=k_F$). Sloppily 
speaking, there are two gapless {\it points} on the Fermi surface. In the 
gapless 2SC phase, however, there are two values of $|{\bf k}|$ between
the original two Fermi momenta, for which the spectrum becomes gapless. 
Consequently, there are two gapless spherical {\it surfaces}. In the 
discussion of spin-one color superconductors, we also encounter 
the situation of gapless {\it lines} on the Fermi surface, 
cf.\ Sec.\ \ref{gapeqsolution}. 

It should be mentioned that the values of crucial quantities such as the 
quark density in the interior of a neutron star, the strange quark mass 
or the strong coupling constant at moderate (= not asymptotically large) 
densities are poorly known. Since all these quantities enter the  
description of the difference in Fermi momenta, the quantitative
value of this difference is absolutely unclear. It is easy to 
understand that there is a limit value for this difference, above which the
superconducting state, with Cooper pairs formed by fermions
with different Fermi momenta, is no longer favored. Consequently, 
in this case, besides the normal-conducting state, only pairing between 
fermions with equal Fermi momenta is
possible. In a dense quark system, this is equivalent to one-flavor
superconductivity, meaning either a system consisting only of 
quarks of a single flavor or a many-flavor system where each flavor
separately forms Cooper pairs. Therefore, in the main part of this thesis, 
we focus on this special case which, as shown in Sec.\ \ref{oneflavor}, 
corresponds to the spin-one phases of color superconductivity. 

Finally, let us quote several different theoretical approaches 
to color superconductivity. Most of the above mentioned works are based on a 
phenomenological Nambu-Jona-Lasinio (NJL) model \cite{alford1,rapp}, which 
is valid at densities of the order of a few times nuclear matter ground
state density. In the following sections of this thesis, color 
superconductivity is studied 
from first principles in the framework of QCD at weak coupling 
\cite{pisarski2,pisarski3,son1,schaeferwil,hong}, which is rigorously valid at asymptotically 
large densities. 
Extrapolating the QCD gap parameter (for two and three flavors) 
to moderate densities yields a value that is in agreement with the 
NJL approach. 
Another approach to color superconductivity is the 
phenomenological Ginzburg-Landau theory, already mentioned in 
Sec.\ \ref{superconductivity}. This approach \cite{iida1,iida2,giannakis}
is valid in temperature regions in the vicinity of the critical temperature 
and has been used for instance for the investigation of vortices in 
(two- and three-flavor) superconducting/superfluid quark matter 
\cite{iida3,blaschke}.

      %Color superconductivity
\chapter{Spin-one color superconductivity} \label{mainpart}

In this chapter, which forms the main part of the thesis, we mainly study the
properties of spin-one color superconductors. In its first section,
Sec.\ \ref{gapeqsolution}, which is based on Refs.\ \cite{schmitt1,schmitt4}, 
we compute the gap parameters, the excitation energies, and the critical 
temperatures for
several phases in a spin-one color superconductor and compare them to the
corresponding values in a spin-zero color superconductor. A derivation of  
the gap equation starting from the QCD partition function is outlined in 
Sec.\ \ref{derivgapeq}. In Sec.\ \ref{exenergies}, the structure of the   
quasiparticle excitation spectra for color superconductors are determined. 
In the case of the spin-one phases, there are anisotropies occurring in these 
spectra, as expected from the discussion in the introduction. 
Moreover, there are phases with a 
two-gap structure, i.e., two excitation energies with different, nonzero 
energy gaps, as it is known from the CFL phase.
The gap equation is solved in Sec.\ \ref{gapsolve}
in order to determine the gap parameter, i.e., the value of
the energy gap at the Fermi surface for zero temperature. This is done
by using a general notation valid for an arbitrary order parameter. 
Consequently, a general result for the gap parameter is obtained, which
can be evaluated for several different phases. First, we recover the well-known
results for the 2SC and CFL phases. Second, we apply the result to
four different spin-one phases.  Finally, in Sec.\ \ref{critictemp}, 
the critical temperatures are determined
for the several phases (again, after deriving a general expression from
the gap equation). We discuss the validity of the BCS relation between the
zero-temperature gap and the critical temperature, 
cf.\ Eq.\ (\ref{BCSrelation}).

In Sec.\ \ref{grouptheory}, we discuss a systematic classification of 
possible order parameters for spin-one color superconductors. 
From the discussion of superfluid
$^3$He we know that an order parameter in the form of a complex 
$3\times 3$ matrix a priori allows for more than one 
superfluid/superconducting phase. Therefore, we use group-theoretical 
arguments in order to list all matrices that lead to a 
superconducting phase. Obviously, a residual subgroup (called $H$ in the
introduction) can be assigned to each of those order parameters. 
From these subgroups one can qualitatively read off several properties of the
corresponding state. One of the most interesting properties is related to the
Anderson-Higgs mechanism, namely, one can read off
whether the phase exhibits a Meissner effect. 

The quantitative discussion of the Meissner effect in spin-one 
color superconductors is presented in Sec.\ \ref{mixingscreening}. 
This section is based on Refs.\ \cite{schmitt2,schmitt3}. 
We present a fundamental derivation of the mixing between gluons and 
photons (which has been explained in simple words in the introduction),
starting from the QCD partition function. From a calculation of the 
gluon and photon polarization tensors we deduce the Meissner masses for 
the gauge bosons. Furthermore, besides magnetic screening, also electric
screening is discussed via a calculation of the Debye masses. The final
results are given for the 2SC and CFL phases (partially already known 
in the literature \cite{litim,meissner2,meissner3}) and for two 
spin-one phases. At the end of this section, we discuss the Meissner effect
in many-flavor systems, where each flavor separately forms spin-one 
Cooper pairs.

In the last section of the main part, Sec.\ \ref{thepressure}, the 
QCD effective potential is considered in order to determine the pressure
of several color-superconducting phases. 
Using the results of the previous sections, 
especially those of Sec.\ \ref{gapeqsolution}, a relatively simple 
calculation shows which of the spin-one phases corresponds to the 
maximum pressure at zero temperature and therefore is expected to be 
the favored one.    

Our convention for the metric tensor is 
$g^{\mu\nu}=\mbox{diag}\{1,-1,-1,-1\}$. 
Our units are $\hbar=c=k_B=1$ (deviating from the nonrelativistic 
convention used in Sec.\ \ref{BCS_He3}). Four-vectors
are denoted by capital letters, 
$K\equiv K^\mu=(k_0,{\bf k})$, 
and $k\equiv|{\bf k}|$, while $\uk\equiv{\bf k}/k$.
We work in the imaginary-time formalism, i.e., $T/V \sum_K \equiv
T \sum_n \int d^3{\bf k}/(2\pi)^3$, where $n$ labels the Matsubara 
frequencies $\omega_n \equiv i k_0$. For bosons, $\omega_n=2n \pi T$,
for fermions, $\omega_n=(2n+1) \pi T$.
 
   \section{The gap equation}
\label{gapeqsolution}

\subsection{Derivation of the gap equation} \label{derivgapeq}

In the following, we outline the derivation of the QCD gap equation 
\cite{rischkerev}. We apply the so-called ``Cornwall-Jackiw-Tomboulis'' (CJT) 
formalism \cite{cjt},
which allows for a derivation of self-consistent Dyson-Schwinger equations
from the QCD partition function.
It is especially useful in the case of spontaneous symmetry breaking, which 
is taken into account via a bilocal source term in the QCD action. We 
start from the partition function
\be \label{QCDpartition}
{\cal Z} = \int {\cal D}A\,{\cal D}\bar{\psi}{\cal D}\psi\,\exp S \,\, ,
\ee
where the action $S$ is composed of three parts,
\be \label{action}
S=S_A + S_F + g\int_X \bar{\psi}(X) \g^\m T_a \psi(X) A_\m^a(X) \,\, .
\ee
The first term is the gluon field part, which will be discussed in detail
in Sec.\ \ref{mixingscreening}. It contains a gauge field term, $S_{F^2}$,
a gauge fixing term, $S_{gf}$, and a ghost term, $S_{FPG}$,   
\be
S_A=S_{F^2} + S_{gf} + S_{FPG} \,\, .
\ee
The second term is the free fermion part in the presence of a chemical 
potential $\m$,
\be \label{fermion}
S_F=\int_X \bar{\psi}(X)\,(i\g\cdot \partial_X + \m\g_0 - m)\psi(X) \,\, .
\ee
Here, $\psi(X)$ and $\bar{\psi}(X)=\psi^\dag(X)\g_0$ are the quark and 
adjoint quark fields, respectively. The space-time integration is 
defined as $\int_X \equiv \int_0^{1/T}d\tau\int_V d^3{\bf x}$, where
$T$ is the temperature and $V$ the volume of the system, and $m$ is the
fermion mass. In principle, $m$ is a mass matrix accounting for different
masses of different quark flavors. 

The third term in Eq.\ (\ref{action}) describes the coupling between 
quarks and gluons. Such a term arises in any gauge theory where the requirement
of gauge invariance leads to a covariant derivative including a gauge field.
Here, the gauge fields $A_\m^a$ $(a=1,\ldots,8)$ are gluon fields, while
the Gell-Mann matrices $T_a$ are the generators of the gauge group 
$SU(3)_c$ in the adjoint representation. As already introduced in Table
\ref{analogy}, $g$ is the strong coupling constant. In this section, 
we restrict our discussion to the strong interaction which is responsible 
for the formation of Cooper pairs. In Sec.\ \ref{mixingscreening}, also the 
photon field is taken into account in order to investigate the mixing 
between photons and gluons. 

\begin{figure}[ht] 
\begin{center}
\vspace{0.5cm}
\includegraphics[width=14cm]{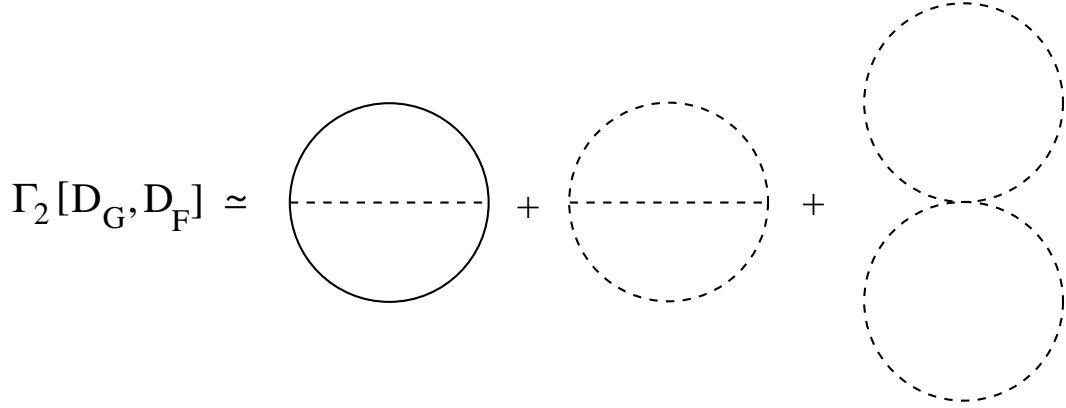}
\vspace{0.5cm}
\caption[Two-loop approximation]{Two-loop approximation of 
$\G_2[D_G,D_F]$ (without ghost contributions). Dashed lines represent
the gluon propagator $D_G$ while full lines represent the quark propagator
$D_F$.}
\label{twoloop}
\end{center}
\end{figure}

In order to implement a bilocal source term into the action, it is
convenient to introduce Nambu-Gor'kov spinors
\be
\Psi = \left(\begin{array}{c} \psi \\ \psi_C \end{array}\right) \,\, ,\qquad
\bar{\Psi} = (\bar{\psi},\bar{\psi}_C) \,\, ,   
\ee
where $\psi_C\equiv C \bar{\psi}^T$ is the charge conjugate spinor, arising
from the original spinor through multiplication with the charge 
conjugation matrix 
$C\equiv i\g^2\g_0$. In the $2N_cN_f$-dimensional Nambu-Gor'kov space, the
fermion action, Eq.\ (\ref{fermion}), can be written as 
\be
S_F = \frac{1}{2} \int_{X,Y} \bar{\Psi}(X)\,{\cal S}_0^{-1}(X,Y)\,\Psi(X) \,\,.
\ee
The factor $1/2$ accounts for the doubling of degrees of freedom. The 
inverse free fermion propagator now has an additional $2\times 2$ structure,
\be \label{inversefermion}
{\cal S}_0^{-1} \equiv \left(\begin{array}{cc}[G_0^+]^{-1} & 0 \\ 
0 & [G_0^-]^{-1}\end{array} \right) \,\, ,
\ee 
where
\be
[G_0^\pm]^{-1}(X,Y)\equiv -i\,(i\g\cdot\partial_X \pm \m\g_0 -m)
\,\d^{(4)}(X-Y) \,\, .
\ee
The interaction term reads in the new basis 
\be
g\int_X \bar{\psi}(X) \g^\m T_a \psi(X) A_\m^a(X) = 
\frac{1}{2}g\int_X \bar{\Psi}(X)\G_a^\m\Psi(X)A_\m^a(X) \,\, ,
\ee
where
\be \label{Gammadef}
\G_a^\m \equiv \left(\begin{array}{cc} \g^\m T_a & 0 \\ 0 & -\g^\m T_a^T\end{array}
\right) \,\, .
\ee
Now we extend the action $S\to S[{\cal K}]$ by adding a bilocal source 
term ${\cal K}(X,Y)$, which is a $2\times 2$ matrix,
\be
{\cal K}\equiv \left(\begin{array}{cc} \sigma^+ & \varphi^- \\ \varphi^+ & \sigma^-
\end{array}\right) \,\, .
\ee
Including this term, the new action $S[{\cal K}]$ is given by
\be
S[{\cal K}] = S + \int_{X,Y} \bar{\Psi}(X)\,{\cal K}(X,Y)\Psi(Y)  \,\, .
\ee
Note that the crucial quantities regarding superconductivity are the 
off-diagonal elements of ${\cal K}$, $\varphi^+$ and $\varphi^-$. Since
they couple two (adjoint) quarks (while the diagonal elements $\sigma^+$ 
and $\sigma^-$ couple quarks with adjoint quarks), a nonzero
value of these elements is equivalent to Cooper pairing, or, in other 
words, to a nonvanishing diquark expectation value $\langle \psi\psi\rangle$.
The four entries of ${\cal K}$ are not independent. They are related
via $\sigma^- = C[\sigma^+]^\dag C^{-1}$ (due to charge conjugation 
invariance) and $\varphi^- = \g_0[\varphi^+]^\dag \g_0$ (since the action has to 
be real-valued). Finally, we arrive at the new QCD partition function
\be
{\cal Z}[{\cal K}] = \int {\cal D}A\,{\cal D}\bar{\Psi}{\cal D}\Psi\,
\exp S[{\cal K}] \,\, .
\ee
At this point, the CJT formalism can be applied. Details can be 
found in Ref.\  \cite{cjt}. It results in 
an effective action $\G$ which, in general, is a functional of one- and 
two-point functions. In the following, we neglect the expectation value of 
the gluon field, 
present in order to ensure color neutrality \cite{dietrich}.
Then, the effective
action is a functional only of two-point functions, namely the gauge boson  
and fermion propagators $D_G$ and $D_F$,
\be \label{effectiveaction}
\G[D_G,D_F] = -\frac{1}{2}\Tr\ln D_G^{-1} - \frac{1}{2} \Tr(\D_0^{-1}D_G -1)
+\frac{1}{2} \Tr\ln D_F^{-1} + \frac{1}{2} \Tr ({\cal S}_0^{-1}D_F -1)
+\G_2[D_G,D_F]\,\, ,
\ee
where $\D_0^{-1}$ is the inverse free gluon propagator and the traces run over
Nambu-Gor'kov, Dirac, flavor, color, and momentum space. The functional
$\G_2[D_G,D_F]$ denotes the sum of all two-particle irreducible diagrams 
without external legs and with internal lines given by the gluon and
quark propagators. The physical situations correspond to the
stationary points of the effective potential, obtained after taking the 
functional derivatives with respect to the gluon and fermion propagators.  
One then obtains  
a set of equations for the stationary point $(D_G,D_F)=(\Delta,{\cal S})$,
\begin{subequations} \label{dysonschwinger}
\bea  
\D^{-1} &=& \D_0^{-1} + \Pi \,\, , \label{dysonschwinger1} \\ 
{\cal S}^{-1} &=& {\cal S}_0^{-1} + \Sigma \,\, , \label{dysonschwinger2}
\eea
\end{subequations}
where we defined the gluon and photon self-energies as the functional
derivatives of $\G_2$ at the stationary point, 
\be \label{selfenergydef} 
\Pi\equiv -2\left.\frac{\d\G_2}{\d D_G}\right|_{(D_G,D_F)=(\D,{\cal S})} 
\,\, , \qquad  
\Sigma \equiv 2\left.\frac{\d\G_2}{\d D_F}
\right|_{(D_G,D_F)=(\D,{\cal S})} \,\, . 
\ee 
In order to find the full propagators, one has to solve the Dyson-Schwinger
equations, Eqs.\ (\ref{dysonschwinger}), self-consistently. To this end, 
we denote the entries of the $2\times 2$ fermion self-energy by
\be \label{sigmanambu} 
\Sigma\equiv\left(\begin{array}{cc}\Sigma^+ & \Phi^- \\ \Phi^+ & \Sigma^- 
\end{array}\right) \,\, ,
\ee
and invert Eq.\ (\ref{dysonschwinger2}) formally 
\cite{manuel2}, which yields the full quark propagator in the form
\be \label{fullquark}
{\cal S} = \left(\begin{array}{cc} G^+ & \Xi^- \\ \Xi^+ & G^- 
\end{array}\right) \,\, ,
\ee   
where the fermion propagators for quasiparticles and charge-conjugate 
quasiparticles are
\be \label{Gpm}
G^\pm = \left\{[G_0^\pm]^{-1} + \Sigma^\pm - 
\Phi^\mp([G_0^\mp]^{-1}+\Sigma^\mp)^{-1}
\Phi^\pm\right\}^{-1} \,\, ,
\ee 
and the so-called anomalous propagators, typical for a superconducting
system, are given by
\be \label{Xpm}
\Xi^\pm = -([G_0^\mp]^{-1} + 
\Sigma^\mp)^{-1}\Phi^\pm G^\pm \,\, .
\ee
At this point, one has to restrict oneself to an approximation for 
$\G_2[D_G,D_F]$.
In Fig.\ \ref{twoloop}, all two-particle irreducible diagrams with two loops
are shown. It is an important property of the CJT formalism that 
truncating the infinite set of diagrams contained in $\G_2$ still yields
a well-defined, self-consistent set of equations. The two-loop approximation
of $\G_2$ is equivalent to a one-loop approximation of the self-energies
$\Pi$ and $\Sigma$.   

\begin{figure}[ht] 
\begin{center}
\vspace{0.5cm}
\includegraphics[width=13cm]{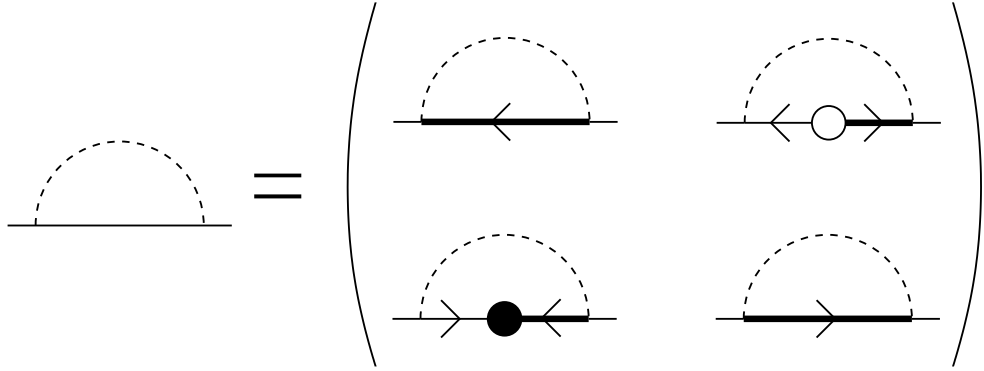}
\vspace{0.5cm}
\caption[Quark self-energy]{Diagrammatic representation of 
Eq.\ (\ref{sigmaeq}). The quark 
self-energy is shown as a $2\times 2$ Nambu-Gor'kov matrix (cf.\ also 
Eqs.\ (\ref{sigmacomponents})). Dashed lines correspond to the gluon 
propagator $\D$. The normal full quasiparticle propagators $G^+$ and $G^-$ 
are denoted
by thick lines with an arrow pointing to the left and right, respectively.
The anomalous propagators $\Xi^\pm$ are drawn according to their structure
given in Eq.\ (\ref{Xpm}): Thin lines correspond to the term 
$([G_0^\mp]^{-1} + \Sigma^\mp)^{-1}$, while the 
full and empty circles denote the gap matrices $\Phi^+$ and $\Phi^-$,
respectively. }
\label{figsigma}
\end{center}
\end{figure}

According to its definition, given in Eq.\ (\ref{selfenergydef}), 
the quark self-energy $\Sigma$ is obtained from $\G_2$ by cutting one
quark line. In the above approximation, this is equivalent to 
cutting a quark line in the left diagram in Fig.\ \ref{twoloop}.
Consequently, in momentum space, we obtain
\be \label{sigmaeq}
\Sigma(K) = -g^2 \int_Q \G^\m_a \, {\cal S}(Q) \, \G_b^\n \, 
\D_{\m\n}^{ab}(K-Q) \,\, .
\ee
Due to the Nambu-Gor'kov structure, this equation actually is a set of four 
equations. With Eqs.\ (\ref{sigmanambu}) and (\ref{fullquark}) these 
equations are 
\begin{subequations} \label{sigmacomponents}
\bea
\Sigma^+(K) &=& -g^2\int_Q \g^\m \, T_a \, G^+(Q) \, \g^\n \, T_b \, 
\D_{\m\n}^{ab}(K-Q) 
\,\, , \\
\Sigma^-(K) &=& -g^2\int_Q \g^\m \, T_a^T \, G^-(Q) \, \g^\n \, T_b^T \, 
\D_{\m\n}^{ab}(K-Q) 
\,\, , \\
\Phi^+(K) &=& g^2\int_Q \g^\m \, T_a^T \, \Xi^+(Q) \, \g^\n \, T_b \, 
\D_{\m\n}^{ab}(K-Q)
\,\, , \label{gapeq21} \\
\Phi^-(K) &=& g^2\int_Q \g^\m \, T_a \, \Xi^-(Q) \, \g^\n \, T_b^T \, 
\D_{\m\n}^{ab}(K-Q)
\,\, . \label{gapeq12}
\eea
\end{subequations}
In Fig.\ \ref{figsigma}, we represent these equations diagrammatically, taking
into account the structure of the anomalous propagators given in 
Eq.\ (\ref{Xpm}). 
 
The last two equations, Eqs.\ (\ref{gapeq21}) and (\ref{gapeq12}), are the
gap equations. The quantities $\Phi^\pm(K)$ are matrices in flavor, color, 
and Dirac space and functions of the quark four-momentum $K$. They occur on 
the 
left-hand side of the equations as well as on the right-hand side, implicitly
present under the integral in the anomalous propagators. Both quantities are 
related via 
\be \label{phiminus}
\Phi^- = \g_0 [\Phi^+]^\dag\g_0 \,\, .
\ee
In the following, we will 
use the term {\it gap matrix} for $\Phi^+(K)$. A complete self-consistent 
solution of the four coupled integral equations is too difficult and certain
approximations have to be made. In 
Secs.\ \ref{gapsolve} and \ref{critictemp}, we use the gap equation to 
determine the value of the gap at the Fermi surface for $T=0$ and the
critical temperature $T_c$. In the next section, Sec.\ \ref{exenergies},
we determine the structure of the quasiparticle excitation energies which 
follows from the ansatz for the gap matrix $\Phi^+$ (without solving 
the gap equation).

\subsection{Excitation energies} \label{exenergies}
 
Before we solve the gap equation, let us first investigate the structure of 
the full quasiparticle propagator $G^+$. From this propagator we can read off
the excitation energy 
of the quasiparticles. This energy spectrum has a nonzero gap in the case of a 
superconductor, as discussed in Sec.\ \ref{superconductivity}.   
In this section, it is shown how the ansatz for the gap matrix, and especially
the form of the order parameter, determines the excitation spectrum.
Here and in all following sections of this thesis, for the sake 
of simplicity,  we consider ultrarelativistic quarks, $m=0$, which is a very
good approximation at least for the $u$ and $d$ quarks. Nonzero 
quark masses cause tremendous technical complications \cite{fugleberg}. In this
case, restricting to Cooper pairing in the even-parity channel, the gap
matrix can be written as
\be \label{gm2SC}
\Phi^+(K)=\sum_{e=\pm}  \phi^e(K)\, {\cal M}_{\bf k} \, \Lambda^e_{\bf k} 
\,\, ,
\ee
where $\phi^e(K)$ is the gap function, ${\cal M}_{\bf k}$ is a matrix
defined by the symmetries of the 
color-superconducting condensate, and 
$\Lambda_{\bf k}^e=(1+e\gamma_0 \vg \cdot \uk)/2$, 
$e=\pm$, are projectors onto states of positive or negative energy. 
In general, ${\cal M}_{\bf k}$ is a matrix in color, flavor, and Dirac
space, and it can be chosen such that 
\be \label{M}
[{\cal M}_{\bf k},\Lambda^e_{\bf k}] = 0 \,\, .
\ee
Let us comment on the relation between the matrix ${\cal M}_{\bf k}$ and 
the order parameter $\D$ discussed in the introduction of this thesis. 
In the matrix ${\cal M}_{\bf k}$ not only the order parameter $\D$ is 
implicitly present but also the basis elements of the special representation 
of the underlying symmetry groups accounting for the (anti-)symmetry of the 
representation. As explained
in the introduction, the representation solely depends on the number 
of flavors, while the choice of the order parameter corresponds to a 
special phase. This will become more transparent below, in Sec.\ 
\ref{exenergies}, when ${\cal M}_{\bf k}$ is specified for several numbers 
of flavors and several phases, see, for instance, Eq.\ (\ref{M2SC}). 
What might appear awkward at first sight, but is technically convenient, is 
that the $4\times 4$ Dirac structure is partially included into 
${\cal M}_{\bf k}$ and partially explicitly written via the energy 
projectors. For details concerning the Dirac structure of the gap matrix, 
cf.\ Ref.\ \cite{rischke2}.   
 
Using Eq.\ (\ref{Gpm}), we can write the quasiparticle propagator as
\bea \label{S11}
G^+ &=& ([G^-_0]^{-1} + \Sigma^-)\left\{([G_0^+]^{-1} + \Sigma^+)
([G^-_0]^{-1} + \Sigma^-) \right. \nonumber \\ 
&& \left. - \; \Phi^-([G_0^-]^{-1}+\Sigma^-)^{-1}
\Phi^+([G_0^-]^{-1}+\Sigma^-)\right\}^{-1} \,\, .
\eea
The free fermion (charge-conjugate) propagator for massless quarks in 
momentum space is 
\be \label{freefermionprop}
G_0^\pm(K)=( \g^{\m}K_{\m} \pm \m \g_0 )^{-1} \,\, ,
\ee 
In Ref.\ \cite{wang}, it was shown that, in order to solve
the gap equation to subleading order, it is 
permissible to approximate the diagonal elements of the quark self-energy by 
\be \label{selfenergy}
\Sigma(K)\equiv \Sigma^+(K)=\Sigma^-(K)\simeq \gamma_0\, \bar{g}^2\,k_0\,
\ln\frac{M^2}{k_0^2}\,\, ,
\ee
where 
\be
\bar{g}\equiv \frac{g}{3\sqrt{2}\pi}\,\, , \qquad M^2\equiv
\frac{3\pi}{4}m_g^2 \,\, , 
\ee
and the zero-temperature gluon mass parameter (squared) is 
\be
m_g^2\equiv\frac{N_f g^2 \mu^2}{6\pi^2}\,\, .
\ee
Then, using the above ansatz for the gap matrix, the second term 
in curly brackets in Eq.\ (\ref{S11}) is   
\be  \label{C2}
\Phi^-([G_0^-]^{-1}+\Sigma^-)^{-1}
\Phi^+([G_0^-]^{-1}+\Sigma^-) = \sum_e |\phi^e(K)|^2 \, L_{\bf k}\, 
\Lambda^{-e}_{\bf k} \,\, ,
\ee
where
\be \label{Ldef}
L_{\bf k}\equiv\gamma_0\, {\cal M}^\dagger_{\bf k}{\cal M}_{\bf k}\, 
\gamma_0\,\, .
\ee
In principle, a notation that includes a ``+'', such as $L^+_{\bf k}$, would 
be appropriate since, in general, the corresponding
matrix $L^-_{\bf k}$ occurring in the expression for the charge-conjugate 
propagator $G^-$ might be different. Indeed, in one of the phases we consider
in this thesis, namely the A phase in a spin-one color superconductor, 
we have $L^+_{\bf k}\neq L^-_{\bf k}$. For more details concerning this 
special case, see Sec.\ \ref{thepressure}. Here and in the following
sections it is sufficient to consider $L_{\bf k}\equiv L^+_{\bf k}$, 
since in the gap equation, only $G^+$ and not $G^-$ occurs, and thus
we simplify the notation by omitting the +.
  
Note that also $L_{\bf k}$ commutes with the energy projectors,
$[L_{\bf k}, \Lambda_{\bf k}^e] =0$. Since $L_{\bf k}$ is hermitian, 
it has real eigenvalues and can be expanded in terms of a complete set of
orthogonal projectors ${\cal P}_{\bf k}^r$,
\be \label{L}
L_{\bf k} = \sum_r \lambda_r \, {\cal P}_{\bf k}^r \,\, ,
\ee 
where $\lambda_r$ are the eigenvalues of $L_{\bf k}$. In this 
decomposition, the sum over $r$ is defined such that all $\l_r$ are 
different, i.e.,
the corresponding projectors ${\cal P}_{\bf k}^r$ project onto   
$n_r$-dimensional eigenspaces, where $n_r\equiv \Tr{\cal P}_{\bf k}^r$
is the degeneracy of the eigenvalue $\l_r$. In general, these 
eigenvalues can depend on the direction of the quark 3-momentum ${\bf k}$.
For the sake of notational convenience, we do not write this $\uk$-dependence
explicitly. Denoting the number of different eigenvalues with 
$n$, the projectors can be computed via
\be \label{projgeneral}
{\cal P}_{\bf k}^r = \prod_{s\neq r}^n \frac{L_{\bf k}-\l_s}{\l_r-\l_s} \,\, .
\ee
For a proof of this relation, see Appendix \ref{proof1}. Obviously, these 
projectors also commute with the energy projectors, 
$[{\cal P}_{\bf k}^r, \Lambda_{\bf k}^e] =0$.
The maximum number 
of different eigenvalues for the color-superconducting phases we 
consider in this thesis is $n=3$. However, we solve the gap equation 
only for special cases in which there are at most two different eigenvalues 
of $L_{\bf k}$. Therefore, let us explicitly write the expression for the 
projectors in this case,
\be \label{proj}
{\cal P}_{\bf k}^{1,2}=\frac{L_{\bf k}
-\lambda_{2,1}}{\lambda_{1,2}-\lambda_{2,1}} \,\, .
\ee     
The next step is to compute the full quasiparticle propagator $G^+$.
The inversion of the term in curly brackets in Eq.\ (\ref{S11}) is now 
particularly simple, because the $2n$ projectors ${\cal P}_{\bf k}^r
\Lambda_{\bf k}^\pm$ are orthogonal and form a complete set in color, 
flavor, and Dirac space. With Eqs.\ (\ref{S11}), (\ref{selfenergy}), 
(\ref{C2}), and (\ref{L})
we obtain
\be \label{fullprop} 
G^+(K)=([G_0^-(K)]^{-1}+\Sigma^-(K))\sum_{e,r}
{\cal P}_{\bf k}^r\, \Lambda_{\bf k}^{-e}\, \frac{1}{\left[k_0/Z(k_0)\right]^2
-\left[\e_{k,r}^e(\phi^e)\right]^2} \,\, ,
\ee
where
\be
Z(k_0)\equiv\left(1+\bar{g}^2\,\ln\frac{M^2}{k_0^2}\right)^{-1}
\ee
is the wave function renormalization factor introduced in Ref.\ 
\cite{manuel2}, and 
\be \label{excite2}
\e_{k,r}^e(\phi^e)\equiv \left[(k-e\mu)^2+\lambda_r|\phi^e|^2\right]^{1/2}
\ee
are the excitation energies for quasiparticles, $e=+$, or 
quasi-antiparticles, $e=-$. They are the relativistic analogues to 
Eq.\ (\ref{excitenonrel}), extended by a possibly nontrivial (multi-)gap 
structure, which also includes possible anisotropies of the 
``true'' gaps $\sqrt{\l_r}\,|\phi^e|$. It was the goal of this section
to show that the investigation of this gap structure 
is equivalent to studying the spectrum of the matrix $L_{\bf k}$.

\subsection{Solution of the gap equation} \label{gapsolve}

In this section, we present a method to compute the value of the gap at the 
Fermi surface at zero temperature from the gap equation.
In the case of the 2SC phase, this value is given
by 
\be \label{phi02SC}
\phi_0^{\rm 2SC}=2 \, \tilde{b} \, b_0'\, \mu\, \exp\left(
-\frac{\pi}{2\, \bar{g}}\right) \,\, ,
\ee
where 
\be \label{constants}
\tilde{b} \equiv 256 \pi^4 \left(\frac{2}{N_f g^2} \right)^{5/2}\,\, , \qquad
b_0' \equiv \exp\left(-\frac{\pi^2+4}{8}\right)
\,\, .
\ee
The term in the exponent of Eq.\ (\ref{phi02SC}) was
first computed by Son \cite{son1}. It arises 
from the exchange of almost static magnetic gluons.
The factor $\tilde{b}$ in front of the exponential originates from 
the exchange of static electric and non-static magnetic gluons
\cite{pisarski3,schaeferwil,hong}.
The prefactor $b_0'$ is due to the quark self-energy \cite{wang,ren}.

In color superconductors with $N_f=1$ and 3 flavors, various other
prefactors may arise \cite{schaefer,schaefer2}, but the exponential
$\exp[-\pi/(2\bar{g})]$ remains the same. 
As will be demonstrated in the following, this is not an accident, but
due to the fact that the leading-order contribution to the
QCD gap equation does not depend on the 
detailed color, flavor, and Dirac structure of the 
color-superconducting order parameter. This structure
only enters at subleading order, and we provide
a simple method to extract these subleading contributions.

Let us briefly recall what the terms
``leading,'' ``subleading,'' and ``sub-subleading order'' mean
in the context of the QCD gap equation \cite{wang}.
Due to the non-analytic dependence of $\phi_0$ on 
the strong coupling constant $g$ 
one cannot apply the naive perturbative counting scheme
in powers of $g$ in order to identify
contributions of different order.
In the QCD gap equation there are also logarithms of the form
$\ln (\m/\phi_0)$, which are $\sim 1/g$ due to
Eq.\ (\ref{phi02SC}) and thus may cancel simple powers of $g$.
A detailed discussion of the resulting, modified power-counting scheme was 
given in the introduction of Ref.\ \cite{wang} and need not be
repeated here. In short,
leading-order contributions in the QCD gap equation are
due to the exchange of almost static magnetic gluons and
are proportional to $g^2\, \phi_0\, \ln^2(\m/\phi_0) \sim \phi_0$.
They determine the argument of the exponential
in Eq.\ (\ref{phi02SC}).
Subleading-order contributions are due to the exchange of
static electric and non-static magnetic gluons and are 
$\sim g^2 \, \phi_0 \, \ln (\m /\phi_0) \sim g \, \phi_0$.
They determine the prefactor of the exponential in Eq.\ (\ref{phi02SC}).
Finally, sub-subleading contributions 
arise from a variety of sources and, at present, cannot be systematically
calculated. They are proportional to $g^2\, \phi_0$ and
constitute $O(g)$ corrections to the prefactor in Eq.\ (\ref{phi02SC}).
It was argued that also gauge-dependent terms
enter at this order \cite{gauge}. This is, of course,
an artefact of the mean-field approximation which was used to derive
the QCD gap equation \cite{rischke2}. On the quasiparticle mass shell,
the true gap parameter is in principle a physical observable and
thus cannot be gauge dependent.

Let us anticipate the most important results of this section. The general 
result 
for the gap para\-meter, which will be proven in the remainder of this 
section, is (if $L_{\bf k}$ has two different eigenvalues $\l_1$, $\l_2$)
\be \label{phi0}
\phi_0 = 2\,\, b \, b_0' \, \m \, \exp\left(- \frac{\pi}{2 \, \bar{g}}
\right)\, \left( \langle\lambda_1\rangle^{a_1} \, 
\langle\lambda_2\rangle^{a_2} \right)^{-1/2}\,\, ,
\ee
where $a_1,a_2$ are positive constants obeying the constraint
\be \label{constraint}
a_1+a_2 = 1 \,\, .
\ee
With $\langle - \rangle$ we denote the angular average 
(in the case of the spin-one phases, the eigenvalues might depend on the
direction of the quark momentum ${\bf k}$),
\be
\langle\l_r\rangle \equiv \int\frac{d\Omega_{\bf k}}{4\pi} \l_r \,\, .
\ee
The constant $b$ is defined as
\be \label{b} 
b\equiv \tilde{b}\,\exp(-d) \,\, ,
\ee
with $\tilde{b}$ from Eq.\ (\ref{constants}), and $d$ a constant
of order one.
The constant $d$ originates from subleading contributions to the gap equation.
For spin-zero condensates, $d=0$, due to an accidental 
cancellation of some of the subleading terms arising
from static electric and non-static magnetic gluon exchange.
In the spin-one cases, this cancellation does not occur 
and, consequently, $d\neq 0$. Actually, in one of the spin-one phases
considered below, $d$ is not a constant but depends on the direction of
the quark momentum, $d=d(\uk)$. Nevertheless, let us keep the simplified 
notation $d$ without explicitly writing a potential $\uk$ dependence 
(as we also do for the eigenvalues $\l_1$, $\l_2$).

From Eqs.\ (\ref{phi02SC}) and (\ref{phi0}) one 
immediately determines $\phi_0$ in units
of the gap in the 2SC phase,
\be \label{ratio}
\frac{\phi_0}{\phi_0^{\rm 2SC}} = \exp(-d) \, 
\left( \langle\lambda_1\rangle^{a_1} \, 
\langle\lambda_2\rangle^{a_2} \right)^{-1/2}\,\, .
\ee 
In this thesis we present a simple method to extract the value
of the constant $d$ without actually solving a gap equation.
This method utilizes the fact that, to subleading order,
the integration over gluon
momenta in the QCD gap equation can be written as a sum
of a few integrals multiplied by constants. Only these
constants depend on the detailed color, flavor, and Dirac structure
of the order parameter. The integrals are generic for all
cases studied here and have to be computed only once.

This is a nontrivial fact. It means that the 
leading contribution to the gap equation is {\em unique}. 
If it were not, then the prefactor
of the gap integral would be different for each case.
In other words, the contribution of almost static magnetic gluons
to the gap equation is universal in the sense that it is
independent of the detailed color, flavor, and Dirac structure
of the color-superconducting order parameter.
Differences between the cases studied here occur
at subleading order. Only at this order the specific
structure of the order parameter 
is important and leads to different values
for the constant $d$ in Eq.\ (\ref{b}).

Let us now start to prove Eq.\ (\ref{phi0}) 
(and consequently also Eq.\ (\ref{ratio}))
using the gap equation, Eq.\ (\ref{gapeq21}). For the right-hand side
of this equation we need the anomalous propagator $\Xi^+(K)$, defined in 
Eq.\ (\ref{Xpm}). Inserting the expression for the propagator $G^+$, 
Eq.\ (\ref{fullprop}), into this definition and using the form of 
the gap matrix $\Phi^+$ given in Eq.\ (\ref{gm2SC}), we obtain 
\be \label{S212SC}
\Xi^+(K)=-\sum_{e,r} \gamma_0 \, {\cal M}_{\bf k}\, 
\gamma_0\, {\cal P}_{\bf k}^r \Lambda_{\bf k}^{-e}
\, \frac{\phi^e(K)}{\left[k_0/Z(k_0)\right]^2-
\left[\e_{k,r}^e(\phi^e)\right]^2} \,\, . 
\ee

To derive the gap equation for the gap function $\phi^e(K)$, we 
insert Eq.\ (\ref{S212SC}) into Eq.\ (\ref{gapeq21}), 
multiply both sides from the right
with ${\cal M}^\dagger_{\bf k}\, \Lambda_{\bf k}^e$
and trace over color, flavor, and Dirac space. 
To subleading order in the gap equation, it is permissible to use the 
gluon propagator in the Hard-Dense-Loop (HDL) approximation 
\cite{rischke6}, where it is diagonal in adjoint color space, 
$\Delta^{\m\n}_{ab}=\delta_{ab}\, \Delta^{\m\n}$. We obtain
\be \label{gap3}
\phi^e(K)=g^2\frac{T}{V}\sum_{Q}\sum_{e',s} 
\frac{\phi^{e'}(Q)}{\left[q_0/Z(q_0)\right]^2-
\left[\e_{q,s}^{e'}(\phi^{e'})\right]^2} \, \Delta^{\m\n}(K-Q) \,
{\cal T}_{\m\n}^{ee',s}({\bf k},{\bf q}) \,\, ,
\ee 
where the sum over $s$ corresponds to the two eigenvalues $\l_1$ 
and $\l_2$, i.e., $s=1,2$, and  
\be \label{T2SC}
{\cal T}_{\m\n}^{ee',s}({\bf k},{\bf q})\equiv -\frac{ {\rm Tr}
\left[\gamma_\m \, T_a^T \, \gamma_0 \,{\cal M}_{\bf q}\, \gamma_0
\,  {\cal P}_{\bf q}^s \, 
\Lambda_{\bf q}^{-e'}\, \gamma_\n \, T_a\, {\cal M}^\dagger_{\bf k} \, 
\Lambda_{\bf k}^e\right]}{{\rm Tr}
\left[{\cal M}_{\bf k}\,  {\cal M}^\dagger_{\bf k} \, 
\Lambda_{\bf k}^e\right]} \,\, .
\ee
The form (\ref{gap3}) of the gap equation holds for all cases considered 
in this thesis. What is different in each case is the structure
of the term ${\cal T}_{\m\n}^{ee',s}({\bf k},{\bf q})$. 
Our computation will be done in pure Coulomb gauge, where
\be
\Delta^{00}(P)=\Delta_\ell(P) \,\, , \,\,\, \Delta^{0i}(P)=0 \,\, ,
\,\,\,  \Delta^{ij}(P)=(\delta^{ij}-\hat{p}^i\hat{p}^j)\, \Delta_t(P)
\,\, ,
\ee
with the longitudinal and transverse propagators $\Delta_{\ell,t}$ and
$P\equiv K-Q$.
Consequently, we only need the 00-component, 
${\cal T}_{00}^{ee',s}({\bf k},{\bf q})$, and the transverse 
projection of the $ij$-components, 
\be \label{Ttransv}
{\cal T}_t^{ee',s}({\bf k},{\bf q})\equiv -(\delta^{ij}-\hat{p}^i\hat{p}^j) 
\, {\cal T}_{ij}^{ee',s}({\bf k},{\bf q}) \,\, ,
\ee
of the tensor (\ref{T2SC}). (The extra minus sign is included for
the sake of notational convenience.)
It will turn out that in all cases studied here
the quantities ${\cal T}_{00,t}^{ee',s}({\bf k},{\bf q})$
are related in the following way:
\be \label{relation2}
\frac{{\cal T}_{00}^{ee',2}({\bf k},{\bf q})}{
{\cal T}_{00}^{ee',1}({\bf k},{\bf q})} 
= \frac{{\cal T}_t^{ee',2}({\bf k},{\bf q})}{
{\cal T}_t^{ee',1}({\bf k},{\bf q})} 
= \mbox{const.} \,\, .
\ee

The right-hand side of Eq.\ (\ref{T2SC}) depends on $k$, $q$, and 
$\uk\cdot\uq$. The latter can be replaced by the square of the 
gluon 3-momentum $p^2$ via $\uk\cdot\uq=(k^2+q^2-p^2)/(2kq)$. 
Thus, the relevant components 
can be written in terms of a power series in $p^2$,
\begin{subequations} \label{T}
\bea
{\cal T}_{00}^{ee',s}({\bf k},{\bf q})&=& a_s \sum_{m=-1}^\infty\,
\eta_{2m}^\ell(ee',k,q)\, \left(\frac{p^{2}}{kq}\right)^m \,\, ,\\
{\cal T}_t^{ee',s}({\bf k},{\bf q})
&=& a_s\sum_{m=-1}^\infty\,
\eta_{2m}^t(ee',k,q)\, \left(\frac{p^{2}}{kq}\right)^m\,\, .
\eea
\end{subequations}
Here, the coefficients $\eta_{2m}^{\ell, t}(ee',k,q)$ no longer
depend on $s$ on account of Eq.\ (\ref{relation2}). 
The overall normalization on the right-hand sides of Eqs.\ (\ref{T})
is still free, and we choose it such that Eq.\ (\ref{constraint})
is fulfilled. This 
uniquely determines the values of the dimensionless coefficients 
$\eta_{2m}^{\ell,t}(ee',k,q)$ and we have the following 
relation for the coefficients $a_1$, $a_2$,
\be \label{a1a2}
a_{1/2}\equiv \frac{{\cal T}_t^{ee',1/2}({\bf k},{\bf q})}
{{\cal T}_t^{ee',1}({\bf k},{\bf q})
+{\cal T}_t^{ee',2}({\bf k},{\bf q})}
= \frac{{\cal T}_{00}^{ee',1/2}({\bf k},{\bf q})}
{{\cal T}_{00}^{ee',1}({\bf k},{\bf q})
+{\cal T}_{00}^{ee',2}({\bf k},{\bf q})} \,\, .
\ee
We now perform the Matsubara sum in Eq.\ (\ref{gap3}), which does
not depend on the detailed structure of the tensor 
${\cal T}_{\m\n}^{ee',s}({\bf k},{\bf q})$. This calculation is 
similar to that of Ref.\ \cite{pisarski3}. The difference is the appearance
of the wave function renormalization factor $Z(q_0)$ \cite{wang}.
To subleading order, this amounts to an extra factor $Z(\e_{q,s}^{e'})$
in the gap equation. Since there are two different excitation energies 
$\e_{q,1}$ and $\e_{q,2}$ on the right-hand side of the gap equation, 
we can put the gap function on the left-hand side on either one of the 
two possible quasiparticle mass shells $k_0=\e_{k,1}$ or $k_0=\e_{k,2}$.
One then obtains
\begin{eqnarray}
\phi^e(\e_{k,r}^e,k) & = & \frac{g^2}{16\pi ^2 k} 
\int_{\mu-\delta}^{\mu+\delta}
dq \, q \sum_{e',s} a_s\,  Z(\e_{q,s}^{e'})\,
\frac{\phi^{e'}(\e_{q,s}^{e'},q)}{\e_{q,s}^{e'}}\,
\tanh\left(\frac{\e_{q,s}^{e'}}{2T}\right) 
\non 
&& \times \sum_m  \int_{|k-q|}^{k+q} dp\,p
 \left(\frac{p^2}{kq}\right)^m\, \left\{\frac{2}{p^2+3m_g^2}\,
\eta_{2m}^\ell + \left[\;\frac{2}{p^2}\, \Theta(p-M)   \right.\right.
\non
&&\left.\left.
+\Theta(M-p)\left(\frac{p^4}{p^6+M^4(\e_{q,s}^{e'}+\e_{k,r}^e)^2}
+\frac{p^4}{p^6+M^4(\e_{q,s}^{e'}-\e_{k,r}^e)^2}\right)\right]
\eta_{2m}^t \right\} \,\,.
\label{a1}
\end{eqnarray}
The first term in braces arises from static 
electric gluons, while the two terms in brackets originate 
from non-static and almost static magnetic gluons, respectively. Various 
other terms which yield sub-subleading contributions to the gap equation 
\cite{pisarski3} have been omitted. 
In deriving Eq.\ (\ref{a1}), the angular integration  
$d\Omega_{\bf q}=\sin\theta d\theta d\varphi$ has been reduced to an 
integration over the modulus of the gluon momentum $p$: 
The integral over the polar angle $\theta$ can be transformed into an 
integral over $p$ by choosing the coordinate system such that 
$\uk\cdot\uq = \cos\theta$ and applying the relation 
$\uk\cdot\uq=(k^2+q^2-p^2)/(2kq)$. The integration over the azimuthal angle
$\varphi$ has already been performed.  
This integration is trivial in the 2SC and CFL 
phases, since in these cases the gap function only depends on the 
modulus of ${\bf q}$ and the eigenvalues
$\l_s$ are constants. However, in the spin-one phases, there might occur 
angular dependent terms, and consequently, the $d\Omega_{\bf q}$ integration 
becomes nontrivial. First, through a potential angular dependence of 
$\l_s$, the
excitation energies in the integrand can be $\uq$-dependent. Second,
the gap function itself may turn out to depend on $\uq$. In order
to proceed, we approximate the angular integral by replacing the eigenvalues
in the excitation energies $\e_{q,s}^{e'}$ by their angular averages 
\be \label{replacelambda}
\l_s \to \langle \l_s\rangle \,\, ,
\ee
and then performing the resulting integral over the azimuthal angle 
and over $p$.
In order to avoid a too complicated notation, we keep the same symbols
for the excitation energies, $\e_{q,s}^{e'}$, but understand them with
the above replacement. This simple approximation is based on the assumption 
that all neglected terms are of sub-subleading order, for which, in principle, 
a rigorous proof is required.      
An angular-dependent gap function $\phi^{e'}(\e_{q,s}^{e'},q)$  
occurs in one of the spin-one phases, namely the polar phase. 
This dependence, entering via an angular dependent function
$d=d(\uq)$, is neglected here, cf.\ also comments in the
discussion of this phase, Sec.\ \ref{results}.

Although the coefficients $\eta_{2m}^{\ell,t}$ depend on $k$ and $q$,
to subleading order in the gap equation we may approximate 
$k\simeq q\simeq \m$. This can be easily proven by power counting.
To this end, it is sufficient to take $k= \m$, 
and write $q = \m + \xi$, where $\xi = q - \m$. 
In weak coupling, the gap function is sharply peaked around the
Fermi surface, and thus the range of integration
in the gap equation can be restricted to a small region of size $2\, \d$
around the Fermi surface. All that is necessary is that
$\d$ is parametrically much larger than $\phi_0$,
but still much smaller than $\m$, $\phi_0 \ll \d \ll \m$
\cite{pisarski3}. It turns out that $\d \sim m_g$ is a convenient choice
\cite{reuter}.
Since the integral over $\x$ is symmetric around $\x = 0$, 
terms proportional to odd powers of $\x$ vanish by symmetry.
Thus, corrections to the leading-order terms are at most $\sim
(\x/\m)^2$. As long as $\d$ is parametrically of the order of $m_g$,
$\x \leq m_g$, and these corrections are $\sim g^2$,
i.e., suppressed by two powers of the coupling
constant. Even for the leading terms in the gap equation the
correction due to terms 
$\sim (\x/\m)^2$ is then only of sub-subleading order and thus
negligible.

Since the coefficients $\eta_{2m}^{\ell,t}$ are dimensionless, 
with the approximation $k \simeq q \simeq \m$ they become pure 
numbers which, as we shall see in the following, are directly 
related to the constant $d$. In all cases considered
here, $\eta_{2m}^{\ell,t}=0$ for 
$m\geq 3$, and the series in Eqs.\ (\ref{T}) terminate after the first few
terms. Moreover, $\eta_{-2}^\ell$ always vanishes and, to subleading order, 
also $\eta_{-2}^{t}=0$.
For the remaining $m$, the $p$ integral in Eq.\ (\ref{a1}) 
can be performed exactly. 
The details of this calculation are deferred to Appendix \ref{AppB}.
We obtain
\begin{eqnarray} 
\phi^e(\e_{k,r}^e,k) & = & \frac{g^2}{16\pi ^2} 
\int_{\mu-\delta}^{\mu+\delta}
dq \sum_{e',s} a_s \, Z(\e_{q,s}^{e'})\,
\frac{\phi^{e'}(\e_{q,s}^{e'},q)}{\e_{q,s}^{e'}}\,
\tanh\left(\frac{\e_{q,s}^{e'}}{2T}\right) 
\non 
&&\times \left[\eta_0^t\, \frac{1}{3}\, 
\ln\frac{M^2}{|(\e_{q,s}^{e'})^2-(\e_{k,r}^e)^2|}
+\eta_0^\ell\, \ln\frac{4\mu^2}{3m_g^2} \right.
\non
&&\left.+\eta_0^t\, \ln\frac{4\mu^2}{M^2}
+4(\eta_2^\ell+\eta_2^t)+8(\eta_4^\ell+\eta_4^t)\right] \; .
\label{a1-1}
\end{eqnarray}
Note that the contribution from almost static magnetic gluons 
only appears in the term  proportional to $\eta_0^t$, 
while non-static magnetic and static electric gluons 
contribute to all other terms.

The antiparticle contribution ($e'=-$)
does not have a BCS logarithm, since $\e_{q,s}^-\simeq q+\m$. 
For the same reason, for antiparticles the logarithm from almost static
magnetic gluons is also only of order 1,
and furthermore there is no large logarithm from the $p$ integrals.
Therefore, the antiparticles contribute at most to sub-subleading
order to the gap equation and can be neglected. In the following,
we may thus set $e=e'=+$ and omit this superscript for the sake
of simplicity. Then, the gap equation for the quasiparticle gap 
function reads 
\begin{equation} \label{gap4}
\phi(\e_{k,r},k)=\bar{g}^2
\int_0^\delta d(q-\mu)\; \sum_s a_s \, Z(\e_{q,s})\,
\frac{\phi(\e_{q,s},q)}{\e_{q,s}}\;
\tanh\left(\frac{\e_{q,s}}{2T}\right)\;\frac{3}{4}\, \eta_0^t\,
\ln\left(\frac{b^2\mu^2}{|\e_{q,s}^2-\e_{k,r}^2|}\right) \,\, ,
\end{equation}
where
\be
b^2=\frac{64\, \m^4}{M^4}\left(\frac{4\m^2}{3m_g^2}
\right)^{3\eta_0^\ell/\eta_0^t}\exp(-2d) \,\, ,
\ee
with
\be \label{d}
 d=-\frac{6}{\eta_0^t}\, \left[\eta_2^\ell+\eta_2^t+2(\eta_4^\ell+\eta_4^t)
\right] \,\, .
\ee
In all cases considered here, $\eta_0^\ell=\eta_0^t$, so that
$b$ assumes the value quoted in Eq.\ (\ref{b}). 
The expression (\ref{d}) is a general formula to compute the 
constant $d$ from the coefficients $\eta_{2m}^{\ell,t}$.
We also find that, for all cases considered here, 
$\eta_0^t=2/3$. This is the uniqueness of the leading-order
contribution
to the gap equation mentioned before. 

Let us, in order to solve Eq.\ (\ref{gap4}),
distinguish between the cases where $a_1=1, \, a_2=0$, and
where both $a_1$ and $a_2$ are nonzero. It turns out that 
the former case corresponds to only one gapped quasiparticle
excitation, i.e., $\l_2=0$. This is expected, since in this case, 
due to $a_2=0$, the ungapped excitation is not present in the gap equation.  
The solution of the gap equation is
well-known for this case. It was discussed in detail in Ref.\ \cite{wang}.
All one has to do is replace the constant $\tilde{b}$
in the calculation of Ref.\ \cite{wang} by the constant
$b = \tilde{b} \exp(-d)$, cf.\ Eq.\ (\ref{b}). The result for
the value of the gap function at the Fermi surface is
Eq.\ (\ref{phi0}), but without the factor  
$(\langle\lambda_1\rangle^{a_1}\, \langle\lambda_2\rangle^{a_2})^{-1/2}$.  
However, in the respective cases this factor simply is 
$\langle\l_1\rangle^{-1/2}$ (remember that $0^0=1$).  
This factor is easily reproduced using the method presented in Ref.\
\cite{wang}: 
One has to multiply both sides of 
Eq.\ (\ref{gap4}) with $\langle\lambda_1\rangle^{1/2}$ in order to 
obtain a gap equation for which the solution of Ref.\ \cite{wang} applies.

%We will show below that in the 2SC phase, $d=0$, and consequently 
%$b= \tilde{b}$, such that the result coincides with Eq.\ (\ref{phi02SC}). 
%In the phases where $d >0$, the gap is reduced as compared
%to the 2SC phase by a factor $\exp(-d)$.

In the case of two gapped quasiparticle excitations, which is 
equivalent to $a_1\neq 0$, $a_2\neq 0$,
the solution of Eq.\ (\ref{gap4}) is more complicated. 
A priori, one has to solve two gap equations, one 
for each quasiparticle mass shell, $k_0=\e_{k,1}$ and $k_0=\e_{k,2}$. 
Therefore, as a function of momentum $k$, there are in principle 
two different gap functions,
$\phi_r(k) \equiv \phi(\e_{k,r},k), \, r = 1,2$. 

In order to proceed with the solution, to subleading order 
we may approximate the logarithm in Eq.\ (\ref{gap4}) in a way 
first proposed by Son \cite{son1},
\be
\frac{1}{2}\,\ln\left(\frac{b^2\m^2}{|\e_{q,s}^2-\e_{k,r}^2|}\right)
\simeq\Theta(\e_{q,s}-\e_{k,r})\, \ln \left(\frac{b\m}{\e_{q,s}}\right)
+\Theta(\e_{k,r}-\e_{q,s})\, \ln \left(\frac{b\m}{\e_{k,r}}\right) \,\, .
\ee
With this approximation and the new variables
\be \label{vartrans}
x_r \equiv \bar{g} \, \ln\left(\frac{2b\m}{k-\m+\e_{k,r}}\right) \quad, \qquad
y_s \equiv \bar{g} \, \ln\left(\frac{2b\m}{q-\m+\e_{q,s}}\right) \,\, ,
\ee
to subleading order the gap equation (\ref{gap4}) transforms into \cite{wang}
\bea 
\phi(x_r)&=&\sum_s a_s \left\{ x_r \int_{x_r}^{x_s^*} dy_s  \, 
(1-2\,\bar{g}\,y_s) \, \tanh \left[\frac{\e(y_s)}{2T}\right]\, \phi(y_s)
\right.
\non
&&\left.
\hspace{1cm} + \int_{x_0}^{x_r}dy_s \, y_s \, (1-2\, \bar{g}\,y_s) \, 
\tanh \left[\frac{\e(y_s)}{2T}\right]\, \phi(y_s)\right\} \,\, .
\label{phixr}
\eea
Here, we denoted the value of $x_s$ at the Fermi surface, i.e,
for $k= \mu$ and $\e_{k,s} = \e_{\mu,s}$, by
\be \label{xsdef}
x_s^* \equiv \bar{g} \, 
\ln\left(\frac{2b\m}{\sqrt{\langle\lambda_s\rangle} 
\, \phi_{0,s}}\right) \,\, ,
\ee
where $\phi_{0,s} \equiv \phi(x_s^*)$ is the value of
the function $\phi(x_s)$ at the Fermi surface. Remember that in the excitation 
energies $\e_{q,s}$ we replaced $\l_s$ by the angular average 
$\langle\lambda_s\rangle$. 
The single point $k= \mu$ in momentum space thus corresponds to two
different points $x_1^*,\, x_2^*$, $x_1^* \neq x_2^*$ ,
in the new variables $x_s$. 
Since we expect $\phi_{0,s}$ to be $\sim \exp(-1/\bar{g})$, 
$x_s^*$ is a constant of order one.  
Furthermore we defined
\be
x_0 \equiv \bar{g} \, \ln\left(\frac{b\m}{\d}\right) \,\, .
\ee
This constant is parametrically of order $O(\bar{g})$.
To subleading order, the relation between the new variable
$y_s$ and the excitation energy is given by \cite{pisarski3},
\be
\e(y_s) = b\, \m \, \exp\left(-\frac{y_s}{\bar{g}}\right)\,\, .
\ee
A consequence of the transformation of variables (\ref{vartrans})
and of neglecting sub-subleading corrections is
that the two equations (\ref{phixr}) for $r=1$ and $r=2$ become
identical. The only difference is the notation for the argument of 
the function $\phi$, which in both cases we may simply call $x$.
Therefore, instead of two separate equations, we only have to 
consider a single equation which determines the function $\phi(x)$.
Moreover, $y_s$ is merely an integration variable, 
and we may set $y_s \equiv y$ in the following.

With Eq.\ (\ref{constraint}), we rewrite Eq.\ (\ref{phixr}) in the form
\begin{eqnarray}
\phi(x) & = & x \int_x^{x_2^*}dy\, (1- 2\, \bar{g} \, y) \,
\tanh\left[\frac{\e(y)}{2T}\right]  \, \phi(y)
+ \int_{x_0}^x dy\, y\, (1- 2\, \bar{g}\, y)\, 
\tanh \left[\frac{\e(y)}{2T}\right]\, \phi(y) \nonumber \\
&  & - a_1\, x \int_{x_1^*}^{x_2^*}dy\, (1- 2\, \bar{g} \, y) \,
\tanh\left[\frac{\e(y)}{2T}\right]\, \phi(y) \,\, .
\label{phix}
\end{eqnarray}
One can also write this equation in a form where
$x_2^*$ is replaced by $x_1^*$ and $a_1$ by $a_2$, respectively.
Equation (\ref{phix}) is an integral equation for the function 
$\phi(x)$, which is solved in the standard manner by converting 
it into a set of differential equations \cite{son1},
\begin{subequations}
\begin{eqnarray} \label{firstder}
\frac{d\phi}{dx} & = & \int_x^{x_2^*}dy\, (1- 2\, \bar{g}\, y)\,
\tanh \left[ \frac{\e(y)}{2T} \right]\, \phi(y)
- a_1 \int_{x_1^*}^{x_2^*} dy\,(1- 2\, \bar{g}\, y)\,
\tanh \left[ \frac{\e(y)}{2T} \right] \, \phi(y)\,\, ,  \\
\frac{d^2\phi}{dx^2} & = & - (1- 2\, \bar{g}\, x)\, 
\tanh \left[ \frac{\e(x)}{2T} \right]\, \phi(x) \, \, .
\label{diffeq}
\end{eqnarray}
\end{subequations}
We now solve the second-order differential equation (\ref{diffeq}) at
zero temperature, $T=0$. One immediately observes that this equation is
identical to Eq.\ (22c) of Ref.\ \cite{wang}, and its solution
proceeds along the same lines as outlined there. The only difference
compared to the previous calculation are the extra terms 
$\sim a_1$ in Eqs.\ (\ref{phix}) and (\ref{firstder}).
To subleading order, we expect $\phi_{0,1}/ \phi_{0,2} \simeq 1$
(we show below that this assumption is consistent with our final result),
such that the difference
\be \label{x2x1}
x_2^* - x_1^* = \bar{g} \, \ln \left( 
\frac{\sqrt{\langle\lambda_1\rangle}\, 
\phi_{0,1}}{\sqrt{\langle\lambda_2\rangle}\, \phi_{0,2}} \right)
\simeq \frac{\bar{g}}{2}\, \ln \left( 
\frac{\langle\lambda_1\rangle}{\langle\lambda_2\rangle} \right)
\ee
is of order $O(\bar{g})$. Consequently, the extra terms $\sim a_1$
are of subleading order, $O(\bar{g}\phi_0)$, and we may
approximate
\be \label{subcorr}
\int_{x_1^*}^{x_2^*} dy\,(1- 2\, \bar{g}\, y)\,  \, \phi(y)
\simeq (x_2^*-x_1^*) \, \phi_{0,2} \,\, .
\ee
Ordering the eigenvalues such that $\langle\lambda_1\rangle >
\langle\lambda_2\rangle$, we have $x_2^* - x_1^* >0$.

The subleading correction (\ref{subcorr}) qualitatively changes
the behavior of the gap function $\phi(x)$ near the Fermi surface.
In the absence of the term $\sim a_1$ in Eq.\ (\ref{firstder}), 
the derivative of the gap function vanishes for $x=x_2^*$,
and the gap function assumes its maximum at this point 
\cite{wang}. 
The subleading correction (\ref{subcorr}) induced by the two-gap
structure causes the derivative (\ref{firstder}) of the function
$\phi(x)$ to be {\it negative\/} at the Fermi surface. Consequently, since 
we still expect $\phi(x)$ to rapidly vanish away from the Fermi surface,
this function assumes its maximum not right {\it at\/} the Fermi surface,
but at a point $x_{\rm max}$ which is close, but not identical to
$x_2^*$. We shall see that $x_2^* - x_{\rm max} \sim O(\bar{g})$.

The subleading correction (\ref{subcorr}) modifies the solution 
of the differential equation (\ref{diffeq}) from the one
given in Ref.\ \cite{wang}. Again, we fix the two unknown constants in the
general solution of the second-order differential equation
(\ref{diffeq}) by matching the solution and its
derivative to the right-hand sides of Eqs.\ (\ref{phix}) and
(\ref{firstder}) at the point $x= x_2^*$.
Introducing the variables 
$z \equiv - (2 \bar{g})^{-2/3} \, (1-2\bar{g}x)$ and 
$z^* \equiv - (2 \bar{g})^{-2/3} \,(1-2\bar{g}x_2^*)$, 
the solution reads
\bea 
\phi(z) &=& \phi_{0,2}  \left\{ \frac{M(|z|)}{M(|{z^*}|)} 
\frac{ \sin \left[\varphi(|z^*|) - \theta(|z|) \right] }{
\sin \left[\varphi(|z^*|) - \theta(|z^*|) \right] } \right.
\non
&&\left. + a_1 \, (x_2^*-x_1^*)\, (2\bar{g})^{-1/3} \,
\frac{M(|z|)}{N(|{z^*}|)}\,
\frac{ \sin \left[\theta(|z^*|) - \theta(|z|) \right] }{ 
\sin \left[\varphi(|z^*|) - \theta(|z^*|) \right] } \right\}\,\, ,
\label{phiz}
\eea
where the functions $M(|z|),\, N(|z|),\, \varphi(|z|),$ and 
$\theta(|z|)$ are related
to the Airy functions ${\rm Ai}(z), \, {\rm Bi}(z)$ and their derivatives
in the standard way \cite{abramowitz}.
The derivative $d\phi(z)/dz$ can be obtained from Eq.\ (\ref{phiz})
simply by replacing $M(|z|)$ and $\theta(|z|)$ by $N(|z|)$ and
$\varphi(|z|)$, respectively.
The difference to the solution for a single gapped quasiparticle
excitation, cf.\ Eq.\ (27) of Ref.\ \cite{wang}, 
is the term proportional to $a_1$.

Finally, we have to determine the value of $\phi_{0,2}$.
To this end, we rewrite Eq.\ (\ref{phix}) at the point $x= x_2^*$
in the form
\be \label{relation3}
\left[ z_0 + (2\bar{g})^{-2/3} \right] \, \frac{d \phi}{d z}(z_0) = 
\phi(z_0) \,\, ,
\ee
where $z_0 \equiv - (2 \bar{g})^{-2/3} \, (1-2\bar{g}x_0)$.
Remarkably, this equation holds in this form also in the case
of a single gapped quasiparticle excitation, 
cf.\ Eq.\ (29) of Ref.\  \cite{wang}.
In weak coupling, the dependence on the variable $z_0$ is spurious.
Inserting the solution (\ref{phiz}) and its derivative for 
$z=z_0$ and expanding 
$M(|z_0|),\, N(|z_0|),\, \varphi(|z_0|),$ and $\theta(|z_0|)$ 
to order $O(\bar{g})$ as demonstrated in Ref.\ \cite{wang},
one derives the condition
\be
x_2^* \simeq \frac{\pi}{2} + \bar{g} \, \frac{\pi^2+4}{8} + 
a_1\, (x_2^*-x_1^*)
\,\, .
\ee
The second term is the $O(\bar{g})$ correction originating from 
the quark self-energy. It leads to the constant $b_0'$ in
Eq.\ (\ref{phi02SC}) and was first derived in Refs.\ \cite{wang,ren}.
The last term $\sim a_1$ is the correction arising from the
two-gap structure to the result (33) of Ref.\ \cite{wang}.
Because of Eq.\ (\ref{x2x1}), this correction is also of order
$O(\bar{g})$.
Using the definition (\ref{xsdef}) of $x_2^*$, as well as the
condition (\ref{constraint}), we 
conclude that the expression for $\phi_{0,2}$ is identical to the
one for $\phi_0$ in Eq.\ (\ref{phi0}).
This is the value of the gap function at the Fermi surface, $k = \mu$,
or $x=x_2^*$, for the quasiparticle excitation branch $\e_{k,2}$. 
The additional factor compared to
the 2SC gap $\phi_0^{\rm 2SC}$ of Eq.\ (\ref{phi02SC}),
which originates from the two-gap structure,
is $\left( \langle\lambda_1\rangle^{a_1}\, \langle\lambda_2\rangle^{a_2} \right)^{-1/2}$.

We can also compute the gap function at the Fermi surface
for the first excitation branch $\e_{k,1}$, i.e., at $x=x_1^*$.
The difference $\phi_{0,2}- \phi_{0,1}$ can be obtained from
Eq.\ (\ref{phix}) as
\be
\phi_{0,2}- \phi_{0,1} = \int_{x_1^*}^{x_2^*} dy \, 
\left[ y-x_1^* - a_1\, (x_2^* - x_1^*) \right]
\, (1- 2\, \bar{g}\, y)\, \phi(y) \,\, .
\ee
An upper bound for the term in brackets is given by setting
$y = x_2^*$, where it assumes the value $a_2 (x_2^*-x_1^*)$
on account of Eq.\ (\ref{constraint}).
Pulling this factor out of the integral, the latter can
be estimated with Eq.\ (\ref{subcorr}). This proves that
the difference $\phi_{0,2}- \phi_{0,1}$ is only of order 
$O(\bar{g}^2\phi_0)$,
which shows that our above assumption
$\phi_{0,1}/\phi_{0,2} \simeq 1$ is consistent up to subleading order.
To this order, we may therefore set
$\phi_{0,1} = \phi_{0,2} \equiv \phi_0$.

We now determine the value of $x_{\rm max}$, 
where the gap function assumes its maximum, by
setting the left-hand side of Eq.\ (\ref{firstder}) equal to zero.
This leads to the condition
\be
\int_{x_{\rm max}}^{x_2^*}dy\, (1- 2\, \bar{g}\, y)\, \phi(y)
= a_1 \int_{x_1^*}^{x_2^*} dy\,(1- 2\, \bar{g}\, y)\, \phi(y)\,\, .
\ee
To order $O(\bar{g}\phi_0)$, one may easily solve this
equation for $x_{\rm max}$, with the result
\be
x_{\rm max} = x_2^* - a_1 \, \frac{\bar{g}}{2}\, 
\ln \left( \frac{\langle\lambda_1\rangle}{\langle\lambda_2\rangle} \right) 
\,\, ,
\ee
i.e., $x_{\rm max}$ is indeed smaller than $x_2^*$ by a term of 
order $O(\bar{g})$, as claimed above.
Obviously, since $a_1 < 1$, from Eq.\ (\ref{x2x1})
we derive the inequality $x_1^* < x_{\rm max} < x_2^*$, i.e.,
the gap function assumes its maximum between the values
$x_1^*$ and $x_2^*$. The value of the gap function at $x_{\rm max}$
can be estimated via a calculation similar to the one for the
difference $\phi_{0,2} - \phi_{0,1}$ above. The result is
$\phi_{max} \simeq \phi_0 \, [1+O(\bar{g}^2)]$. 
This means that the gap function
is fairly flat over a region of size $O(\bar{g})$ (in the variable
$x$) in the vicinity of the Fermi surface.

\subsection{The critical temperature} \label{critictemp}

In this section, we compute the critical temperature $T_c$ for the 
color-superconducting phase transition. Again, let us present the final result
before we prove it. We find for the ratio of the critical temperature
and the gap parameter
\be
\label{Tc}
\frac{T_c}{\phi_0}=\frac{e^\gamma}{\pi} \, 
\left( \langle\lambda_1\rangle^{a_1} \, \langle\lambda_2\rangle^{a_2} 
\right)^{1/2} 
\simeq 0.57 \, \left( \langle\lambda_1\rangle^{a_1} \, 
\langle\lambda_2\rangle^{a_2} \right)^{1/2}\,\, ,
\ee
where $\gamma\simeq 0.577$ is the Euler-Mascheroni constant, already 
introduced in Eq.\ (\ref{BCSrelation}). 
In the case of 
$(\langle\lambda_1\rangle^{a_1} \, \langle\lambda_2\rangle^{a_2})^{1/2} = 1$,
we recover the relation (\ref{BCSrelation}) from BCS theory.

Note that, although a factor 
$(\langle\lambda_1\rangle^{a_1} \, \langle\lambda_2\rangle^{a_2})^{1/2}\neq 1$
renders the ratio $T_c/\phi_0$ different from BCS theory,
the absolute value of $T_c$ is not affected by this factor. If the
energy scale is set by $\phi_0^{\rm 2SC}$, then
\be \label{Tcabs}
\frac{T_c}{\phi_0^{\rm 2SC}} = \exp(-d)
\ee
because the factor
$(\langle\lambda_1\rangle^{a_1}\, \langle\lambda_2\rangle^{a_2})^{-1/2}$
in Eq.\ (\ref{ratio}) simply cancels the factor
$(\langle\lambda_1\rangle^{a_1}\, \langle\lambda_2\rangle^{a_2})^{1/2}$ in 
Eq.\ (\ref{Tc}). As mentioned above, in one of the cases we study,
namely the polar phase, the gap function $\phi_0$ depends on the direction 
of the quark momentum. This dependence enters via the function 
$d=d({\bf k})$, cf.\ Sec.\ \ref{results}. In this case, Eq.\ (\ref{Tcabs})
has to be replaced by
\be \label{Tcabspolar}
\frac{T_c}{\phi_0^{\rm 2SC}} = \langle\exp(-d)\rangle \,\, .
\ee

Let us now prove Eq.\ (\ref{Tc}).
In the case where $a_1=1$ and $a_2=0$,
the calculation of Ref.\ \cite{wang} applies. Therefore, let us focus
on the more complicated case where both $a_1$ and $a_2$ are nonzero.
The calculation follows the line
of arguments presented in Ref.\ \cite{wang}, taking into account
the additional term $\sim a_1$ in Eq.\ (\ref{phix}).
As in Refs.\ \cite{pisarski3,wang} we assume that, to leading
order, the effect of temperature is a change of the magnitude
of the gap, but not of the shape of the gap function,
\be
\phi(x,T) \simeq \phi(T)\, \frac{\phi(x,0)}{\phi_0} \;,
\ee
where $\phi(T)\equiv\phi(x_2^*,T)$ is the value of the gap 
at the Fermi surface at temperature $T$,
$\phi(x,0)$ is the zero-temperature gap function $\phi(x)$ computed in the
last section, cf.\ Eq.\ (\ref{phiz}),
and $\phi_0 \equiv \phi_{0,2} = \phi(x_2^*,0)$. 
With this assumption, Eq.\ (\ref{phix}) reads at the Fermi surface
\begin{eqnarray}
1 & = & \int_{x_0}^{x_\k} dy\, y\, (1- 2\, \bar{g} \, y) \,
\tanh\left[\frac{\e(y)}{2T}\right]  \, \frac{\phi(y,0)}{\phi_0}
+ \int_{x_\k}^{x_2^*}dy\, y\, (1- 2\, \bar{g} \, y) \,
\tanh\left[\frac{\e(y)}{2T}\right]  \, \frac{\phi(y,0)}{\phi_0}
\nonumber \\
&   & - a_1\, x_2^*  \int_{x_1^*}^{x_2^*}dy\, (1- 2\, \bar{g} \, y) \,
\tanh \left[ \frac{\e(y)}{2T} \right]\, \frac{\phi(y,0)}{\phi_0}
\nonumber \\
& \equiv & {\cal I}_1 + {\cal I}_2 + {\cal I}_3
\,\, ,
\label{gapeqtemp}
\end{eqnarray}
where we divided the second integral in Eq.\ (\ref{phix}) 
into two integrals: ${\cal I}_1$ which runs from 
$x_0$ to $x_\kappa$, with $x_\kappa \equiv
x_2^*-\bar{g}\, \ln(2\kappa)$, $\kappa \gg 1$, and ${\cal I}_2$ which runs 
from $x_\kappa$ to $x_2^*$ \cite{pisarski3}. 
We now compute the integrals ${\cal I}_1$ through ${\cal I}_3$
separately to subleading accuracy, i.e., to order $O(\bar{g})$.

In the first integral ${\cal I}_1$, which runs over a region far from the
Fermi surface, $\e(y) \gg T$, and we may approximate the $\tanh$ by 1.
This integral can be formally solved by integration by parts using the
differential equation (\ref{diffeq}),
\be
{\cal I}_1 = \frac{1}{\phi_0} \, \left[ \phi(x_\k,0) - x_\k \,
\frac{d\phi}{dx}(x_\k,0) \right] \,\,,
\ee
where we exploited the condition (\ref{relation3}).
Expanding the functions on the right-hand side around $x_2^*$ we
obtain to subleading order
\be \label{I1}
{\cal I}_1 = 1 - \frac{\pi}{2} \,\left[ \bar{g}\, \ln (2 \k) - a_1 \,
(x_2^*-x_1^*) \right]\,\, .
\ee
This estimate is similar to the one made in Eq.\ (36) of Ref.\
\cite{wang}. The main difference to that calculation is the term
$\sim a_1$ which appears because the first derivative of the
gap function no longer vanishes at the Fermi surface, cf.\ the
discussion in the previous section.

In the second integral ${\cal I}_2$, 
which only contributes to order $O(\bar{g})$
to the right-hand side of Eq.\ (\ref{gapeqtemp}), to subleading order
we may set $\phi(y,0)/\phi_0 \simeq 1$ and $y \simeq  x_2^* \simeq
\pi/2$.
Reverting the transformation of variables (\ref{vartrans}) 
we obtain
\be
{\cal I}_2 = \frac{\pi}{2} \bar{g} \int_0^{\sqrt{\langle\lambda_2\rangle}
\k\phi_0} 
\frac{d(q-\m)}{\e_{q,2}}\, \tanh  \left( \frac{\e_{q,2}}{2T} \right) \,\, .
\ee

The last integral in Eq.\ (\ref{gapeqtemp}), ${\cal I}_3$, 
also contributes a term of order $O(\bar{g})$, 
and may thus be approximated by an argument
similar to that leading to Eq.\ (\ref{subcorr}),
\be
{\cal I}_3 = a_1\, x_2^* (x_2^* - x_1^*) \, 
\tanh \left[ \frac{\phi(T)}{2T} \right] \,\, .
\ee
At the critical temperature $T_c$, where $\phi(T_c) = 0$, 
this term vanishes. 
Putting everything together, at $T=T_c$ Eq.\ (\ref{gapeqtemp}) becomes
\be \label{gapeqtemp2}
\bar{g} \int_0^{\sqrt{\langle\lambda_2\rangle}\k\phi_0} 
d(q-\m)\, \left[ \frac{1}{q-\m}\, \tanh 
\left( \frac{q-\m}{2T_c} \right) - 
\frac{1}{\sqrt{(q-\m)^2 + \langle\lambda_2\rangle \phi_0^2}} \right] = 
- a_1 \, (x_2^*- x_1^*) \,\, ,
\ee
where the term $\ln (2\k)$ in Eq.\ (\ref{I1}) was expressed
in terms of an integral according to Eq.\ (96) of Ref.\ \cite{pisarski3}.
In the integral on the left-hand side, we may send $\k \rightarrow
\infty$ \cite{pisarski3}. This allows us to perform it analytically,
which yields the result 
$\ln[ e^\gamma \sqrt{\langle\lambda_2\rangle}\phi_0/(\pi T_c)]$.
If the right-hand side of Eq.\ (\ref{gapeqtemp2}) 
were zero, for $\langle\lambda_2\rangle = 1$
this would then lead to the BCS relation $T_c/\phi_0 = e^\gamma/\p$.
However, using Eq.\ (\ref{x2x1}) we now obtain Eq.\ (\ref{Tc}).
The last factor on the right-hand side of this equation
is exactly the inverse of the additional 
factor in Eq.\ (\ref{phi0}). 
This factor violates the BCS relation $T_c/\phi_0 = e^\gamma/\pi$.

\subsection{Results: The 2SC, CFL, polar, planar, A, and CSL phases} 
\label{results}

In the previous sections we derived general expressions for the gap
parameter and the critical temperature. The final expressions are given
in Eqs.\ (\ref{phi0}) and (\ref{Tc}), respectively. In this section, 
we apply these 
expressions to spin-zero color superconductors in the 2SC and CFL phases, 
as well as to spin-one color superconductors in four different phases.
For each case, we need to determine the following quantities:

\begin{itemize}

\item The eigenvalues $\l_s$ (and their degeneracies) of the matrix 
$L_{\bf k}$. This leads
to the excitation spectrum, cf.\ Eq.\ (\ref{excite2}).

\item The projectors ${\cal P}_{\bf k}^s$, which project onto the 
corresponding eigenspaces. They are needed for

\item the traces ${\cal T}_{00,t}^{ee',s}({\bf k},{\bf q})$, 
defined in Eq.\ (\ref{T2SC}) and occurring in the gap equation.

\item The constants $a_1$, $a_2$, introduced in Eqs.\ (\ref{T}) and 
computed via Eq.\ (\ref{a1a2}). 

\item The coefficients $\eta_{2m}^{\ell,t}$, which lead to 

\item the number $d$, entering the prefactor of the gap, cf.\ Eqs.\ 
(\ref{phi0}) and (\ref{b}).

\end{itemize}

\subsubsection{The 2SC phase} \label{2SC}

For $N_f=2$, the spin-zero condensate is a singlet in flavor and an 
antitriplet in color space \cite{bailin}, cf.\ Sec.\ \ref{2SCCFL}. 
As explained in the introduction, the order parameter of the 
2SC phase is $\D_i = \d_{i3}$. Consequently, the matrix 
${\cal M}_{\bf k}$ reads
\be \label{M2SC}
{\cal M}_{\bf k}= \D_i J_i\,\tau_2 \, \gamma_5  = 
J_3\,\tau_2 \, \gamma_5 \,\, ,
\ee
where $\gamma_5$ takes into account that we restrict our discussion
to the even-parity channel. The antisymmetric $3\times 3$ matrices 
$(J_i)_{jk}=-i\e_{ijk}$, $i,j,k=1,2,3$, 
form a basis of the color antitriplet $[{\bf\bar{3}}]^a_c$. They are, up to 
a factor $\pm 2$ , identical to three of the Gell-Mann matrices, $2\,T_2=J_3$, 
$2\,T_5=-J_2$, and $2\,T_7=J_1$. The antisymmetric singlet 
structure in flavor space is represented by the second 
Pauli matrix $(\tau_2)_{fg}=-i\,\e_{fg}$, $f,g=1,2$.  
The matrix ${\cal M}_{\bf k}$ obviously fulfills the condition 
(\ref{M}). From Eq.\ (\ref{Ldef}) we construct the matrix
\be \label{L2SC}
\left(L_{\bf k}\right)_{ij}^{fg} = (J_3^2)_{ij} \, (\tau_2^2)^{fg} = (\d_{ij}-
\d_{i3}\d_{j3})\, \d^{fg} \,\, .
\ee
In this case, $L_{\bf k}$ does not depend on ${\bf k}$, and consists of
a unit matrix in flavor space and a projector onto the first two colors
in color space. In principle, it also consists of a unit matrix in
Dirac space, which we disregard on account of the spin-zero nature of
the condensate.

The eigenvalues of $L_{\bf k}$ are (cf.\ Appendix \ref{AppA})
\be \label{EV2SC}
\lambda_1=1 \quad (\mbox{4-fold}) \quad , \qquad \lambda_2=0 \quad 
(\mbox{2-fold}) \,\, .
\ee
{}From Eq.\ (\ref{excite2}) we conclude that there are four gapped and
two ungapped excitations. (Taking into account the $4\times 4$ Dirac structure,
the degeneracies are 16 and 8, respectively.) 

The projectors ${\cal P}^r_{\bf k}$ follow from Eq.\ (\ref{proj}),
\be \label{p2SC}
{\cal P}_{\bf k}^1=J_3^2 \quad , \qquad {\cal P}_{\bf k}^2=1-J_3^2
\,\, .
\ee
They have the property that $J_3{\cal P}_{\bf k}^1=J_3$ and
$J_3{\cal P}_{\bf k}^2=0$. Consequently, the tensor 
${\cal T}_{\m\n}^{ee',2}({\bf k},{\bf q})$ vanishes trivially. Therefore,
as expected, the ungapped excitation branch does not enter the
gap equation. For $s=1$ we obtain
\begin{subequations} \label{T2SC2}
\bea
{\cal T}_{00}^{ee',1}({\bf k},{\bf q})&=&\frac{1}{3}\,\left(1+
ee'\,\uk\cdot\uq\right) \,\, ,\\
{\cal T}_t^{ee',1}({\bf k},{\bf q})
&=&\frac{1}{3}\, \left[3-ee'\,\uk\cdot\uq-\frac{(ek-e'q)^2}{p^2}\,
\left(1+ee'\,\uk\cdot\uq\right)\right] \,\, .
\eea
\end{subequations}
We now match this result to the expansion in terms of $p^2$, Eq.\ (\ref{T}).
Since ${\cal T}_{\m\n}^{ee',2}({\bf k},{\bf q})=0$
and because of Eq.\ (\ref{constraint}) we have 
\be \label{a2SC}
a_1=1 \quad , \qquad a_2=0 \,\, .
\ee
This uniquely fixes the coefficients $\eta_{2m}^{\ell,t}(ee',k,q)$. 
To subleading order we only require their values for
$e=e'=+$ and $k\simeq q\simeq \m$, 
\be \label{eta2SC}
\eta_0^\ell=\frac{2}{3} \quad , \qquad 
\eta_2^\ell=-\frac{1}{6} \quad , \qquad 
\eta_4^\ell=0 \quad , \qquad
\eta_0^t=\frac{2}{3} \quad, \qquad  
\eta_2^t=\frac{1}{6} \quad , \qquad 
\eta_4^t=0 \,\, .
\ee
This result implies that the contributions from static electric and 
non-static magnetic gluons to the constant $d$ defined in Eq.\ (\ref{d})
cancel, and consequently $d=0$.

\subsubsection{The CFL phase} \label{CFL}

In the CFL phase, the spin-zero condensate is 
a flavor antitriplet locked with 
a color antitriplet \cite{alford2}, cf.\ Sec.\ \ref{2SCCFL}. The order
parameter is $\D_{ij}=\d_{ij}$, and therefore,  
\be \label{MCFL}
{\cal M}_{\bf k}=\D_{ij} J_i\,I_j \, \gamma_5 = {\bf J} \cdot {\bf I} \, 
\gamma_5 \,\, ,
\ee
where, as above, ${\bf J}=(J_1,J_2,J_3)$ represents the antitriplet in color 
space. The vector ${\bf I}$ represents
the antitriplet in flavor space and is defined analogously. Consequently,
$({\bf J}\cdot{\bf I})_{ij}^{fg}=-\delta_i^f\,\delta_j^g+\delta_i^g\,
\delta_j^f$. 

{}From Eq.\ (\ref{Ldef}) we obtain the matrix
\be
(L_{\bf k})_{ij}^{fg}=\left[({\bf J} \cdot {\bf I})^2\right]_{ij}^{fg}
=\delta_i^f\,\delta_j^g+\delta_{ij}\,\delta^{fg} \,\, .
\ee
As in the 2SC case, the operator $L_{\bf k}$ is independent of 
${\bf k}$, and we omitted its trivial Dirac structure. 
It can be expanded in terms of its eigenvalues and 
projectors as in Eq.\ (\ref{L}), with (cf.\ Appendix \ref{AppA})
\be \label{EVCFL}
\lambda_1=4 \quad (\mbox{1-fold}) \quad , \qquad \lambda_2=1 \quad 
(\mbox{8-fold}) \,\, ,
\ee
and 
\be
({\cal P}_{\bf k}^1)_{ij}^{fg}=\frac{1}{3}\,\delta_i^f\,\delta_j^g
\quad , \qquad 
({\cal P}_{\bf k}^2)_{ij}^{fg}=\delta_{ij}\delta^{fg}-\frac{1}{3}\,
\delta_i^f\,\delta_j^g
\,\, ,
\ee
where ${\cal P}_{\bf k}^1$ and ${\cal P}_{\bf k}^2$ correspond to
the singlet and octet projector, also used in Refs.\ \cite{shovkovy3,zarembo}.

We now compute the relevant components of the tensor  
${\cal T}_{\m\n}^{ee',s}({\bf k},{\bf q})$. 
Since the Dirac structure of ${\cal M}_{\bf k}$ 
is the same as in the 2SC case, the dependence
on ${\bf k}$ and ${\bf q}$ is identical to the one in Eq.\ (\ref{T2SC2}). 
However, since the color-flavor structure is different, 
we obtain a non-trivial
result both for $s=1$ and $s=2$, with different prefactors,
\begin{subequations} \label{TCFL}
\bea
{\cal T}_{00}^{ee',1}({\bf k},{\bf q})=
\frac{1}{2} \, {\cal T}_{00}^{ee',2}({\bf k},{\bf q})&=&\frac{1}{9}\,\left(1+
ee'\,\uk\cdot\uq\right) \,\, ,\\
{\cal T}_t^{ee',1}({\bf k},{\bf q})=  
\frac{1}{2} \, {\cal T}_t^{ee',2}({\bf k},{\bf q})
&=&\frac{1}{9}\, \left[3-ee'\,\uk\cdot\uq-\frac{(ek-e'q)^2}{p^2}\,
\left(1+ee'\,\uk\cdot\uq\right)\right] \,\, .
\eea
\end{subequations}
Obviously, the condition (\ref{relation2}) is fulfilled.
The coefficients $\eta_{2m}^{\ell,t}$ remain the same as 
in Eq.\ (\ref{eta2SC}), which again yields $d=0$. However, the
two-gap structure leads to the constants 
\be \label{aCFL}
a_1=\frac{1}{3} \quad , \qquad a_2=\frac{2}{3} \,\, .
\ee

In our treatment we have so far neglected the color-sextet,
flavor-sextet gap which is induced by condensation in the 
color-antitriplet, flavor-antitriplet channel \cite{pisarskisym}. 
Such a color-flavor symmetric structure is generated 
in the anomalous propagator $\Xi^+$, even for the
completely antisymmetric order parameter of Eq.\ (\ref{MCFL}).
(This does not happen in the 2SC case, where the color-flavor structure
of $\Xi^+$ remains completely antisymmetric.)
Consequently, it also appears on the right-hand side of the gap equation.    
The reason why it disappeared in our calculation is that we projected 
exclusively onto the antisymmetric color-flavor channel when we multiplied 
both sides of Eq.\ (\ref{gapeq21}) with 
${\cal M}^\dagger_{\bf k}\,\Lambda_{\bf k}^e$ and traced over color, 
flavor, and Dirac space.
To be consistent, one should have started with an order parameter
which includes both the symmetric and the antisymmetric color-flavor
structures. In weak coupling, however, the symmetric gap is suppressed by
an extra power of the strong coupling constant $g$ \cite{schaefer2}.
This fact by itself is not sufficient to neglect the symmetric gap in
the weak-coupling solution of the gap equation because this could still 
lead to a subleading correction which modifies the prefactor of the 
(antisymmetric) gap. A more detailed investigation
of this problem, however, is outside the scope of this thesis.

\subsubsection{The spin-one phases}

In the following, we discuss four different phases in spin-one (= one-flavor)
color superconductors, namely the polar, planar, A, and color-spin-locked
(CSL) phases. The former three are termed according to their analogues
in superfluid $^3$He (remember that in spin-one color 
superconductors as well as in superfluid $^3$He the order parameter is
a complex $3\times 3$ matrix). The latter corresponds to the B phase
in $^3$He. While in the B phase angular momentum is locked with 
(nonrelativistic) spin (cf.\ Sec.\ \ref{he3}), the CSL phase 
locks color with (total) spin in a similar way (and with the same
order parameter $\D$). A detailed discussion of the symmetry 
breaking patterns in spin-one color superconductors is presented in 
Sec.\ \ref{grouptheory}, where it is also argued in which sense the
above mentioned four phases are the most important ones. Here, we simply
insert the different order parameters in order to determine the 
quasiparticle excitation spectrum and the gap parameter. 
 
The general form of the matrix ${\cal M}_{\bf k}$ in the case
of a spin-one color superconductor is
\be \label{Mk}
{\cal M}_{\bf k} = \sum_{i,j=1}^3 J_i\, \D_{ij}\left[\a\,\hat{k}_j + 
\b\,\g_\perp^j({\bf k})\right] \,\, .
\ee
Contrary to the spin-zero phases, the flavor structure
is trivial. Instead, the spin-triplet structure has to be taken into
account. This is done by the terms in the angular brackets, which are 
components of 3-vectors and therefore serve as a basis for the spin-triplet
representation $[{\bf 3}]_J$. The first
term, proportional to $\hat{k}_j$, describes pairing of quarks with the
same chirality, since it commutes with the chirality projector 
${\cal P}_{r,\ell}=(1\pm\g_5)/2$. The second one, proportional to
\be 
\g_\perp^j({\bf k})\equiv \g_j - \hat{k}_j\vg\cdot\uk \,\, ,\qquad j=1,2,3
\,\, ,
\ee
corresponds to pairing of quarks of opposite chirality, since commuting
this term with the chirality projector flips the sign of chirality. 
In the above ansatz for the gap matrix we allow for a general linear 
combination 
of these two terms, determined by the real coefficients $\a$ and $\b$ with 
\be \label{normalize}
\a^2 + \b^2 = 1 \,\, .
\ee
This normalization was chosen in Ref.\ \cite{schaefer} in order to 
introduce a single parameter $\vartheta$ and $\a\to\sin\vartheta$,
$\b\to\cos\vartheta$. Since our results are not essentially 
simplified by this redefinition, we will keep the coefficients $\a$ and $\b$.
In Refs.\ \cite{pisarski3,schmitt1} the special cases $(\a,\b)=(1,0)$ and 
$(\a,\b)=(0,1)$ were 
termed ``longitudinal'' and ``transverse'' gaps, respectively. We 
will also use these terms in the following. (In Ref.\ \cite{schaefer},
the LL and RR gaps correspond to the longitudinal and the LR and RL gaps
to the transverse gaps.) The reason why both cases can be studied 
separately is that a purely 
longitudinal gap matrix on the right-hand side of the gap equation 
does not induce a transverse gap on the left-hand side and
vice versa. More precisely, inserting the matrix ${\cal M}_{\bf k}$ from
Eq.\ (\ref{Mk}) with $\b=0$ into the anomalous propagator from Eq.\
(\ref{S212SC}), and the result into the right-hand side of the gap equation 
(\ref{gapeq21}), we realize that the Dirac structure still commutes with 
$\g_5$ and thus preserves chirality. The analogous argument holds
for the transverse gap, $\a=0$.
For the case of an equal admixture of longitudinal and transverse gaps, i.e., 
$\a=\b=1/\sqrt{2}$, let us use the term ``mixed'' gap. In the following,
we determine the eigenvalues of $L_{\bf k}$ for a general linear 
combination of longitudinal and transverse gaps, i.e., for arbitrary
coefficients $\a$, $\b$. But in the calculation
of the gap parameter, 
for simplicity, we focus on the three special cases of a longitudinal, a 
mixed, and a transverse gap (except for the polar phase, where a general
treatment is presented). 

Note that the ansatz for the matrix ${\cal M}_{\bf k}$ given in Eq.\ (\ref{Mk})
fulfills the condition (\ref{M}), due to the fact
that $\Lambda_{\bf k}^e$ commutes with 
$\gperp({\bf k})$. Had we used $\vg$ in Eq.\ (\ref{Mk}), like
in Ref.\ \cite{schaefer}, this condition would have been violated
and the general discussion presented above would not be applicable 
to the spin-one phases. However, both choices are equivalent as was shown
in Appendix C of Ref.\ \cite{schmitt1}.

\subsubsection{The polar phase} \label{polar}

In the polar phase, the order parameter is given by $\D_{ij}=\d_{i3}\d_{j3}$
\cite{vollhardt,schaefer,schmitt1}. Inserting this into Eq.\ (\ref{Mk})
yields 
\be \label{Mkpolar}
{\cal M}_{\bf k} = J_3\,\left[\a\,\hat{k}_3 + \b\,\g_\perp^3({\bf k})\right] \,\, .
\ee
From Eq.\ (\ref{Ldef}) we then conclude 
\be 
L_{\bf k} = J_3^2\left[\b^2+(\a^2-\b^2)\,\cos^2\theta\right] \,\, ,
\ee
where $\theta$ is the angle between the direction of the quark momentum
${\bf k}$ and the $z$-axis, $\cos\theta = \hat{k}_3$. Note that $L_{\bf k}$
is a $12\times 12$ matrix in color and Dirac space. However, color and
Dirac parts factorize and the Dirac part is trivial, i.e., proportional
to the unit matrix. Therefore, it is obvious that, for arbitrary 
coefficients $\a$, $\b$, four eigenvalues of the matrix $L_{\bf k}$
are zero, since the color part is given by $J_3^2={\rm diag}(1,1,0)$. 
Physically, this means that, as in the 2SC phase, there are ungapped
excitation branches in the polar phase. The eigenvalues are
\be
\l_1 = \b^2 + (\a^2-\b^2)\,\cos^2\theta \qquad (\mbox{8-fold})\quad, 
\qquad \l_2 = 0 \qquad (\mbox{4-fold}) \,\, .
\ee 
Consequently, in general, the first eigenvalue depends on $\hat{k}_3$;
nevertheless, the projectors do not. They are the same as in the 2SC phase,
\be \label{ppolar}
{\cal P}_{\bf k}^1=J_3^2 \quad , \qquad {\cal P}_{\bf k}^2=1-J_3^2
\,\, .
\ee
Therefore, we immediately conclude 
${\cal T}_{00,t}^{ee',2}({\bf k},{\bf q})=0$, and thus, on account of
Eq.\ (\ref{a1a2}),
\be \label{apolar}
a_1=1 \quad , \qquad a_2=0 \,\, .
\ee
Because of their length, we do not show the explicit results for the traces 
${\cal T}_{00,t}^{ee',1}({\bf k},{\bf q})$ and the coefficients
$\eta_{2m}^{\ell,t}(ee',k,q)$ in the general case (= keeping $\a$ and $\b$). 
We find 
\be \label{dpolar}
d = \frac{3}{2}\frac{(3\b^2-4\a^2)\,\cos^2\theta -3\,\b^2}
{(\b^2-\a^2)\,\cos^2\theta - \b^2} \,\, , 
\ee
The function $d=d(\uk)$ is constant with respect to $\uk$ in the cases of a 
longitudinal and a transverse gap, where it assumes the values $d=6$ 
and $d=9/2$, respectively. In all other cases, the factor 
$\exp(-d)$ causes an anisotropy of the function $\phi_0$. 

In order to elaborate on the anisotropies of the polar phase,
let us discuss the special case of a mixed gap in somewhat more detail.
In this case, i.e., for $\a=\b=1/\sqrt{2}$, we find
\be \label{eigenpolar}
\lambda_1=\frac{1}{2} \quad (\mbox{8-fold}) \quad , \qquad \lambda_2=0 \quad 
(\mbox{4-fold}) \,\, .
\ee
Note that these eigenvalues differ from those of Ref.\ \cite{schmitt1} by
a factor 2. The reason for this is the choice of the coefficients 
$\a$ and $\b$.
While in Ref.\ \cite{schmitt1}, $\a=\b=1$, here we choose $\a=\b=1/\sqrt{2}$,
in order to fulfill the normalization given in Eq.\ (\ref{normalize}).
Therefore, let us check which are the implications of a rescaling with
a constant factor $c$, i.e.,
\be \label{factorc}
{\cal M}_{\bf k} \to c\,{\cal M_{\bf k}} \,\, .
\ee
We find
\be \label{rescaling}
\l_r \to c^2 \l_r \,\, , \qquad d \to d  \,\, , \qquad
\phi_0 \to \frac{1}{c}\phi_0 \,\, . 
\ee
Consequently, the physically relevant quantities are not influenced by the
rescaling,
\be
\sqrt{\l_r}\,\phi_0 \to \sqrt{\l_r}\,\phi_0 \,\, , \qquad 
 T_c \to T_c  \,\, .
\ee
(Remember that in the quasiparticle excitation energies $\phi_0$ is 
multiplied by $\sqrt{\l_r}$.)

Since the projectors do not depend on $\a$ and $\b$, Eq.\ (\ref{ppolar})
also holds for the mixed gap. This yields
\begin{subequations} \label{Tpolar}
\bea
{\cal T}_{00}^{ee',1}({\bf k},{\bf q})&=&
\frac{1}{3}\,\left\{ 
\left( 1 + ee'\,\uk\cdot\uq \right)  \, 
\left[ 1 + (1+ee') \, \hat{k}_3 \, \hat{q}_3 \right]
 - (e\hat{k}_3 + e'\hat{q}_3)^2 \right\} \,\, ,\\
{\cal T}_t^{ee',1}({\bf k},{\bf q})&=&
\frac{1}{3}\, \Big(
2\, \hat{k}_3\, \hat{q}_3 \, \left(1-e e' \uk \cdot \uq\right)
+ \left[ 1-\frac{(ek-e'q)^2}{p^2}\right]  \nonumber \\
&   &  \times 
\left\{ \left( 1+ee'\,\uk\cdot\uq \right) \left[ 1 + (1+ee')\,
\hat{k}_3 \, \hat{q}_3\right]-(e\hat{k}_3+e'\hat{q}_3)^2
\right\} \Big) \,\, .
\eea
\end{subequations}
There is a marked difference between the expressions 
(\ref{Tpolar}) and the 
corresponding ones for the previously discussed spin-zero cases. 
In contrast to these cases, there are two independent, fixed
spatial directions, that chosen by the order parameter and that of the
vector ${\bf k}$. Since we 
already aligned the order parameter with the $z$-direction, we are
no longer free to choose ${\bf k}=(0,0,k)$ for the 
$d^3 {\bf q}$-integration. Without loss of generality, however,
we may assume ${\bf k}$ to lie in the $xz$-plane, i.e.,
${\bf k}=k\,(\sin\theta,0,\cos\theta)$.
Then, we rotate the coordinate frame for the 
$d^3{\bf q}$-integration by the angle $\theta$ around the $y$-axis,
such that the rotated $z$-direction aligns with ${\bf k}$.   
The quantities $\uk\cdot\uq$, $\hat{q}_3$, and $\hat{k}_3$
appearing in Eqs.\ (\ref{Tpolar}) are expressed in terms of the new 
spherical coordinates $(q,\theta',\varphi')$ and the rotation angle 
$\theta$ as follows:
\be 
\uk\cdot\uq=\cos\theta' \quad, \qquad \hat{q}_3=\cos\theta' \, \cos\theta 
-\sin\theta' \, \sin\theta \, \cos\varphi' \quad, \qquad 
\hat{k}_3=\cos\theta \,\, .
\ee 
In the new coordinates the angle between ${\bf k}$ and ${\bf q}$
is identical with $\theta'$, and thus
$p$ becomes independent of $\varphi'$. Still, the 
$\varphi'$-integral is not trivial because of the potential 
$\varphi'$ dependence of the gap function. At this point we can
only proceed by assuming the gap function to be independent of $\varphi'$.
With this assumption, the $\varphi'$-integration becomes elementary,
and we are finally able to read off the coefficients $\eta_{2m}^{\ell,t}$,
which now depend on $\theta$,
\be \label{etaCSLpolar}
\eta_0^\ell=\frac{2}{3} \; , \quad 
\eta_2^\ell=-\frac{2+\cos^2\theta}{6} \; , \quad 
\eta_4^\ell=\frac{1+\cos^2\theta}{24} \; , \quad 
\eta_0^t=\frac{2}{3} \; , \quad 
\eta_2^t=-\frac{2-\cos^2\theta}{6} \; , \quad 
\eta_4^t=\frac{1-3\cos^2\theta}{24} \,\, .
\ee
{}From this and Eq.\ (\ref{d}) we compute 
\be \label{dpolarmixed}
d=\frac{3(3+\cos^2\theta)}{2} \,\, ,
\ee
which is in agreement with the general formula, Eq.\ (\ref{dpolar}),
setting $\a=\b$. 

Let us now comment on our assumption that the gap function is independent 
of $\varphi'$. As shown above, the value
of the gap function at the Fermi surface, $\phi_0$, 
is proportional to $\exp(-d)$. The angular dependence of $d$ then implies
a similar dependence of the gap itself. If ${\bf k}$ points in the same
direction as the order parameter, $\theta=0$, we find $d=6$, while for
${\bf k}$ being orthogonal to the order parameter, $\theta=\pi/2$,
one obtains $d=9/2$. These two cases have also been discussed in Refs.\ 
\cite{schaefer,ren}, with the same results for the constant $d$.
Our results surpass the previous ones in that they interpolate between these
two limiting cases. 

However, the angular dependence of $\phi_0$ causes the following problem,
already mentioned below Eq.\ (\ref{replacelambda}).
Since the gap function $\phi(\e_{k,1},{\bf k})$ is proportional 
to $\phi_0$, it also depends on $\theta$. Under
the $d^3{\bf q}$-integral on the right-hand side of the gap equation,
this dependence translates into a $\varphi'$ dependence of 
$\phi(\e_{q,1},{\bf q})$. Our previous assumption, which was necessary 
in order to perform the $\varphi'$-integral, precisely neglected this
dependence. Therefore, this approximation is in principle inconsistent.
Nevertheless, the agreement of our results with the ones of Refs.\
\cite{schaefer,ren} suggest that the 
$\varphi'$-dependence of the gap function could be a sub-subleading
effect.

\subsubsection{The planar phase} \label{planar}

In the planar phase, the order parameter is given by 
$\D_{ij}=\d_{ij} - \d_{i3}\d_{j3}$ \cite{vollhardt,schaefer,schmitt4}. 
This leads to 
\be \label{Mkplanar}
{\cal M}_{\bf k} = J_1[\a\,\hat{k}_1 + \b\,\g_\perp^1({\bf k})]+
J_2[\a\,\hat{k}_2 + \b\,\g_\perp^2({\bf k})] \,\, 
\ee
and
\bea \label{Lplanar}
L_{\bf k} &=& J_1^2\,[(\a^2-\b^2)\,\hat{k}^2_1 + \b^2] + 
J_2^2\,[(\a^2-\b^2)\,\hat{k}^2_2 + \b^2] + \{J_1,J_2\}(\a^2-\b^2)\hat{k}_1
\hat{k}_2  \\ \nonumber 
&& + [J_1,J_2] \, \b\left\{\a\left[\hat{k}_2\,\g_\perp^1({\bf k})-  
\hat{k}_1\,\g_\perp^2({\bf k})\right] - \b\left[\g_\perp^1({\bf k})\,\g_\perp^2({\bf k})-
\hat{k}_1\hat{k}_2\right]\right\} \,\, ,
\eea
where $\{-,-\}$ denotes the anticommutator and $[-,-]$ the commutator.
For the eigenvalues of $L_{\bf k}$ we find ($\theta$ defined as above)
\be \label{eigenplanar}
\l_1=\a^2\sin^2\theta+\b^2(1 + \cos^2\theta)\qquad 
(\mbox{8-fold})\quad, \qquad \l_2 = 0 \qquad (\mbox{4-fold}) \,\, . 
\ee
In Appendix \ref{AppA} we present a proof for this result. Then, with 
Eq.\ (\ref{proj}), the projectors obviously are
\be \label{pplanar}
{\cal P}_{\bf k}^1=\frac{1}{\l_1}L_{\bf k} \quad , \qquad 
{\cal P}_{\bf k}^2=1-\frac{1}{\l_1}L_{\bf k}
\,\, .
\ee
The fact that there is an ungapped excitation branch again leads to
${\cal T}_{\m\n}^{ee',2}({\bf k},{\bf q})=0$, and
\be \label{aplanar}
a_1=1 \quad , \qquad a_2=0 \,\, .
\ee
For simplicity, we do not consider the general case of arbitrary coefficients
$\a$, $\b$, but focus on the three cases of a longitudinal, a mixed, and a 
transverse gap. In all three cases, there is a gapped excitation branch
with degeneracy 8 and an ungapped one with degeneracy 4.
Also, we do not show the traces ${\cal T}_{00,t}^{ee',1}({\bf k},{\bf q})$
explicitly, but rather quote the final result for the constant $d$.   
Contrary to the polar phase, $d$ is not only constant for longitudinal and 
transverse gaps but also for $\a=\b$. We find 
\be
d = \left\{\begin{array}{cl} 6 & (\mbox{longitudinal})\,\, , \\ & \\ 
\frac{21}{4} & (\mbox{mixed})\,\, , \\ & \\
\frac{9}{2} & (\mbox{transverse})\,\, . \end{array} \right.
\ee

\subsubsection{The A phase} \label{A}

The A phase in superfluid $^3$He has been discussed in the 
introduction, cf.\ Sec.\ \ref{he3} and phase diagrams in Figs.\ 
\ref{Hephase1} and \ref{Hephase2}. In spin-one color superconductors,
we use the term A phase for the analogous phase, i.e., the phase
which is described by the same order parameter, 
$\D_{ij}=\d_{i3}(\d_{j1} + i\d_{j2})$ \cite{schaefer}, cf.\ also 
Eq.\ (\ref{heA}). Consequently,
\be \label{MkA}
{\cal M}_{\bf k} = J_3\,\left[\a\,(\hat{k}_1+i\hat{k}_2)+\b\,(\g_\perp^1({\bf k})
+i\g_\perp^2({\bf k}))\right] \,\, .
\ee
This leads to 
\be \label{LA}
L_{\bf k}=J_3^2\left(\a^2(1-\hat{k}^2_3)+\b^2(1+\hat{k}^2_3) 
2i\b\left\{\a\left[\hat{k}_2\,\g_\perp^1({\bf k})-  
\hat{k}_1\,\g_\perp^2({\bf k})\right] - \b\left[\g_\perp^1({\bf k})\,\g_\perp^2({\bf k})-
\hat{k}_1\hat{k}_2\right]\right\}\right) \,\, .
\ee
In Appendix \ref{AppA}, we show that the eigenvalues of $L_{\bf k}$ are
\bea \label{eigenA}
\l_{1/2}&=&\a^2\sin^2\theta+\b^2(1 + \cos^2\theta) \pm
2\,\b\,\sqrt{\a^2\sin^2\theta+\b^2(1 + \cos^2\theta) - \b^2} \qquad 
(\mbox{4-fold}\,\,{\rm each}) \,\, ,  \nonumber \\
\l_3 &=& 0  \qquad (\mbox{4-fold})\,\, .
\eea
For the corresponding projectors see Eqs.\ (\ref{pA}) in Appendix \ref{AppA}.
Contrary to all previously discussed phases, there are, in general, three 
different eigenvalues and thus three different excitation branches, one
of which is ungapped.

Again, for the determination of the gap parameter we focus on three special 
cases: In the case of a longitudinal gap, 
$\l_1 = \l_2 = \sin^2\theta$, i.e., there is one gapped branch 
with an 8-fold degeneracy and an ungapped branch with a 4-fold degeneracy.
In this case, we find 
\be
d=6 \qquad (\mbox{longitudinal}) \,\, .
\ee
For a mixed gap, 
there is also one gapped and one ungapped branch. But since in this case 
$\l_1 = 2$ and $\l_2 = 0$, the 
degeneracy of the gapped energies is 4, while the one of the ungapped energies
is 8. Here we have 
\be
d=\frac{21}{4} \qquad (\mbox{mixed}) \,\, ,
\ee
as in the planar phase. In the case of 
a transverse gap, there are two different gapped 
branches, $\l_{1/2} = (1\pm |\cos\theta|)^2$.
In principle, for this case one needs three constants $a_1$, $a_2$, $a_3$,
instead of two as in Eqs.\ (\ref{T}). One finds 
\be \label{aA}
a_1=a_2=\frac{1}{2} \,\, , \qquad  a_3=0 \,\, .
\ee
However, as in all other cases with one ungapped excitation branch, the
corresponding coefficient (here: $a_3$) vanishes. Therefore, the
additional third eigenvalue does not cause any complications for the
solution of the gap equation. The structure of the gap equation is similar
to the one in the CFL phase, where also two different nonzero eigenvalues
are present, and thus the above presented general formalism applies also 
for the transverse A phase. One finds that 
\be
d=\frac{9}{2} \qquad (\mbox{transverse}) \,\, .
\ee

\subsubsection{The CSL phase} \label{CSL}

The CSL phase is the analogue of the B phase in superfluid $^3$He. Therefore,
the order parameter is given by $\D_{ij}=\d_{ij}$, cf.\ Eq.\ (\ref{heB}).
We obtain
\be \label{MkCSL}
{\cal M}_{\bf k} = {\bf J}\cdot\left[\a\,\uk+\b\,\gperp({\bf k})\right] \,\, ,
\ee
Denoting color indices by $i$, $j$, we find
\be \label{LgeneralCSL}
(L_{\bf k})^{ij} = (\a^2+2\b^2)\,\d^{ij} - \left[\a\,\hat{k}^j+
\b\,\g_\perp^j({\bf k})\right]\left[\a\,\hat{k}^i-\b\,\g_\perp^i({\bf k})\right] \,\, .
\ee
Note that $i$, $j$ indeed indicate the components in color space, although
they appear with the vectors $\uk$ and $\gperp({\bf k})$. This is 
the consequence of color-spin locking, where each spatial component 
is assigned to a direction in color space, 
$(x,y,z)\to ({\rm red},{\rm green},{\rm blue})$. For more details about this
interesting symmetry breaking pattern see Sec.\ \ref{grouptheory}.
The eigenvalues of $L_{\bf k}$ are given by \cite{schaefer}
\be
\l_{1/2} = \frac{1}{2}\a^2+2\b^2\pm\frac{1}{2}\a\sqrt{\a^2+8\b^2}\qquad 
(\mbox{4-fold}\,\,{\rm each}) \quad, \qquad
\l_3 = \a^2  \qquad (\mbox{4-fold})\,\, .
\ee 
Here, in general three different nonzero gaps are possible. Note that
for all $\a$, $\b$, the eigenvalues are independent of $\uk$. 
Let us elaborate on the cases of a longitudinal, a mixed, and a transverse
gap. In these cases, at most two of the eigenvalues $\l_1$, $\l_2$, $\l_3$ 
assume a nonzero value.

In the CSL phase with longitudinal gaps only, the matrix 
${\cal M}_{\bf k}$ can be read off from the general case in 
Eq.\ (\ref{LgeneralCSL}),
\be \label{LCSLlong}
(L_{\bf k})^{ij}=\d^{ij}-\hat{k}^i\,\hat{k}^j 
\,\, .
\ee
This matrix is a projector onto the subspace orthogonal to $\hat{\bf k}$.
However, due to color-spin locking, the indices $i,j$ run over  
fundamental colors and not over spatial dimensions, and thus, amusingly, 
this projection actually occurs in color space. 
Since $L_{\bf k}$ is a projector, the eigenvalues are 
(also obvious from the general ones given above)
\be \label{EVlong}
\lambda_1=1 \quad (\mbox{8-fold}) \quad , \qquad \lambda_2=0 \quad 
(\mbox{4-fold}) \,\, . 
\ee
The projectors ${\cal P}_{\bf k}^{1,2}$ follow from Eq.\ (\ref{proj}),
\be \label{pCSLlong}
{\cal P}_{\bf k}^1=L_{\bf k} \quad , \qquad {\cal P}_{\bf k}^2=1-L_{\bf k}
\,\, .
\ee
The peculiar feature of Eq.\ (\ref{pCSLlong}) is that the projector 
${\cal P}_{\bf k}^1$ belongs to the eigenvalue corresponding to 
quasiparticle excitations with a longitudinal gap, but it actually 
projects onto the subspace orthogonal to $\hat{\bf k}$. This is,
however, not a contradiction, since the projection occurs in color
space, while the gap is longitudinal (parallel to $\hat{\bf k}$)
in real space.

For $s=2$, the quantities ${\cal T}_{00,t}^{ee',s}({\bf k},{\bf q})$ vanish
because ${\bf J}\cdot\uk \, {\cal P}_{\bf k}^2 =0$.
For $s=1$, we obtain
\begin{subequations} \label{TCSLlong}
\bea
{\cal T}_{00}^{ee',1}({\bf k},{\bf q})&=&\frac{1}{3}\,\uk\cdot\uq \, \left(1+
ee'\,\uk\cdot\uq\right) \,\, ,\\
{\cal T}_t^{ee',1}({\bf k},{\bf q})
&=&\frac{1}{3}\,\uk\cdot\uq\, \left[3-ee'\,\uk\cdot\uq-\frac{(ek-e'q)^2}{p^2}\,
\left(1+ee'\,\uk\cdot\uq\right)\right] \,\, ,
\eea
\end{subequations}
which only differ by an overall factor $\uk\cdot\uq$ from those of
Eq.\ (\ref{T2SC2}). While the constants $a_r$ are the same as in the 2SC 
case, see Eq.\ (\ref{a2SC}), this factor substantially 
changes the coefficients $\eta_{2m}^{\ell,t}$,
\be \label{etaCSLlong}
\eta_0^\ell=\frac{2}{3} \quad , \qquad 
\eta_2^\ell=-\frac{1}{2} \quad , \qquad 
\eta_4^\ell=\frac{1}{12} \quad , \qquad 
\eta_0^t=\frac{2}{3} \quad , \qquad 
\eta_2^t=-\frac{1}{6} \quad , \qquad 
\eta_4^t=-\frac{1}{12} \,\, .
\ee
This leads to $d=6$.

In the mixed CSL phase, we find from Eq.\ (\ref{LgeneralCSL}) with
$\a=\b=1/\sqrt{2}$ (and denoting Dirac indices with $a,b,c$)
\be \label{LCSL}
(L_{\bf k})_{ab}^{ij}=\delta^{ij}\, \delta_{ab} +
\left[\hat{k}^i\, \d_{ac}+\gamma_{\perp ac}^i({\bf k})\right]
\left[\hat{k}^j\, \d_{cb}-\gamma_{\perp cb}^j({\bf k})\right] \,\, .
\ee
The eigenvalues are
\be \label{EVCSL}
\lambda_1=2 \quad (\mbox{4-fold}) \quad , \qquad \lambda_2=\frac{1}{2} \quad 
(\mbox{8-fold}) \,\, . 
\ee
Again, the eigenvalues differ by a factor of two from those of
Ref.\ \cite{schmitt1}, cf.\ remarks below Eq.\ (\ref{eigenpolar}).

The projectors follow from Eq.\ (\ref{proj}),
\begin{subequations}
\bea
({\cal P}^1_{\bf k})_{ab}^{ij}&=&\frac{1}{3} \, 
\left[\hat{k}^i\, \d_{ac}+\gamma_{\perp ac}^i({\bf k})\right]
\left[\hat{k}^j\, \d_{cb}-\gamma_{\perp cb}^j({\bf k})\right] 
\,\, ,\\
({\cal P}^2_{\bf k})_{ab}^{ij}&=&\delta^{ij}\, \delta_{ab} -
\frac{1}{3} \,
\left[\hat{k}^i\, \d_{ac}+\gamma_{\perp ac}^i({\bf k})\right]
\left[\hat{k}^j\, \d_{cb}-\gamma_{\perp cb}^j({\bf k})\right] \,\, .
\eea
\end{subequations}
Inserting these projectors into Eq.\ (\ref{T2SC}) we obtain
\begin{subequations} \label{TCSL}
\bea
\frac{1}{2}\,{\cal T}_{00}^{ee',1}({\bf k},{\bf q})&=&
{\cal T}_{00}^{ee',2}({\bf k},{\bf q})=
\frac{1}{27} \, \left(1+ee'\,\uk\cdot\uq\right)
\left[1+(1+ee')\,\uk\cdot\uq\right] \,\, ,\\
\frac{1}{2}\,{\cal T}_t^{ee',1}({\bf k},{\bf q})&=&
{\cal T}_t^{ee',2}({\bf k},{\bf q})=
\frac{1}{27} \, \Big\{  2 \, \uk \cdot \uq \left( 1 - e e'
\, \uk \cdot \uq \right) + \left[ 1- \frac{(ek-e'q)^2}{p^2} \right] \nonumber\\
&& \hspace*{2.5cm}  
\times\left(1+ee'\,\uk\cdot\uq\right)\left[1+(1+ee')\,
\uk\cdot\uq\right]\Big\} \,\, .
\eea
\end{subequations}
Comparing this to Eq.\ (\ref{TCFL}), the prefactor 1/2 now accompanies
${\cal T}_{00,t}^{ee',1}$ instead of ${\cal T}_{00,t}^{ee',2}$.
Consequently, the constants $a_1$ and $a_2$ exchange their
roles compared to the CFL case, Eq.\ (\ref{aCFL}),
\be  \label{aCSL}
a_1=\frac{2}{3} \quad , \qquad a_2=\frac{1}{3}  
\ee
and, to subleading order,
\be \label{etaCSL}
\eta_0^\ell=\frac{2}{3} \quad , \qquad 
\eta_2^\ell=-\frac{7}{18} \quad , \qquad 
\eta_4^\ell=\frac{1}{18} \quad , \qquad 
\eta_0^t=\frac{2}{3} \quad , \qquad 
\eta_2^t=-\frac{5}{18} \quad , \qquad 
\eta_4^t=0 \,\, .
\ee
According to Eq.\ (\ref{d}), this yields $d=5$. 

As in the CFL case, another condensate 
with a symmetric color structure is
induced. This condensate belongs to the color-sextet representation and, 
for $N_f=1$, necessarily carries spin zero. To identify this induced 
condensate, one has to explicitly 
analyze the color structure of $\Xi^+$. By analogy to the CFL case, we 
expect this condensate to be suppressed by a power of $g$ compared
to the primary spin-one, color-antitriplet condensate. 
Its contribution to the gap equation could be of sub-subleading order,
if there is a cancellation of the leading terms involving the
spin-zero gap in the gap equation for the spin-one gap. A more
detailed investigation, however, is beyond the scope of this thesis.

For transverse gaps in the CSL phase we obtain
\be \label{LCSLtrans}
(L_{\bf k})_{ab}^{ij}=2\,\hat{k}^i\,\hat{k}^j\, \d_{ab}
-\g_{\perp ac}^i({\bf k}) \, \g_{\perp cb}^j ({\bf k})
\,\, .
\ee
The eigenvalues of this matrix are 
\be \label{EVtrans}
\lambda_1=2 \quad (\mbox{8-fold}) \quad , \qquad \lambda_2=0 \quad 
(\mbox{4-fold}) \,\, . 
\ee
The projectors ${\cal P}_{\bf k}^{1,2}$ are given by
\be \label{pCSLtrans}
{\cal P}_{\bf k}^1=\frac{1}{2}\,L_{\bf k} \quad , \qquad 
{\cal P}_{\bf k}^2=1-\frac{1}{2}\,L_{\bf k}
\,\, .
\ee
Although ${\bf J}\cdot\gperp({\bf k})\, {\cal P}_{\bf k}^s \neq 0$ for both 
$s=1$ and $s=2$, the final result for
${\cal T}_{00,t}^{ee',2}({\bf k},{\bf q})$ is nevertheless zero.
To see this, however, one has to explicitly perform the trace 
in Eq.\ (\ref{T2SC}). For $s=1$, we obtain
\begin{subequations} \label{TCSLtrans}
\bea
{\cal T}_{00}^{ee',1}({\bf k},{\bf q})&=&\frac{1}{6} \, \left(1+
ee'\,\uk\cdot\uq\right)^2 \,\, ,\\
{\cal T}_t^{ee',1}({\bf k},{\bf q})
&=&\frac{1}{6}\, \left(1+ee'\,\uk\cdot\uq\right)^2 \, 
\left[1-\frac{(ek-e'q)^2}{p^2}\right] \,\, .
\eea
\end{subequations}
The constants $a_r$ are the same as in the 2SC and longitudinal CSL phases, 
see Eq.\ (\ref{a2SC}). 
The coefficients $\eta_{2m}^{\ell,t}$ are
\be \label{etaCSLtrans}
\eta_0^\ell=\frac{2}{3} \quad , \qquad 
\eta_2^\ell=-\frac{1}{3} \quad , \qquad 
\eta_4^\ell=\frac{1}{24} \quad , \qquad 
\eta_0^t=\frac{2}{3} \quad , \qquad 
\eta_2^t=-\frac{1}{3} \quad , \qquad 
\eta_4^t=\frac{1}{24} \,\, .
\ee
This gives $d=9/2$.

\subsubsection{Summary of the results}

\begin{figure}[ht] 
\begin{center}
\includegraphics[width=13cm]{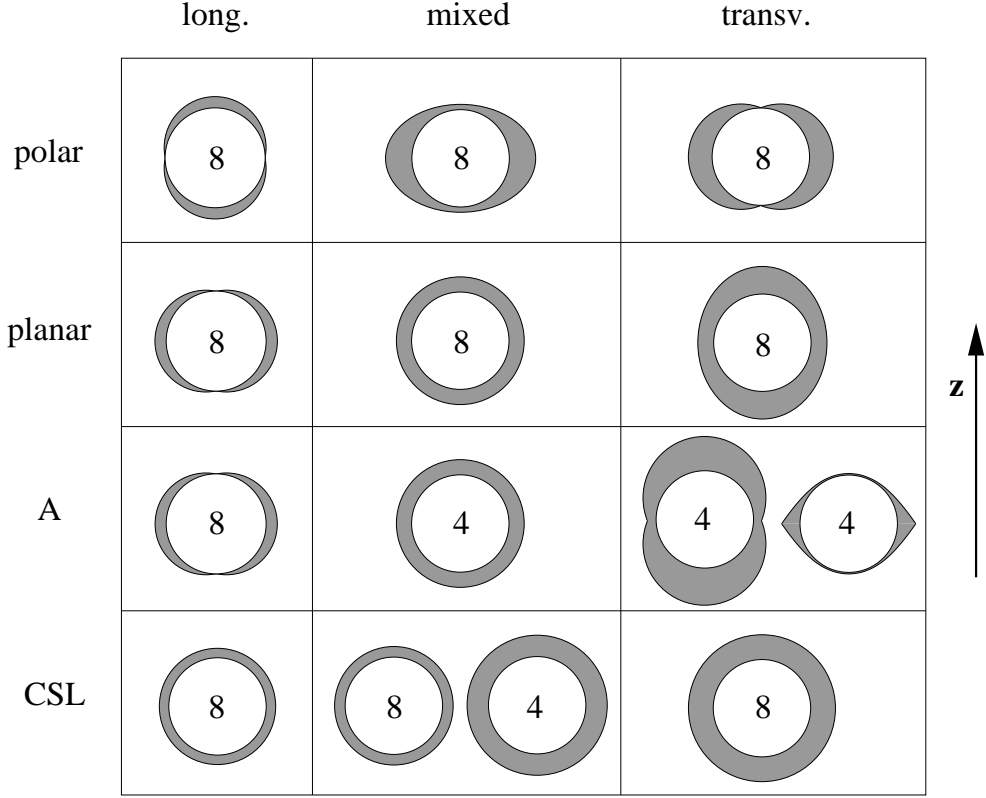}
\vspace{0.5cm}
\caption[Gap functions in a spin-one color superconductor]
{Schematic representation of the functions 
$\sqrt{\l_r}\phi_0$ occurring in the quasiparticle excitation energies,
Eq.\ (\ref{excite2}),   
for longitudinal, mixed, and transverse gaps in four different spin-one
color-superconducting phases. All functions are symmetric with respect to 
rotations around the $z$-axis. The numbers correspond to the degeneracies
of the respective excitation branch. In each case there are 12 excitation
branches, from which we only show the gapped ones. 
}
\label{gapfigure}
\end{center}
\end{figure}

We summarize our results in Table \ref{tablegaps}, Fig.\ \ref{gapfigure},
and Table \ref{tableTc}.

\begin{table} 
\begin{center}
\begin{tabular}[t]{|c||r|c|r|}
\hline 
$\sqrt{\l_r}\phi_0/\phi_0^{\rm 2SC}$ & longitudinal & mixed & transverse 
$\qquad\,\,\,\;\;\;$   \\ \hline\hline
polar & $3^{1/2}|\cos\theta|\, e^{-6}$ & $e^{-3(3+\cos^2\theta)/2}$ & 
$\left(\frac{3}{2}\right)^{1/2}|\sin\theta|\,e^{-9/2}$ \\ \hline
planar & $\left(\frac{3}{2}\right)^{1/2}|\sin\theta|\, e^{-6}$ & $e^{-21/4}$ & 
$\left(\frac{3}{4}\right)^{1/2}\sqrt{1+\cos^2\theta}\,e^{-9/2}$ \\
\hline
A & $\left(\frac{3}{2}\right)^{1/2}|\sin\theta| \, e^{-6}$ & $e^{-21/4}$ & 
$\left(\frac{3}{\sqrt{7}}\right)^{1/2}(1\pm|\cos\theta|)\,e^{-9/2}$ \\
\hline
CSL & $e^{-6}$ & $2^{(-1\pm 3)/6}\, e^{-5}$ & $e^{-9/2}$ \\ \hline

\end{tabular}
\caption[Gap functions]{The ratio $\sqrt{\l_r}\phi_0/\phi_0^{2SC}$ 
for four different 
spin-one color superconductors and longitudinal, mixed, and transverse gaps.
The angle $\theta$ is the angle between the quark momentum and the $z$-axis. 
(Fig.\ \ref{gapfigure} shows the same results graphically.) In the cases
of the transverse A phase and the mixed CSL phase, there are two gapped
excitation branches, $\l_1\neq 0$, $\l_2\neq 0$.
}
\label{tablegaps} 
\end{center}
\end{table}

In Table \ref{tablegaps}, we show the ratio $\sqrt{\l_r}\phi_0/\phi_0^{2SC}$
for all spin-one phases that were discussed in detail. There are three 
essential contributions to that ratio. 
First, through the eigenvalue $\l_r$
an angular dependence enters. The eigenvalue is included into the table, 
since it also multiplies the gap in the quasiparticle energies, 
Eq.\ (\ref{excite2}). Only for the CSL phase, the eigenvalues are 
constants in all three cases (longitudinal,
mixed, and transverse gap) . In all other 
phases, the eigenvalues for the longitudinal and transverse gap depend
on $\uk$. However, for an
equal admixture of longitudinal and transverse gaps, $\a=\b$ in 
Eq.\ (\ref{Mk}), no angular dependence of the eigenvalues is present in any
phase. 

The second contribution is the
factor $\exp(-d)$, typical for all spin-one gaps (for spin-zero condensates,
$d=0$). From the table it is obvious that all longitudinal gaps carry 
the same suppression factor $e^{-6}\simeq 2.5\cdot 10^{-3}$. In 
Appendix \ref{applonggap}, we prove that this is indeed a universal result 
(=\,independent of the order parameter $\Delta$). 
Also for the transverse gaps a common factor $e^{-9/2}\simeq 1.1\cdot 10^{-2}$
is found. Consequently, the magnitude of the gap is larger in the case 
of the transverse gaps. In the case of the mixed phases, the factor 
$e^{-d}$ differs from phase to phase. The polar phase is the only one 
where this factor is angular dependent. In all phases, the mixed gap
is larger than the longitudinal, but smaller than the transverse gap.
(Note that $e^{-21/4}\simeq 5.2\cdot 10^{-3}$ and 
$e^{-5}\simeq 6.7\cdot 10^{-3}$.) 
In conclusion,
the factor $e^{-d}$, arising in all spin-one phases, reduces the
magnitude of the gap compared to the spin-zero phases by two to three
orders of magnitude. Assuming that the gap in the 2SC phase is of the order of
\mbox{10 - 100 MeV}, the gap in spin-one color superconductors is of the 
order of 20 - 400 keV.

The third factor listed in Table \ref{tablegaps} arises from a potential
two-gap structure. It is absent in all cases with only
one nonzero eigenvalue $\l_1$. In these cases, there is one gapped and one 
ungapped excitation branch. The factor is nontrivial (and of order
one) in the cases with two 
different nonzero eigenvalues $\l_1$, $\l_2$. In three of the cases 
we consider, this situation occurs: The first case is the CFL phase (not
included into Table \ref{tablegaps}). In this phase, an additional factor
$(\l_1^{a_1}\l_2^{a_2})^{-1/2} = 2^{-1/3}$ arises compared to the 2SC gap. 
Including the factor $\sqrt{\l_r}$, as in Table \ref{tablegaps},
we obtain the two factors $\sqrt{\l_{1,2}}\,2^{-1/3}= 2^{(1\pm3)/6}$.
The second case
is the mixed CSL phase, which, in this respect, is very similar to the
CFL phase. While the eigenvalues are identical in both cases (up to an
irrelevant factor, cf.\ discussion below Eq.\ (\ref{eigenpolar})), 
the difference is the reversed order of the constants 
$a_1$, $a_2$ (cf.\ Eqs.\ (\ref{aCFL}) and (\ref{aCSL})). Therefore, as 
can be seen in the last row of Table \ref{tablegaps}, the factor
arising from the two-gap structure is slightly different than in the
CFL phase. The third 
phase with two nonzero eigenvalues of the matrix $L_{\bf k}$ is the
transverse A phase. In this case, 
$(\langle\l_1\rangle^{a_1}\langle\l_2\rangle^{a_2})^{-1/2} = 
(3/\sqrt{7})^{1/2}$, 
where we used Eqs.\ (\ref{eigenA}) and (\ref{aA}).

The results of Table \ref{tablegaps} are shown graphically in 
Fig.\ \ref{gapfigure}. In this figure, we show the angular dependence
of the gap schematically. We
also indicate the effect of the suppression factor, which leads to 
larger gaps in the transverse cases than in the longitudinal ones. 

In the first line of Fig.\ \ref{gapfigure}
we show the angular structure of the gap in the polar phase.
In the longitudinal case (first column), the gap
vanishes at the equator of the Fermi surface since 
$\sqrt{\l_1}=|\cos\theta|$, i.e., there is a nodal 
line. In the mixed phase,
$\l_1$ does not depend on the angle $\theta$. Nevertheless, there is 
an anisotropy due to the exponential factor $\exp(-d)$, cf.\ Eqs.\ 
(\ref{phi0}) and (\ref{dpolarmixed}). 
This anisotropy is also indicated schematically in Fig.\ \ref{gapfigure}. 
In the case of a transverse gap, $\b=1$, the gap function has nodal points 
at the north and south pole of the Fermi surface since 
$\sqrt{\l_1}=|\sin\theta|$.   

In the planar phase, the longitudinal gap 
has point nodes, since $\sqrt{\l_1}=|\sin\theta|$, while in the 
transverse case the gap is nonvanishing for all angles, 
$\sqrt{\l_1}=\sqrt{(1+\cos^2\theta)}$. In the mixed case, the gap is 
isotropic, $\l_1 = 1$.  

The results for the A phase are illustrated in the third row of 
Fig.\ \ref{gapfigure}.
In the longitudinal case, the gap has two point nodes, identical to 
the planar phase. It is not surprising that this structure can also 
be found in the A phase of superfluid $^3$He (cf.\ Fig.\ \ref{figgaps}
in the introduction). The reason for this is the similarity of the
matrix ${\cal M}_{\bf k}$, Eq.\ (\ref{Mk}), in the case $\b=0$ with the 
corresponding ansatz for the gap matrix in superfluid $^3$He \cite{vollhardt}.
Therefore, all longitudinal gaps reproduce the nonrelativistic limit.
The 3-vector $\gperp({\bf k})$, accounting for pairing of quarks with 
opposite chirality, is not present in nonrelativistic theories,
wherefore the results presented in the second and third column of 
Fig.\ \ref{gapfigure} can be considered as relativistic extensions
of the first column. 

The mixed gap in the A phase is isotropic (note that this case is 
the only one where the ungapped excitation has degeneracy eight), 
whereas the transverse 
case exhibits two different gap structures, one having point nodes at
the north and south pole of the Fermi surface, one nonvanishing for all
directions of the quark momentum. Of course, these structures cannot be 
seen in the nonrelativistic theory of the A phase in $^3$He.

In the CSL phase, the gaps in all three cases are isotropic. Again, 
the nonrelativistic case reproduces the result for the B phase 
in $^3$He, shown in Fig.\ \ref{figgaps}.  

\begin{table} 
\begin{center}
\begin{tabular}[t]{|c||c|c|c|}
\hline 
$T_c/\phi_0^{2SC}$ & longitudinal & mixed & transverse  \\ \hline\hline
polar & $e^{-6}$ & $1.4\,e^{-21/4}$ & 
$e^{-9/2}$ \\ \hline
planar & $e^{-6}$ & $e^{-21/4}$ & $e^{-9/2}$ \\
\hline
A & $e^{-6}$ & $e^{-21/4}$ & $e^{-9/2}$ \\
\hline
CSL & $e^{-6}$ & $e^{-5}$ & $e^{-9/2}$ \\ \hline

\end{tabular}
\caption[Critical Temperatures]{The critical temperature $T_c$ 
in units of $\phi_0^{\rm 2SC}$
for four different 
spin-one color superconductors and longitudinal, mixed, and transverse gaps.
}
\label{tableTc} 
\end{center}
\end{table}

In Table \ref{tableTc}, we present the absolute value of the critical 
temperature in units of the 2SC gap, $\phi_0^{\rm 2SC}$.
Then, $T_c$ simply is given by $\exp(-d)$, cf.\
Eq.\ (\ref{Tcabs}). In the polar phase with a mixed gap, the critical 
temperature is given by the angular average of this term, cf.\ 
Eq.\ (\ref{Tcabspolar}). With 
\mbox{$\phi_0^{\rm 2SC}\sim$ 10 - 100 MeV}, the critical temperature in a 
spin-one color superconductor is therefore of the order of 10 - 400 keV.
Finally, we remark that the BCS relation (\ref{BCSrelation}) for the
critical temperature is violated
in the cases of the mixed CSL and the transverse A phases, which can 
be seen from Eq.\ (\ref{Tc}). This violation,
also present in the CFL phase, originates from the two-gap structure, i.e., 
two excitation branches with different, non-zero energy gaps.         %General gap equation, results (anisotropies)
   \section{Patterns of symmetry breaking}  \label{grouptheory}

\renewcommand{\labelenumii}{(\roman{enumii})}

In this section, we discuss possible symmetry breaking patterns in a spin-one
color superconductor. In other words, we present a systematic classification 
of theoretically possible superconducting phases. As explained in 
the introduction, a superconducting
phase is mathematically described by a nonvanishing order parameter $\D$ that 
spontaneously breaks the original symmetry group $G$ of the system. 
Remember that $\D$ is an element of a representation of $G$. In the case
of a spin-one color superconductor, this representation is given by the 
tensor product of the antisymmetric color antitriplet $[\bar{\bf 3}]_c^a$ 
and the symmetric spin triplet $[{\bf 3}]_J^s$, cf.\ Eq.\ \ref{colorchannels}
and Sec.\ \ref{oneflavor},
\be
\D \in  [\bar{\bf 3}]_c^a \otimes [{\bf 3}]_J^s \,\, .
\ee
Therefore, $\D$ is, as in the case of superfluid $^3$He, a complex $3\times 3$
matrix. There is no nontrivial contribution from the flavor 
structure since we consider systems with only one quark flavor.
In Sec.\ \ref{he3}, it has been discussed that the order parameter 
breaks down the group $G$ to a residual (proper) subgroup $H\subseteq G$.
This means that any transformation $g\in H$ leaves the order parameter 
invariant,
\be \label{invariance0}
g(\D) = \D \,\, .
\ee
In the following, we investigate this invariance condition in order to 
determine all possible order parameters $\D$ and the corresponding 
residual groups $H$. The method we use in this section is motivated 
by the analogous one for the case of superfluid $^3$He \cite{vollhardt}.
Note that in the previous section $\D$ already has been introduced as a part 
of the gap matrix, cf.\ Eqs.\ (\ref{gm2SC}) and (\ref{Mk}). The specific
structure of $\D$ has been used for four different spin-one 
color-superconducting phases, namely the polar, planar, A, and CSL
phases. Therefore, we expect to recover these four phases (among others) in 
the following discussion.

Let us first elaborate on the mathematical contents of 
Eq.\ (\ref{invariance0}).
From Sec.\ \ref{oneflavor} we know that the relevant symmetry group 
in the case of a spin-one color superconductor is 
\be
G=SU(3)_c\times SU(2)_J \times U(1)_{em} \times U(1)_B \,\, ,
\ee
where $SU(3)_c$
is the (local) color gauge group, $SU(2)_J$ the (global) spin
group, $U(1)_{em}$ the (local) electromagnetic gauge group, and $U(1)_B$
the (global) baryon number conservation group. Here, we put ``local'' and
``global'' in parentheses because these properties of the symmetries are
not essential for the following classification. They are of relevance
only for the interpretation of the results. Without loss of 
generality, we can restrict ourselves to the group
\be
G=G_1\times G_2\times G_3 \,\, ,
\ee
where 
\be
G_1 = SU(3)_c \,\, , \qquad G_2 = SU(2)_J \,\, ,\qquad G_3 = U(1) \,\, .
\ee
Since in the one-flavor case the transformations of both $U(1)_{em}$ 
and $U(1)_B$ correspond to a multiplication with a phase factor, we can 
reduce our following arguments to a single $U(1)$. The extension to both 
$U(1)$'s can be done straightforwardly at the end.   
After specifying the group $G$, we have to specify how $G$ acts on the order 
parameter in Eq.\ (\ref{invariance0}). To this end, we write an arbitrary 
group element $g\in G$ as (in this section, no confusion with the 
strong coupling constant $g$ is possible)
\be
g=(g_1,g_2,g_3) \,\, ,
\ee
where
\be \label{g1g2g3}
g_1=\exp(-ia_mT_m^T) \,\, , \qquad g_2=\exp(ib_nJ_n) \,\, , \qquad 
g_3=\exp(2ic\bf{1}) \,\, , 
\ee
with real coefficients $a_m$ ($m=1,\ldots ,8$), $b_n$ ($n=1,2,3$), and $c$. 
The Gell-Mann matrices $T_m$ generate the group $SU(3)_c$, and we 
have taken into account that the color representation is 
an {\it anti}triplet.
The matrices $J_n$ are the same as in Eq.\ (\ref{Mk}), where they served
as the basis of the color antitriplet. Here, they are used as the 
generators of the spin group 
$SU(2)_J$. For the generator of $U(1)$ we choose $2\cdot{\bf 1}$, where
${\bf 1}$ is the $3\times 3$ unit matrix. The factor 2 accounts for the
diquark nature of the order parameter. 
The transformation $g$ acts on the matrix ${\cal M}_{\bf k}$, defined
in Eq.\ (\ref{Mk}), in the following way (abbreviating the 3-vector
$\kappa_j\equiv [\a\,\hat{k}_j + \b\,\g_\perp^j({\bf k})]$, such that 
${\cal M}_{\bf k} = J_i\D_{ij}\k_j$)
\bea
g({\cal M}_{\bf k}) &=& g_3 \,  (g_1J_i)\,\D_{ij}\, (g_2\kappa_j) \non
&=&g_3 \, g_1^{ik}\, J_k \,\D_{ij} \,g_2^{j\ell} \,\kappa_\ell 
\,\, .
\eea
Therefore, since $J_i\otimes \kappa_j$ is a basis of the representation 
$[\bar{\bf 3}]_c^a \otimes [{\bf 3}]_J^s$, the matrix $\D$ transforms as
\be
g(\D_{ij}) = g_3 \, g_1^{ki} \, \D_{k\ell} \, g_2^{\ell j} \,\, .
\ee
Then, using Eqs.\ (\ref{g1g2g3}), the infinitesimal transformations of 
$\D$ by $G$ are given by
\be
g(\D) \simeq \D  - a_mT_m\D + b_n\D J_n + 2c\,\D\,\, ,
\ee
where $T_m\D$ as well as $\D J_n$ are matrix products. 
The invariance condition for the order parameter
(\ref{invariance0}) is thus equivalent to
\be  \label{invariance}
- a_mT_m\D + b_n\D J_n + 2c\,\D = 0\,\, .
\ee
This matrix equation can be written as a system of nine equations for
the nine complex entries $\D_{11},\ldots,\D_{33}$ of the matrix $\D$. 
In principle,
one can find all possible symmetry breaking patterns and corresponding order
parameters by setting the determinant of the coefficient matrix to zero. Then,
each possibility to make the determinant vanish yields a set of 
conditions for the coefficients $a_m$, $b_n$, $c$, and it can be checked if 
these conditions correspond to a residual subgroub $H$. But since this is 
much too complicated, we proceed
via investigating possible subgroups explicitly. In the following, we 
focus only on the continuous subgroups of $G$. 

Let us start with subgroups 
$H$ that contain the smallest possible continuous group, $U(1)$, i.e., 
\be \label{H'}
H=U(1)\times H' \,\, ,
\ee
where $H'$ is a direct product of Lie groups.
The residual $U(1)$ must be generated by a $3\times 3$ matrix $U$ which is 
a linear 
combination of the generators of $G$, i.e., in general,
\be
U=a_mT_m + b_nJ_n + 2c{\bf 1} \,\, .
\ee
Let us restrict to linear combinations that involve one generator of each 
group $G_1$, $G_2$, $G_3$, for instance  
\be  \label{generator1}
U = a_8T_8 + b_3J_3 + 2c{\bf 1} \,\, .
\ee 
In order to find all possible order parameters, it is necessary to consider
at least one more combination, namely $U=a_2T_2 + b_3J_3 + 2c{\bf 1}$. We 
comment on these two choices of the residual generator below. 
With Eq.\ (\ref{generator1}), the invariance condition
\be
e^{iU}(\D)=\D
\ee
results in a system of nine equations, which can be discussed explicitly,
\begin{subequations} \label{setofequations}
\bea
\left(-\frac{a_8}{2\sqrt{3}} + 2c\right)\D_{11} + ib_3\D_{12} &=& 0\,\, , \\
\left(-\frac{a_8}{2\sqrt{3}} + 2c\right)\D_{12} - ib_3\D_{11} &=& 0\,\, , \\
\left(-\frac{a_8}{2\sqrt{3}} + 2c\right)\D_{13} &=& 0 \,\, , \\
\left(-\frac{a_8}{2\sqrt{3}} + 2c\right)\D_{21} + ib_3\D_{22} &=& 0\,\, , \\
\left(-\frac{a_8}{2\sqrt{3}} + 2c\right)\D_{22} - ib_3\D_{21} &=& 0\,\, , \\
\left(-\frac{a_8}{2\sqrt{3}} + 2c\right)\D_{23} &=& 0 \,\, , \\
\left(-\frac{a_8}{\sqrt{3}} + 2c\right)\D_{31} + ib_3\D_{32} &=& 0\,\, , \\
\left(-\frac{a_8}{\sqrt{3}} + 2c\right)\D_{32} - ib_3\D_{31} &=& 0\,\, , \\
\left(-\frac{a_8}{\sqrt{3}} + 2c\right)\D_{33} &=& 0 \,\, .
\eea
\end{subequations}
The corresponding coefficient 
matrix $A$ exhibits a block structure and the determinant thus factorizes
into four sub-determinants. Therefore, we have to consider the equation 
\be 
0 = {\rm det} A = {\rm det} A_1\,{\rm det} A_2\,{\rm det} A_3\,
{\rm det}A_4 \,\, ,
\ee  
where
\begin{subequations}
\bea
{\rm det}A_1 &=& \left[\left(-\frac{a_8}{2\sqrt{3}} + 
2c\right)^2-b_3^2\right]^2    \,\, , \label{det1}\\
{\rm det}A_2 &=& \left(-\frac{a_8}{2\sqrt{3}}+2c\right)^2 \,\, ,\label{det2}\\
{\rm det}A_3 &=& \left(\frac{a_8}{\sqrt{3}}+2c\right)^2 - b_3^2 \,\, ,
\label{det3}\\
{\rm det}A_4 &=& \frac{a_8}{\sqrt{3}}+2c \,\, .\label{det4}
\eea
\end{subequations}
Now, one can systematically list all possibilities that yield a zero 
determinant of the coefficient matrix and thus allow for a nonzero order 
parameter. For convenience, let us introduce a normalization for the order
parameter,
\be \label{normalize2}
\Tr(\D\D^\dag)=1 \,\, .
\ee

\begin{enumerate}
\item ${\rm det}A_1 = 0$.

Here we distinguish between the cases (i) where the two terms in the angular 
brackets of Eq.\ (\ref{det1}) cancel each other and (ii) where they 
separately vanish.

\begin{enumerate}

\item $a_8$, $c$ arbitrary, $b_3 = -a_8/(2\sqrt{3}) + 2c$.

Inserting these conditions for the coefficients into Eqs.\ 
(\ref{setofequations})), one obtains for the order parameter matrix
\be
\D= \frac{1}{N}\left(\begin{array}{ccc} \D_1&i\D_1&0\\ \D_2&i\D_2&0\\0&0&0 
\end{array}\right) \,\, , \label{order1i} 
\ee
where the factor $1/N$ with  
$N=(2|\D_1|^2+2|\D_2|^2)^{1/2}$ accounts for the normalization 
(\ref{normalize2}). In this case, the order parameter contains 
two independent parameters $\D_1$ and $\D_2$. 
%A slight subtlety of this case is that the single condition 
%$b_3 = -a_8/(2\sqrt{3}) + c$ allows for the choice of two coefficients to 
%be arbitrary. As far as these arbitrary coefficients are nonzero, the order
%parameter does not depend on this choice (i.e., $a_8$, $b_3$ arbitrary, and
%$c=a_8/(2\sqrt{3}) + b_3$ leads to the same $\D$). 
From this form of the order parameter, we can now determine the group $H'$
in Eq.\ (\ref{H'}). 
Inserting $\D$ into Eq.\ (\ref{invariance}) and using the fact that the 
parameters $\D_1$, $\D_2$ are independent of each other, one obtains 
the conditions
\be \label{conditions1i}
a_1 = \ldots = a_7 = b_1 = b_2 = 0 \,\, ,\quad 
\frac{1}{2\sqrt{3}}a_8 + b_3 -2c = 0 \,\,.
\ee
Consequently, 
\be \label{H1i}
H = U(1)\times U(1) \,\, ,
\ee 
since a vanishing coefficient in Eq.\ (\ref{conditions1i}) translates to 
a ``broken dimension''
of $G$. For instance, $a_1=0$ means that $T_1$ does not occur in the generators
of $H$, etc. The dimensions of the residual Lie group can be counted with
the help of the number of the conditions for the coefficients. 
Since ${\rm dim}\,G={\rm dim}\,G_1 + {\rm dim}\,G_2 + {\rm dim}\,G_3 
= 8+3+1=12$, 
and the number of conditions in Eqs.\ (\ref{conditions1i}) is 10, we
conclude ${\rm dim}\,H=2$, which is in agreement with Eq.\ (\ref{H1i}).
Or, in other words, there is an additional $U(1)$, i.e., $H'=U(1)$ in 
Eq.\ (\ref{H'}) because the equation relating
the three coefficients $a_8$, $b_3$, $c$ allows for two linear independent 
generators $U$ and $V$ which are linear combinations of the 
generators $T_8$, $J_3$, ${\bf 1}$. Note that $U$ and $V$ are not uniquely 
determined. One possible choice is
\be \label{gen1i}
U = T_8 - \frac{1}{2\sqrt{3}}J_3 \,\, , \qquad V = 2J_3 + {\bf 1} \,\, .
\ee

Different order parameters are obtained from two subcases: 

First, 
one can impose the additional relation $c=-a_8/(2\sqrt{3})$ between the
two coefficients that have been arbitrary above. Then,
$b_3=-a_8\sqrt{3}/2$. These two conditions yield
\be
\D= \frac{1}{N}\left(\begin{array}{ccc} \D_1&i\D_1&0\\ \D_2&i\D_2&0\\0&0&\D_3 
\end{array}\right) \,\, , \label{order1iadd} 
\ee
where $N=(2|\D_1|^2+2|\D_2|^2+|\D_3|^2)^{1/2}$. In this case, 
Eq.\ (\ref{invariance}) leads to 11 conditions for the coefficients 
$a_m$, $b_n$, $c$, which leaves a subgroup
\be
H=U(1) \,\, ,
\ee
generated by a linear combination of generators of all three original 
subgroups $G_1$, $G_2$, $G_3$, 
\be
U= T_8 - \frac{\sqrt{3}}{2}J_3 - \frac{1}{2\sqrt{3}}{\bf 1} \,\, .
\ee

Second, one can set one of the coefficients $a_8$, $c$ to zero. 
The condition $c=0$ does not yield a new case. But $a_8=0$, and 
consequently $b_3=2c$, has to be treated separately. 
In this case, Eqs.\ (\ref{setofequations}) yield
\be
\D= \frac{1}{N}\left(\begin{array}{ccc} \D_1&i\D_1&0\\ \D_2&i\D_2&0\\ \D_3&
i\D_3 &0 
\end{array}\right) \,\, , \label{order1ii} 
\ee
where $N=(2|\D_1|^2+2|\D_2|^2+2|\D_3|^2)^{1/2}$. The residual group is given 
by
\be
H = U(1) \,\, ,
\ee
generated by
\be
U = 2J_3 + {\bf 1} \,\, .
\ee 

\item $c=a_8/(4\sqrt{3})$, $b_3=0$. 

Here, one obtains
\be
\D= \frac{1}{N}\left(\begin{array}{ccc} \D_1&\D_2&\D_3\\ \D_4&\D_5&\D_6\\ 
0& 0 &0  
\end{array}\right) \,\, , \label{order1iv} 
\ee
where $N=(\sum_{i=1}^6|\D_i|^2)^{1/2}$. Again, the residual group is
one-dimensional,
\be
H=U(1) \,\,  ,
\ee
generated by
\be
U = T_8 + \frac{1}{4\sqrt{3}}{\bf 1} \,\, .
\ee

\end{enumerate}

\item ${\rm det}A_2 = 0$.

This determinant vanishes in the following cases:

\begin{enumerate}
\item $b_3$ arbitrary, $c=a_8/(4\sqrt{3})$.

With Eqs.\ (\ref{setofequations}), one obtains
\be
\D= \frac{1}{N}\left(\begin{array}{ccc} 0&0&\D_1\\ 0&0&\D_2\\ 0&0&0 
\end{array}\right) \,\, , \label{order2i} 
\ee
where $N=(|\D_1|^2+|\D_2|^2)^{1/2}$. Inserting $\D$ into 
Eq.\ (\ref{invariance}) yields
\be \label{conditions2i}
a_1 = \ldots = a_7 = b_1 = b_2 = 0 \,\, ,\quad 
c=\frac{1}{4\sqrt{3}}a_8  \,\,.
\ee
As for the order parameter (\ref{order1i}), the residual group
is given by  
\be \label{H2i}
H = U(1)\times U(1) \,\, .
\ee 
However, the corresponding generators differ from those in Eqs.\ (\ref{gen1i}),
\be
U = T_8 + \frac{1}{4\sqrt{3}} {\bf 1} \,\, ,\qquad V=J_3 \,\, .
\ee 

\item $b_3$ arbitrary, $a_8=c=0$.

In this case,
\be 
\D= \frac{1}{N}\left(\begin{array}{ccc} 0&0&\D_1\\ 0&0&\D_2\\ 0&0&\D_3 
\end{array}\right) \,\, , \label{order2ii} 
\ee
where $N=(|\D_1|^2+|\D_2|^2+|\D_3|^2)^{1/2}$. The residual group is
\be 
H=U(1)\,\, ,
\ee
which is a subgroup of the spin group $G_2=SU(2)_J$, since it is
generated by 
\be
U=J_3 \,\, .
\ee
\end{enumerate}

\item ${\rm det}A_3 = 0$.
 
\begin{enumerate}
\item $a_8$, $c$ arbitrary, $b_3=a_8/\sqrt{3} + 2c$.

In this case, we find with Eqs.\ (\ref{setofequations}),
\be
\D= \frac{1}{\sqrt{2}}\left(\begin{array}{ccc} 0&0&0\\ 0&0&0\\ 1&i&0 
\end{array}\right) \,\, . \label{orderA} 
\ee
This matrix differs from all previously discussed order parameters in that 
it is uniquely determined.
Here, as in all cases above, we omitted a possible phase factor which 
could multiply $\D$ without violating the normalization. In 
Eq.\ (\ref{orderA}) we recover the A phase, which already has been
introduced in Eq.\ (\ref{heA}) for superfluid $^3$He and in Eq.\ (\ref{MkA})
for a spin-one color superconductor.   
Inserting $\D$ into Eqs.\ (\ref{invariance}) yields the following relations,
\be \label{conditions3i}
a_4 = \ldots = a_7 = b_1 = b_2 = 0 \,\, ,\quad 
\frac{1}{\sqrt{3}}a_8-b_3+2c=0 \,\,.
\ee
Consequently, besides the relation between $a_8$, $b_3$, and $c$, 
there are only 6
additional conditions. Thus, the dimension of the residual group is 
$12-7=5$. We obtain
\be
H = SU(2)\times U(1)\times U(1) \,\, ,
\ee
where $SU(2)$ is generated by $2T_1$, $2T_2$, $2T_3$, and thus is a 
subgroup of the color gauge group $G_1=SU(3)_c$. The factor 2 accounts
for the validity of the $SU(2)$ commutation relations,
\be \label{SU2commutation}
[J_i,J_j]=i\,\e_{ijk} J_k \,\, , \qquad i,j,k \le 3 \,\, .
\ee  
For the generators of the two $U(1)$'s one can choose 
\be \label{gen3i}
U=T_8-\frac{1}{2\sqrt{3}}{\bf 1} \,\, , \qquad V=J_3+{\bf 1} \,\, .
\ee
As in case 1.(i), there is a subcase that produces an additional 
order parameter. Namely, if we require the condition $c=a_8/(4\sqrt{3})$, 
which yields $b_3=\sqrt{3}\,a_8/2$, we obtain
\be
\D= \frac{1}{N}\left(\begin{array}{ccc} 0&0&\D_2\\ 0&0&\D_3\\ \D_1&i\D_1&0 
\end{array}\right) \,\, , \label{order3i1} 
\ee
where $N=(2|\D_1|^2+|\D_2|^2+|\D_3|^2)^{1/2}$. From Eq.\ (\ref{setofequations})
we conclude that all other coefficients vanish. Consequently,
\be
H=U(1) \,\, ,
\ee
with the generator
\be
U=T_8 + \frac{\sqrt{3}}{2}J_3 + \frac{1}{4\sqrt{3}}{\bf 1} \,\, .
\ee

\item $b_3 = 0$, $c=-a_8/(2\sqrt{3})$.

With Eqs.\ (\ref{setofequations}) one obtains
\be \label{order3ii}
\D=\frac{1}{N}\left(\begin{array}{ccc} 0&0&0\\0&0&0\\ \D_1&\D_2&\D_3 
\end{array}\right) \,\, ,  
\ee
where $N$ is defined as in Eq.\ (\ref{order2ii}). From Eq.\ (\ref{invariance})
we conclude in this case 
\be
a_4 = \ldots = a_7= b_1=b_2=b_3=0 \,\, , \quad c=-\frac{a_8}{2\sqrt{3}} \,\, .
\ee
Hence, the residual group is
\be
H = SU(2)\times U(1) \,\, ,
\ee
generated by $2T_1$, $2T_2$, $2T_3$, and
\be
U = T_8 - \frac{1}{2\sqrt{3}}{\bf 1} \,\, .
\ee

\end{enumerate}

\item ${\rm det}A_4 = 0$.

There are two cases in which ${\rm det}A_4=0$:

\begin{enumerate}
\item $b_3$ arbitrary, $c=-a_8/(2\sqrt{3})$. 

These relations lead to 
\be
\D=\left(\begin{array}{ccc} 0&0&0\\0&0&0\\0&0&1 
\end{array}\right) \,\, , \label{orderpolar} 
\ee
As in the A phase, Eq.\ (\ref{orderA}), the order parameter is uniquely 
determined. It describes the polar phase, which has been studied in the
previous section. Inserting the order parameter of the polar phase 
into the invariance condition, Eq.\ (\ref{invariance}), yields
\be
a_4 = \ldots = a_7 = b_1 = b_2 = 0 \,\, ,\quad 
c=-\frac{1}{2\sqrt{3}}a_8 \,\,.
\ee
Therefore, the residual group is
\be 
H=SU(2)\times U(1)\times U(1)
\ee
with the generators $2T_1$, $2T_2$, $2T_3$, and 
\be \label{genpolar}
U=T_8-\frac{1}{2\sqrt{3}}{\bf 1} \,\, , \qquad V=J_3 \,\, .
\ee
Thus, the symmetry breaking pattern in the polar phase is similar to that
in the A phase, cf.\ (\ref{gen3i}). But while in the A phase both 
residual $U(1)$'s are combinations of the original symmetries, 
in the polar phase one of them is a subgroup of $G_2=SU(2)_J$.

\item $b_3$ arbitrary, $a_8=c=0$.

This case is identical to case 2.(ii).

\end{enumerate}
\end{enumerate}

As mentioned above, one
also has to take into account generators of a residual $U(1)$ where $T_8$ 
in Eq.\ (\ref{generator1})
is replaced by $T_2$, i.e., 
\be  \label{generator2}
U = a_2T_2 + b_3J_3 + 2c{\bf 1} \,\, .
\ee 
The reason why $T_2$ plays a special role is that we 
used the generator $J_3$ of the spin group, which is proportional to
$T_2$. Consequently, we expect to find additional residual groups that 
connect the color group with the spin group
(meaning a residual $U(1)$ generated by a combination of a color and a 
spin generator). The calculations with this generator are completely 
analogous to the ones with the ansatz (\ref{generator1}). Therefore, without 
presenting the details, let us mention 
that the ansatz (\ref{generator2}) yields one new {\it unique} order parameter, 
namely
\be \label{orderplanar}
\D=\frac{1}{\sqrt{2}}\left(\begin{array}{ccc} 1&0&0\\0&1&0\\0&0&0 
\end{array}\right) \,\, .
\ee
This order parameter corresponds to the planar phase, introduced in 
Eq.\ (\ref{Mkplanar}). In this case, Eq.\ (\ref{invariance}) yields
\be 
a_1 = 0 \,\, , \quad  a_3 = a_4 = \ldots = a_7 = b_1 = b_2 = 0 \,\, ,\quad 
a_2 = 2b_3\,\, , \quad c=\frac{1}{4\sqrt{3}}a_8  \,\,.
\ee
Thus, the residual group is 
\be
H=U(1)\times U(1) \,\, ,
\ee
with the generators
\be
U=2T_2 + J_3 \,\, ,\qquad V=T_8+\frac{1}{4\sqrt{3}} \,{\bf 1} \,\, .
\ee 

Let us now turn to possible groups $H$ that do not contain any $U(1)$ but
solely consist of higher-dimensional Lie groups, say 
\be \label{H'2}
H=SU(2)\times H' \,\, .
\ee
Let $U_1$, $U_2$, $U_3$ be the generators of the residual $SU(2)$. They 
are linear combinations of the generators of $G$,
\begin{subequations}
\bea
U_1&=&a^1_mT_m + b^1_nJ_n + 2c^1{\bf 1} \,\, , \\
U_2&=&a^2_mT_m + b^2_nJ_n + 2c^2{\bf 1} \,\, , \\
U_3&=&a^3_mT_m + b^3_nJ_n + 2c^3{\bf 1} \,\, .
\eea
\end{subequations}
Since they must fulfill
the $SU(2)$ commutation relations, Eq.\ (\ref{SU2commutation}), 
they must not contain the generator of 
$G_3=U(1)$, the unit matrix, i.e., $c^1=c^2=c^3=0$ . Therefore, there are 
three possibilities.
First, each $U_i$ is a combination of color and spin 
generators. Second and third, each $U_i$ is composed solely of color or 
spin generators, respectively. The simplest options to realize these 
cases are
\begin{subequations} \label{SU2gen}
\bea 
&& (a)\qquad U_i=T_i'+J_i \,\, , \\
&& (b)\qquad U_i=T_i' \,\, ,    \\
&& (c)\qquad U_i=J_i \,\, ,   
\eea
\end{subequations}
where $(T_1',T_2',T_3')$ is either given by $(2T_1,2T_2,2T_3)$ or  
$(2T_7,-2T_5,2T_2)$, which both fulfill the required commutation relations. 
Using the options $(a)$, $(b)$, $(c)$, let us first show that 
$H'$ in Eq.\ (\ref{H'2}) cannot be a second $SU(2)$. To this end, assume
that $H'=SU(2)$ with generators $V_1$, $V_2$, $V_3$, which have the same form
as the generators $U_i$ in Eqs.\ (\ref{SU2gen}). Then, since
the Lie algebra of $H$ is a direct sum of the constituent Lie algebras,
cf.\ Sec.\ \ref{he3}, we have to require
\be
[U_i,V_j] = 0\,\, .
\ee
This condition reduces all options to one, namely
\be
U_i = T_i' \,\, , \qquad V_i = J_i 
\ee
(or vice versa). However, now the invariance equation for the order parameter yields
\be \label{exclude}
\D J_i=0 \,\, ,
\ee
for all $i=1,2,3$, which does not allow for a nonzero order parameter $\D$.
Therefore, $H'=SU(2)$ is forbidden. Since the cases with $H'=U(1)$ and
$H'=U(1)\times U(1)$ were already covered in the above discussion,
the only possibility that is left is $H'={\bf 1}$ and thus $H=SU(2)$. 
 
From Eqs.\ (\ref{SU2gen}), case $(c)$ can be immediately excluded since it 
also leads to Eq.\ (\ref{exclude}). The same 
argument excludes case $(b)$ with $(T_1',T_2',T_3')=(2T_7,-2T_5,2T_2)$.
Case $(b)$ with $(T_1',T_2',T_3')=(2T_1,2T_2,2T_3)$ leads to two 
order parameters already considered above, namely the A phase, 
Eq.\ (\ref{orderA}), and the polar phase, Eq.\ (\ref{orderpolar}). 
In case $(a)$, only $(T_1',T_2',T_3')=(2T_7,-2T_5,2T_2)$ is possible. With
\be
U_i\D = -T_i'\D + \D J_i = 0 \,\, ,
\ee
one finds
\be \label{orderCSL}
\D=\frac{1}{\sqrt{3}}\left(\begin{array}{ccc} 1&0&0\\0&1&0\\0&0&1 
\end{array}\right) \,\, . \\      
\ee 
Indeed, it can be checked with Eq.\ (\ref{invariance}) that this order
parameter leads to
\be
a_1=a_3=a_4=a_6=a_8=c=0\,\, , \quad a_2 = 2b_3  \,\, , \quad a_5 = -2b_2
\,\, , \quad a_7 = 2b_1 \,\, ,
\ee
which corresponds to $H=SU(2)$, consisting of joint rotations in color and 
spin space. This is the CSL phase introduced for a spin-one color 
superconductor in Eq.\ (\ref{MkCSL}) and for superfluid $^3$He (therein
called B phase) in Eq.\ (\ref{heB}).

Finally, we give an argument why an even larger subgroup, i.e., an 
$SU(3)$, cannot occur in the residual group $H$. Assume that there 
are eight generators $W_1,\ldots ,W_8$ of this $SU(3)$. Then, as 
for the $SU(2)$ subgroup above, there can be no contribution to 
$W_1,\ldots ,W_8$ from the 
$G_3$ generator due to the $SU(3)$ commutation relations for the generators,
\be \label{SU3commutation}
[W_i,W_j] = i\,f_{ijk} W_k \,\, , \qquad i,j,k \le 8 \,\, ,
\ee
where $f_{ijk}$ are the $SU(3)$ structure constants.
Also, $W_i=T_i$ is excluded 
because in this case the invariance condition yields 
$T_8\D=0$ and thus $\D=0$. Therefore, at least one of the spin 
generators has to be included. For instance, choose $W_8=T_8 + b_3 J_3$.
Then, from the commutation relation \mbox{$[W_4,W_5]\sim W_8$} we conclude
that also $J_1$ and $J_2$ must be included via $W_4=T_4 + b_1 J_1$ and   
$W_5=T_5 + b_2 J_2$. But now the three equations $W_4\D=W_5\D=W_8\D=0$
lead to $\D=0$. Therefore, we conclude that
there is no residual group $H$ that contains an $SU(3)$. For a more 
rigorous proof one has to take into account more complicated linear 
combinations of the original generators.

Before we summarize and interpret the results, let us add two remarks.

First, we explain how the above discussion can be extended to two 
original $U(1)$ symmetries, one accounting for electromagnetism (local)
and one for baryon number conservation (global). Denoting the generators of 
these two groups by $2q\,{\bf 1}_{em}$ and $2\cdot{\bf 1}_B$, respectively, 
where $q$ is 
the electric charge of the quark, it is very simple to
generalize the results to this situation. For instance, in the polar phase 
there are not only two residual $U(1)$'s generated by 
$U=T_8-1/(2\sqrt{3}){\bf 1}$ and $V=J_3$, cf.\ Eqs.\ (\ref{genpolar}),  
but three $U(1)$'s generated by 
$U_1=T_8-1/(2q\sqrt{3}){\bf 1}_{em}$ (local!),  
$U_2=T_8-1/(2\sqrt{3}){\bf 1}_{B}$ (global!), and $V=J_3$. The extension 
in the other phases works analogously. Note that the same problem in 
a two- or three-flavor color superconductor is less trivial. In these cases,
the generator of the electromagnetic gauge group is not proportional to 
the unit matrix since it involves different electric charges of the 
respective quark flavors.    

The second remark concerns the difference between the uniquely determined 
order parameters, such as those for the polar, planar, A, and CSL phases,
and the ones that depend on one or more free parameters.  
Without elaborating on the group-theoretical details, let us refer to the
analogous situation in $^3$He.
In this case, all order parameters of the first 
kind lead to so-called ``inert'' states \cite{vollhardt}. 
All experimentally known states of superfluid 
$^3$He belong to this class of order parameters. 
Mathematically speaking, these
order parameter matrices play a special role due to a theorem 
(``Michel's Theorem'') \cite{vollhardt,michel}, which ensures that these 
order parameters correspond to a stationary point of any $G$-invariant 
functional of $\D$ (for instance the effective potential). 
This is the reason why in Secs.\ \ref{gapeqsolution}, \ref{mixingscreening},
and \ref{thepressure} we focus on these order parameters.

\begin{figure}[ht] 
\begin{center}
\includegraphics[width=11cm]{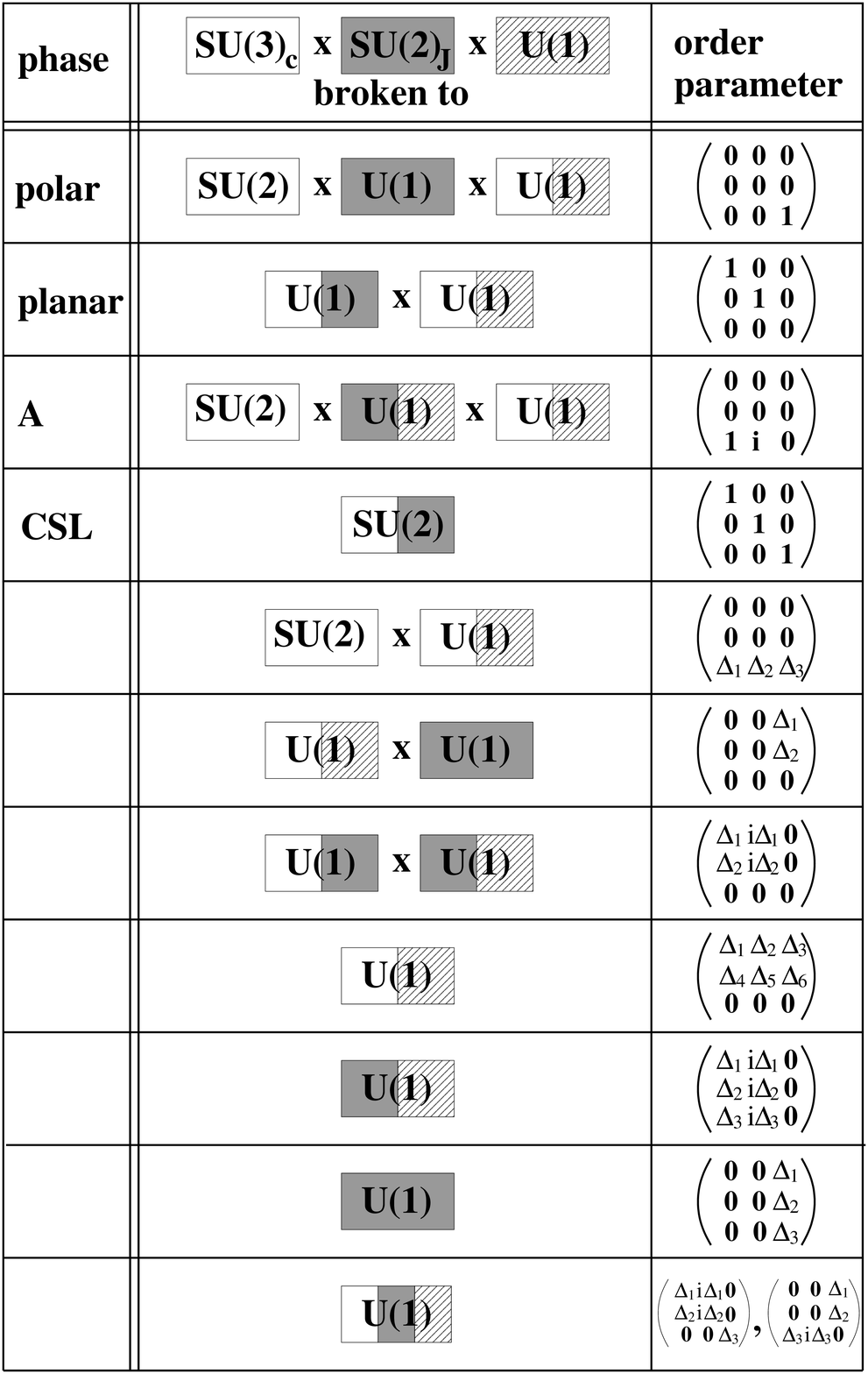}
\vspace{0.5cm}
\caption[Symmetry breaking patterns in a spin-one color superconductor]
{Symmetry breaking patterns and 
order parameters for a 
spin-one color superconductor. The original symmetry group (first line) 
is given 
by the color gauge group $SU(3)_c$ (blank background), the spin group 
$SU(2)_J$ (grey background), and a $U(1)$ (hatched background). The latter 
can be regarded as the electromagnetic 
gauge group or the baryon number conservation group. The backgrounds of the 
residual groups illustrate the symmetry breaking pattern. For
instance, the blank $SU(2)$ occuring in the residual group of the
polar phase is generated solely by generators of the original $SU(3)_c$ 
while the blank/hatched $U(1)$ in the same line is generated by a 
linear combination of the generators of the original $SU(2)_J$ and $U(1)$ 
groups. For the explicit expression of these generators, see text.
}
\label{symmetries}
\end{center}
\end{figure}

In Fig.\ \ref{symmetries}, we summarize our results in a list 
of all superconducting phases that we have found in the above 
discussion. It should be mentioned that this list is not complete, since
for the generators of the residual $U(1)$'s we have restricted ourselves to 
two special forms given in Eqs.\ (\ref{generator1}) and (\ref{generator2}).
Therefore, there are certainly more (at least noninert) order parameters 
that lead to an allowed symmetry breaking. The 
inert states are listed in the first four lines. Each of these four states 
has its analogue in superfluid $^3$He \cite{vollhardt}. Note that the
${\rm A}_1$ phase, which is experimentally observed in $^3$He in the 
presence of an external magnetic field (see the phase diagram 
in Fig.\ \ref{Hephase1}), does not lead to an allowed symmetry breaking
in the case of a spin-one color superconductor \cite{schaefer}. 
To see this, one inserts the corresponding order parameter from 
Eq.\ (\ref{heA1}) into Eq.\ (\ref{invariance}). One obtains
\be
a_1=a_3=b_1=b_2=0\,\, ,\quad a_4=a_7\,\, ,\quad a_5=a_6\,\, , \quad
\frac{1}{2}a_2+\frac{1}{2\sqrt{3}}a_8-b_3-2c=0 \,\, .
\ee
These 7 conditions suggest that ${\rm dim}\,H=12-7=5$ and thus 
$H=SU(2)\times U(1)\times U(1)$. However, there is no possibility to construct
three generators from the above conditions that fulfill the commutation 
relations (\ref{SU2commutation}). For instance, assume that two of these 
generators
are given by $U_1=T_4 + T_7$ and $U_2=T_5 +T_6$. Then, with 
Eq.\ (\ref{SU3commutation}), $[U_1,U_2]\sim T_3$. But since $a_3=0$, 
the third generator $U_3$ cannot be proportional to $T_3$. 
Consequently, there is no ${\rm A}_1$ phase in a spin-one color superconductor.
  
Below the four inert states we list the eight noninert states which 
have been found above. Note that one of these noninert phases, given in
Eq.\ (\ref{order3ii}), has a larger residual symmetry group than the planar
and CSL phases. 

There are several properties of the spin-one phases which can easily
be read off from Fig.\ \ref{symmetries}. First, consider  
the spin group $SU(2)_J$. This symmetry accounts for the rotational symmetry
in real space of the normal-conducting phase (remember that, in the case of 
the spin-one representation, one can equivalently consider $SO(3)_J$ instead 
of $SU(2)_J$). The list shows that spatial symmetry is broken in each case. 
For instance, in the polar phase, $SU(2)_J$ is broken to its subgroup $U(1)_J$.
Therefore, the superconducting phase is invariant under rotations around
one fixed axis in real space. This axis (the 3-axis) is chosen by the 
spins of the Cooper pairs which all are aligned in one direction. In the
previous section it has been shown that this breakdown of the 
rotational symmetry leads to an anisotropy of the energy gap. In most
of the other cases, the breaking of the spatial rotation symmetry is more
subtle: For instance in the planar phase, the superconducting state is 
invariant under a special joint rotation in color and real space. 
Note that in the planar phase, unlike the polar phase, the special
role of the third spatial direction originates from the fact that the 
spins are aligned perpendicular to the 3-axis. The most interesting
breakdown of spatial symmetries is present in the CSL phase. 
Here, any rotation in real space leaves the system invariant as long as one 
simultaneously performes the same rotation in fundamental color space 
which is spanned by the three directions red, green, and blue. This special
form of locking
of color and spatial indices is responsible for the fact that the
CSL phase (and the B phase in $^3$He) is ``pseudoisotropic''.  
In fact, in the previous section it has been shown that there is no
anisotropy in the gap function, cf.\ Fig.\ \ref{gapfigure}.

Next, let us read off some properties concerning the color symmetry.
It is obvious that in none of the cases the 
full color symmetry is preserved. In this sense, it is justified to call
each phase a color superconductor. 
In three of the cases, there is a residual 
color subgroup $SU(2)$, namely in the polar phase, the A phase, and the
noninert phase given by Eq.\ (\ref{order3ii}). Mathematically speaking, 
this residual group originates from
the fact that the order parameter has only nonzero elements in its third
row. Therefore, the third direction in fundamental color space is preferred.
Physically, this means that the Cooper pairs carry color charge anti-blue, 
or, in other words, only red and green quarks form Cooper pairs. Of course,
the choice of the anti-blue direction is convention; more generally speaking,
quarks of one color remain unpaired. Remember that this is also true 
for the 2SC phase in a two-flavor color superconductor, 
cf.\ Sec.\ \ref{2SCCFL}. 

From Sec.\ \ref{higgs} we know that a spontaneously
broken gauge symmetry is equivalent to massive gauge bosons. In the case
of a color superconductor, these masses are the magnetic screening masses 
of the gluons. Therefore, in the cases where there is a residual color
subgroup $SU(2)$, we expect a Meissner effect for five of the eight 
gluons. Three of the gluons, however, namely those corresponding to 
the generators $T_1$, $T_2$, $T_3$, do not attain a Meissner mass. This is 
also obvious from the fact 
that these gluons do not see the (anti-)blue color charge which is
carried by the Cooper pairs. For a more detailed and quantitative 
discussion of the color Meissner effect, see the next section.
Also with respect to the breakdown of the color symmetry, the CSL 
phase is exceptional. Although there is a residual $SU(2)$, where three
of the color generators are involved, we expect all eight gluons to 
attain a Meissner mass. To this end, note that this residual $SU(2)$ is 
a global symmetry and 
therefore all dimensions of the gauge group have to be considered as broken.   
This is analogous to the CFL phase introduced in Sec.\ \ref{2SCCFL}. In this
three-flavor color-superconducting phase, color is locked to flavor which 
results in a color Meissner effect for all eight gluons.

In order to discuss the question whether the color superconductors in Fig.\ 
\ref{symmetries} are also electromagnetic superconductors, one regards 
the $U(1)$ of the original symmetry (hatched background in the figure)
as the electromagnetic gauge group $U(1)_{em}$. From 
ordinary superconductors, cf.\ Sec.\ \ref{superconductivity}, we know
that this symmetry is spontaneously broken below the critical temperature.
Therefore, a simple conclusion is that all states in the list without a 
hatched background occurring in the residual group are electromagnetic
superconductors. Obviously, this is the case for the CSL phase and for one of 
the noninert states, defined in Eq.\ (\ref{order2ii}). This is a 
remarkable result because electromagnetic superconductivity is 
absent in both spin-zero superconductors that were discussed in the 
introduction. Moreover, 
the presence of the electromagnetic Meissner effect in a spin-one color 
superconductor probably leads to observable implications for the
properties of a neutron star \cite{schmitt2}. 

In most of the 
phases listed in Fig.\ \ref{symmetries}, the question of the 
electromagnetic Meissner effect is less trivial. Although in none of the 
cases the electromagnetic gauge group is untouched, there are residual groups
that involve the $U(1)_{em}$ generator. Here we have to distinguish between
two cases. First, $U(1)_{em}$ may only be coupled to the global spin group 
which results in a global residual (sub)group. In this case we expect the
photon to attain a magnetic screening mass since this situation is analogous
to the CFL (and CSL) phase(s), where a coupling of the gluons to the global
flavor (spin) symmetry leads to a color Meissner effect for all gluons. 
Second, $U(1)_{em}$ may couple to the color gauge group. This happens 
for instance in the polar, planar, and A phases. Since this 
scenario exactly corresponds to the cases of the 2SC and CFL phases, we 
expect a ``rotated'' magnetic photon (cf.\ Sec.\ \ref{higgs}) that can 
penetrate the superconductor (= {\it no} electromagnetic Meissner effect). 
Besides the
color Meissner effect, also the electromagnetic Meissner effect is treated
quantitatively in Sec.\ \ref{mixingscreening}. In this section, we also
discuss the case of a many-flavor system in which 
each quark flavor separately forms spin-one Cooper pairs in the polar phase.
It turns 
out that in this case there is an electromagnetic Meissner effect, in spite
of the apparently contradicting conclusion from Fig.\ \ref{symmetries}.      
 
Finally, let us comment on the superfluidity of the phases. To this end,
regard the hatched $U(1)$ as the particle number conservation 
symmetry $U(1)_B$. As discussed in the introduction, a spontaneous breakdown
of this symmetry leads to superfluidity (= a stable superflow and vortices
in a rotating system). Again, the only phases in the list, where the 
breakdown of this symmetry is obvious, are the CSL phase and the phase 
from Eq.\ (\ref{order2ii}). In all other cases, the question of 
superfluidity is nontrivial. It is comparable to the A phase in $^3$He,
see discussion in Sec.\ \ref{he3}. Remember that in the CFL phase, $U(1)_B$
is completely broken, cf.\ Eq.\ (\ref{HCFL}), which renders this phase
a superfluid, contrary to the 2SC phase, cf.\ Eq.\ (\ref{H2SC}), 
which behaves similar to most of the spin-one phases (e.g., the polar phase)
and is no superfluid. In Sec.\ \ref{neutronstars}, it has been indicated
that also the question of superfluidity is of great physical relevance, 
since superfluid vortices (together with magnetic flux tubes) play
an important role for possible explanations of the properties of a 
neutron star.       

       %Group Theory, Classification
   \section{Mixing and screening of photons and gluons} \label{mixingscreening} 

From the previous section, where symmetry breaking
patterns in a spin-one color superconductor have been studied, we know
that different color-superconducting phases have different properties
with respect to electromagnetic and color fields, i.e., with respect to
photons and gluons. With the help of simple group-theoretical arguments 
it has been shown that for instance the CSL phase exhibits an electromagnetic
Meissner effect. This is in contrast to the 2SC and CFL phases in a 
spin-zero color superconductor. In these phases there is a {\it color} Meissner
effect for five and eight gluons, respectively, but due to a locking
of the color and electromagnetic gauge groups there is no {\it electromagnetic}
Meissner effect. In Sec.\ \ref{higgs}, it was mentioned
that this absence of the Meissner effect is somehow related to a mixing
of the gauge boson fields and that this is analogous to what happens 
in the Weinberg-Salam model of electroweak interactions. 

It is the goal of this section to clarify this mixing between photons and 
gluons, which is also expected to be present in several spin-one phases,
cf.\ Fig.\ \ref{symmetries}. Furthermore, we not only present a calculation
of the magnetic screening masses (= Meissner masses) in order to 
quantitatively verify the group-theoretical predictions from the last 
section, but also compute the electric screening masses (= Debye masses)
for all gauge bosons. The calculations are shown for four different 
color-superconducting phases. We consider the 2SC and CFL phases as well 
as the polar and CSL phases. More precisely, 
for the spin-one phases we focus on the 
``mixed polar'' and the ``mixed CSL'' phases, cf.\ explanation below 
Eq.\ (\ref{Mk}), but refer to them shortly as ``polar'' and ``CSL'' phases  
 
This section is organized as follows. 
In Sec.\ \ref{propagators} we start from 
the QCD partition function for massless quarks in the presence of 
gluon and photon gauge fields in order to derive 
the gauge boson propagators in a color superconductor. 
This fundamental study provides the 
basis for the following sections, since it can be applied for arbitrary 
phases in color superconductivity. It is also general in the sense that 
we consider gauge fields for arbitrary four-momenta $P$ while 
afterwards we focus on the limiting case $P\to 0$ in order to discuss screening
of static and homogeneous electric and magnetic fields. 

In Sec.\ \ref{mixing}, we present a general definition of the mixing angle.
This mixing angle is the analogue of the Weinberg angle, cf.\ Table 
\ref{analogy}, and quantitatively describes the mixing between photons and 
gluons. Its definition is based on the results of Sec.\ \ref{propagators}.
Making use of the results of the group-theoretical study in 
Sec.\ \ref{grouptheory}, the explicit form of the mixing angle in terms
of the strong and electromagnetic coupling constants $g$ and $e$ is 
presented for the above mentioned four color-superconducting phases. 
 
In the next sections we extend the results of Sec.\ \ref{mixing}
by performing a calculation of the Debye and Meissner masses for
all gauge bosons. These masses correspond to the inverse screening lengths
for electric and magnetic fields, respectively. They are obtained 
from the longitudinal and transverse components of the polarization 
tensors in the zero-energy, low-momentum limit.
In the first of these sections, Sec.\ \ref{calc}, we perform all calculations
that do not depend on the special color-superconducting phase. A part
of these calculations, 
namely the integrals over the absolute value of the quark momentum, 
are deferred to Appendix \ref{Appquark}.  In 
Sec.\ \ref{mixingresults}, we specify the results for the above 
mentioned phases, i.e., we derive the 
particular expressions for the relevant components of the polarization tensors.
Readers who are not interested in the technical details may skip this
section. All results and a discussion are given in Sec.\ \ref{discussion}.
We present all screening masses for photons and gluons which yield the
mixing angles and the masses for the (physically relevant)
mixed, or ``rotated'', gauge bosons. One of the points we discuss in this 
section is in which cases these mixing angles obtained from the 
calculation of the screening masses differ from the ones 
obtained in Sec.\ \ref{mixing}.

\subsection{Gluon and photon propagators} \label{propagators}

In the following, we provide the theoretical basis for the calculation 
of the Debye and Meissner masses in a color superconductor. The results of
this section are the final form for gluon and photon propagators, 
Eq.\ (\ref{phogluprop}), and the definitions for the ``rotated'' Debye and 
Meissner masses, Eqs.\ (\ref{defmasses}). 

As in Sec.\ \ref{derivgapeq}, where the QCD gap equation has been derived,
we start from the QCD partition function for massless quarks.
(In Sec.\ \ref{derivgapeq}, the quark mass $m$ was kept for the derivation of 
the gap equation, but was set to zero before the explicit calculations were 
performed.) 
Besides the gluon fields $A_a^\mu$, however, we also include
a photon field $A^\mu$,
\be \label{QCDpartition2}
{\cal Z}=\int {\cal D}A\,e^{S_A}{\cal Z}_q[A] \,\, .
\ee 
Since the emphasis of this section is put on the gauge fields, we choose a
slightly different notation compared to Eq.\ (\ref{QCDpartition}).
The action for gluon and photon 
fields is given by $S_A$. It consists of three parts, 
cf.\ also Eq.\ (\ref{action}),
\be
S_A=S_{F^2} + S_{gf} + S_{FPG} \,\, ,
\ee
where $S_{gf}$ and $S_{FPG}$ are the gauge fixing and 
ghost terms, respectively, and 
\be 
S_{F^2}\equiv-\frac{1}{4}\int_X (F_a^{\m\n}F^a_{\m\n}+F^{\m\n}F_{\m\n})
\ee
is the gauge field part. The space-time integration is defined as in Sec.\
\ref{derivgapeq}.
The field strength tensors $F^a_{\m\n}=
\partial_\m A_\n^a-\partial_\n A_\m^a + g f^{abc}A_\m^b A_\n^c$ 
correspond to the gluon fields $A_\m^a$, while $F_{\m\n}=
\partial_\m A_\n-\partial_\n A_\m$ corresponds to the photon 
field $A_\m$. The functional ${\cal Z}_q[A]$ is the grand
partition function of massless quarks in the presence of
gluon and photon fields and a chemical potential $\m$. It is given by
\be \label{quarkpartition}
{\cal Z}_q[A]=\int{\cal D}\bar{\psi}\,{\cal D}\psi\exp\left[\int_X
\bar{\psi}\,(i\,\gamma^\mu\partial_\mu+\mu\gamma_0+g\,\gamma^\mu A_\mu^aT_a
+e\,\gamma^\mu A_\mu Q)\, \psi\right] \,\, ,
\ee
where $T_a$ are the generators in the fundamental representation 
of the gauge group of strong
interactions, $SU(3)_c$, and $Q$ is the generator of the
electromagnetic $U(1)_{em}$. Here, $Q$ is not necessarily proportional 
to the unit matrix (as it was the case in Sec.\ \ref{grouptheory}), 
since in the current general treatment the number 
of flavors is not specified. The coupling constants for the
strong interaction and electromagnetism are denoted by $g$ and
$e$, respectively. The quark fields $\psi$ are spinors in color, flavor,
and Dirac space. 

In order to take into account Cooper pairing of the quarks, we 
include a diquark source term. This has been done for instance
in Refs.\ \cite{bailin,shovkovy}. We generalize this treatment by
also taking into account the photon field.
Then, after integrating out the fermion fields, the partition function
is given by \cite{shovkovy}
\be \label{log}
{\cal Z}=\int {\cal D}A\exp\left[S_A+\frac{1}{2}
{\rm Tr}\,\ln({\cal S}^{-1}+A_\mu^a\hat{\Gamma}^\mu_a)\right] \,\, ,
\ee
where the trace runs over space-time, Nambu-Gor'kov, color, flavor, and Dirac
indices.  The sum over $a$ now runs from 1 to 9, where $A_\mu^9\equiv A_\m$ 
is the photon field. 
${\cal S}\equiv {\cal S}(X,Y)$ is the quasiparticle propagator in 
Nambu-Gor'kov space. We also defined the $2\times 2$ Nambu-Gor'kov matrices 
\be \label{vertex}
\hat{\Gamma}_a^\m\equiv {\rm diag}\left(\Gamma_a^\m,
\overline{\Gamma}_a^\m\right)
\equiv \left\{\begin{array}{lll} {\rm diag}(g\,\g^\m T_a,-g\,\g^\m T_a^T) & 
\mbox{for} & \quad a=1,\ldots,8 \,\, , \\ \\
{\rm diag}(e\, \g^\m Q,-e\,\g^\m Q) & \mbox{for} & \quad a=9 \,\, .
\end{array} \right.    
\ee
For convenience, we included 
the coupling constants into $\Gamma_a^\m$ and $\overline{\Gamma}_a^\m$,
although it differs from the notation introduced in Sec.\ \ref{gapeqsolution},
see Eq.\ (\ref{Gammadef}).

 In Eq.\ (\ref{log}), we did not explicitly keep the
fluctuations of the order parameter field, cf.\ Eq.\ (26) of Ref.\ 
\cite{shovkovy}. This is possible, since we are only interested in
the gauge field propagator. Nevertheless, it should be kept in mind
that, like in any theory with spontaneously broken gauge symmetry, 
these fluctuations mix with the (4-)longitudinal (unphysical) 
components of the gauge field. As usual, using a suitable choice of
't Hooft gauge, the fluctuations can be decoupled from the gauge field.
Then, in unitary gauge, one finds that the gauge field propagator is explicitly
(4-)transverse \cite{shovkovy}, 
in accordance with general principles \cite{lee,lebellac}.

Expanding the logarithm in Eq.\ (\ref{log}) to second order in the 
gauge fields, we obtain
\be
\frac{1}{2}{\rm Tr}\,\ln({\cal S}^{-1}+A_\mu^a\hat{\Gamma}^\mu_a)
\simeq\frac{1}{2}{\rm Tr}\,\ln{\cal S}^{-1}
+\frac{1}{2}\,{\rm Tr}[A_\mu^a\,\hat{\Gamma}_a^\mu\,{\cal S}]
-\frac{1}{4}{\rm Tr}[A_\mu^a\,\hat{\Gamma}_a^\mu\,{\cal S}
\,A_\nu^b\,\hat{\Gamma}_b^\nu\,{\cal S}] \,\, .
\ee
Following Ref.\ \cite{shovkovy}, the sum of all terms
which are quadratic in the gauge fields will be denoted by $S_2$. 
$S_2$ does not only contain the 
pure gluon and photon terms but also two terms
which mix gluon and photon fields. In order to perform
the trace over space-time, we introduce the Fourier transforms
\begin{subequations}
\bea
{\cal S}(X,Y)&=&\frac{T}{V}\sum_K e^{-iK\cdot (X-Y)} {\cal S}(K) \,\, , \\
A_\mu^a(X)&=&\sum_P e^{-iP\cdot X} A_\m^a(P) \,\, ,
\eea
\end{subequations}
where we used translational invariance,
${\cal S}(X,Y)={\cal S}(X-Y)$. 
Then we obtain
\bea \label{S2}
S_2 & = & -\frac{1}{4}\int_{X,Y}{\rm Tr}[A_\mu^a(X)\,\hat{\Gamma}_a^\mu
\,{\cal S}(X,Y)\,A_\nu^b(Y)\,\hat{\Gamma}_b^\nu\,{\cal S}(Y,X)] \nonumber \\
& = & -\frac{1}{4}\sum_{K,P}{\rm Tr}[A_\mu^a(-P)\,\hat{\Gamma}_a^\mu
\,{\cal S}(K)\,A_\nu^b(P)\,\hat{\Gamma}_b^\nu\,{\cal S}(K-P)] \nonumber \\ 
& = & -\frac{1}{2}\frac{V}{T}\sum_P A_\mu^a(-P) \, \Pi_{ab}^{\m\n}(P)
\, A_\n^b(P) \,\, ,
\eea
where the trace now runs over Nambu-Gor'kov, color, flavor, and Dirac
indices and where we defined the polarization tensor
\be \label{poltensors} 
\Pi_{ab}^{\m\n}(P)\equiv\frac{1}{2}\frac{T}{V}\sum_K{\rm Tr}
[\hat{\Gamma}_a^\mu\,{\cal S}(K)\,\hat{\Gamma}_b^\nu\,{\cal S}(K-P)] \,\, .
\ee
The Nambu-Gor'kov quasiparticle propagator in momentum space ${\cal S}(K)$
is defined as in Eq.\ (\ref{fullquark}),
\be 
{\cal S}(K) = \left(\begin{array}{cc} G^+(K) & \Xi^-(K) \\ \Xi^+(K) & G^-(K) 
\end{array}\right) \,\, .
\ee   
In the following, we may put the regular self-energies $\Sigma^\pm$ 
to zero since, to the order we are computing, they do not influence 
the results for the polarization tensors. Consequently,
\begin{subequations}
\bea
G^\pm(K)&=&\left\{[G_0^\pm]^{-1}(K)-\Phi^\mp(K)\, G_0^\mp(K) \,\Phi^\pm(K)
\right\}^{-1} \,\, , \\
\Xi^\pm(K) &=& -G_0^\mp(K)\Phi^\pm(K) G^\pm(K) \,\, ,
\eea
\end{subequations}
where $G_0^\pm(K)=(\g_\m K^\m\pm\m\g_0)^{-1}$ is the free
propagator. The gap matrices $\Phi^\pm(K)$ are given by 
Eqs.\ (\ref{phiminus}) and (\ref{gm2SC}).
The quasiparticle propagators are 
\be  \label{prop}
G^\pm(K)=[G_0^\mp(K)]^{-1}\sum_{e=\pm}\sum_{r=1,2}{\cal P}^r_{\bf k}
\, \Lambda_{\bf k}^{\mp e}\, \frac{1}{k_0^2-\left[\e_{k,r}^e(\phi^e)\right]^2}
\,\, .
\ee
In Sec.\ \ref{gapeqsolution}, only $G^+$ was needed, 
cf.\ Eq.\ (\ref{fullprop}). In order to derive the corresponding expression 
for $G^-$, we used the identity 
$\g_0{\cal M}_{\bf k}^\dagger {\cal M}_{\bf k} \g_0 = {\cal M}_{\bf k}
{\cal M}_{\bf k}^\dagger$.
This identity is valid in the phases we consider,
i.e., the 2SC, CFL, polar, and CSL phases. (However, in the A phase it does
not hold, which causes a slight modification of the above result, cf.\
Sec.\ \ref{thepressure}.) All quantities in Eq.\ (\ref{prop}) are defined
as in Sec.\ \ref{gapeqsolution}, i.e.,  
${\cal P}^r_{\bf k}$ are projectors onto the eigenspaces
of the hermitian matrix 
$L_{\bf k} = \g_0{\cal M}_{\bf k}^\dag{\cal M}_{\bf k}\g_0$. In all
phases considered here, there are two eigenvalues of this matrix,
denoted by $\l_1$ and $\l_2$. They appear in the quasiparticle 
excitation energies
\be \label{excite}
\e_{k,r}^e(\phi^e)\equiv \left[(\m-ek)^2+\l_r|\phi^e|^2\right]^{1/2} \,\, .
\ee
The projectors are given by Eq.\ (\ref{proj}).
The anomalous propagators can be written as 
\begin{subequations} \label{aprop}
\bea 
\Xi^+(K) &=& -\sum_{e,r}\g_0{\cal M}_{\bf k}\g_0 \, {\cal P}^r_{\bf k} \,
\Lambda_{\bf k}^{-e}\, \frac{\phi^e(K)}{k_0^2-\left[\e_{k,r}^e(\phi^e)
\right]^2}
\,\, , \\
\Xi^-(K) &=& -\sum_{e,r}{\cal M}_{\bf k}^\dagger \, {\cal P}^r_{\bf k} \,
\Lambda_{\bf k}^e\, \frac{\phi^{e\,*}(K)}{k_0^2-\left[\e_{k,r}^e(\phi^e)
\right]^2}\,\, .
\eea
\end{subequations}
As in Ref.\ \cite{shovkovy},  we introduce a set of complete, 
orthogonal projectors for each 4-vector $P^\m=(p_0,{\bf p})$,
\be \label{defABE}
{\cal Q}_1^{\m\n}\equiv g^{\m\n}-{\cal Q}_2^{\m\n}-{\cal Q}_3^{\m\n} 
\,\, , \qquad
{\cal Q}_2^{\m\n}\equiv\frac{N^\m N^\n}{N^2} \,\, , \qquad
{\cal Q}_3^{\m\n}\equiv\frac{P^\m P^\n}{P^2}\,\, .
\ee
With $N^\m\equiv(p_0p^2/P^2,p_0^2{\bf p}/P^2)$, the projector
${\cal Q}_2^{\m\n}$ projects onto the one-dimensional subspace that is 
\mbox{(4-)orthogonal} to $P^\m$ but (3-)parallel to ${\bf p}$. The
operator ${\cal Q}_3^{\m\n}$ projects onto the one-dimensional 
subspace parallel to $P^\m$.  
Consequently, ${\cal Q}_1^{\m\n}$ projects onto a two-dimensional subspace
that is \mbox{(4-)orthogonal} to both $P^\m$ and $N^\m$. Furthermore,
this subspace is (3-)orthogonal to ${\bf p}$. With the additional 
tensor
\be \label{defC}
{\cal Q}_4^{\m\n}=N^\m P^\n + P^\m N^\n
\ee
we can decompose the polarization tensor \cite{lebellac},
\be \label{decompose}
\Pi^{\m\n}_{ab}(P)=\sum_i\Pi^i_{ab}(P)\,{\cal Q}_i^{\m\n} \,\, .   
\ee
(In  the notation of Refs.\ \cite{shovkovy,lebellac}, ${\cal Q}_1\equiv 
{\rm A}, {\cal Q}_2\equiv {\rm B},{\cal Q}_3\equiv {\rm E},
{\cal Q}_4\equiv {\rm C} $.) 
Since for $i=1,2,3$, ${\cal Q}_i^{\m\n} {\cal Q}_{4\n\m}=0$, 
the coefficients $\Pi^i_{ab}(P)$ of the 
projection operators are given by 
\be \label{components}
\Pi^i_{ab}(P) = \frac{\Pi^{\m\n}_{ab}(P){\cal Q}_{i\n\m}}{{\cal Q}_{i\l}^\l} 
\quad, \qquad i=1,2,3 \,\, .
\ee
The remaining coefficient corresponding to ${\cal Q}_4$ is
\be \label{Pi4}
\Pi^4_{ab}(P) = \frac{ \Pi_{ab}^{\m\n}(P){\cal Q}_{4\n\m}}{{\cal Q}_4^{\l\s}
{\cal Q}_{4\s\l}}  \,\, .
\ee
The explicit forms for $\Pi^i_{ab}(P)$ are given
in Ref.\ \cite{shovkovy}, Eqs.\ (42) and (43). 
Employing the decomposition of the polarization tensor in Eq.\ (\ref{S2}),
we obtain
\bea 
S_2&=&-\frac{1}{2}\frac{V}{T}\sum_P\sum_i A_\mu^a(-P)
{\cal Q}_i^{\m\n}\Pi_{ab}^i(P)A_\n^b(P) \,\, .
\eea 
Now we add the free gauge field term
\be
S^{(0)}_{F^2}=-\frac{1}{2}\frac{V}{T}\sum_PA_\m^a(-P)
(P^2g^{\m\n}-P^\m P^\n)A_\n^a(P)=-\frac{1}{2}\frac{V}{T}\sum_P 
A_\m^a(-P)P^2({\cal Q}_1^{\m\n}+{\cal Q}_2^{\m\n})A_\n^a(P) \,\,.
\ee
We obtain
\be \label{before}
S_2 + S^{(0)}_{F^2} =
-\frac{1}{2}\frac{V}{T}\sum_P A_\m^a(-P)\left\{ \sum_{i=1}^2 {\cal Q}_i^{\m\n}
\left[\d_{ab}P^2 + \Pi_{ab}^i(P)\right] + \sum_{i=3}^4 {\cal Q}_i^{\m\n}
\Pi_{ab}^i(P)\right\} A_\n^b(P) \,\, .
\ee
Since we finally want to read off the gluon and photon propagators, we have
to transform this expression in two ways.
First, concerning the Dirac structure it is necessary to get rid of 
the term proportional to ${\cal Q}_4$ which mixes the longitudinal
mode (3-parallel to ${\bf p}$) with the unphysical mode (4-parallel to
$P^\m$). Then, inverting the inverse propagator becomes trivial, because
it is just a linear combination
of the complete, orthogonal projectors ${\cal Q}_1, {\cal Q}_2, {\cal Q}_3$.
Second, in order to obtain the physical modes we have to diagonalize 
the resulting $9\times 9$ matrices
which, after eliminating ${\cal Q}_4$, will replace  
$\d_{ab}P^2 + \Pi_{ab}^i(P)$, $i=1,2$, and 
$\Pi_{ab}^i(P)$, $i=3,4$ in Eq.\ (\ref{before}). 

We first write $S_2 + S^{(0)}_{F^2}$ as 
\bea \label{withoutgauge}
S_2 + S^{(0)}_{F^2} 
&=&-\frac{1}{2}\frac{V}{T}\sum_{P}\Big\{\sum_{i=1}^2 A_i^{a\m}(-P) 
\left[\d_{ab}P^2 + \Pi_{ab}^i(P)\right] A^b_{i\m}(P) + A_3^{a\m}(-P) 
 \Pi_{ab}^3(P) A^b_{3\m}(P) \nonumber \\
&& \hspace{2.1cm} +\; A_2^{a\m}(-P) N_\m \Pi^4_{ab}(P) P_\n A_3^{b\n}(P)
\nonumber \\
&& \hspace{2.1cm} +\;A_3^{a\m}(-P) P_\m \Pi^4_{ab}(P) N_\n A_2^{b\n}(P) \Big\} 
\,\, ,
\eea
where $A_i^{a\m}(P)\equiv {\cal Q}_i^{\m\n} A_\n^a(P)$ are the gauge fields
projected on the subspace corresponding to ${\cal Q}_i$. 
Now one can ``unmix'' the  
fields $A_2^\m(P)$ and $A_3^\m(P)$ by the following transformation
of the unphysical field component $A^a_{3\m}(P)$,
which does not affect the final result since we integrate over all fields
in the partition function,
\begin{subequations} \label{shift}
\bea 
A_3^{a\m}(-P) &\longrightarrow& A_3^{a\m}(-P) - 
A_2^{c\n}(-P)N_\n \Pi^4_{cb}(P)\left[\Pi^3(P)\right]^{-1}_{ba}
\, P^\m \,\, ,\\
A^a_{3\m}(P) &\longrightarrow& A^a_{3\m}(P) - 
P_\m \left[\Pi^3(P)\right]^{-1}_{ab}
\Pi^4_{bc}(P)\,N_\n A_2^{c\n}(P) \,\, .
\eea
\end{subequations}
After this transformation,
one is left with quadratic expressions in the projected fields.
The transformation modifies the term corresponding to $i=2$, 
\bea 
S_2 + S^{(0)}_{F^2} &=& 
-\frac{1}{2}\frac{V}{T}\sum_P A_\m^a(-P)\Big({\cal Q}_1^{\m\n}
\left[\d_{ab}P^2 + \Pi_{ab}^1(P)\right] \nonumber \\ 
&& + \;{\cal Q}_2^{\m\n}
\left\{\d_{ab}P^2 + \Pi_{ab}^2(P)-P^2N^2\Pi^4_{ac}(P)\,
\left[\Pi^3(P)\right]^{-1}_{cd}\,\Pi^4_{db}(P)\right\} 
\nonumber \\
&&+\;{\cal Q}_3^{\m\n}
\Pi^3_{ab}(P)\Big) A^b_\n(P) \,\, . \label{S_2+S_F}
\eea
Before we do the diagonalization in the 9-dimensional 
gluon-photon space, we add the gauge fixing term $S_{gf}$. We choose the
following gauge with gauge parameter $\l$,
\be \label{gf}
S_{gf}=-\frac{1}{2\l}\frac{V}{T}\sum_P A_\m^a(-P)P^\m  P^\n A_\n^a(P)
=-\frac{1}{2\l}\frac{V}{T}\sum_P A_\m^a(-P)P^2{\cal Q}_3^{\m\n}
A_\n^a(P) \,\, .
\ee
This looks like a covariant gauge but, including the fluctuations 
of the order parameter, which we did not write explicitly, 
it is actually some kind of 't Hooft gauge, cf.\ Eq.\ (50) of
Ref.\ \cite{shovkovy}. Moreover, had we fixed the gauge {\em prior\/}
to the shift (\ref{shift}) of the gauge fields, we would have
to start with an expression which is non-local and also involves the 
(3-)longitudinal components $A_2^{a\m}$ of the gauge field.

Adding the gauge fixing term (\ref{gf}) to Eq.\ (\ref{S_2+S_F}) leads to 
\be
S_2 + S^{(0)}_{F^2} +S_{gf} =  
-\frac{1}{2}\frac{V}{T}\sum_P \sum_{i=1}^3 A^a_\m(-P){\cal Q}_i^{\m\n}
\Theta^i_{ab}(P)A^b_\n(P) \,\, ,
\ee
with
\be \label{thetahat}
\Theta^i_{ab}(P) \equiv \left\{ \begin{array}{cl}
\d_{ab}P^2 + \Pi_{ab}^1(P)  & 
\mbox{for}\quad i=1  \,\, ,
\\ \\
\d_{ab}P^2 + \Pi_{ab}^2(P)-P^2N^2\Pi^4_{ac}(P)\,
\left[\Pi^3(P)\right]^{-1}_{cd}\,\Pi^4_{db}(P) & \mbox{for}\quad i=2 \,\, ,
\\ \\
\d_{ab}\frac{1}{\l}P^2+\Pi^3_{ab}(P) & \mbox{for}\quad i=3 \,\, .
\end{array} \right. 
\ee

In order to obtain the physical modes, we have to diagonalize 
the $9\times 9$ matrices $\Theta^i_{ab}(P)$. Since 
$\Theta^i_{ab}(P)$ is real and symmetric, 
diagonalization is achieved via an orthogonal 
transformation with a $9\times 9$ matrix ${\cal O}_i(P)$, 
\be \label{rotatedfields}
S_2 + S^{(0)}_{F^2} + S_{gf}=-\frac{1}{2}\frac{V}{T}\sum_P\sum_{i=1}^3      
\tilde{A}_{\mu,i}^a(-P){\cal Q}_i^{\m\n}
\tilde{\Theta}_{aa}^i(P)\tilde{A}_{\n,i}^a(P) \,\, ,
\ee
where 
\be
\tilde{A}_{\mu,i}^a(P)={\cal O}_i^{ab}(P)A_\m^b(P)
\ee
are the rotated gauge fields and
\be
\tilde{\Theta}_{aa}^i(P)={\cal O}_i^{ab}(P)
\Theta^i_{bc}(P){\cal O}_i^{ac}(P) 
\ee
are diagonal matrices.
The index $i$ in $\tilde{A}^a_{\m,i}$
has a different origin than in $A^a_{i\m}$ introduced in 
Eq.\ (\ref{withoutgauge}). For $\tilde{A}^a_{\m,i}$, it indicates 
that for each $i=1,2,3$ one has to perform a separate diagonalization. 
For $A^a_{i\m}$ it denoted the
projection corresponding to the projector ${\cal Q}_i^{\m\n}$. 

Note that the orthogonal matrix ${\cal O}_i(P)$ depends on $P^\m$.
For energies and momenta much larger than the superconducting gap parameter
the polarization tensor is explicitly (4-)transverse, $\Pi^3=0$,
and diagonal in $a, b$. In this case, ${\cal O}_i(P)\to {\bf 1}$. Consequently,
the gauge fields are not rotated. However, in the limit $p_0=0$, $p\to 0$
it is known that gluons and photons mix at least in the 2SC and CFL
phases, cf.\ Sec.\ \ref{higgs} and references therein, 
${\cal O}_i(P)\neq {\bf 1}$. 
Thus, the mixing angle between
gluons and photons, which will be discussed
in Sec.\ \ref{mixing}, is in general a function of $P^\m$ and interpolates
between a nonvanishing value at $p_0=0$, $p\to 0$ and zero when $p_0,p\to
\infty$. Note also that the orthogonal matrix ${\cal O}_i(P)$ depends on
$i$, i.e., it may be different for longitudinal and transverse modes. We
comment on this in more detail below.

{}From Eq.\ (\ref{rotatedfields}), we can immediately read off the inverse 
propagator for gluons and photons. It is
\be
{\Delta^{-1}}_{aa}^{\m\n}(P) =
\sum_{i=1}^3{\cal Q}_i^{\m\n}\tilde{\Theta}_{aa}^i(P) \,\, .
\ee
{}From the definition of $\tilde{\Theta}_{aa}^i(P)$ we conclude
\be  \label{diaglimit}
\tilde{\Theta}^i_{aa}(P) \equiv \left\{ \begin{array}{cl} P^2+
\tilde{\Pi}_{aa}^i(P) & \mbox{for}\quad i=1,2  \,\, ,
\\ \\
\frac{1}{\l}P^2+\tilde{\Pi}^3_{aa}(P) & \mbox{for}\quad i=3 \,\, ,
\end{array} \right. 
\ee
where 
\be \label{diag}
\tilde{\Pi}_{aa}^i(P)=\left\{ \begin{array}{cl}{\cal O}_i^{ab}(P)
\Pi^i_{bc}(P){\cal O}_i^{ac}(P) & \mbox{for}\quad i=1,3  \,\, ,
\\ \\
{\cal O}_2^{ab}(P)\left\{
\Pi_{bc}^2(P)-P^2N^2\Pi^4_{bd}(P)\,
\left[\Pi^3(P)\right]^{-1}_{de}\,\Pi^4_{ec}(P)\right\}{\cal O}_2^{ac}(P) 
& \mbox{for}\quad i=2 \,\, .
\end{array} \right.  
\ee
In the case $p_0=0$, using Eqs.\ (42) and (43)
of Ref.\ \cite{shovkovy} one realizes that
the extra term involving $\Pi^3$ and $\Pi^4$ 
for $i=2$ vanishes. Thus, in this case
one only has to diagonalize the original polarization tensors $\Pi^i_{ab}$.

Finally, we end up with the gauge boson propagator 
\be \label{phogluprop}
\Delta_{aa}^{\m\n}(P)=\frac{1}{P^2+\tilde{\Pi}_{aa}^1(P)}{\cal Q}_1^{\m\n}+ 
\frac{1}{P^2+\tilde{\Pi}_{aa}^2(P)}{\cal Q}_2^{\m\n}+ 
\frac{\l}{P^2+\l\tilde{\Pi}_{aa}^3(P)}{\cal Q}_3^{\m\n} \,\, .
\ee
Setting the gauge parameter $\l=0$, we are left with the transverse and 
(3-)longitudinal modes, in accordance with general principles
\cite{lee,lebellac}. The static color and electromagnetic properties
of the color superconductor are characterized by the Debye masses
$\tilde{m}_{D,a}$ and the Meissner masses $\tilde{m}_{M,a}$ which are 
defined as
\begin{subequations} \label{defmasses}
\bea 
\tilde{m}^2_{D,a}&\equiv& -\lim\limits_{p\to 0}\tilde{\Pi}_{aa}^2(0,{\bf p})
=-\lim\limits_{p\to 0}\tilde{\Pi}_{aa}^{00}(0,{\bf p}) \,\, ,\\
\tilde{m}^2_{M,a}&\equiv& -\lim\limits_{p\to 0}\tilde{\Pi}_{aa}^1(0,{\bf p})
=\frac{1}{2}\lim\limits_{p\to 0}
(\d^{ij}-\hat{p}^i\hat{p}^j)\tilde{\Pi}_{aa}^{ij}(0,{\bf p})
 \,\, . 
\eea
\end{subequations} 
Since the orthogonal matrices ${\cal O}_i(P)$ are regular in the limit 
$p_0= 0$, $p\to 0$, the masses can also be obtained by first computing
$\lim_{p\to 0}\Pi_{ab}^{\m\n}(0,{\bf p})$ and then
diagonalizing the resulting $9\times 9$ mass matrix. In Sec.\
\ref{calc}, we use this method to compute $\tilde{m}^2_{D,a}$
and $\tilde{m}^2_{M,a}$, since the diagonalization of the matrix
$\Pi_{ab}^{\m\n}(P)$ for arbitrary $P^\m$ is too difficult.

\subsection{The mixing angle} \label{mixing}

In this section, we investigate the structure of the orthogonal 
matrices ${\cal O}_i(P)$ which diagonalize the gauge field part of
the grand partition function. 
In general, the matrices ${\cal O}_i(P)$ mix all gluon components among 
themselves and with the photon. However, in the limit $p_0=0$, $p\to 0$
it turns out that 
the only non-zero off-diagonal elements of the tensor
$\Pi_{ab}^i\equiv \lim_{p\to 0} \Pi_{ab}^i(0,{\bf p})$ are 
$\Pi_{89}^i=\Pi_{8\g}^i=\Pi_{\g 8}^i$. 
Physically speaking, gluons do not mix among themselves and 
only the eighth gluon mixes with the photon. In this case, Eq.\ (\ref{diag})
reduces to the diagonalization of a $2\times 2$ matrix.   
Consequently, the diagonalization is determined by only one
parameter $\theta_i$ and the (nontrivial part of the) transformation 
operator reads
\be
{\cal O}_i=\left(\begin{array}{cc}
\cos\theta_i&\sin\theta_i\\
-\sin\theta_i&\cos\theta_i\end{array}\right) \,\, .
\ee
The new fields are
\begin{subequations} \label{newfields}
\bea 
\tilde{A}^8_{\m,i}&=&\cos\theta_i\,A^8_\m
+\sin\theta_i\,A_\m \,\, ,\\
\tilde{A}_{\m,i}&=&-\sin\theta_i\,A^8_\m
+\cos\theta_i\,A_\m 
\,\, .
\eea
\end{subequations}
The new polarization functions (eigenvalues of $\Pi_{ab}^i$) are 
\begin{subequations} \label{eigenval}
\bea
\tilde{\Pi}_{88}^i&=&\Pi_{88}^i\cos^2\theta_i+2\Pi_{8\g}^i
\sin\theta_i\cos\theta_i+\Pi^i_{\g\g}\sin^2\theta_i \,\, ,\\
\tilde{\Pi}^i_{\g\g}&=&\Pi_{88}^i\sin^2\theta_i-2\Pi_{8\g}^i
\sin\theta_i\cos\theta_i+\Pi^i_{\g\g}\cos^2\theta_i \,\, .
\eea
\end{subequations}
The mixing angle $\theta_i$ is given by
\be \label{theta}
\tan 2\theta_i=\frac{2\Pi_{8\g}^i}{\Pi_{88}^i-\Pi^i_{\g\g}} \,\, .
\ee
If $[\Pi_{8\g}^i]^2=\Pi_{88}^i\Pi^i_{\g\g}$, the determinant
of $\Pi_{ab}^i$ is zero, which means that there is a vanishing eigenvalue.
 In this case, we have
\be \label{cossintheta}
\cos^2\theta_i=\frac{\Pi_{88}^i}{\Pi_{88}^i+\Pi_{\g\g}^i} \,\, ,
\ee
and the new polarization tensors, Eqs.\ (\ref{eigenval}),
have the simple form 
\be \label{detvanish}
\tilde{\Pi}_{88}^i=\Pi_{88}^i+\Pi^i_{\g\g} \quad ,\qquad 
\tilde{\Pi}^i_{\g\g}=0 \,\, .
\ee
Physically, a vanishing polarization tensor for the rotated 
photon corresponds to the absence of the electromagnetic Meissner effect 
for $i=1$, or the absence of Debye screening for $i=2$.

Next, we show how to determine the specific mixing angles from the results of 
Sec.\ \ref{grouptheory}. In an ordinary superconductor,
the Meissner effect and
thus a non-vanishing Meissner mass $m_{M,\g}$ originate from a non-zero 
electric charge of a Cooper pair.
Besides this magnetic screening there is also  
electric screening of photons described by the photon Debye mass 
$m_{D,\g}$. Of course, in a color superconductor, in addition to the 
electric charge we also have to take into account the color charge.
The group-theoretical method from Sec.\ \ref{grouptheory} 
allows us to investigate whether there exists a (new) charge which generates
an unbroken symmetry. In this case, a Cooper pair is neutral with respect 
to this charge, and consequently one expects that there is
neither a Meissner effect nor Debye screening.
The new charge is a linear combination of electric and color charges.
Correspondingly, the associated gauge field is a linear combination of the 
photon and the eighth gluon field, which, in turn, defines the mixing
angle. The group-theoretical method only allows to identify a new charge, 
and thus does not distinguish between electric Debye or magnetic Meissner 
screening. Consequently, the mixing angles for longitudinal (electric) and 
transverse (magnetic) modes deduced by this method are identical, 
$\theta_1=\theta_2\equiv\theta$. 

From Sec.\ \ref{grouptheory} we know that in the polar phase (and in several
other spin-one phases) there is a
residual symmetry group $U(1)$ generated by a linear combination of 
the generators $Q\equiv 2q$ of $U(1)_{em}$ and $T_8$ of $SU(3)_c$.
Let us denote this residual group by $\tilde{U}(1)_{em}$ and the corresponding 
generator by $\tilde{Q}$. In general, the new charge generator is given by
\be \label{Qtilde}
\tilde{Q}=Q+\eta \, T_8 \,\, ,
\ee
where $\eta$ is a real coefficient. For instance, in the polar phase,
cf.\ Eqs.\ (\ref{genpolar}), $\eta = -2\sqrt{3}q$. 
In Table \ref{tablesymmetry}, we list the relevant quantities regarding
the spontaneous breakdown of the gauge symmetries $SU(3)_c\times U(1)_{em}$.
The results for the polar and CSL phases are taken 
from Sec.\ \ref{grouptheory}. Note that in the case of the CSL phase, there
is no nontrivial residual gauge group. The only residual symmetry is global,
cf.\ Fig.\ \ref{symmetries}. The corresponding results for the spin-zero
phases can be found using exactly the same method as presented in 
Sec.\ \ref{grouptheory}, see Refs.\ 
\cite{alford3,gorbar,litim,manuel,schmitt3}. The respective charge generators
$Q={\rm diag}(q_1,\ldots ,q_{N_f})$ are defined by the electric charges of 
the quark flavors. With $q_1=2/3$, $q_2=q_3=-1/3$ for $u$, $d$, and 
$s$ quarks, the charge generators shown in the table
apply to a system with $u$ and $d$ quarks (2SC) and a system with $d$, $s$, and
$u$ quarks (CFL).

\begin{table}  
\begin{center}
\begin{tabular}{|c||c|c|c|}
\hline
      & Q & generators of & $\eta$ 
\\ 
 & & residual gauge group & 
\\ \hline\hline
2SC   & 
      diag$(2/3,-1/3)$ & $T_1,T_2,T_3,Q+\eta T_8$ & $-1/\sqrt{3}$ 
\\ \hline
CFL   &  diag$(-1/3,-1/3,2/3) \,\, $ & $Q+\eta T_8$ & $2/\sqrt{3}$ 
\\ \hline
CSL   & $2q$ & 0 & -- 
\\ \hline 
polar & $2q$ & 
$\,\, T_1,T_2,T_3,Q+\eta T_8 \,\, $ & $\,\, -2\sqrt{3}\, q \,\,$ 
\\ \hline
\end{tabular}
\caption[Residual gauge groups]{Charge
generators $Q$, generators of the residual gauge group, and 
coefficients $\eta$ for all color-superconducting phases considered 
in this section. 
For the CSL and polar phases, the electric charge of the quark is denoted
by $q$.}
\label{tablesymmetry}
\end{center}
\end{table}

In order to deduce the mixing angle $\theta$, we rewrite the covariant 
derivative, $D_\m = \partial_\m - igA_\m^a T_a - ieA_\m Q$, in terms
of the new charge generator $\tilde{Q}$ and the  
linearly independent generator  
\be
\tilde{T}_8 = T_8 + \l Q \,\, ,
\ee
which belongs to the broken part of the group; $\l$ is a real 
constant to be determined below. We also have to replace
the gauge fields $A_\m^8$ and $A_\m$ by the rotated fields 
(\ref{newfields}) and the associated coupling constants $g$, $e$ by 
new coupling constants $\tilde{g}$, $\tilde{e}$. Thus, we demand the 
identity \cite{alford3,gorbar,manuel}
\be
gA_\m^8 T_8 + eA_\m Q  = \tilde{g}\tilde{A}_\m^8\tilde{T}_8 
+ \tilde{e}\tilde{A}_\m \tilde{Q} \,\, .
\ee
Inserting Eqs.\ (\ref{newfields}) for the rotated fields and the definitions
of $\tilde{Q}$ and $\tilde{T}_8$ we determine   
\be
\tilde{g}=g\cos\theta \quad , \qquad \tilde{e} = e\cos\theta \quad , \qquad
\l=-\eta\frac{e^2}{g^2} \,\, ,
\ee
and the mixing angle
\be \label{allangles}
\cos^2\theta=\frac{g^2}{g^2+\eta^2 e^2} \,\, .
\ee
Since $g\gg e$, the mixing angle in all three cases is small. Thus,
according to Eqs.\ (\ref{newfields}), the gluon almost remains a gluon
and the photon almost remains a photon and therefore it is justified to
call the rotated gauge bosons the new gluon and the new photon.

In the 2SC, CFL, and polar phases, the 
new charge is neither Debye- nor Meissner-screened. We therefore
expect that the polarization tensor for the rotated photon vanishes 
in the zero-energy, zero-momentum limit,
\be \label{zero}
\tilde{\Pi}_{\g\g}^{1,2}=0 \,\, .
\ee
As we shall see in the following section, this argument is not quite correct 
for the 2SC and polar phases, as it neglects the effect of the
unpaired blue quarks on the screening of electric and magnetic fields.
In the CSL phase, on the other hand, there is no residual 
$\tilde{U}(1)_{em}$ symmetry and we conclude that all gauge fields
experience the Meissner effect and Debye screening, since in this
case there does not exist any charge with respect to which a Cooper 
pair is neutral.
Concerning color screening, one concludes from Table \ref{tablesymmetry}, that
in the 2SC phase and the polar phase only the gluons 4-8 are
screened while in the CFL and CSL phases all gluons are screened 
(see also discussion in Sec.\ \ref{grouptheory}).

\subsection{Calculation of the polarization tensors} \label{calc}

In this section, we calculate the polarization tensors 
$\Pi_{ab}^{\m\n}(P)$ given in Eq.\ (\ref{poltensors}) in the limit 
\mbox{$p_0=0$}, \mbox{$p\to 0$}. In this case, the Debye and Meissner masses, 
cf.\ Eqs.\ (\ref{defmasses}), are obtained from the coefficients 
of the first two projectors in the decomposition (\ref{decompose}). 
They will be calculated in the next section.
Here, we first derive a general expression for the polarization tensor 
that holds for all different phases. Then we insert the order parameters
of the 2SC, CFL, polar, and CSL phases, and show the results for
each phase separately.

We start from Eq.\ (\ref{poltensors}) and first perform the trace
over Nambu-Gor'kov space,
\bea \label{nambu}
\Pi_{ab}^{\m\n}(P)&=&\frac{1}{2}\frac{T}{V}\sum_K
\left\{{\rm Tr}[\Gamma_a^\m \, G^+(K) \, \Gamma_b^\n G^+(K-P)]
+{\rm Tr}[\overline{\Gamma}_a^\m \, G^-(K) \, \overline{\Gamma}_b^\n \, G^-(K-P)]
\right. 
\nonumber \\
&&\left.+{\rm Tr}[\Gamma_a^\m \, \Xi^-(K) \, \overline{\Gamma}_b^\n \, 
\Xi^+(K-P)]
+{\rm Tr}[\overline{\Gamma}_a^\m \, \Xi^+(K)\, \Gamma_b^\n \, \Xi^-(K-P)] 
\right\} \,\, ,
\eea
where the traces now run over color, flavor, and Dirac space. 
In the following, we first consider the traces with the quark propagators
$G^\pm$ and afterwards investigate the traces containing the anomalous 
propagators $\Xi^\pm$. 

In order to find the results for the former, we first perform the
Matsubara sum. This is completely analogous to the calculation of
Ref.\ \cite{meissner2}. The only difference is our more compact
notation with the help of the projectors ${\cal P}_{\bf k}^r$, cf.\
Eq.\ (\ref{prop}). Thus, abbreviating $K_1\equiv K$, $K_2\equiv K-P$, and
$k_i\equiv |{\bf k}_i|$ for $i=1,2$, we conclude 
\begin{subequations} \label{matsub1}
\bea 
&&T\,\sum_{k_0}{\rm Tr}
\left[\Gamma_a^\m G^+(K_1)\Gamma_b^\n G^+(K_2)\right] 
\non
&&\hspace{2cm} = \sum_{e_1,e_2}\sum_{r,s}{\rm Tr}\left[\Gamma_a^\m\,\g_0\,
{\cal P}_{{\bf k}_1}^r\,\Lambda_{{\bf k}_1}^{-e_1} \,  
\Gamma_b^\n \,\g_0\,{\cal P}_{{\bf k}_2}^s\,
\Lambda_{{\bf k}_2}^{-e_2}\right]\, v_{e_1e_2}^{rs,+}
(k_1,k_2,p_0) \,\, , \\
&&T\,\sum_{k_0}{\rm Tr}
\left[\overline{\Gamma}_a^\m G^-(K_1)\overline{\Gamma}_b^\n G^-(K_2)\right] 
\non
&&\hspace{2cm} =
\sum_{e_1,e_2}\sum_{r,s}{\rm Tr}\left[\overline{\Gamma}_a^\m\,\g_0\,
{\cal P}_{{\bf k}_1}^r\,\Lambda_{{\bf k}_1}^{e_1} \,  
\overline{\Gamma}_b^\n \,\g_0\,{\cal P}_{{\bf k}_2}^s\,
\Lambda_{{\bf k}_2}^{e_2}\right]\, v_{e_1e_2}^{rs,-}
(k_1,k_2,p_0) \,\, ,
\eea
\end{subequations}
where (cf.\ Eq.\ (40) of Ref.\ \cite{meissner2})
\begin{subequations} \label{defv}
\bea 
v_{e_1e_2}^{rs,+}(k_1,k_2,p_0) &\equiv&
-\left(\frac{n_{1,r}(1-n_{2,s})}{p_0+\e_{1,r}+\e_{2,s}}-\frac{(1-n_{1,r})n_{2,s}}
{p_0-\e_{1,r}-\e_{2,s}}\right)(1-N_{1,r}-N_{2,s}) \nonumber \\
&&-\left(\frac{(1-n_{1,r})(1-n_{2,s})}{p_0-\e_{1,r}+\e_{2,s}}-\frac{n_{1,r}
 n_{2,s}}
{p_0+\e_{1,r}-\e_{2,s}}\right)(N_{1,r}-N_{2,s}) \,\, ,  \\
v_{e_1e_2}^{rs,-}(k_1,k_2,p_0) &\equiv&
-\left(\frac{(1-n_{1,r})n_{2,s}}{p_0+\e_{1,r}+\e_{2,s}}-\frac{n_{1,r}(1-n_{2,s})}
{p_0-\e_{1,r}-\e_{2,s}}\right)(1-N_{1,r}-N_{2,s}) \nonumber \\
&&-\left(\frac{n_{1,r} n_{2,s}}{p_0-\e_{1,r}+\e_{2,s}}-\frac{(1-n_{1,r})(1-n_{2,s})}
{p_0+\e_{1,r}-\e_{2,s}}\right)(N_{1,r}-N_{2,s}) \,\, .
\eea
\end{subequations}
Here, we abbreviated
\be \label{abbreviations}
\e_{i,r}\equiv \e^{e_i}_{k_i,r} \quad,\qquad
n_{i,r}\equiv\frac{\e_{i,r}+\m-e_ik_i}{2\e_{i,r}}
\quad,\qquad N_{i,r}\equiv\frac{1}{\exp(\e_{i,r}/T)+1} \qquad
(i=1,2) \,\, .
\ee 
Note that, for $p_0=0$, we have
\be \label{spurious}
v_{e_1e_2}^{rs,+}(k_1,k_2,0)=
v_{e_1e_2}^{rs,-}(k_1,k_2,0)\equiv 
v_{e_1e_2}^{rs}(k_1,k_2,0) \,\, .
\ee

Next we discuss the traces containing the anomalous propagators. Again
the Matsubara sum is completely analogous to the calculation in 
Ref.\ \cite{meissner2}.
Therefore, using Eq.\ (\ref{aprop}), we obtain 
\begin{subequations} \label{matsub2}
\bea 
&&T\, \sum_{k_0}{\rm Tr}\left[\Gamma_a^\m\Xi^-(K_1)\,\overline{\Gamma}_b^\n\, 
\Xi^+(K_2)\right]
\non
&&\hspace{1cm}=\sum_{e_1,e_2}\sum_{r,s}
{\rm Tr}\left[\Gamma_a^\m {\cal M}_{{\bf k}_1}^\dag
{\cal P}_{{\bf k}_1}^r\Lambda_{{\bf k}_1}^{e_1}\overline{\Gamma}_b^\n
\g_0 {\cal M}_{{\bf k}_2} \g_0
{\cal P}_{{\bf k}_2}^s\Lambda_{{\bf k}_2}^{-e_2}\right]\, 
w_{e_1e_2}^{rs}(k_1,k_2,p_0) \,\, , \\
&&T\, \sum_{k_0}{\rm Tr}\left[\overline{\Gamma}_a^\m\Xi^+(K_1)\,\Gamma_b^\n\, 
\Xi^-(K_2)\right]
\non
&&\hspace{1cm}=\sum_{e_1,e_2}\sum_{r,s}
{\rm Tr}\left[\overline{\Gamma}_a^\m \g_0 {\cal M}_{{\bf k}_1} \g_0
{\cal P}_{{\bf k}_1}^r\Lambda_{{\bf k}_1}^{-e_1}\Gamma_b^\n
{\cal M}_{{\bf k}_2}^\dag
{\cal P}_{{\bf k}_2}^s\Lambda_{{\bf k}_2}^{e_2}\right]\, 
w_{e_1e_2}^{rs}(k_1,k_2,p_0) \,\, ,
\eea
\end{subequations}
where (cf.\ Eq.\ (93) of Ref.\ \cite{meissner2})
\bea  \label{defw}
w_{e_1e_2}^{rs}(k_1,k_2,p_0)&\equiv&
\frac{\phi_{1,r}\phi_{2,s}}{4\e_{1,r}\e_{2,s}}
\left[\left(\frac{1}{p_0+\e_{1,r}+\e_{2,s}}-\frac{1}{p_0-\e_{1,r}-\e_{2,s}}\right)
(1-N_{1,r}-N_{2,s}) \right. \nonumber \\
&&\left.-\left(\frac{1}{p_0-\e_{1,r}+\e_{2,s}}-\frac{1}{p_0+\e_{1,r}-\e_{2,s}}
\right)(N_{1,r}-N_{2,s})\right] \,\,. 
\eea
Here,
\be
\phi_{i,r}\equiv\phi^{e_i}(\e_{i,r},{\bf k}_i)
\ee
is the gap function on the quasiparticle mass shell given by the 
excitation branch $k_0=\e_{i,r}$. 
In the derivation of Eqs.\ (\ref{matsub1}) and (\ref{matsub2}), we 
assumed that, in the 2SC and CFL phases, the chemical potentials 
for all $N_f$ quark flavors are the 
same, $\m_1=\ldots=\m_{N_f}\equiv\m$. In this case, the functions $v$ and $w$
only depend on the single chemical potential $\m$ and can thus be factored
out of the flavor trace. In the cases where quarks of the 
same flavor form Cooper pairs, i.e., in the polar and CSL phases,
our formalism also allows for the treatment of a system with $N_f>1$ and
different chemical potentials $\m_n$, $n=1,\ldots,N_f$. 
Then, $v$ and $w$ depend on the quark flavor through $\m_n$ and have to 
be included into the trace over flavor space. 
 
Inserting Eqs.\ (\ref{matsub1}) and
(\ref{matsub2}) into Eq.\ (\ref{nambu}), we obtain for the
general polarization tensor
\begin{samepage}
\bea \label{general}
\Pi_{ab}^{\m\n}(P)&=&\frac{1}{2}\int\frac{d^3\bf k}{(2\p)^3} 
\sum_{e_1,e_2}\sum_{r,s} \Big\{
{\rm Tr}\left[\Gamma_a^\m\,\g_0\,
{\cal P}_{{\bf k}_1}^r\,\Lambda_{{\bf k}_1}^{-e_1} \,  
\Gamma_b^\n \,\g_0\,{\cal P}_{{\bf k}_2}^s\,
\Lambda_{{\bf k}_2}^{-e_2}\right]\, v_{e_1e_2}^{rs,+}
(k_1,k_2,p_0) \nonumber \\
&& \hspace*{2cm}+\,{\rm Tr}\left[\overline{\Gamma}_a^\m\,\g_0\,
{\cal P}_{{\bf k}_1}^r\,\Lambda_{{\bf k}_1}^{e_1} \,  
\overline{\Gamma}_b^\n \,\g_0\,{\cal P}_{{\bf k}_2}^s\,
\Lambda_{{\bf k}_2}^{e_2}\right]\, v_{e_1e_2}^{rs,-}
(k_1,k_2,p_0)  \nonumber \\
&& \hspace*{2cm}+\,{\rm Tr}\left[\Gamma_a^\m{\cal M}_{{\bf k}_1}^\dag
{\cal P}_{{\bf k}_1}^r\Lambda_{{\bf k}_1}^{e_1}\overline{\Gamma}_b^\n
\g_0{\cal M}_{{\bf k}_2}\g_0
{\cal P}_{{\bf k}_2}^s\Lambda_{{\bf k}_2}^{-e_2}\right]\, 
w_{e_1e_2}^{rs}(k_1,k_2,p_0)  \nonumber \\
&&\hspace*{2cm}+\,{\rm Tr}\left[\overline{\Gamma}_a^\m 
\g_0{\cal M}_{{\bf k}_1}\g_0
{\cal P}_{{\bf k}_1}^r\Lambda_{{\bf k}_1}^{-e_1}\Gamma_b^\n
{\cal M}_{{\bf k}_2}^\dag
{\cal P}_{{\bf k}_2}^s\Lambda_{{\bf k}_2}^{e_2}\right]\, 
w_{e_1e_2}^{rs}(k_1,k_2,p_0) \Big\} \,\, .
\eea
\end{samepage}
In the following, we focus on the special case $p_0=0$ and $p\to 0$,
i.e, ${\bf k}_2\to{\bf k}_1\equiv {\bf k}$.  
In this limit, the traces only depend on $\uk$
and the functions $v$ and $w$ only on $k\equiv |{\bf k}|$.  
Thus, the $d^3{\bf k}$ integral factorizes into an angular 
and a radial part. With the abbreviations
\begin{subequations} \label{VWdef}
\bea 
{\cal V}_{ab,e_1e_2}^{\m\n,rs}
&\equiv&\frac{1}{2}\int\frac{d\Omega_{\bf k}}{(2\p)^3}
\left\{ {\rm Tr}\left[\Gamma_a^\m\,\g_0\,{\cal P}_{\bf k}^r\,
\Lambda_{\bf k}^{-e_1}
\,\Gamma_b^\n\,\g_0\,{\cal P}_{\bf k}^s\,\Lambda_{\bf k}^{-e_2}\right] 
\right.\non
&& \hspace{1.5cm}\left.+\;{\rm Tr}\left[\overline{\Gamma}_a^\m\,
\g_0\,{\cal P}_{\bf k}^r\,\Lambda_{\bf k}^{e_1}
\,\overline{\Gamma}_b^\n\,\g_0\,{\cal P}_{\bf k}^s\,\Lambda_{\bf k}^{e_2}
\right] \right\} \,\, , 
\\ 
{\cal W}_{ab,e_1 e_2}^{\m\n,rs}&\equiv&
\frac{1}{2}\int\frac{d\Omega_{\bf k}}{(2\p)^3}
\left\{
{\rm Tr}\left[\Gamma_a^\m\,{\cal M}_{\bf k}^\dag\,
{\cal P}_{\bf k}^r
\,\Lambda_{\bf k}^{e_1}\,\overline{\Gamma}_b^\n\,\g_0{\cal M}_{\bf k}\g_0
\,{\cal P}_{\bf k}^s\,\Lambda_{\bf k}^{-e_2}\right] \right.\nonumber \\  
&& \hspace{1.5cm} \left.+ \; {\rm Tr}\left[\overline{\Gamma}_a^\m\,
\g_0{\cal M}_{\bf k}\g_0\,{\cal P}_{\bf k}^r
\,\Lambda_{\bf k}^{-e_1}\,\Gamma_b^\n\,{\cal M}_{\bf k}^\dag
\,{\cal P}_{\bf k}^s\,\Lambda_{\bf k}^{e_2}\right] \right\} \,\, ,
\eea
\end{subequations}
we can write the polarization tensor as
\bea 
&&\Pi_{ab}^{\m\n}(0) \equiv \lim_{p\to 0} \Pi_{ab}^{\m\n}(0,{\bf p})=
\sum_{e_1,e_2}
\sum_{r,s}\left[{\cal V}_{ab,e_1e_2}^{\m\n,rs}
\, \lim\limits_{p\to 0}\int dk\,k^2 v_{e_1e_2}^{rs}(k_1,k_2,0) 
\right.\non
&& \left. \hspace{5.3cm}+\,{\cal W}_{ab,e_1e_2}^{\m\n,rs}
\,\lim\limits_{p\to 0}\int dk\,k^2 w_{e_1e_2}^{rs}(k_1,k_2,0)\right] \,\, .
\label{polshort}
\eea
Note that only the angular integrals over the color, flavor, and Dirac traces, 
defined by ${\cal V}$ and ${\cal W}$, depend on the symmetries
of the order parameter and thus have to be calculated separately for each 
phase. Therefore, we first consider the $dk$ integrals which are the same 
for all cases. 

In order to see how the two different quasiparticle excitations 
branches (labelled by $r$, $s$), 
as well as normal and
anomalous propagation of the respective excitations (represented by the 
functions $v$, $w$) contribute in the final expressions for the polarization 
tensors in the zero-energy, zero-momentum limit, it is convenient 
to define the quantities
\begin{subequations} \label{kintegrals}
\bea
v^{rs}&\equiv& \frac{1}{\m^2}\lim\limits_{p\to 0}\int dk\,k^2 \left[
v_{++}^{rs}(k_1,k_2,0) + v_{--}^{rs}(k_1,k_2,0)\right] \,\, , \\
\bar{v}^{rs}&\equiv& \frac{1}{\m^2}\lim\limits_{p\to 0}\int dk\,k^2 \left[
v_{+-}^{rs}(k_1,k_2,0) + v_{-+}^{rs}(k_1,k_2,0)\right] \,\, , \\
w^{rs}&\equiv& \frac{1}{\m^2}\lim\limits_{p\to 0}\int dk\,k^2 \left[
w_{++}^{rs}(k_1,k_2,0) + w_{--}^{rs}(k_1,k_2,0)\right] \,\, , \\ 
\bar{w}^{rs}&\equiv& \frac{1}{\m^2}\lim\limits_{p\to 0}\int dk\,k^2 \left[
w_{+-}^{rs}(k_1,k_2,0) + w_{-+}^{rs}(k_1,k_2,0)\right] \,\, .
\eea
\end{subequations}
These quantities are dimensionless since 
$v_{e_1e_2}^{rs}(k_1,k_2,0)$ and $w_{e_1e_2}^{rs}(k_1,k_2,0)$
have the dimension [1/energy] (cf.\ the definitions in Eqs.\ (\ref{defv}) and
(\ref{defw})).
Combining particle-particle ($e_1=e_2=+$) and antiparticle-antiparticle
($e_1=e_2=-$), as well as particle-antiparticle ($e_1=-e_2=\pm$) 
contributions is possible since the corresponding integrals ${\cal V}, 
{\cal W}$ multiplying these
terms turn out to be the same.
In the definitions of $v^{rs}$, $\bar{v}^{rs}$
and $w^{rs}$, $\bar{w}^{rs}$ we divided by the square of the
quark chemical potential $\m^2$ in order to make these quantities 
independent of the quark flavor. This will be 
convenient for the results of the spin-one phases, 
where the formation of Cooper pairs 
is possible for different chemical potentials for each quark flavor.   
In Appendix \ref{Appquark}, we present the calculation of the relevant 
integrals defined in Eqs.\ (\ref{kintegrals}).
As in Refs.\ \cite{meissner2,meissner3}, we neglect the
antiparticle gap, $\phi^-\simeq 0$, and compute the integrals up to
leading order. In Table \ref{tablevw} we collect all results.
Some integrals vanish since they are proportional to a
vanishing gap. This is the case for $\bar{w}^{rs}=0$ (for all $r,s$), 
since these integrals are proportional to at least one antiparticle gap. 
Clearly, for $T\ge T_c$, the gap vanishes in all phases and thus 
$w^{rs}=\bar{w}^{rs}=0$.   

\begin{table} 
\begin{center}
\begin{tabular}{|c|c||c|c|c|c|c|c|c|c|}
\hline
 & & $v^{11}$ &$v^{22}$ &$v^{12}$ & $v^{21}$ 
&$\bar{v}^{11}$ &$\bar{v}^{22}$ &$\bar{v}^{12}$ & $\bar{v}^{21}$
\\
\hline
\hline
$T=0$ & 2SC, polar & $-\frac{1}{2}$ & $-1$ & 
\multicolumn{2}{c}{$-\frac{1}{2}$}\vline &  
\multicolumn{4}{c}{$\frac{1}{2}$}\vline \\
\cline{2-10}
& CFL, CSL & \multicolumn{4}{c}{$-\frac{1}{2}$}\vline &
\multicolumn{4}{c}{$\frac{1}{2}$}\vline \\
\hline
$T\ge T_c$ & all phases & \multicolumn{4}{c}{$-1$}\vline &  
\multicolumn{4}{c}{$\frac{1}{2}$}\vline \\
\hline
\hline
 & & $w^{11}$ &$w^{22}$ &$w^{12}$ & $w^{21}$ 
&$\bar{w}^{11}$ &$\bar{w}^{22}$ &$\bar{w}^{12}$ & $\bar{w}^{21}$ \\
\hline
\hline
$T=0$ & 2SC, polar &  $\frac{1}{2}$ & 
\multicolumn{3}{c}{-}\vline &  
0 & \multicolumn{3}{c}{-}\vline \\ 
\cline{2-10}
& CFL, CSL & $\frac{1}{8}$ & 
$\frac{1}{2}$ & \multicolumn{2}{c}{$\frac{1}{3}\ln 2$}\vline
 & \multicolumn{4}{c}{0}\vline \\
\hline
$T\ge T_c$ & all phases &  \multicolumn{8}{c}{0} 
\vline\\
\hline
\end{tabular}
\caption[Integral values]{Leading order results  
for the integrals defined in Eqs.\ (\ref{kintegrals}).  
The indices 1 and 2 correspond to the two gaps of each phase (the second
one vanishing in the 2SC and the polar phase), while $v$ 
corresponds to the trace over quasiparticle propagators $G^\pm$ and
$w$ to the trace over anomalous propagators $\Xi^\pm$. The fields with no
entry indicate that these values do not occur in the calculations.}
\label{tablevw}
\end{center}
\end{table} 

In order to discuss the traces and the angular integral in 
Eq.\ (\ref{polshort}), we have to distinguish between the several
phases since the special form of the gap matrix, Eq.\ (\ref{gm2SC}),
is explicitly involved. Therefore, in the following sections we discuss
the 2SC, CFL, polar, and CSL phases separately
and compute the Debye and Meissner masses for photons and gluons in each 
phase. 
For the 2SC phase \cite{meissner2} and the CFL phase \cite{meissner3,son},
the results for the gluons are already known. Also the masses of
the rotated gauge bosons, where the rotation is given by the new 
generator $\tilde{Q}$, cf.\ Eq.\ (\ref{Qtilde}), were considered for these
two phases \cite{litim}. Nevertheless, we will briefly discuss also these 
cases, since we first want to establish our notation and second, we will show
that for the 2SC phase, there is actually no mixing between electric
gluons and photons, i.e., the longitudinal mixing angle $\theta_2$ is 
zero. Consequently, there is no rotated electric photon.

%\newpage
\subsection{The 2SC, CFL, polar, and CSL phases}\label{mixingresults}

In Table \ref{tableMP}, we collect the relevant color-flavor-Dirac matrices 
for the considered phases. They have been introduced in 
Sec.\ \ref{gapeqsolution}. In the case of the spin-one phases, 
we consider the mixed gaps, i.e., $\a=\b$ in Eq.\ (\ref{Mk}). For
simplicity, we choose $\a=\b=1$. Although this choice of the 
coefficients $\a$, $\b$ violates the normalization 
(\ref{normalize}), it does not change the results for the 
Debye and Meissner masses compared to the case $\a=\b=1/\sqrt{2}$.
To this end, note that a rescaling
\be
{\cal M}_{\bf k} \to c\,{\cal M}_{\bf k}
\ee
leads to
\begin{subequations} \label{VWrescale}
\bea 
{\cal V}&\to& {\cal V} \,\, ,\qquad
\hspace{0.5cm} v^{rs} \, ,\;\bar{v}^{rs} \to v^{rs} \, , 
\; \bar{v}^{rs} \,\, , \\
{\cal W} &\to& c^2\,{\cal W}
\,\, ,\qquad w^{rs} \, , \; \bar{w}^{rs} \to 
\frac{1}{c^2}\,w^{rs}\, ,\;\frac{1}{c^2}\,\bar{w}^{rs} \,\, ,
\eea
\end{subequations}
where the results from Appendix \ref{Appquark} have been used. Therefore,
the result for the polarization tensor (\ref{polshort}) is not affected by this
rescaling.

\begin{table}
\begin{center} 
\begin{tabular}{|c||c|c|c|} 
\hline
 & ${\cal M}_{\bf k}$ & ${\cal P}^1_{\bf k}$ & ${\cal P}^2_{\bf k}$ \\
\hline\hline
2SC & $J_3\t_2\g_5$ & $J_3^2$ & $1-J_3^2$ 
\\
\hline
CFL & ${\bf J}\cdot{\bf I}\,\g_5$ & $\frac{1}{3}[({\bf J}\cdot{\bf I})^2
-1)]$ & $\frac{1}{3}[4-({\bf J}\cdot{\bf I})^2]$ \\
\hline
polar &  $\quad J_3[\hat{k}^z+\g_\perp^z({\bf k})]\quad$ & 
 $J_3^2$ & $1-J_3^2$ \\
\hline
CSL & $\quad{\bf J}\cdot[\uk+\gperp({\bf k})]\quad$ & 
$\quad\frac{1}{3}[\uk+\gperp({\bf k})][\uk-\gperp({\bf k})]\quad$ & 
$1-{\cal P}^1_{\bf k}$ \\
\hline
\end{tabular}
\caption[Gap matrices and projectors]{Relevant color-flavor-Dirac matrices 
for the calculation of
the Debye and Meissner masses in a given color-superconducting phase.
The matrix ${\cal M}_{\bf k}$ reflects the symmetries of the 
various gap matrices. For the definition of the projectors 
${\cal P}_{\bf k}^{1,2}$ see Eq.\ (\ref{proj}).
In color space, we use the matrices $(J_i)_{jk} = -i\e_{ijk}$, 
($i,j,k=1,2,3$); in flavor space, we use $(I_i)_{jk} = -i\e_{ijk}$ 
and the second Pauli matrix $\t_2$. In Dirac space, we defined 
$\gperp({\bf k})\equiv\vg-\uk\,\vg\cdot\uk$. }\label{tableMP}
\end{center}
\end{table}

\subsubsection{The 2SC phase} \label{2SCphase}

\begin{center}
{\it 1. Gluon polarization tensor ($a,b\le 8$)}
\end{center}

Inserting the matrices given in the second line of Table \ref{tableMP} into 
Eqs.\ (\ref{VWdef}), we obtain 
\begin{subequations} \label{VW2SCgluon}
\bea
{\cal V}_{ab,e_1e_2}^{\m\n,rs}&=&
g^2\left\{{\rm Tr}[T_a\,{\cal P}^r\,T_b\,{\cal P}^s]
\;{\cal T}_{-e_1,-e_2}^{\m\n}
 + {\rm Tr}[T_a^T\,{\cal P}^r\,T_b^T\,{\cal P}^s]\;{\cal T}_{e_1,e_2}^{\m\n}
\right\} \,\, ,
 \\
{\cal W}_{ab,e_1e_2}^{\m\n,rs}&=&g^2\left\{{\rm Tr}[T_a\,J_3\,{\cal P}^r\,T_b^T
\,J_3\,{\cal P}^s]\;{\cal U}_{e_1,-e_2}^{\m\n} + 
{\rm Tr}[T_a^T\,J_3\,{\cal P}^r\,T_b\,J_3\,{\cal P}^s]
\;{\cal U}_{-e_1,e_2}^{\m\n}\right\} \,\, , 
\eea
\end{subequations}
where the traces only run over color space and where we defined
\begin{subequations}\label{TUdef} 
\bea 
{\cal T}_{e_1,e_2}^{\m\n}&\equiv&\int\frac{d\Omega_{\bf k}}{(2\p)^3}
{\rm Tr}\left[\g^\m\,\g_0\,\Lambda_{\bf k}^{e_1}
\,\g^\n\,\g_0\,\Lambda_{\bf k}^{e_2}\right] \,\, , \\
{\cal U}_{e_1,e_2}^{\m\n}&\equiv&\int\frac{d\Omega_{\bf k}}{(2\p)^3}
{\rm Tr}\left[\g^\m\,\g^5\,\Lambda_{\bf k}^{e_1}\,\g^\n
\,\g^5\,\Lambda_{\bf k}^{e_2}\right] \,\, .
\eea
\end{subequations}
Here, the traces only run over Dirac space.
We used the fact that the projectors 
${\cal P}^{r,s}\equiv{\cal P}_{\bf k}^{r,s}$
do not depend on the quark momentum ${\bf k}$ and that 
the color and Dirac traces factorize. 
Furthermore, the trivial flavor trace was already performed, yielding a 
factor 2. The angular integrals over the Dirac traces, Eqs. (\ref{TUdef}), are
easily evaluated,
\begin{subequations} \label{TU}
\bea
{\cal T}^{00}_{e_1,e_2}&=&-{\cal U}^{00}_{e_1,-e_2}=\frac{1}{2\p^2}
(1+e_1e_2) 
\,\, , \label{TU00}\\
{\cal T}^{0i}_{e_1,e_2}&=&{\cal T}^{i0}_{e_1,e_2}=
{\cal U}^{0i}_{e_1,-e_2}={\cal U}^{i0}_{e_1,-e_2}=0 
\,\, , \label{TU0i}\\
{\cal T}^{ij}_{e_1,e_2}&=&{\cal U}^{ij}_{e_1,-e_2}=\frac{1}{2\p^2}\d_{ij}
(1-\frac{1}{3}e_1e_2)     \,\, , \label{TUij}
\eea
\end{subequations}
where $i,j=1,2,3$. 
For the evaluation of the color traces note that
$J_3{\cal P}^1=J_3$ and $J_3{\cal P}^2=0$. 
We find that ${\cal V}$ and ${\cal W}$, given in Eqs.\ (\ref{VW2SCgluon}),
are diagonal in the adjoint color indices $a$ and $b$.

\begin{enumerate}
\renewcommand{\labelenumi}{(\alph{enumi})}
\item $\m=\n=0$.
With Eq.\ (\ref{TU00}) we obtain after performing the color traces
\be \label{2SCgluon00}
\Pi^{00}_{ab}(0)=\d_{ab}\frac{2g^2\m^2}{\p^2}\left\{
\begin{array}{cl} \frac{1}{2}v^{11}+\frac{1}{2}w^{11} &
\quad \mbox{for} \quad a=1,2,3 \,\, , \\ \\
\frac{1}{4}(v^{12}+v^{21}) & \quad \mbox{for} \quad 
a=4-7 \,\, ,\\ \\
\frac{1}{6}v^{11}+\frac{1}{3}v^{22}-\frac{1}{6}w^{11} &
\quad \mbox{for} \quad a=8 \,\, . 
\end{array}
\right. 
\ee 

\item $\m=0,\n=i$ and $\m=i,\n=0$.
Due to Eq.\ (\ref{TU0i}), 
\be
\Pi_{ab}^{0i}(0) = \Pi_{ab}^{i0}(0) = 0 \,\, .
\ee 

\item $\m=i,\n=j$.
Due to Eq.\ (\ref{TUij}), the polarization tensor is diagonal
in spatial indices $i$, $j$. We obtain 
\be \label{2SCgluonij}
\Pi^{ij}_{ab}(0)=\d_{ab}\d^{ij}\frac{2g^2\m^2}{3\p^2}
\left\{
\begin{array}{cl} \frac{1}{2}(v^{11}+2\bar{v}^{11}) - 
\frac{1}{2}(w^{11}+2\bar{w}^{11}) &
\quad \mbox{for} \quad a=1,2,3 \,\, ,\\ \\
\frac{1}{4}[(v^{12}+v^{21})+2(\bar{v}^{12}+\bar{v}^{21})]
 & \quad \mbox{for} \quad 
a=4-7 \,\, ,\\ \\
\frac{1}{6}(v^{11}+2\bar{v}^{11})+\frac{1}{3}(v^{22}
+2\bar{v}^{22})+\frac{1}{6}(w^{11}+2\bar{w}^{11}) &
\quad \mbox{for} \quad a=8 \,\, . 
\end{array}
\right. 
\ee

\end{enumerate}

\begin{center}
{\it 2. Mixed polarization tensor ($a\le 8, b=9$ and $a=9, b\le 8$)}
\end{center}

For a system in the color-superconducting 2SC phase where quarks
with the electric charges $q_1$ and $q_2$ form Cooper pairs,
the electric charge
generator, introduced in Eq.\ (\ref{quarkpartition}), is given by
$Q={\rm diag}(q_1,q_2)$. Thus, we obtain
\begin{subequations} \label{VW2SCmixed}
\bea
{\cal V}_{a\g,e_1e_2}^{\m\n,rs}&=& 
\frac{1}{2}\,eg(q_1+q_2)\left\{ {\rm Tr}[T_a\,{\cal P}^r\,
{\cal P}^s]\,{\cal T}_{-e_1,-e_2}^{\m\n} +{\rm Tr}[T_a^T\,{\cal P}^r\,
{\cal P}^s]\,{\cal T}_{e_1,e_2}^{\m\n} \right\} \,\, , \\
{\cal W}_{a\g,e_1e_2}^{\m\n,rs}&=&
\frac{1}{2}\,eg(q_1+q_2)\left\{ {\rm Tr}[T_a\,J_3\,{\cal P}^r\,
J_3\,{\cal P}^s]\,
{\cal U}_{e_1,-e_2}^{\m\n} + {\rm Tr}[T_a^T\,J_3\,{\cal P}^r\,J_3\,
{\cal P}^s]\,{\cal U}_{-e_1,e_2}^{\m\n}\right\}
\,\, . 
\eea
\end{subequations}
It is not difficult to show that the polarization tensor 
is symmetric under the exchange of photon and gluon indices, 
\be \label{symmetric}
\Pi_{a\g}^{\m\n}(0)=
\Pi_{\g a}^{\m\n}(0) \,\, .
\ee
Using $J_3^3=J_3$, all color traces in Eq.\ (\ref{VW2SCmixed}) reduce 
to ${\rm Tr}[T_a J_3^2]=\d_{a8}/\sqrt{3}$. Therefore, we obtain
for the various Dirac components of the tensor:

\begin{enumerate}
\renewcommand{\labelenumi}{(\alph{enumi})}
\item $\m=\n=0$.
\be \label{2SCmixed00}
\Pi_{a\g}^{00}(0)=(q_1+q_2)\frac{eg\,\m^2}{\sqrt{3}\p^2}\left\{
\begin{array}{cl} 0 &  \quad \mbox{for} \quad a=1-7 \,\, ,\\ \\
v^{11}-v^{22}-w^{11} &  \quad \mbox{for} \quad a=8 \,\, .\\
\end{array} \right.
\ee

\item $\m=0,\n=i$ and $\m=i,\n=0$.
\be
\Pi_{a\g}^{0i}(0) = \Pi_{a\g}^{i0}(0) = 0 \,\, .
\ee 

\item $\m=i,\n=j$.
\be \label{2SCmixedij}
\Pi_{a\g}^{ij}(0)=\d_{ij}(q_1+q_2)\frac{eg\,\m^2}{3\sqrt{3}\p^2}\left\{
\begin{array}{cl} 0 &  \quad \mbox{for} \quad a=1-7 \,\, ,\\ \\
(v^{11}+2\bar{v}^{11})-(v^{22}+2\bar{v}^{22})+
(w^{11}+2\bar{w}^{11}) &  \quad \mbox{for} \quad a=8 \,\, .\\
\end{array} \right.
\ee

\end{enumerate}

\newpage
\begin{center}
{\it 3. Photon polarization tensor ($a=b=9$)}
\end{center}

In this case, the tensors ${\cal V}$, ${\cal W}$ are
\begin{subequations} \label{VW2SCphoton}
\bea
{\cal V}_{\g\g,e_1e_2}^{\m\n,rs}&=& 
\frac{1}{2}\,e^2(q_1^2+q_2^2)\,{\rm Tr}[{\cal P}^r\,
{\cal P}^s]\,\left( {\cal T}_{-e_1,-e_2}^{\m\n} + {\cal T}_{e_1,e_2}^{\m\n}
\right) \,\, ,
\\
{\cal W}_{\g\g,e_1e_2}^{\m\n,rs}&=&
e^2q_1q_2\,{\rm Tr}[J_3\,{\cal P}^r\,J_3\,{\cal P}^s]
\,\left( {\cal U}_{e_1,-e_2}^{\m\n} +{\cal U}_{-e_1,e_2}^{\m\n} \right) 
\,\, . 
\eea
\end{subequations}
Here, we performed the flavor traces ${\rm Tr}[Q^2]=q_1^2+q_2^2$ and 
${\rm Tr}[Q\t_2Q\t_2]=2q_1q_2$.
After performing the color traces and the sums over $e_1$, $e_2$
and $r$, $s$, the results for the different components are as follows. 

\begin{enumerate}
\renewcommand{\labelenumi}{(\alph{enumi})}
\item $\m=\n=0$.
\be \label{2SCphoton00}
\Pi_{\g\g}^{00}(0)=\frac{e^2\m^2}{\p^2}
\left[(q_1^2+q_2^2)\,(2v^{11}+v^{22})
-4q_1q_2w^{11}\right]
\,\, .
\ee

\item $\m=0,\n=i$ and $\m=i,\n=0$.
\be
\Pi_{\g\g}^{0i}(0) = \Pi_{\g\g}^{i0}(0) = 0 \,\, .
\ee 

\item $\m=i,\n=j$.
\be  \label{2SCphotonij}
\Pi_{\g\g}^{ij}(0)=\d_{ij}
\frac{e^2\m^2}{3\p^2}
\left\{(q_1^2+q_2^2)\,\left[2(v^{11}+2\bar{v}^{11})+
(v^{22}+2\bar{v}^{22})\right]+
4q_1q_2\,(w^{11}+2\bar{w}^{11})\right\}  \,\, .
\ee

\end{enumerate}

\subsubsection{The CFL phase} \label{CFLphase}

\begin{center}
{\it 1. Gluon polarization tensor ($a,b\le 8$)}
\end{center}

With the matrix ${\cal M}_{\bf k}$ and the projectors 
${\cal P}_{\bf k}^{1,2}$ for the CFL phase, given in Table \ref{tableMP},
Eqs.\ (\ref{VWdef}) become
\begin{subequations} \label{VWCFLgluon}
\bea
{\cal V}_{ab,e_1e_2}^{\m\n,rs}&=&\frac{1}{2}\,g^2\left\{ 
{\rm Tr}[T_a\,{\cal P}^r\,T_b\,{\cal P}^s]\,{\cal T}_{-e_1,-e_2}^{\m\n} +  
{\rm Tr}[T_a^T\,{\cal P}^r\,T_b^T\,{\cal P}^s]\,{\cal T}_{e_1 e_2}^{\m\n}
\right\} \,\, ,
\\
{\cal W}_{ab,e_1e_2}^{\m\n,rs}&=&
\frac{1}{2}\,g^2\left\{ {\rm Tr}\left[T_a\,{\bf J}\cdot{\bf I}
\,{\cal P}^r\,T_b^T\,{\bf J}\cdot{\bf I}\,{\cal P}^s\right]
\,{\cal U}_{e_1,-e_2}^{\m\n} \right. \non
&&\hspace{0.6cm} \left. + \;{\rm Tr}\left[T_a^T\,{\bf J}\cdot{\bf I}
\,{\cal P}^r\,T_b\,{\bf J}\cdot{\bf I}\,{\cal P}^s\right]
\,{\cal U}_{-e_1,e_2}^{\m\n} \right\} \,\, . 
\eea
\end{subequations}
As in the 2SC phase, the projectors ${\cal P}^r$, ${\cal P}^s$ do not
depend on momentum. Consequently, the angular integrals defined in 
Eq.\ (\ref{TUdef}) also appear in the CFL phase. Therefore,
also in the CFL phase the $(0i)$ and $(i0)$ components of the
polarization tensor vanish, and the $(ij)$ components are proportional
to $\d^{ij}$. Contrary to the 2SC
phase, color and flavor traces cannot be performed separately, since the
projectors ${\cal P}^r$, ${\cal P}^s$ are nontrivial matrices both
in color and in flavor space. 
In order to perform the color-flavor trace,
one uses the relations ${\rm Tr}[T_a{\cal P}^1T_b{\cal P}^1]=0$ and 
${\bf J}\cdot{\bf I}\,{\cal P}^1=-2\,{\cal P}^1$.  
The polarization tensor is not only diagonal in color, but also 
has the same value for all eight gluons. One finally obtains the 
following expressions for the gluon polarization tensors.
\begin{enumerate}
\renewcommand{\labelenumi}{(\alph{enumi})}
\item $\m=\n=0$.

\be \label{CFLgluon00}
\Pi_{ab}^{00}(0)=\d_{ab}\frac{g^2\m^2}{6\p^2}
\left[(v^{12}+v^{21})+7v^{22}+2(w^{12}+w^{21})
+2w^{22}\right] \,\, .
\ee

\item $\m=0,\n=i$ and $\m=i,\n=0$.
\be
\Pi_{ab}^{0i}(0) = \Pi_{ab}^{i0}(0) = 0 \,\, .
\ee 

\item $\m=i,\n=j$.
\bea \label{CFLgluonij}
\Pi_{ab}^{ij}(0)&=&\d_{ij}\d_{ab}\frac{g^2\m^2}{18\p^2}
\left[(v^{12}+v^{21})+2(\bar{v}^{12}+\bar{v}^{21})+
7(v^{22}+2\bar{v}^{22}) \right.\nonumber \\
&&\left. \hspace{1.7cm} -2(w^{12}+w^{21})
-4(\bar{w}^{12}+\bar{w}^{21})
-2(w^{22}+2\bar{w}^{22})\right] \,\, .
\eea
\end{enumerate}

%\clearpage
\begin{center}
{\it 2. Mixed polarization tensor ($a\le 8, b=9$ and $a=9, b\le 8$)}
\end{center}

To compute the mixed polarization tensors, we need the electric charge
generator $Q$. Since we consider a system with three quark flavors of electric 
charges $q_1$, $q_2$, $q_3$, we have $Q={\rm diag}(q_1,q_2,q_3)$. We shall insert 
the charges for $u$, $d$, and $s$ quarks in the final result.
We obtain
\begin{subequations} \label{VWCFLmixed}
\bea
{\cal V}_{a\g,e_1e_2}^{\m\n,rs}&=&\frac{1}{2}\,eg\,\left\{ 
{\rm Tr}[T_a\,{\cal P}^r\,Q\,{\cal P}^s]\,{\cal T}_{-e_1,-e_2}^{\m\n} + 
{\rm Tr}[T_a^T\,{\cal P}^r\,Q\,{\cal P}^s]\,{\cal T}_{e_1 e_2}^{\m\n} \right\}
 \,\, ,\\
{\cal W}_{a\g,e_1e_2}^{\m\n,rs}&=&
\frac{1}{2}\,eg\,\left\{{\rm Tr}\left[T_a\,{\bf J}\cdot{\bf I}
\,{\cal P}^r\,Q\,{\bf J}\cdot{\bf I}\,{\cal P}^s\right]
\,{\cal U}_{e_1,-e_2}^{\m\n} \right. \non
&&\hspace{0.6cm} \left. + \;{\rm Tr}\left[T_a^T\,{\bf J}\cdot{\bf I}
\,{\cal P}^r\,Q\,{\bf J}\cdot{\bf I}\,{\cal P}^s\right]
\,{\cal U}_{-e_1,e_2}^{\m\n} \right\}\,\, . 
\eea
\end{subequations}
First we note that Eq.\ (\ref{symmetric}) also holds for the CFL phase.
With the help of the relations 
\be
{\rm Tr}[T_a{\cal P}^1Q{\cal P}^1]=0 \quad, \qquad
{\rm Tr}[T_aQ]={\rm Tr}[T_a]{\rm Tr}[Q]=0 \,\, , 
\ee
and 
\bea
{\rm Tr}[T_a{\bf J}\cdot{\bf I}\,Q\,{\bf J}\cdot{\bf I}]&=&
3\,{\rm Tr}[T_a{\cal P}^1Q]=
3\,{\rm Tr}[T_a{\cal P}^1Q\,{\bf J}\cdot{\bf I}] \non 
&=&\d_{a3}\frac{1}{2}(q_1-q_2)
+\d_{a8}\frac{1}{2\sqrt{3}}(q_1+q_2-2q_3) 
\eea
we obtain the following results.

\begin{enumerate}
\renewcommand{\labelenumi}{(\alph{enumi})}
\item $\m=\n=0$.
\be \label{CFLmixed00}
\Pi_{a\g}^{00}(0)=\frac{eg\,\m^2}{6\p^2}\left\{
\begin{array}{cl} 0 &  \quad \mbox{for} \quad a=1,2,4-7 \,\, ,\\ \\
(q_1-q_2)\left[(v^{12}+v^{21})-2v^{22}  \right. &\\
\left.+\,2(w^{12}+w^{21})-7w^{22})\right] &  \quad \mbox{for} \quad a=3 \,\, ,\\ \\
\frac{1}{\sqrt{3}}(q_1+q_2-2q_3)
\left[(v^{12}+v^{21})-2v^{22} \right. &\\ 
\left.+\,2(w^{12}+w^{21})-7w^{22})\right] &  \quad \mbox{for} \quad a=8 \,\, .\\
\end{array} \right.
\ee

\item $\m=0,\n=i$ and $\m=i,\n=0$.
\be
\Pi_{a\g}^{0i}(0) = \Pi_{a\g}^{i0}(0) = 0 \,\, .
\ee 

\item $\m=i,\n=j$.
\be \label{CFLmixedij}
\Pi_{a\g}^{ij}(0)=\d^{ij}\frac{eg\,\m^2}{18\p^2}\left\{
\begin{array}{cl} 0 &  \quad \mbox{for} \quad a=1,2,4-7 \,\, ,\\ \\
(q_1-q_2)
\left[(v^{12}+v^{21})+2(\bar{v}^{12}+\bar{v}^{21}) \right. & \\
\left. -\,2v^{22}-4\bar{v}^{22}-2(w^{12}+w^{21}) \right. & \\
\left.-\,4(\bar{w}^{12}+\bar{w}^{21})
+7w^{22}+14\bar{w}^{22}\right] &  \quad \mbox{for} \quad a=3 \,\, ,\\ \\
\frac{1}{\sqrt{3}}(q_1+q_2-2q_3)
\left[(v^{12}+v^{21})+2(\bar{v}^{12}+\bar{v}^{21}) \right. & \\
\left. -\,2v^{22}-4\bar{v}^{22}-2(w^{12}+w^{21}) \right. & \\
\left.-\,4(\bar{w}^{12}+\bar{w}^{21})
+7w^{22}+14\bar{w}^{22}\right] &  \quad \mbox{for} \quad a=8 \,\, .\\
\end{array} \right.
\ee
\end{enumerate}

\begin{center}
{\it 3. Photon polarization tensor ($a=b=9$)}
\end{center}

In this case, the tensors ${\cal V}$, ${\cal W}$ read
\begin{subequations} \label{VWCFLphoton}
\bea
{\cal V}_{\g\g,e_1e_2}^{\m\n,rs}&=&\frac{1}{2}\,e^2 
{\rm Tr}[Q\,{\cal P}^r\,Q\,{\cal P}^s]\,\left({\cal T}_{-e_1,-e_2}^{\m\n} + 
{\cal T}_{e_1,e_2}^{\m\n} \right) \,\, ,\\
{\cal W}_{\g\g,e_1e_2}^{\m\n,rs}&=&
\frac{1}{2}\,e^2{\rm Tr}\left[Q\,{\bf J}\cdot{\bf I}
\,{\cal P}^r\,Q\,{\bf J}\cdot{\bf I}\,{\cal P}^s\right]
\,\left({\cal U}_{e_1,-e_2}^{\m\n} + {\cal U}_{-e_1,e_2}^{\m\n}\right)\,\, . 
\eea
\end{subequations}
In order to perform the color-flavor traces, we abbreviate
the following sums over the three quark charges,
\be
x\equiv\sum_{n,m=1}^3 q_nq_m \quad, \qquad y\equiv\sum_{n=1}^3 q_n^2 \,\, .
\ee
Then, with ${\rm Tr}[Q\,{\cal P}^1Q\,{\cal P}^1]=x/9$, 
${\rm Tr}[Q^2{\bf 1}_c]=9\,{\rm Tr}[Q\,{\cal P}^1Q]=3y$ (where ${\bf 1}_c$
is the unit matrix in color space), and 
${\rm Tr}[Q\,{\bf J}\cdot{\bf I}\,Q\,{\bf J}\cdot{\bf I}]=
3{\rm Tr}[Q\,{\bf J}\cdot{\bf I}\,{\cal P}^1Q\,{\bf J}\cdot{\bf I}]=2(x-y)$,
we obtain the following results.

\begin{enumerate}
\renewcommand{\labelenumi}{(\alph{enumi})}
\item $\m=\n=0$.
\bea \label{CFLphoton00}
\Pi_{\g\g}^{00}(0)&=&\frac{e^2\m^2}{9\p^2}
\left[x\,v^{11}+(3y-x)(v^{12}+v^{21})+(21y+x)v^{22}
\right. \nonumber \\
&&\left. \hspace{0.8cm}- 4x\,w^{11}+(6y-2x)(w^{12}+w^{21})
+(6y-10x)w^{22}   \right]
\,\, .
\eea
In a system of $d$, $s$, and $u$ quarks we have the electric charges
$q_1=q_2=-1/3$ and $q_3=2/3$. In this case,
\be \label{uds}
x=0 \quad, \qquad y=\frac{2}{3} \,\, .
\ee
Inserting these values into Eq.\ (\ref{CFLphoton00}), the result 
becomes proportional to the gluon polarization tensor, 
Eq.\ (\ref{CFLgluon00}),
\be 
\Pi_{\g\g}^{00}(0)=\frac{4}{3}\frac{e^2}{g^2}\Pi_{aa}^{00}(0) \,\, .
\ee

\item $\m=0,\n=i$ and $\m=i,\n=0$.
\be
\Pi_{\g\g}^{0i}(0) = \Pi_{\g\g}^{i0}(0) = 0 \,\, .
\ee 

\item $\m=i,\n=j$.
\bea \label{CFLphotonij}
\Pi_{\g\g}^{ij}(0)&=&\d_{ij}\frac{e^2\m^2}{27\p^2}
\left\{x\,(v^{11}+2\bar{v}^{11})+(3y-x)\left[v^{12}+v^{21}
+2(\bar{v}^{12}+\bar{v}^{21})\right] \right.\nonumber \\
&&\left.\hspace{1.5cm} +\,(21y+x)(v^{22}+2\bar{v}^{22})
+ 4x\,(w^{11}+2\bar{w}^{11})
\right. \nonumber \\
&&\left. \hspace{1.5cm} -\,(6y-2x)\left[
w^{12}+w^{21}+2(\bar{w}^{12}+\bar{w}^{21})\right] \right. \non
&&\left.\hspace{1.5cm} -\,(6y-10x)(w^{22}+2\bar{w}^{22})   \right\}
\,\, .
\eea
Again, using the quark charges of $d$, $s$, and $u$ quarks that lead to
Eq.\ (\ref{uds}), we obtain a result proportional to that given in 
Eq.\ (\ref{CFLgluonij}),
\be
\Pi_{\g\g}^{ij}(0)=\frac{4}{3}\frac{e^2}{g^2}
\Pi_{aa}^{ij}(0) \,\, .
\ee

\end{enumerate}

\subsubsection{The polar phase} \label{polarphase}

In the following we consider
a system of quarks with $N_f$ different flavors where each quark flavor
forms Cooper pairs separately. Each quark flavor has a separate electric
charge, $q_1,\ldots,q_{N_f}$, and chemical potential, $\m_1,\ldots,\m_{N_f}$.

\begin{center}
{\it 1. Gluon polarization tensor ($a,b\le 8$)}
\end{center}

In this case, we have
\begin{subequations} \label{VWpolargluon}
\bea
{\cal V}_{ab,e_1e_2}^{\m\n,rs}&=&\frac{1}{2}\,g^2
\left\{{\rm Tr}[T_a\,{\cal P}^r\,T_b\,{\cal P}^s]
\,{\cal T}_{-e_1,-e_2}^{\m\n} + {\rm Tr}[T_a^T\,{\cal P}^r\,T_b^T\,{\cal P}^s]
\,{\cal T}_{e_1,e_2}^{\m\n}\right\} \,\, ,\\
{\cal W}_{ab,e_1e_2}^{\m\n,rs}&=&\frac{1}{2}\,g^2
\left\{{\rm Tr}[T_a\,J_3\,{\cal P}^r\,T_b^T
\,J_3\,{\cal P}^s]\,\hat{{\cal U}}_{e_1,-e_2}^{\m\n} +
{\rm Tr}[T_a^T\,J_3\,{\cal P}^r\,T_b
\,J_3\,{\cal P}^s]\,\hat{{\cal U}}_{-e_1,e_2}^{\m\n} \right\} \,\, , 
\eea
\end{subequations}
where
\be
\hat{{\cal U}}_{e_1,e_2}^{\m\n}\equiv -\int\frac{d\Omega_{\bf k}}{(2\p)^3}
{\rm Tr}[\g^\m\,(\hat{k}^z-\g^z_\perp(\uk))\,\Lambda_{\bf k}^{e_1}
\,\g^\n\,(\hat{k}^z-\g^z_\perp(\uk))\,\Lambda_{\bf k}^{e_2}] \,\, .
\ee
Since 
\be
{\rm Tr}[\g^\m\,(\hat{k}^z-\g^z_\perp(\uk))\,\Lambda_{\bf k}^{e_1}
\,\g^\n\,(\hat{k}^z-\g^z_\perp(\uk))\,\Lambda_{\bf k}^{-e_2}]=
{\rm Tr}[\g^\m\,\Lambda_{\bf k}^{e_1}\,\g^\n\,\Lambda_{\bf k}^{-e_2}] \,\, ,
\ee
we find
\be \label{2SCpolar}
\hat{{\cal U}}_{e_1,e_2}^{\m\n}={\cal U}_{e_1,e_2}^{\m\n} \,\, .
\ee
Consequently, the gluon polarization tensor in
the polar phase is almost identical to that in the
2SC phase. The only difference is the flavor trace
which here yields a factor $\sum_{n=1}^{N_f}\m_n^2$. 
Therefore, Eqs.\ (\ref{2SCgluon00}) -- 
(\ref{2SCgluonij}) hold also for the polar phase 
after replacing $2\m^2$ with $\sum_{n=1}^{N_f}\m_n^2$.

\begin{center}
{\it 2. Mixed polarization tensor ($a\le 8, b=9$ and $a=9, b\le 8$)}
\end{center}

Obviously, also for the mixed polarization tensor, the Dirac and 
color part is identical to the 2SC case. Consequently, in 
Eqs.\ (\ref{2SCmixed00}) -- (\ref{2SCmixedij}) one has to replace 
the factor $(q_1+q_2)\m^2$ by $\sum_{n=1}^{N_f} q_n\m_n^2$ in order to 
obtain the results for the polar phase. 

\newpage
\begin{center}
{\it 3. Photon polarization tensor ($a=b=9$)}
\end{center}

Analogously, for the photon polarization tensor, the 2SC results
given in Eqs.\ (\ref{2SCphoton00}) -- (\ref{2SCphotonij}) are
valid for the polar phase with the following modifications. 
Since the 2SC phase has a nontrivial flavor structure 
of the matrix ${\cal M}_{\bf k}$, there are two different flavor 
factors in Eqs.\ (\ref{2SCphoton00}) -- (\ref{2SCphotonij}), 
namely $(q_1^2+q_2^2)\m^2$ and $2q_1q_2\m^2$. 
Replacing each of these factors
by the common factor $\sum_{n=1}^{N_f}q_n^2\m_n^2$ yields the corresponding
results for the polar phase.

\subsubsection{The CSL phase} \label{CSLphase}

As for the polar phase, we consider a system of $N_f$ quark flavors,
each flavor forming Cooper pairs separately. 

\begin{center}
{\it 1. Gluon polarization tensor ($a,b\le 8$)}
\end{center}

Inserting the matrices ${\cal M}_{\bf k}$ and ${\cal P}_{\bf k}^{1,2}$
from Table \ref{tableMP} into Eq.\ (\ref{VWdef}) yields   
\begin{subequations} \label{VWCSLgluon}
\bea 
{\cal V}_{ab,e_1e_2}^{\m\n,rs}
&=&\frac{g^2}{2}\,\int\frac{d\Omega_{\bf k}}{(2\p)^3}
\Big\{{\rm Tr}\left[\g^\m\,\g_0\,T_a\,{\cal P}_{\bf k}^r\,
\Lambda_{\bf k}^{-e_1}\,\g^\n\,\g_0\,T_b\,{\cal P}_{\bf k}^s
\Lambda_{\bf k}^{-e_2}\right] \nonumber \\
&&\hspace{2.2cm}+ \left(e_{1,2}\to -e_{1,2},\; 
T_{a,b}\to T_{a,b}^T \right) \Big\} \,\, , \\
{\cal W}_{ab,e_1e_2}^{\m\n,rs}
&=&-\frac{g^2}{2}\, \int\frac{d\Omega_{\bf k}}{(2\p)^3}
\Big\{ {\rm Tr}\left[\g^\m\,T_a\,{\bf J}\cdot(\uk-\gperp(\uk))
\,{\cal P}_{\bf k}^r
\,\Lambda_{\bf k}^{e_1}\,\g_\n\,T_b^T\,{\bf J}\cdot(\uk-\gperp(\uk))\,
{\cal P}_{\bf k}^s\,\Lambda_{\bf k}^{-e_2}\right] \nonumber \\
&&\hspace{2.2cm}+ \left(e_{1,2}\to -e_{1,2}, \;
T_{a,b}\to T_{a,b}^T \right) \Big\} \,\, .
\eea
\end{subequations}
The trace over the $12\times 12$ color-Dirac
matrices is more complicated than in all previously discussed cases. 
For the $(00)$ components, both ${\cal V}$ and 
${\cal W}$ are diagonal in color space. The diagonal elements
are collected in Table \ref{tableCSL1}. They are divided into two
parts, one corresponding to the symmetric Gell-Mann matrices,
$a=1,3,4,6,8$, the other corresponding to the antisymmetric Gell-Mann
matrices, $a=2,5,7$. 
\begin{table}
\begin{center}
\begin{tabular}{|c||c|c|c|}
\hline
$a$ & ${\cal V}_{aa,e_1e_2}^{00,11}$ & 
${\cal V}_{aa,e_1e_2}^{00,12}={\cal V}_{aa,e_1e_2}^{00,21}$ & 
${\cal V}_{aa,e_1e_2}^{00,22}$ \\
\hline
\hline
$\,\,1,3,4,6,8\,\,$ & 0 & $(1+e_1e_2)/(12\p^2)$ & 
$(1+e_1e_2)/(12\p^2)$ \\
\hline
$2,5,7$ & $\,\,(1+e_1e_2)/(18\p^2)\,\,$ & $\,\,(1+e_1e_2)/(36\p^2)\,\,$ & 
$\,\,5(1+e_1e_2)/(36\p^2)\,\,$ \\
\hline 
\hline
 & ${\cal W}_{aa,e_1e_2}^{00,11}$ & 
${\cal W}_{aa,e_1e_2}^{00,12}={\cal W}_{aa,e_1e_2}^{00,21}$ & 
${\cal W}_{aa,e_1e_2}^{00,22}$ \\
\hline
\hline
$\,\,1,3,4,6,8\,\,$ & 0 & $(1+e_1e_2)/(6\p^2)$ 
 & $-(1+e_1e_2)/(12\p^2)$ \\
\hline
$2,5,7$ & $\,\,2(1+e_1e_2)/(9\p^2)\,\,$ & 
$\,\,-(1+e_1e_2)/(18\p^2)\,\,$ & $\,\,5(1+e_1e_2)/(36\p^2)\,\,$ \\
\hline
\end{tabular}
\caption[CSL tensors, I]{$(00)$ components of the tensors ${\cal V}$, 
${\cal W}$ for the
gluon polarization tensor in the CSL phase, defined in 
Eqs.\ (\ref{VWCSLgluon}). All tensors are diagonal in color
space.  }
\label{tableCSL1}
\end{center}
\end{table}
For the $(i0)$ and $(0i)$ components, we find (omitting the indices $r$, $s$
and $e_1$, $e_2$)
\be \label{VW0i}
{\cal V}_{ab}^{i0}={\cal V}_{ab}^{0i}=
{\cal W}_{ab}^{i0}={\cal W}_{ab}^{0i}=0 \,\, .
\ee
Therefore, also for the CSL phase, the $(0i)$ and $(i0)$ components 
of the gluon polarization tensor vanish.

For the $(ij)$ components, we find that ${\cal V}$
and ${\cal W}$ are neither diagonal with respect to
color $a$, $b$ nor with respect to spatial indices. 
Nevertheless, after inserting the $k$ integrals from
Table \ref{tablevw}, the gluon polarization tensor becomes diagonal,
i.e., $\Pi_{ab}^{ij}(0)\sim\d_{ab}\d^{ij}$. The reason
for the cancellation of all non-diagonal elements are the following properties
of ${\cal V}$ and ${\cal W}$,
\begin{subequations} \label{VWijgluon}
\bea
\sum_{rs}{\cal V}^{ij,rs}_{ab,e_1e_2}&\sim&\d_{ab}\d^{ij}\frac{1}{6\p^2} 
\label{Vijgluon} \,\, ,\\
{\cal W}_{ab,e_1e_2}^{ij,rs}&\sim& 1-e_1e_2 \qquad \mbox{for ($i\neq j$;
$a$, $b$ arbitrary), and for ($i=j$; $a\neq b$)} \,\, .\label{Wijgluon}
\eea
\end{subequations}
Taking into account the trace over flavor space, which is the same as in the
polar phase, we obtain
\begin{enumerate}
\renewcommand{\labelenumi}{(\alph{enumi})}
\item $\m=\n=0$.
\be
\Pi_{ab}^{00}(0)=\d_{ab}\frac{g^2}{18\p^2}
\sum_{n=1}^{N_f}\m_n^2
\left\{
\begin{array}{cl}  3(v^{12}+v^{21})+3v^{22}  & \\
+6(w^{12}+w^{21})-3w^{22} &  \quad \mbox{for} \quad 
a=1,3,4,6,8 \,\, ,\\ \\
2v^{11}+(v^{12}+v^{21})+5v^{22} & \\
+8w^{11}-2(w^{12}+w^{21})+5w^{22}
&  \quad \mbox{for} \quad a=2,5,7 \,\, .\\
\end{array} \right.
\ee

\item $\m=0,\n=i$ and $\m=i,\n=0$.
\be
\Pi_{ab}^{0i}(0) = \Pi_{ab}^{i0}(0) = 0 \,\, .
\ee 

\item $\m=i,\n=j$. The general result (keeping all functions $v$, $w$)
is a complicated $24\times 24$ matrix and therefore not shown here.
However, as stated above, the polarization tensor 
is diagonal after inserting the values for the functions $v$ and $w$, 
and the result for the Debye and Meissner masses will be given in 
the next section. 

\end{enumerate}

\begin{center}
{\it 2. Mixed polarization tensor ($a\le 8, b=9$ and $a=9, b\le 8$)}
\end{center}

Here we have 
\begin{subequations} \label{VWCSLmixed}
\bea 
{\cal V}_{a\g,e_1e_2}^{\m\n,rs}
&=&\frac{eg}{2}\,\int\frac{d\Omega_{\bf k}}{(2\p)^3}
\Big\{ {\rm Tr}\left[\g^\m\,\g_0\,T_a\,{\cal P}_{\bf k}^r\,
\Lambda_{\bf k}^{-e_1}\,\g^\n\,\g_0\,Q\,{\cal P}_{\bf k}^s
\Lambda_{\bf k}^{-e_2}\right] \nonumber \\
&& \hspace{2.2cm} + \left(e_{1,2}\to -e_{1,2},\; 
T_a\to T_a^T \right) \Big\}\,\, , \\
{\cal W}_{a\g,e_1e_2}^{\m\n,rs}&=&-\frac{eg}{2}\,
\int\frac{d\Omega_{\bf k}}{(2\p)^3}
\Big\{{\rm Tr}\left[\g^\m\,T_a\,{\bf J}\cdot(\uk-\gperp(\uk))
\,{\cal P}_{\bf k}^r\,\Lambda_{\bf k}^{e_1}\,\g_\n\,Q\,{\bf J}
\cdot(\uk-\gperp(\uk))\,{\cal P}_{\bf k}^s\,\Lambda_{\bf k}^{-e_2}
\right] \nonumber \\
&& \hspace{2.2cm} + \left(e_{1,2}\to -e_{1,2},\; T_a\to T_a^T \right)
\Big\}  \,\, ,
\eea
\end{subequations}
where $Q={\rm diag}(q_1,\ldots,q_{N_f})$.
All $(00)$, $(i0)$, and $(0i)$ components of these integrals vanish, 
\be \label{VWmixedvanish} 
{\cal V}^{00}_{a\g}={\cal V}^{0i}_{a\g}={\cal V}^{i0}_{a\g}
={\cal W}^{00}_{a\g}={\cal W}^{i0}_{a\g}={\cal W}^{0i}_{a\g}=0 \,\,.
\ee
(Here, we again omitted the indices $r$, $s$ and $e_1$, $e_2$.)
The $(ij)$  components of ${\cal V}$ and ${\cal W}$ are nonvanishing and, 
as for the gluonic CSL case,
we do not show them explicitly. However, the final result, i.e. the 
polarization tensor in the considered limit, vanishes
because of the following relations,
\begin{subequations} \label{VWijmixed} 
\bea
\sum_{rs}{\cal V}^{ij,rs}_{a\g,e_1,e_2}&=&0  \,\, , \label{Vijmixed} \\
{\cal W}_{a\g,e_1e_2}^{ij,rs}&\sim& 1-e_1e_2 \,\, . \label{Wijmixed}
\eea
\end{subequations}

\begin{enumerate}
\renewcommand{\labelenumi}{(\alph{enumi})}
\item $\m=\n=0$.
According to Eq.\ (\ref{VWmixedvanish}), we have
\be \label{CSLmixed00}
\Pi_{a\g}^{00}(0)=0\,\, .
\ee
\item $\m=0,\n=i$ and $\m=i,\n=0$.
\be
\Pi_{a\g}^{0i}(0) = \Pi_{a\g}^{i0}(0) = 0 \,\, .
\ee 
\item $\m=i,\n=j$. This polarization tensor  
has a complicated structure in terms of $v$, $w$ and is thus not given here.
However, as stated above, the final result is zero. 

\end{enumerate}

\begin{center}
{\it 3. Photon polarization tensor ($a=b=9$)}
\end{center}

For the photon polarization tensor in the CSL phase we need
\begin{subequations} \label{VWCSLphoton}
\bea 
{\cal V}_{\g\g,e_1e_2}^{\m\n,rs}
&=&\frac{e^2}{2}\,\int\frac{d\Omega_{\bf k}}{(2\p)^3}
\Big\{ {\rm Tr}\left[\g^\m\,\g_0\,Q\,{\cal P}_{\bf k}^r\,
\Lambda_{\bf k}^{-e_1}\,\g^\n\,\g_0\,Q\,{\cal P}_{\bf k}^s
\Lambda_{\bf k}^{-e_2}\right] + \left(e_{1,2}\to -e_{1,2}\right) 
\Big\}\,\, , \\
{\cal W}_{\g\g,e_1e_2}^{\m\n,rs}&=&
-\frac{e^2}{2}\,\int\frac{d\Omega_{\bf k}}{(2\p)^3}
\Big\{ {\rm Tr}\left[\g^\m\,Q\,{\bf J}\cdot(\uk-\gperp(\uk))\,
{\cal P}_{\bf k}^r\,\Lambda_{\bf k}^{e_1}\,\g_\n\,Q\,{\bf J}
\cdot(\uk-\gperp(\uk))\,{\cal P}_{\bf k}^s\,\Lambda_{\bf k}^{-e_2}\right] 
\nonumber \\
&& \hspace{2.2cm} + \left(e_{1,2}\to -e_{1,2}\right) 
\Big\} \,\, .
\eea
\end{subequations}
The results for the $(00)$ and $(ij)$ 
components are given in Tables \ref{tableCSL2} and \ref{tableCSL3}.
The $(0i)$ and $(i0)$ components vanish (for all $r$, $s$, $e_1$, $e_2$),
\be
{\cal V}_{\g\g}^{i0}={\cal V}_{\g\g}^{0i}=
{\cal W}_{\g\g}^{i0}={\cal W}_{\g\g}^{0i}=0 \,\, .
\ee
\begin{table}
\begin{center}
\begin{tabular}{|c|c|c|}
\hline
${\cal V}_{\g\g,e_1e_2}^{00,11}$ & ${\cal V}_{\g\g,e_1e_2}^{00,12}=
{\cal V}_{\g\g,e_1e_2}^{00,21}$ & ${\cal V}_{\g\g,e_1e_2}^{00,22}$ \\
\hline
\hline
$\,\,(1+e_1e_2)/(2\p^2)\,\,$ & 0 & $(1+e_1e_2)/\p^2$ \\
\hline
\hline
${\cal W}_{\g\g,e_1e_2}^{00,11}$ & 
$\,\,{\cal W}_{\g\g,e_1e_2}^{00,12}={\cal W}_{\g\g,e_1e_2}^{00,21}\,\,$ & 
${\cal W}_{\g\g,e_1e_2}^{00,22}$ \\
\hline
\hline
$\,\,-2(1+e_1e_2)/\p^2\,\,$ & 0 & $\,\,-(1+e_1e_2)/\p^2\,\,$ \\
\hline
\end{tabular}
\caption[CSL tensors, II]{$(00)$ components of the tensors 
${\cal V}$, ${\cal W}$ for the
photon polarization tensor in the CSL phase, defined in 
Eqs.\ (\ref{VWCSLphoton}).  }
\label{tableCSL2}
\end{center}
\end{table}
\begin{table}
\begin{center}
\begin{tabular}{|c|c|c|}
\hline
${\cal V}_{\g\g,e_1e_2}^{ij,11}$ & ${\cal V}_{\g\g,e_1e_2}^{ij,12}=
{\cal V}_{\g\g,e_1e_2}^{ij,21}$ & ${\cal V}_{\g\g,e_1e_2}^{ij,22}$ \\
\hline
\hline
$\d^{ij}(11+7e_1e_2)/(54\p^2)$ & $8\,\d^{ij}(1-e_1e_2)/(27\p^2)$  & 
$\d^{ij}(19-e_1e_2)/(27\p^2)$ \\
\hline
\hline
${\cal W}_{\g\g,e_1e_2}^{ij,11}$ & 
${\cal W}_{\g\g,e_1e_2}^{ij,12}={\cal W}_{\g\g,e_1e_2}^{ij,21}$ & 
${\cal W}_{\g\g,e_1e_2}^{ij,22}$ \\
\hline
\hline
$\,\,2\,\d^{ij}(7+11e_1e_2)/(27\p^2)\,\,$ & 
$\,\,16\,\d^{ij}(1-e_1e_2)/(27\p^2)\,\,$ & 
$\,\,-\d^{ij}(1-19e_1e_2)/(27\p^2)\,\,$ \\
\hline
\end{tabular}
\caption[CSL tensors, III]{$(ij)$ components of the tensors 
${\cal V}$, ${\cal W}$ for the
photon polarization tensor in the CSL phase, defined in 
Eqs.\ (\ref{VWCSLphoton}).  }
\label{tableCSL3}
\end{center}
\end{table}
One obtains the following photon polarization tensor.
\begin{enumerate}
\renewcommand{\labelenumi}{(\alph{enumi})}
\item $\m=\n=0$. 
\be \label{CSLphoton00}
\Pi_{\g\g}^{00}(0)=\sum_{n=1}^{N_f}q_n^2\m_n^2\,
\frac{e^2}{\p^2}(v^{11}+2v^{22}-4w^{11}-2w^{22}) \,\, .
\ee
\item $\m=0,\n=i$ and $\m=i,\n=0$.
\be
\Pi_{\g\g}^{0i}(0) = \Pi_{\g\g}^{i0}(0) = 0 \,\, .
\ee 
\item $\m=i,\n=j$.
\bea 
\Pi_{\g\g}^{ij}(0)&=&\d^{ij}\sum_{n=1}^{N_f}q_n^2\m_n^2\,
\frac{e^2}{27\p^2}\left[9v^{11}+2\bar{v}^{11}+16(\bar{v}^{12}
+\bar{v}^{21})+18v^{22}+20\bar{v}^{22} \right. \nonumber \\
&&\left. \hspace{2cm} +\,36w^{11}-8\bar{w}^{11}+32(\bar{w}^{12}+\bar{w}^{21})
+18w^{22}-20\bar{w}^{22}\right] \label{CSLphotonij} \,\, .
\eea
\end{enumerate}

\subsection{Results and discussion} \label{discussion}

In this section, we use the results of the previous section to calculate
the Debye and Meissner masses. We insert 
the numbers from Table \ref{tablevw} into the results for the
polarization tensors $\Pi_{ab}^{00}(0)$ and $\Pi_{ab}^{ij}(0)$ given
in Sec.\ \ref{mixingresults}
and use the definitions of the screening masses given in 
Sec.\ \ref{propagators} (cf.\ Eqs. (\ref{defmasses}) and comment below this
equation). 

We distinguish between the normal-conducting and the superconducting
phase. The screening properties of the normal-conducting phase are obtained
with the numbers given in Table \ref{tablevw} for temperatures larger
than the critical temperature for the superconducting phase transition,
$T\ge T_c$. For all $a,b\le 9$,  they lead to a 
vanishing Meissner mass, i.e., as 
expected, there is no Meissner effect in the normal-conducting state.
However, there is electric screening for temperatures larger than $T_c$.
Here, the Debye mass solely depends on the number of quark flavors
and their electric charge. We find (with $a\le 8$)
\be \label{normaldebye}
T\ge T_c: \qquad m_{D,aa}^2=3\,N_f\frac{g^2\m^2}{6\p^2} \quad, \qquad m_{D,a\g}^2=0
\quad,\qquad m_{D,\g\g}^2=18\sum_n q_n^2 \frac{e^2\m_n^2}{6\p^2} \,\, .
\ee
Consequently, the $9\times 9$ Debye mass matrix is already diagonal. 
Electric gluons and electric photons are screened. 

The masses in the superconducting phases are more interesting.
The results for all phases are collected in Table \ref{tabledebye}
(Debye masses) and Table \ref{tablemeissner} (Meissner masses). The physically
relevant, or ``rotated'', masses are obtained after a 
diagonalization of the
$9\times 9$ mass matrices. We see from Tables \ref{tabledebye} and
\ref{tablemeissner} that the special situation discussed in 
Sec.\ \ref{mixing} applies to all considered phases; namely, all 
off-diagonal gluon masses, $a,b\le 8$, as well as all mixed masses
for $a\le 7$ vanish. Furthermore, in all cases where the mass matrix
is not diagonal, we find $m_{8\g}^2=m_{88}m_{\g\g}$. Therefore, 
the rotated masses $\tilde{m}_{88}$ and $\tilde{m}_{\g\g}$ (which are the
eigenvalues of the mass matrix) are determined by Eqs.\ (\ref{detvanish}).
The electric and magnetic mixing angles $\theta_D\equiv\theta_2$ and 
$\theta_M\equiv\theta_1$ are 
obtained with the help of Eqs.\ (\ref{cossintheta}). Remember that the indices
1 and 2 originate from the spatially transverse and longitudinal
projectors defined in Eqs.\ (\ref{defABE}). They were associated with
the Meissner and Debye masses in Eqs.\ (\ref{defmasses}). We collect
the rotated masses and mixing angles for all phases in 
Table \ref{tablemixing}.

\begin{table}
\begin{center}
\begin{tabular}{|c||cccccccc|cc|c|}
\hline
& \multicolumn{8}{c}{$m^2_{D,aa}$}\vline & 
\multicolumn{2}{c}{$m^2_{D,a\g}=m^2_{D,\g a}$}\vline
&  $m_{D,\g\g}^2$ \\
\hline\hline
$a$ & 1 & 2 & 3 & 4 & 5 & 6 & 7 & 8 & 1-7 & 8 & 9 \\ 
\hline\hline
2SC & \multicolumn{3}{c}{0}\vline & 
\multicolumn{4}{c}{$\frac{3}{2}g^2$}
\vline &  $3\,g^2$ & 0  & 0  &  $2\,e^2$ \\
\hline
CFL  & \multicolumn{8}{c}{$3\,\z\,g^2$}\vline  
 & 0 &  $-2\,\sqrt{3}\,\z\,eg$ &  $4\,\z\,e^2$ \\
\hline
polar & \multicolumn{3}{c}{0}\vline & 
\multicolumn{4}{c}{$\frac{3}{2}g^2$}\vline 
& $3\,g^2$ & 0 & 0 &  $18\,q^2 e^2$ \\
\hline   
CSL & $ 3\,\b g^2$ & $ 3\,\a g^2$ & $ 3\,\b g^2$ 
& $3\,\b g^2$
 & $ 3\,\a g^2$ & $ 3\,\b g^2$ & $ 3\,\a g^2$ & 
$ 3\,\b g^2$ & 0 &  0 
& $18\,q^2 e^2$ \\
\hline 
\end{tabular} 
\caption[Debye masses]{Zero-temperature Debye masses. All masses are 
given in units of 
$N_f\m^2/(6\p^2)$, where $N_f=2$ in the 2SC phase, $N_f=3$ in the CFL
phase, and $N_f=1$ in the polar and CSL phases. We use
the abbreviations $\z\equiv (21-8\ln 2)/54$, 
$\a\equiv (3+4\ln 2)/27$, and
$\b\equiv (6-4\ln 2)/9$. 
}\label{tabledebye}
\end{center}
\end{table}

\begin{table}
\begin{center}
\begin{tabular}{|c||cccccccc|cc|c|}
\hline
& \multicolumn{8}{c}{$m^2_{M,aa}$}\vline & 
\multicolumn{2}{c}{$m^2_{M,a\g}=m^2_{M,\g a}$}\vline
&  $m_{M,\g\g}^2$ \\
\hline\hline
$a$ & 1 & 2 & 3 & 4 & 5 & 6 & 7 & 8 & 1-7 & 8 & 9 \\ 
\hline\hline
2SC  & \multicolumn{3}{c}{0}\vline & 
\multicolumn{4}{c}{$\frac{1}{2}g^2$}\vline  
 & $\frac{1}{3}g^2$ & 0 &  $\frac{1}{3\sqrt{3}}eg $ &  
$ \frac{1}{9}e^2$ \\
\hline
CFL  & \multicolumn{8}{c}{$\z\,g^2$}\vline & 0  &  
$-\frac{2}{\sqrt{3}}\,\z\,eg$  &  $\frac{4}{3}\,\z\,e^2$ \\
\hline
polar & \multicolumn{3}{c}{0}\vline & 
\multicolumn{4}{c}{$\frac{1}{2}g^2$}\vline & 
 $\frac{1}{3}g^2$ & 0 &   $\frac{2}{\sqrt{3}}\,q\,eg$ &  
$4\,q^2 e^2$ \\
\hline   
CSL & $\,\,\b g^2\,\,$ & $\,\,\a g^2\,\,$ & $\,\,\b g^2\,\,$ & $\,\,\b g^2\,\,$ & $\,\,\a g^2\,\,$ & $\,\,\b g^2\,\,$ & $\,\,\a g^2\,\,$ & $\,\,\b g^2\,\,$ & 0 &  0 
& $6 \, q^2 e^2$ \\
\hline 
\end{tabular} 
\caption[Meissner masses]{Zero-temperature Meissner masses. All results 
are given in the same units as the 
Debye masses in Table \ref{tabledebye}. The abbreviations of Table
\ref{tabledebye} are used.}
\label{tablemeissner}
\end{center}
\end{table}

\begin{table}
\begin{center}
\begin{tabular}{|c||c|c|c|}
\hline
 & $\tilde{m}_{D,88}^2$ & $\tilde{m}_{D,\g\g}^2$ & $\cos^2\theta_D$  \\
\hline
\hline
2SC & $3\,g^2$ & $2\,e^2$ & 1  \\
\hline
CFL & $\quad (4\,e^2+3\,g^2)\,\z\quad $ & 0 & $\quad 3g^2/(3g^2+4e^2)\quad$ \\
\hline
polar & $3\,g^2$ & $\quad 18\,q^2e^2\quad$ & 1 \\
\hline
CSL & $3\,\b\,g^2$ & $18\,q^2e^2$ & 1 \\
\hline
\hline
 &  $\tilde{m}_{M,88}^2$ &$\tilde{m}_{M,\g\g}^2$ & $\cos^2\theta_M$ \\
\hline
\hline
2SC & $\frac{1}{3}g^2+\frac{1}{9}e^2$ & 0 
 & $3g^2/(3g^2+e^2)$ \\
\hline
CFL &  
$\quad (\frac{4}{3}e^2+g^2)\,\z\quad $ & 0 & $3g^2/(3g^2+4e^2)$ \\
\hline
polar &  
$\quad \frac{1}{3}g^2+4\,q^2e^2\quad$ & 0 
& $\quad g^2/(g^2+12\,q^2e^2)\quad$ \\
\hline
CSL & $\b\,g^2$ & $\quad 6\,q^2e^2\quad$ 
& 1 \\
\hline
\end{tabular}
\caption[Rotated Debye and Meissner masses]{Zero-temperature rotated 
Debye and Meissner masses in units of $N_f\m^2/(6\p^2)$
and mixing angles for electric and magnetic gauge bosons. The constants 
$\z$, $\a$, and $\b$ are defined as in Tables \ref{tabledebye} and
\ref{tablemeissner}.}
\label{tablemixing}
\end{center}
\end{table}

Let us first discuss the spin-zero cases, 2SC and CFL.   
In the 2SC phase, the gluon masses are obtained from 
Eqs.\ (\ref{2SCgluon00}) and (\ref{2SCgluonij}). 
Due to a cancellation of the normal and anomalous
parts, represented by $v$ and $w$, the Debye and Meissner masses for the
gluons 1,2, and 3 vanish. Physically, this is easy to understand. Since
the condensate picks one color direction, all quarks of the third color, say
blue, remain unpaired. The first three gluons only interact with red and 
green quarks and thus acquire neither a Debye nor a Meissner mass.
We recover the results of Ref.\ \cite{meissner2}.
For the mixed and photon masses we inserted the electric charges 
for $u$ and $d$ quarks, i.e., in 
Eqs.\ (\ref{2SCmixed00}), (\ref{2SCmixedij}), (\ref{2SCphoton00}), and
(\ref{2SCphotonij}) we set $q_1=2/3$ and $q_2=-1/3$.
Here we find the remarkable result that the mixing angle for the Debye
masses is different from that for the Meissner masses, $\theta_D\neq\theta_M$.
The Meissner mass matrix is not diagonal. By a rotation with the angle 
$\theta_M$, given in Table \ref{tablemixing}, we diagonalize this matrix 
and find a vanishing 
mass for the new photon. Consequently, there is no electromagnetic Meissner
effect in this case. This fact is well-known \cite{alford3,gorbar,manuel}. 
The Debye mass matrix, however, is diagonal. The off-diagonal elements 
$m_{D,8\g}^2$ vanish, since the contribution of the ungapped modes, 
corresponding to 
the blue quarks, $v^{22}$, cancels the one of the gapped modes, 
$v^{11}-w^{11}$, cf.\ Eq.\ (\ref{2SCmixed00}). Consequently,
the mixing angle is zero, $\theta_D=0$. Physically, this means that 
not only the color-electric eighth gluon but also the electric photon
is screened. Had we considered only the gapped quarks, i.e., 
$v^{22}=0$ in Eqs.\ (\ref{2SCgluon00}), (\ref{2SCmixed00}), 
and (\ref{2SCphoton00}), we would have found the same mixing angle as for
the Meissner masses and a vanishing Debye mass for the new photon. This
mixing angle is the same as predicted from simple group-theoretical 
arguments, Eq.\ (\ref{allangles}). The photon Debye mass in the 
superconducting 2SC phase differs from that of the normal phase, 
Eq.\ (\ref{normaldebye}), which, for $q_1=2/3$ and $q_2=-1/3$ is
$m_{D,\g\g}^2=5\,N_fe^2\m^2/(6\p^2)$. 

In the CFL phase, all eight gluon Debye and Meissner masses are equal. This 
reflects the symmetry of the condensate where there is no preferred
color direction. For the mixed and photon masses, 
we used Eq.\ (\ref{uds}), i.e., 
we inserted the electric charges for $u$, $d$, and $s$ quarks into 
the more general expressions given in Sec.\ \ref{CFLphase}.
The results in Tables \ref{tabledebye} and \ref{tablemeissner} show
that both Debye and Meissner mass matrices have nonzero off-diagonal
elements, namely $m_{8\g}^2=m_{\g 8}^2$. Diagonalization yields a zero
eigenvalue in both cases. This means that neither electric nor magnetic
(rotated) photons are screened. Or, in other words, there is a charge
with respect to which the Cooper pairs are neutral. Especially, there
is no electromagnetic Meissner effect in the CFL phase, either. 
Note that the CFL phase is the only one considered in this paper in which 
electric photons are not screened.
Unlike the 2SC phase, both electric and magnetic gauge fields are rotated
with the same mixing angle $\theta_D=\theta_M$. This angle is well-known 
\cite{alford3,gorbar,litim,manuel}. Remember that, due to the spectrum of
the matrix $L_{\bf k}=({\bf J}\cdot{\bf I})^2$, there are two gapped 
branches. Unlike the
2SC phase, there is no ungapped quasiparticle excitation branch. This is the
reason why both angles $\theta_D,\theta_M$ coincide with the one 
predicted in Eq.\ (\ref{allangles}).

Let us now discuss the spin-one phases, i.e., the polar and CSL phases.
For the sake of simplicity, all results in Tables \ref{tabledebye}, 
\ref{tablemeissner}, and \ref{tablemixing} refer to a single quark system,
$N_f=1$, where the quarks carry the electric charge $q$. After discussing
this most simple case, we will comment on the situation where $N_f>1$
quark flavors separately form Cooper pairs. The results for the gluon 
masses show that, up to a factor $N_f$, there is no difference between
the polar phase and the 2SC phase regarding screening of color fields. This
was expected since also in the polar phase the blue quarks remain 
unpaired. Consequently, the gluons with adjoint color index $a=1,2,3$ 
are not screened. Note that, due to Eq.\ (\ref{2SCpolar}), the spatial
$z$-direction picked by the spin of the Cooper pairs has no effect on the
screening masses. As in the 2SC phase, electric gluons do not mix with 
the photon. There is electromagnetic Debye screening, which, in this case,
yields the same photon Debye mass as in the normal phase, 
cf.\ Eq.\ (\ref{normaldebye}). The Meissner mass matrix is diagonalized by 
an orthogonal transformation defined by the mixing angle which equals 
the one in Eq.\ (\ref{allangles}). 

In the CSL phase, we find a special pattern of the gluon Debye and Meissner
masses. In both cases, there is a difference between the gluons corresponding
to the symmetric Gell-Mann matrices with $a=1,3,4,6,8$ and the ones
corresponding to the antisymmetric matrices, $a=2,5,7$. The reason for this
is, of course, the residual symmetry group $SO(3)_{c+J}$ that describes 
joint rotations
in color and real space and which is generated by a combination of the 
generators of the spin group $SO(3)_J$ and the antisymmetric Gell-Mann
matrices, $T_2$, $T_5$, and $T_7$. The remarkable property of the CSL phase
is that both Debye and Meissner mass matrices are diagonal. 
In the case of the Debye masses, the mixed entries of the matrix, $m_{D,a\g}$,
are zero because of the vanishing traces, Eq.\ (\ref{VWmixedvanish}), 
indicating that pure symmetry reasons are responsible for this fact 
(remember that, in the 2SC phase, the reason for the same fact was a
cancellation of the terms originating from the gapped and ungapped
excitation branches). There is a nonzero photon Debye mass which is identical
to that of the polar and the normal phase which shows that electric photons 
are screened in the CSL phase. Moreover, and only in this phase, 
also magnetic photons are screened. This means that 
there is an electromagnetic Meissner effect. Consequently, there is no
charge, neither electric charge,  nor color charge, nor any combination
of these charges, with respect to which the Cooper pairs are neutral. 
This was also shown in Sec.\ \ref{grouptheory} where we argued that in the
CSL phase there is no nontrivial residual $\tilde{U}(1)_{em}$, 
cf.\ Fig.\ \ref{symmetries} and Table \ref{tablesymmetry}. 

Finally, let us discuss the more complicated situation of a many-flavor 
system, $N_f>1$, which is in a superconducting state with spin-one
Cooper pairs. In both polar and CSL phases, this extension of the system 
modifies the results in Tables \ref{tabledebye}, \ref{tablemeissner}, and
\ref{tablemixing}. We have to include several different electric quark charges,
$q_1,\ldots,q_{N_f}$, and chemical potentials, $\m_1,\ldots,\m_{N_f}$, 
in a way explained below Eq.\ (\ref{2SCpolar}) and shown in the 
explicit results of the CSL phase in Sec.\ \ref{CSLphase}. In the CSL phase,
these modifications will change the numerical values of all masses, but 
the qualitative conclusions, namely that there is no mixing and electric as
well as magnetic screening, remain unchanged. In the case of the polar 
phase, however, a many-flavor system might change the conclusions 
concerning the Meissner masses. While in the one-flavor case, diagonalization
of the Meissner mass matrix leads to a vanishing photon Meissner mass, this
is no longer true in the general case with arbitrary $N_f$. There is 
only a zero eigenvalue if the determinant of the matrix vanishes, i.e.,
if $m_{M,8\g}^2=m_{M,88}\,m_{M,\g\g}$. Generalizing the results 
from Table \ref{tablemeissner}, this condition can be written as
\be \label{meissnersurface} 
\sum_{m,n}q_n(q_n-q_m)\,\m_n^2\m_m^2 =0\,\, .
\ee
Consequently, in general and for fixed charges $q_n$, there is a hypersurface
in the $N_f$-dimensional space spanned by the quark chemical potentials
on which there is a vanishing eigenvalue of the Meissner mass matrix 
and thus no electromagnetic Meissner effect. All remaining points in 
this space correspond to a situation where the (new) photon Meissner 
mass is nonzero (although, of course,
there might be a mixing of the eighth gluon and the photon). 
Eq.\ (\ref{meissnersurface}) is trivially fulfilled when the electric 
charges of all quarks are equal. Then we have no electromagnetic 
Meissner effect in the polar phase which is plausible since, regarding
electromagnetism, this situation is similar to the one-flavor case.
For specific values of the electric charges we find very simple 
conditions for the chemical potentials. 
In a two flavor system with $q_1=2/3$, $q_2=-1/3$, 
Eq.\ (\ref{meissnersurface}) reads
\be
\m_1^2\m_2^2=0 
\ee
while, in a three-flavor system with $q_1=-1/3$, $q_2=-1/3$, and $q_3=2/3$, 
we have
\be
(\m_1^2+\m_2^2)\m_3^2=0 \,\, .
\ee
Consequently, these systems {\it always}, i.e., for all combinations
of the chemical potentials $\m_n$,  exhibit the electromagnetic Meissner effect
in the polar phase except when they reduce to the above discussed
simpler cases (same electric charge of all quarks or a one-flavor system).

       %Mixing and screening
   \section{The pressure} \label{thepressure}

In Sec.\ \ref{grouptheory}, it was argued that, 
a priori (= from pure symmetry 
arguments), a multitude of phases is theoretically possible in a spin-one
color superconductor. 
In Secs.\ \ref{gapeqsolution} and \ref{mixingscreening}, some of these phases
were selected in order to derive their physical properties such as 
the magnitude of their gap, their transition temperature, and their
behavior in external electric and magnetic fields. However, 
so far we have not discussed which of the phases we expect to be realized 
in nature. This question will be approached in the following. To this end,
we investigate the effective potential $V_{\rm eff}$, because we expect the 
preferred color-superconducting state to minimize this potential. 
Since the thermodynamic pressure is the negative of the effective
potential (at its stationary point) 
\be \label{pandV}
p = -V_{\rm eff} \,\, ,
\ee
this is equivalent to finding the state with the largest pressure.
It will turn out that the formalism introduced in Sec.\ \ref{gapeqsolution}
provides the means for a fundamental and straightforward derivation of a 
general expression for the pressure. After the
general form is derived, we will specify
the value of the pressure at zero temperature in the polar, planar, A, and
CSL phases for the cases of a longitudinal, mixed, and transverse gap.
  
We start from the effective action
given in Eq.\ (\ref{effectiveaction}), cf.\ also 
Refs.\ \cite{rischkerev,ruester,abuki},
\bea
\G[D_G,D_F] &=& -\frac{1}{2}\Tr\ln D_G^{-1} - \frac{1}{2} \Tr(\D_0^{-1}D_G -1)
\nonumber \\
&&+\,\frac{1}{2} \Tr\ln D_F^{-1} + \frac{1}{2} \Tr ({\cal S}_0^{-1}D_F -1)
+\G_2[D_G,D_F]\,\, .
\eea
The effective potential is defined as 
\be \label{VandG}
V_{\rm eff}[D_G,D_F] = -\frac{T}{V} \G[D_G,D_F] \,\, .
\ee
As in the derivation of the gap equation,  
we restrict ourselves to the two-loop approximation 
for $\G_2[D_G,D_F]$, which, for the fermionic degrees of
freedom, is equivalent to taking into account only the left diagram 
in Fig.\ \ref{twoloop}. Omitting the gluonic part, in this 
approximation we have
\be
\G_2[\D,{\cal S}] \simeq \frac{1}{4}\Tr(\Sigma\, {\cal S}) \,\, .
\ee
Remember that the stationary point of the effective action 
$(D_G,D_F)=(\Delta,{\cal S})$ is given by the Dyson-Schwinger equations 
(\ref{dysonschwinger}). The fermionic part of the effective action at the 
stationary point thus can be written as
\be \label{potgeneral}
\G[{\cal S}] = \frac{1}{2}\Tr\ln {\cal S}^{-1} - 
\frac{1}{4}\Tr(1-{\cal S}_0^{-1}{\cal S}) 
\,\, .
\ee
After performing the trace over Nambu-Gor'kov space, this expression reads
\bea \label{afternambu}
\G[{\cal S}] &=& \frac{1}{2}\Tr\ln \left\{
([G_0^+]^{-1} + \Sigma^+)\,
([G_0^-]^{-1} + \Sigma^-) - 
\Phi^-\,([G_0^-]^{-1} + \Sigma^-)^{-1}\, 
\Phi^+\,([G_0^-]^{-1} + \Sigma^-)\right\}
\non
&&+\;\frac{1}{4}\Tr\left\{2-G^+[G_0^+]^{-1}-G^-
[G_0^-]^{-1}\right\} \,\, .
\eea
This equation is derived as follows: For the inverse full fermion
propagator ${\cal S}^{-1}$ we used the Dyson-Schwinger equation 
(\ref{dysonschwinger2}), which allows us to replace ${\cal S}^{-1}$ with
${\cal S}_0^{-1} + \Sigma$. Then, we use Eqs.\ (\ref{inversefermion})
and (\ref{sigmanambu}) to obtain 
\be
{\cal S}_0^{-1} + \Sigma= \left(\begin{array}{cc} [G_0^+]^{-1} + \Sigma^+ & 
\Phi^- \\ \Phi^+ & [G_0^-]^{-1} +\Sigma^- 
\end{array}\right) \,\, .
\ee
Now note that $\Tr\ln({\cal S}_0^{-1}+ \Sigma) = 
\ln{\rm det}\,({\cal S}_0^{-1}+ \Sigma)$ and that for arbitrary matrices 
${\cal A}$, ${\cal B}$, ${\cal C}$, and an invertible matrix ${\cal D}$, 
\be  
{\rm det} \left(\begin{array}{cc}
{\cal A} & {\cal B} \\ {\cal C} & {\cal D} \end{array}\right)
 = {\rm det} ({\cal A}\,{\cal D} -{\cal B}\,{\cal D}^{-1}{\cal C}\,{\cal D})
\,\, .
\ee 
This yields the first term in Eq.\ (\ref{afternambu}). The second term
is straightforwardly derived with the definitions (\ref{inversefermion})
and (\ref{fullquark}). Note that the anomalous propagators $\Xi^\pm$, 
occuring in the full quark propagator ${\cal S}$, do not enter the result
(\ref{afternambu}).  

In Eq.\ (\ref{afternambu}), the free (charge-conjugate) 
propagator for massless quarks $G_0^\pm$ is given by 
Eq.\ (\ref{freefermionprop}). The gap matrices $\Phi^\pm$ are defined in 
Eqs.\ (\ref{phiminus}) and (\ref{gm2SC}). For the regular quark 
self-energies we use the approximation (\ref{selfenergy}), and for the
full (charge-conjugate) quark propagators we use 
\be 
G^\pm = ([G_0^\mp]^{-1} + \Sigma^\mp)\sum_{e,r}
{\cal P}_{{\bf k},r}^\pm\, \L_{\bf k}^{\mp e} 
\frac{1}{\left[k_0/Z^2(k_0)\right] - [\e_{{\bf k},r}^e]^2} 
\quad ,\qquad (r=1,2) \,\, .
\ee
This form of the quark propagator generalizes the ones from the previous 
sections, Eqs.\ (\ref{fullprop}) and (\ref{prop}). Here, we introduced
{\it two} sets of projectors ${\cal P}_{{\bf k},r}^+$ and 
${\cal P}_{{\bf k},r}^-$; 
they project onto the eigenspaces of the matrices
\be \label{defL}
L^+_{\bf k}\equiv \g_0\,{\cal M}_{\bf k}^\dag\, {\cal M}_{\bf k}\, \g_0
\qquad \mbox{and} \qquad  L^-_{\bf k}\equiv {\cal M}_{\bf k}\,
{\cal M}_{\bf k}^\dag \,\, ,
\ee
respectively. In the notation of the previous sections, 
\be
L_{\bf k}\equiv L^+_{\bf k} \,\, , \qquad 
{\cal P}_{\bf k}^r\equiv {\cal P}_{{\bf k},r}^+ \,\, .
\ee 
As mentioned below Eq.\ (\ref{prop}), in the A phase, 
$L^+_{\bf k}\neq L^-_{\bf k}$. However, in all cases,  
$L_{\bf k}^+$ and $L_{\bf k}^-$ have the same spectrum, given by the 
eigenvalues $\l_r$,
\be \label{spectrum}
L_{\bf k}^\pm = \sum_r \l_r {\cal P}_{{\bf k},r}^\pm \,\, .
\ee
Putting everything together, and using the identity 
\be \label{help}
[G_0^\mp]^{-1}[G_0^\pm]^{-1} = \sum_e[k_0^2-
(\m-e\, k)^2]\L_{\bf k}^{\pm e} \,\, ,
\ee
we find for the first term on the 
right-hand side of Eq.\ (\ref{afternambu}),
\bea 
\frac{1}{2}\Tr\ln {\cal S}^{-1} &=& \frac{1}{2}\Tr\ln 
\sum_e\left[\frac{k_0^2}{Z^2(k_0)}-(\m-ek)^2
-\phi_e^2 L_{\bf k}^+\right] 
\L_{\bf k}^{-e}   \nonumber  \\
&=& \frac{1}{2}\,\sum_{e,r} \sum_K \Tr[{\cal P}_{{\bf k},r}^+\L_{\bf k}^{-e}]
\,\ln\left[\frac{k_0^2}{Z^2(k_0)} - (\e_{k,r}^e)^2\right] \,\, .
\eea
Now we make use of the relation
\be \label{matsu1}
\sum_{k_0}\ln \frac{\e_k^2 - k_0^2}{T^2} = \frac{\e_k}{T} + 2\ln\left[
1 + \exp\left(-\frac{\e_k}{T}\right)\right] \,\, ,
\ee
which is proven in Appendix \ref{matsubara} via performing the 
Matsubara sum in terms of a contour integration in the complex plane.
With this relation, applied to the modified variable
$k_0 \to k_0/Z(k_0)$, one obtains    
\be
\label{firstterm}
\frac{1}{2}\Tr\ln {\cal S}^{-1} =  \frac{1}{2}\frac{V}{T}\sum_{e,r}
\int\frac{d^3{\bf k}}{(2\pi)^3}
\Tr[{\cal P}_{{\bf k},r}^+\L_{\bf k}^{-e}]
\left\{\tilde{\e}_{k,r}^e + 
2\,T\,\ln\left[1 + \exp\left(-\frac{\tilde{\e}_{k,r}^e}{T}\right)
\right]\right\}
\,\, ,
\ee
where the trace now runs only over color, flavor, and 
Dirac space. Employing the notation of Ref.\ \cite{wang}, we defined 
the modified excitation energy
$\tilde{\e}_{{\bf k},r}^e\equiv Z(\e_{{\bf k},r}^e)\,\e_{{\bf k},r}^e$,
which includes the effect of the regular quark self-energy.

For the second term on the right-hand side of Eq.\ (\ref{afternambu})
one obtains
\be 
\frac{1}{4}\Tr(1-{\cal S}_0^{-1}{\cal S}) = 
-\frac{1}{4}\sum_{e,r}\sum_K\, \Tr[{\cal P}_{{\bf k},r}^+
\L_{\bf k}^e + {\cal P}_{{\bf k},r}^-\L_{\bf k}^{-e}]\, Z^2(k_0)\,
\frac{\l_r\,\phi_e^2}{k_0^2-Z^2(k_0)\,[\e_{k,r}^e]^2} \,\, . 
\ee
In this case, the Matsubara sum can be performed with the help of the relation 
\be \label{matsu2}
\sum_{k_0}\frac{\varphi(k_0)}{k_0^2-\e_k^2} = -\frac{\varphi(\e_k)}{2\e_k}\,
\tanh\frac{\e_k}{2T} \,\, ,
\ee
where $\varphi$ is an even function of $k_0$. This relation is also 
proven in Appendix \ref{matsubara}. Then, we obtain
\bea \label{secondterm}
\frac{1}{4}\Tr(1-{\cal S}_0^{-1}{\cal S})&=& \frac{1}{4}\frac{V}{T}
\sum_{e,r}\int\frac{d^3{\bf k}}{(2\pi)^3}\,\Tr[{\cal P}_{{\bf k},r}^+
\L_{\bf k}^e + {\cal P}_{{\bf k},r}^-\L_{\bf k}^{-e}] 
\non 
&&\hspace{2cm}\times\;Z^2(\tilde{\e}_{k,r}^e)\, \frac{\l_r\,
\phi_e^2(\tilde{\e}_{k,r}^e,k)}{2\,\tilde{\e}_{k,r}}
\,\tanh\frac{\tilde{\e}_{k,r}^e}{2\,T} \,\, .
\eea
From the results of Sec.\ \ref{gapeqsolution} we conclude
that for all phases we consider, 
\be \label{traces}
\frac{1}{2}\Tr[{\cal P}_{\bf k}^r] 
=\Tr[{\cal P}_{{\bf k},r}^+\L_{\bf k}^e]
=\Tr[{\cal P}_{{\bf k},r}^-\L_{\bf k}^e] \,\, .
\ee
For the case of the A phase, where 
${\cal P}_{{\bf k},r}^+ \neq {\cal P}_{{\bf k},r}^-$, one uses 
Eqs.\ (\ref{pA}) from Appendix \ref{AppA} to prove these relations. 
Therefore, the final result for the pressure $p$, obtained
by putting together Eqs.\ (\ref{firstterm}) and (\ref{secondterm}), is
\bea 
p &=& \frac{1}{4}\sum_{e,r}\int\frac{d^3{\bf k}}{(2\pi)^3} \,
\Tr[{\cal P}_{\bf k}^r] \,  
\left\{\tilde{\e}_{k,r}^e + 
2\,T\,\ln\left[1 + \exp\left(-\frac{\tilde{\e}_{k,r}^e}{T}\right)
\right] \right.\non 
&&\left. \hspace{2cm} - Z^2(\tilde{\e}_{k,r}^e)\, \frac{\l_r\,
\phi_e^2(\tilde{\e}_{k,r}^e,k)}{2\,\tilde{\e}^e_{k,r}}
\,\tanh\frac{\tilde{\e}_{k,r}^e}{2\,T}
\right\} \,\, .\label{Tnonzero}
\eea
In the following, we leave the general treatment and consider the 
simpler 
situation of zero temperature, $T=0$. We also neglect the effect of the regular
quark self-energy, $Z^2\simeq 1$, $\tilde{\e}\simeq\e$. Furthermore, we
neglect the antiparticle gap, $\phi_-\simeq 0$, and thus denote 
$\phi\equiv\phi_+$. In this case, Eq.\ (\ref{Tnonzero}) becomes
\be
p = \frac{1}{4}\sum_{r}\int\frac{d^3{\bf k}}{(2\pi)^3} \,
\Tr[{\cal P}_{\bf k}^r] \,  
\left[\e_{k,r}^+ + \e_{k,r}^- 
 - \frac{\l_r\, \phi^2(\e_{k,r}^+,k)}{2\,\e^+_{k,r}}
\right] \,\, .
\ee
In order to evaluate the integral over the absolute 
value of the quark momentum, we use 
\be
\int_0^\d d\xi \, \left(\sqrt{\xi^2+\phi^2}-\frac{1}{2}
\frac{\phi^2}{\sqrt{\xi^2+\phi^2}}\right) = \frac{1}{2}\,\d\,\sqrt{\d^2+\phi^2}
= \frac{1}{2}\d^2 + \frac{1}{4} \phi^2  + O\left(\frac{\phi^4}{\d^2}\right)
\,\, .
\ee
Consequently, the pressure of the superconducting phase compared to 
the normal-conducting phase is 
\be \label{pressure}
\Delta p = \frac{\m^2}{16\,\pi^2}\langle\phi_0^2\,\Tr[L_{\bf k}]\rangle \,\, ,
\ee
where $\phi_0$ is the (angular-dependent) value of the gap at the
Fermi surface for $T=0$, obtained by solving the gap equation, and
the angular integration is abbreviated by $\langle - \rangle$, as 
introduced
above. Inserting the result for $\phi_0$ from Eq.\ (\ref{ratio}), we obtain
\be \label{pressure2}
\Delta p=\frac{\mu^2(\phi_0^{\rm 2SC})^2}{16\,\pi^2}\;
\frac{\langle e^{-2d}\,\Tr[L_{\bf k}]\rangle}{\langle\l_1\rangle^{a_1}
\langle\l_2\rangle^{a_2}}\,\, .
\ee
In the 2SC phase, we have $d=0$, $\l_1=1$, $\l_2=0$, $a_1=1$, and $a_2=0$.
Moreover, $\Tr [L_{\bf k}]=16$, since the degeneracy of the eigenvalues is 
16 and
8, respectively (here we have to include the Dirac structure, unlike in 
Eqs.\ (\ref{EV2SC})). Consequently,
\be
\Delta p_{\rm 2SC} = \frac{\m^2 (\phi_0^{\rm 2SC})^2}{\pi^2} \,\, ,
\ee
in accordance with Refs.\ \cite{miransky,ellis}.

The result (\ref{pressure2}) is plausible 
since $\Tr[L_{\bf k}]$, sloppily speaking, 
just counts the gaps. More precisely, $\Tr[L_{\bf k}]=\sum_r n_r \l_r$,
where $n_r=\Tr[{\cal P}_{\bf k}^r]$ is the degeneracy of the corresponding
eigenvalue $\l_r$. In the case of a constant $d$ and 
only one excitation branch, i.e., $\l_2=0$ (and $a_1=1$, $a_2=0$) the 
nonzero eigenvalue $\l_1$ cancels (even if it is angular-dependent). Then,
the pressure is solely determined by the degeneracy of the gapped 
excitation branch $n_r$ and the factor $\exp(-2d)$. Consequently,
as one expects, the more gapped excitation branches there are, 
the more favorable is the phase. 

The results for the polar, planar, A, and CSL phases for 
longitudinal, mixed, and transverse gaps are collected in 
Table \ref{tablepressure}. They are obtained by inserting the specific
values of these phases, cf.\ Sec.\ \ref{results}, into Eq.\ (\ref{pressure2}).

\begin{table}  
\begin{center}
\begin{tabular}{|c||c|c|c|}
\hline
$\Delta p/(p_{\rm 2SC}/2)$ & longitudinal & mixed & transverse \\ \hline\hline
polar & $\,\,e^{-12}\,\,$ & $0.5 \cdot e^{-9}$ & $e^{-9}$ \\ \hline
planar & $e^{-12}$ & $e^{-21/2}$ & $e^{-9}$ \\ \hline
A & $e^{-12}$ & $\,\,\frac{1}{2} \cdot e^{-21/2}\,\,$ & $\,\,
\frac{4}{\sqrt{7}}\cdot e^{-9}\,\,$ 
\\ \hline
CSL & $e^{-12}$ & $3\cdot 2^{-4/3}\cdot e^{-10}$ & $e^{-9}$ \\ \hline
\end{tabular}
\caption[Zero-temperature pressure]{Zero-temperature pressure $\Delta p$ in 
units of 
$\mu^2(\phi_0^{\rm 2SC})^2/(2\pi^2)=p_{\rm 2SC}/2$ for 
longitudinal, mixed, and transverse gaps in four different spin-one phases.
Note that $e^{-12}\simeq 6.1\cdot 10^{-6}$, $e^{-21/2}\simeq 2.8\cdot 10^{-5}$,
$e^{-10}\simeq 4.5\cdot 10^{-5}$, and $e^{-9}\simeq 1.2\cdot 10^{-4}$.}
\label{tablepressure}
\end{center}
\end{table}

The pressure is identical for all phases in the case of a longitudinal gap,
because, in this case, $d$ is constant, and $\l_2=0$. 
In the case of a mixed gap,
the pressure depends on the phase. In the polar phase, the angular-dependent
value of $d$ causes a factor
\be
\langle e^{-2d}\rangle = \frac{\sqrt{3\pi}}{6}\,{\rm erf}(\sqrt{3})
\,e^{-9} \simeq 0.5 \, e^{-9} \,\, ,
\ee
where erf is the error function. In the A phase, 
the gapped excitation is only four-fold degenerate. Therefore, it differs
by a factor $1/2$ from the planar phase (for the degeneracies, see for 
instance Fig.\ \ref{gapfigure}). In the mixed CSL phase, the two-gap
structure produces an additional factor 
\be
\frac{1}{8}\,\frac{n_1\l_1 + n_2\l_2}{\l_1^{a_1} \l_2^{a_2}} = 
3\cdot 2^{-4/3} \,\, .
\ee 
The pressure of all mixed phases is larger than that of the longitudinal 
phases. However, the largest pressure is found in the transverse phases, 
which originates from the smallest value of $d$. Indeed, 
in the case of the polar phase, where $d$ has been computed for arbitrary 
combinations of longitudinal and transverse gaps, cf.\ Sec.\ \ref{results},
we find that 
the pressure is a monotonously increasing function of $\b$, assuming its
minimum for a longitudinal gap, $\b=0$, and its maximum for 
a transverse gap, $\b=1$. Comparing the results for the transverse gaps, 
one observes that the A phase exhibits the largest pressure. It is 
larger by a factor $4/\sqrt{7}\simeq 1.5$, which originates from the 
two-gap structure,
\be
\frac{1}{8}\,\frac{n_1\langle\l_1\rangle + n_2\langle\l_2\rangle}
{\langle\l_1\rangle^{a_1} 
\langle\l_2\rangle^{a_2}} = \frac{4}{\sqrt{7}} \,\, ,
\ee 
where, according to Eqs.\ (\ref{eigenA}), $\langle\l_1\rangle=7/3$, and
$\langle\l_2\rangle=1/3$ has been used. Therefore, we conclude that 
the transverse A phase has the largest pressure of all spin-one
color-superconducting phases we consider. Of course, since
in the spin-zero phases, $d=0$, 
the pressure in the 2SC and CFL phases is larger by four orders of
magnitude. However, this result is only valid in the case where the chemical
potentials of all quark flavors are the same. We already have commented on 
the reliability of this assumption in Sec.\ \ref{more}. 

       %Pressure
   \chapter{Conclusions \& outlook}

\section{Summary and discussion}

In the introduction we have argued that, in order to understand 
systems of cold and dense quark matter in nature, it might be important 
to study spin-one color superconductors as an alternative to spin-zero color
superconductors and to normal-conducting quark matter. In this section,
we discuss this statement in more detail, taking into account 
the results of the main part of this thesis. 

We have, in the weak-coupling limit of QCD, derived 
several properties of spin-one color superconductors that are relevant in order
to decide if the considered phases may exist in the interior of compact
stellar objects, such as neutron stars. Some of these properties 
are directly related to experimentally accessible observables of neutron 
stars. To summarize, we have investigated the following quantities and 
effects:
\begin{itemize}
\item angular dependence of the gap,

\item magnitude of the gap,

\item magnitude of the critical temperature,

\item electromagnetic Meissner effect,

\item thermodynamic pressure.
\end{itemize}  
Furthermore, besides the questions regarding the phenomenological 
implications, also problems mainly of theoretical and fundamental 
significance have been treated, such as
\begin{itemize}
\item universality of the gap equation,

\item group-theoretical classification of spin-one phases,

\item fundamental derivation of ``photon-gluon mixing'',

\item similarities/differences to ordinary condensed-matter physics.
\end{itemize}

Let us recapitulate these points without repeating any technical details. 
First, remember that a mismatch of the 
Fermi momenta of different quark flavors does, if too large, not allow for 
the usual BCS pairing. No mismatch is present in the case of pairing of
quarks with the same flavor (provided that the color-chemical potentials are
small). However, such a one-flavor color 
superconductor must be built of Cooper pairs with total spin-one. 
Learning from the analogy between a spin-one color superconductor and the 
well-known theory of superfluid $^3$He, we have argued that a thorough 
treatment requires 
the discussion of more than one order parameter. This is in contrast to 
ordinary superconductivity and to a two-flavor color superconductor. 
In both cases only one phase, i.e., one symmetry breaking pattern, 
is possible. We have shown that, as in the 
cases of $^3$He and a three-flavor color superconductor, the order parameter 
in a 
spin-one color superconductor mathematically corresponds to a complex
$3\times 3$ matrix. Therefore, we have discussed possible forms of this
matrix that lead to different superconducting phases. 

It has turned
out that there are similarities but also differences to the analogous 
situation in $^3$He. Two of the phases shown in the $^3$He phase diagrams
in the introduction, namely the A and the B phase, have their analogues in 
a spin-one color superconductor. Following Ref.\ \cite{schaefer}, we termed
the analogue of the B phase ``color-spin-locked (CSL) phase''.  
It has been
shown that there is no phase corresponding to the ${\rm A}_1$ phase, which 
occurs in $^3$He in the case of an external magnetic field. The reason 
for this is the different structure of the color gauge group compared to the
(nonrelativistic) spin group. After systematically listing 
different order parameters for a spin-one color superconductor, we have 
seen that they can be divided into two classes. We have found 
order parameters that are uniquely defined and order parameters that depend
on several parameters, leading to more than one gap function. Following 
the similar situation in $^3$He, we selected the order parameters of the 
former class for a detailed investigation. In $^3$He, they correspond 
to stationary points of the free energy and therefore are called ``inert'' 
states. In our case, these are the polar, planar, A, and CSL phases. However,
we did not present a rigorous argument which disfavors 
the other phases. Note, for instance, that the residual group of one of the 
``non-inert'' states is larger than the residual group of 
the planar and CSL phases. Therefore, a study going beyond our simple 
argument might reveal a difference to the situation in $^3$He, leading to a 
stable state containing more than one gap function. 

\bigskip
Using our results of
the group-theoretical classification, it is possible to discuss qualitatively
some physical properties without performing any calculations.
Such a discussion has been presented at the end of Sec.\ \ref{grouptheory}.
One of the main results is the fact that the CSL phase is obviously not only
a color superconductor but also an electromagnetic superconductor, i.e., 
this phase exhibits an electromagnetic Meissner effect. This result is of
physical relevance since the (ordinary) magnetic field is one of the 
observables of neutron stars. It also discriminates a spin-one color 
superconductor from spin-zero color superconductors, where an electromagnetic
Meissner effect is absent due to a mixing of the photon and the eighth gluon.
Motivated by this remarkable result, we have presented a quantitative
and fundamental investigation
of the Meissner effect and the photon-gluon mixing that goes far
beyond the group-theoretical analysis. 

Starting from first principles of 
QCD, we have explained how mixing between the gauge bosons can be 
understood. We have shown that the crucial role in this problem is played
by the polarization tensor. A priori, there are $9\times 9 =81$ 
polarization tensors, accounting for the eight gluons and the photon. 
In other words, there are not only 9 tensors for 9 gauge fields, but 
the mixed terms, including every possible combination of two of the gauge 
fields, have to be taken into account in the superconducting phase. 
In order to derive the gauge 
field propagators, this four-momentum-dependent $9\times 9$ polarization 
matrix has to be diagonalized. This is done by an orthogonal transformation,
which, applied to the original
gauge fields, produces a mixture between these fields. 
Consequently, in the superconducting phase, the physical fields are
not necessarily identical with the original gluon and photon fields, but 
may be complicated combinations of these. Indeed, in the superconducting
phases we have considered, there are nonvanishing off-diagonal 
elements in the polarization matrix. We have not computed them for 
arbitrary four-momenta, but rather
focussed on the zero-energy, low-momentum limit in order to compute 
the electric and magnetic screening masses, which are obtained from 
the longitudinal and transverse modes, respectively. (Note that, keeping the 
four-momentum in the 2SC phase,
there is a mixing between the fourth and fifth as well as the sixth and 
seventh gluon \cite{meissner2}.) Consequently, our 
calculations yield $9\times 9$ ``mass matrices''. The eigenvalues
of these matrices are the screening masses (= Debye and Meissner masses)
for the rotated fields. 
In this limit, for all phases we considered, there is no mixing among the 
gluons themselves. The only nonvanishing off-diagonal elements combine 
the eighth gluon with the photon. Therefore, the orthogonal transformation
matrix reduces to a $2\times 2$ matrix, determined by one single parameter,
the mixing angle. We have shown that the group-theoretical arguments
can be used to determine this mixing angle, making use of the 
explicit form of the generator of the residual gauge group $\tilde{U}(1)_{em}$.
It is also determined by an explicit calculation of the screening mass matrix,
which has been done in Sec.\ \ref{mixingscreening}. 

In the case of the CSL phase, all off-diagonal elements vanish both
in the electric and the magnetic case. This means
that there is no mixing between the gauge bosons, and thus the 
physical fields in the superconducting phase are the original gauge fields. 
Furthermore, all diagonal elements are nonzero, implying electric and magnetic
screening for all eight gluons and the photon. The 
nonzero value for the magnetic photon mass is equivalent to a finite
penetration depth for magnetic fields, as predicted from group theory.

In the case of the polar and 2SC phases, there is a qualitative difference
between the electric and magnetic screening mass matrix, which 
shows that the group-theoretical result has to be interpreted with some care. 
For the 
magnetic modes, we have found a mixing between the eighth gluon and the photon,
corresponding to a mixing angle also predicted by group theory. Moreover, 
the eigenvalues are nonzero for five of the gluons (four ``original'' gluons 
and the ``new'' one) but vanish for the three remaining gluons and 
the new photon. Therefore, there is no electromagnetic Meissner effect in 
these cases. However, we have argued that in the case of a many-flavor system,
where each quark flavor separately forms Cooper pairs in the polar phase, 
the magnetic mass matrix does not have any zero eigenvalue (if the
quark flavors differ in their electric charges, which is the case for 
instance in a two-flavor system with $u$ and $d$ quarks). Thus, in this
case (although the fields are mixed!) there is an electromagnetic Meissner 
effect.
             
For the electric modes in the polar and 2SC phases, however, we have found 
a diagonal mass matrix. This is the 
reason why, in these cases,  there is no mixing between the electric gauge 
bosons, i.e., the mixing angle is zero. 
Six of the Debye masses are nonzero, implying an electric 
screening of five gluons and the photon. Three of the gluon Debye masses 
vanish. These belong to the gluons that do not see the (anti-)blue color
of the Cooper pairs. Therefore, in the low-momentum region, where the 
individual quarks of the pair are not resolved, these gluons are not screened. 
Note that for the 
2SC phase, in Ref.\ \cite{litim} the diagonal matrix of the electric screening
masses has been rotated using 
the mixing angle predicted from group theory. Therefore, a different 
value for the 
photon Debye mass (therein, contrary to our calculation, a ``rotated''
photon mass) is obtained. Trying to clarify this subtlety of the 2SC 
(and polar) phase, we have argued in Sec.\ \ref{discussion}, that the 
ungapped blue quarks play a special role in this problem. 
As obvious from 
our results, the contribution of the ungapped quarks cancels the one
of the gapped quarks in the off-diagonal elements. Considering
only the gapped quarks would yield the same mixing angle 
as in the magnetic case. 

\bigskip
After summarizing the theoretical aspects of the electromagnetic 
superconductivity of spin-one color superconductors, we should comment on 
their physical implications. However, before discussing the effect of
spin-one color-superconducting quark matter in the interior of a compact
star, it has to be clarified if the conditions in these astrophysical 
objects are such that the favored state may indeed be a spin-one 
color-superconductor. To this end, we make use of the results of Sec.\ 
\ref{gapeqsolution}. To summarize the conclusion in one sentence: Since
we have estimated the critical temperature of spin-one 
color superconductors to be of the order of tens to hundreds of keV, and the 
temperature of an old neutron star is in the same range or below, the 
core of the star could very well consist of quark matter in a spin-one 
color-superconducting state. Let us recapitulate our estimate in some more 
detail.

The critical temperature, $T_c$, for the transition from the normal-conducting
to the color-super\-conducting phase has been calculated with the help of the 
QCD gap equation for ultrarelativistic quarks. Due to asymptotic freedom, 
asymptotically large densities 
correspond to the weak coupling limit of QCD, and a systematic solution
of the QCD gap equation is possible via extracting leading and subleading 
contributions to the gap. 
We have shown that the gap equation has a universal form applicable to 
arbitrary color-superconducting phases. Due to this fact, we could formalize 
the solution of the gap equation in the sense that the calculation 
of certain terms, directly related to the order parameter, leads 
to the value of the gap $\phi_0$ at the Fermi surface for $T=0$. 
This formalism has been applied to the computation of the gaps for the 
2SC and CFL phases, which had already been known in the literature. 
In the case of a spin-one color superconductor, we computed the 
gap in the polar, planar, A, and CSL phases. Moreover, in each phase, 
we distinguished between the cases of a longitudinal, a mixed, and a 
transverse gap, arriving at twelve different gap parameters. This 
discrimination
has its origin in the special structure of the spin-one gap matrix, which,
due to the vectorial structure of the spin triplet, contains two terms.
The first term, proportional to the quark momentum $\uk$, 
describes Cooper pairing of two quarks with the same 
chirality, while the second, proportional to $\gperp({\bf k})$, accounts   
for pairing of quarks with opposite chirality. We have termed gap matrices
containing solely one of these terms ``longitudinal'' and ``transverse'',
respectively and have shown that these gaps are consistent with the 
gap equation, i.e., they do not induce each other. Note that the longitudinal 
gap can be considered as the nonrelativistic limit. An equal admixture
of longitudinal and transverse gaps was termed ``mixed gap''. 
  
Two technical complications enter the solution of the gap equation in the 
spin-one phases. The first, also present in the CFL phase, is caused by
a two-gap structure in the mixed CSL and the transverse A phases. This
feature actually produces two gap equations. We have discussed
the solution of these two gap equations, which can be reduced to one, 
in detail. It has turned out (and was already expected from the results
of the CFL phase), that the two-gap structure produces a nontrivial factor 
multiplying the gap parameter. Due to our general treatment we have found an
expression of this factor (which is of order one) applicable to all 
color-superconducting phases that exhibit two different gaps. Using the 
notation introduced in Sec.\ \ref{gapeqsolution}, this factor is 
\be \label{twogapfactors}
(\langle\l_1\rangle^{a_1}\langle\l_2\rangle^{a_2})^{-1/2}
 = \left\{\begin{array}{cl} 2^{-1/3} & {\rm CFL} \\ & \\ 2^{-2/3} & 
{\rm CSL}\,{\rm (mixed)} \\ & \\ \left(\frac{3}{\sqrt{7}}\right)^{1/2} 
& {\rm A} 
\,{\rm  (transverse)} 
\end{array} \right.
\ee
The second complication enters through the potentially angular-dependent
excitation energies. This angular dependence
results in a complicated angular integration in the gap equation. To 
perform this integration, we have used a simple approximation, assuming
the neglected terms to be sub-subleading. In principle, a more rigorous 
treatment of the angular integration is required to prove this 
approximation. 

Let us summarize the results for the gap $\phi_0$. Besides the factor 
originating from the two-gap structure (mainly of theoretical interest, since
of order one), there is another factor occurring in the gap parameter, 
which we have denoted by $\exp(-d)$.
Since $d=0$ in the 2SC phase (and the CFL phase), we found the following
ratio of $\phi_0$ and $\phi_0^{\rm 2SC}$,
\be 
\frac{\phi_0}{\phi_0^{\rm 2SC}} = \exp(-d) \, 
\left( \langle\lambda_1\rangle^{a_1} \, 
\langle\lambda_2\rangle^{a_2} \right)^{-1/2}\,\, .
\ee 
We have shown that in the spin-one phases, $4.5\le d\le 6$. In all 
longitudinal phases, $d=6$, which has been proven for arbitrary order 
parameters. Also, in all transverse phases, $d$ assumes 
the same value, namely $d=4.5$. In the mixed phases, the value of $d$ 
depends on the order parameter.
In conclusion, the value of the gap in the spin-one phases is smaller
by 2 -- 3 orders of magnitude compared to the spin-zero phases, which 
leads to a gap of 20 -- 400 keV, assuming the 2SC gap to be of the order of
\mbox{10 -- 100 MeV}. 

From these results, it is easy to conclude the values of $T_c$. To this end,
we have derived a simple relation between $T_c$ and $\phi_0$, which 
deviates by the (inverse of the) above mentioned factor from the 
corresponding relation in BCS theory,
\be
\frac{T_c}{\phi_0}= 0.57 \, \left( \langle\lambda_1\rangle^{a_1} \, 
\langle\lambda_2\rangle^{a_2} \right)^{1/2}\,\, .
\ee
Consequently, the critical temperature in a spin-one color superconductor 
is in the range of 10 -- 400 keV, and therefore 
the existence of these phases in an old neutron star is not ruled out by the
star's temperature.

\bigskip
Remember from the introduction, and references therein, that the 
conventional picture of a neutron star
includes a core of superconducting protons and superfluid neutrons. Based 
on the assumption that the protons form a type-II superconductor, the 
star's magnetic field can penetrate into the interior of the star through
flux tubes. Moreover, since the star is rotating, vortices are formed due
to the superfluidity of the neutron matter. Since the neutron superflow
surrounding these vortices entrains protons, the vortices are magnetized.
Consequently, the vortices are strongly interacting with the magnetic 
flux tubes. This results in a complicated scenario during the spinning down
of the neutron star, because a decreasing rotation frequency is 
accompanied by an expanding vortex array, entraining the magnetic flux 
tubes \cite{ruderman,link}. However, it has been argued \cite{link} that 
this scenario is inconsistent with observed precession periods of one year
in the case of isolated pulsars. 
A solution to this problem might be a core of color-superconducting quark 
matter. 
While spin-zero color superconductors could be of type II
at small $\mu$ \cite{iida3,blaschke2},
a spin-one color superconductor is most likely always of type I,
because the ratio of the penetration depth to the coherence length
is of order 
\mbox{$\sim \phi_0/(e \mu) \sim 10^{-3}\, [100\, {\rm MeV}/(e\mu)] \ll 1$}.
Consequently, the magnetic field is completely expelled from the 
core of a compact stellar object, if it is a spin-one color-superconductor.
The magnetic field is expelled unless it exceeds the critical field
strength
for the transition to the normal conducting state. The magnetic
field in neutron stars is typically of the order of $10^{12}$ Gauss.
This is much smaller than the critical magnetic field which,
from the results of Ref.\ \cite{iida3}, we estimate to be of
the order of $10^{16}$ Gauss. Consequently, the magnetic field strength of 
a neutron star is consistent with a spin-one color-superconducting core. 

Remember also from the introduction that the superfluid vortices
provide an explanation for the observed glitches, i.e., sudden jumps in the 
rotation frequency of the star. Therefore, also the question of
superfluidity in a spin-one color superconductor should be addressed. We
have not studied this question in detail, and thus it remains  
for further studies. However, from our group-theoretical 
analysis, we can immediately read off which of the phases are expected
to contain stable superfluid vortices. We have discussed this question 
at the end of Sec.\ \ref{grouptheory}. To repeat the main conclusions, let 
us point out that, in the case of a spin-one color superconductor, all 
phases that exhibit an electromagnetic Meissner effect 
(due to pure symmetry reasons), are also superfluid. This simple 
conclusion can be drawn for one-flavor systems. Consequently, restricting 
to the ``inert'' phases, 
only the CSL phase is a superfluid. A two-flavor color superconductor, 
as can be seen from the discussion in the introduction, is neither superfluid
nor an electromagnetic superconductor. A three-flavor color 
superconductor, however, is not an electromagnetic superconductor but
a superfluid.      

\bigskip
In order to implement the properties of quark matter into
the theory of neutron stars, it is necessary to know the ground state.
In the introduction it has been argued that for sufficiently cold and dense 
systems, this ground state is a color superconductor. However, the 
multitude of possible phases and the uncertainty of the values for the
density, strong coupling constant etc.\ render the question for 
the true ground state extremely difficult. Therefore, in the present thesis,
we have focussed on the spin-one color superconductors and have discussed 
the ground state among these phases. A comparison with other phases has 
been discussed on a qualitative level in Sec.\ \ref{more}. 
We have computed the thermodynamic pressure, in order
to find the favored state at zero temperature and zero magnetic field.
It has turned out that all transverse phases are favored compared to 
the mixed and longitudinal phases. The reason is the factor 
$\exp(-d)$ that also occurs in the value of the gap parameter. 
The square of this factor, i.e., $\exp(-2d)$, enters the value for the 
pressure. Since $d$ assumes its minimum, $d=4.5$, in the transverse 
phases, these are preferred compared to all other phases where $d>4.5$.
The exponential factor dominates all other contributions that discriminate
between different spin-one phases. 

One of these other contributions is,
roughly speaking, the number of gaps. More precisely, the pressure 
is enhanced by every excitation branch that is gapped. In the 
normal-conducting phase, there are
twelve ungapped excitation branches, resulting from the color (three) and Dirac
(four) components of the quark fields. In the color-superconducting 
phases, these branches become partially gapped. There is one case, namely 
the mixed CSL phase, where all branches are gapped (with two different 
gaps). However, due to the exponential factor, this fact does not help
to prefer this phase over the transverse phases, even though the latter have 
only eight gapped quasiparticle excitation branches. 

Consequently, in order to determine the favored phase, the several transverse 
phases have to be compared. It has turned out that the pressure (and 
the effective potential) is identical in the transverse polar, planar, and
CSL phases. In the transverse A phase, the pressure is enhanced by a factor 
$1.5$. This additional factor originates from the two-gap structure and 
the angular structure of the gap function,
\be
\frac{p_{\rm A}}{p_{\rm polar}} = 
\frac{1}{8}
\frac{n_1\langle\l_1\rangle + n_2\langle\l_2\rangle}{\langle\l_1\rangle^{a_1} 
\langle\l_2\rangle^{a_2}} = \frac{4}{\sqrt{7}}\simeq 1.5 \,\, .
\ee 
In conclusion, among the spin-one phases that we have considered, the
transverse A phase has, at $T=0$ and without external magnetic field, 
the maximal pressure (and, thus, corresponds to the minimal value of the 
effective potential). However, it should be mentioned that this conclusion
depends on our simple approximation concerning the angular integration
in the gap equation and thus has to be treated with some care. 

\bigskip
Finally, let us comment on parallels or differences to other physical 
systems, such as ordinary superconductors/superfluids.
Of special interest in this context are two features that we encountered 
in the treatment of spin-one color superconductivity, namely the 
two-gap structure and the angular dependence of the gap. We have 
demonstrated how both effects yield contributions to the values of the 
gap parameter, the critical temperature, and the pressure. Therefore, 
it is obvious that they strongly affect the physics of the superconductor. 
We have indicated above that the famous BCS
relation, $T_c/\phi_0 = 0.57$, is violated in a superconducting system
where there are two excitation branches with different energy gaps. 
We have discussed one case where one gap is twice the other gap. This
situation occurs in the CSL phase and was also found in the CFL case 
(however, a subtle difference between these two cases yield two different 
factors violating the BCS relation, cf.\ Eq.\ (\ref{twogapfactors})).
In another case, namely the A phase, we found two gaps that differ in their
angular structure. The idea of a two-gap (or multi-gap)
structure in a superconductor is not new and has been discussed 
in the condensed-matter literature \cite{suhl}. Also, experimental evidence 
for two different gaps in ${\rm MgB}_2$ has been reported and theoretically
interpreted  \cite{twogaps}. Similar to our discussion, a deviation 
from the expected values of the gaps, using weak-coupling BCS theory,
has been found. The origin of the two-gap structure 
in this system, however, is a complicated structure of the Fermi surface,
actually separating into two Fermi surfaces. This is in contrast to 
a spin-one color superconductor in the CSL and A phase, where a two-gap 
structure is present even with a single Fermi sphere. More relevant to 
quark matter could be another study about two-gap systems \cite{babaev2}, 
where the type of the superconductor and consequences for the
vortex structure is investigated. 

Let us add that 
there is no two-gap structure in superfluid $^3$He, although the order 
parameters of the CSL and A phases in a spin-one color superconductor 
are identical to those of the B and A phases in $^3$He. The reason, of course, 
is the different structure of the relativistic gap matrix. In the 
nonrelativistic limit, i.e., in the case of a longitudinal gap, we have 
reproduced the single-gap structure of the $^3$He phases.

Also angular-dependent gap functions are known in 
ordinary condensed-matter physics. Before we discuss some parallels, we
briefly summarize what we have found for spin-one color superconductors. 
We have presented a formalism in which the eigenvalues of a certain 
matrix, basically the square of the gap matrix, determine the structure 
of the quasiparticle excitation energy. 
In particular, the dependence of these eigenvalues on the direction
of the quasiparticle momentum $\uk$ determines the angular dependence of the 
gap (in one of the cases we considered, namely the polar phase, an additional
angular dependence enters via the gap parameter $\phi_0$). Since
the eigenvalues are constant numbers in spin-zero color superconductors,
the gap function is isotropic. In a spin-one
color superconductor we have encountered the following cases 
(illustrated in Fig.\ \ref{gapfigure}):
\begin{itemize}
\item isotropic gap: {\it all CSL, mixed planar, mixed A phases};
\item anisotropic gap without nodes: {\it mixed polar, transverse planar, transverse 
A phases};
\item anisotropic gap with two nodal points at north and south pole of Fermi
sphere: {\it transverse polar, longitudinal planar, longitudinal and 
transverse A phases};
\item anisotropic gap with a nodal line at equator of Fermi sphere: 
{\it longitudinal polar phase}.
\end{itemize}
The transverse A phase occurs twice in this enumeration, because
it has two gaps with different structures. 
Nodes of the gap can only occur for a purely longitudinal or transverse
gap. For all admixtures (especially for the ``mixed'' gap), the gap 
is nonvanishing for all directions of the quark momentum. As  
mentioned in the introduction, nodes of the gap are well-known from 
the A and ${\rm A}_1$ phases in superfluid $^3$He. 
In a spin-one color superconductor, we find, due
to the transverse term (= Cooper pairs with quarks of opposite
chirality), an even richer structure of anisotropies. The most interesting
case is certainly the transverse A phase, where two gaps with different 
angular structures are present. Since this phase is likely to be realized in 
nature (remember the maximal pressure of this phase), further studies 
regarding the physical implications of this unusual gap structure 
should be tackled. The physical consequence of gap nodes 
is, for instance, a modified temperature-dependence of the 
specific heat, as we know from $^3$He. It is plausible that nodes of the 
gap influence the thermodynamic properties of the superconductor/superfluid, 
since thermal excitations of quasiparticles become possible even at small 
temperatures. The implications of the nodes certainly depend on their
geometric dimension. Remember from the introduction that
not only superconductors with nodal points or lines but also with 
two-dimensional, spherical, nodes (``gapless 2SC phase'') are known.

Besides superfluid $^3$He, also unconventional superconductors,
such as high-$T_c$ superconductors, are candidates
for gaps that vanish for special directions of the fermion momentum. 
The anisotropies occur for similar reasons as in $^3$He or 
spin-one color superconductors. In several high-$T_c$ 
superconductors, line nodes and point nodes of the gap have been
studied originating from the spin-singlet, $d$-wave nature of the
Cooper pair. For instance, see Ref.\ \cite{tsuei} for the discussion
of line nodes in cuprate superconductors, and Ref.\ \cite{izawa} 
for point nodes in ${\rm YNi}_2{\rm B}_2{\rm C}$. 
Another example of gap nodes has been found in 
superconducting ${\rm Sr}_2{\rm RuO}_4$ \cite{mackenzie} 
where the order parameter is 
assumed to be of the same form as in the A phase of $^3$He, describing
a condensate in the spin-triplet, $p$-wave channel.

    %Summary and discussion
   \section{Open questions}

In the previous section, we have summarized the main results of this
thesis and have discussed their theoretical and phenomenological meaning.
In that discussion, we have already indicated some open questions that 
remain to be tackled in future work. In this section, we discuss these
questions in more detail and extend the 
list of possible future projects by some additional points.

From the results of this thesis and the introductory remarks about 
the research field of color superconductivity, it is clear that many 
important questions concerning cold and dense matter are unanswered, or
only understood at a relatively elementary level. We simply do not know
which is the form of matter in the core of a neutron star. We do not know
if it is deconfined quark matter, and, if yes, if it is color-superconducting,
and, if yes again, which color-superconducting phase is in fact realized. 
These questions, besides the not less exciting fundamental theoretical 
aspects, are the motivation for a continuing activity in this field. 
One of the goals, of course, should be to find a microscopical model of
matter inside a neutron star that is in agreement with the macroscopical
astrophysical observations. We have argued that a key to a deeper understanding
of cold and dense quark matter is the magnetic field of the star and related 
observables such as its rotation frequency. We have also argued 
that spin-one color superconductors exhibit some striking properties 
concerning magnetism. We have discussed that, in spite of the smaller gap 
compared to spin-zero color superconductors, spin-one phases could be 
relevant in realistic systems. Therefore, let us now point out some 
future projects that could be added to the present status of the theory
of spin-one color superconductors.

First, there are some obvious, mostly technical, extensions and completions
to the calculations of this thesis. For instance, we have not computed the
Debye and Meissner masses for the A phase. Since our results suggest
that the transverse A phase is favored, this is definitely a task of physical 
relevance. With the formalism that we have developed, this calculation 
reduces to the calculation of some traces in color-spin space and can be 
done without conceptual difficulties. Of special interest in connection with
this question could be the general behavior of purely longitudinal 
and transverse gaps regarding magnetic fields. Remember that the calculations
in Sec.\ \ref{mixingscreening} have focussed on the mixed polar and 
CSL phases. 
In none of these phases, the gap function has nodes, which is the case for 
instance in the transverse A phase. It could be discussed, if and how the 
nodes affect a possible Meissner effect, e.g., it can be expected that the 
penetration depth of the magnetic field depends on the spatial direction.
For the transverse A phase, one does not expect an electromagnetic  
Meissner effect in a 
one-flavor system, since there is a residual gauge group 
$\tilde{U}(1)_{em}$. However, it is very likely that, as discussed for the 
polar phase, for many-flavor systems a Meissner effect will be present.

Other questions, directly related to the thesis' calculations, are the 
following. We have proven that all longitudinal gaps are suppressed by 
a factor $e^{-6}$ compared to the spin-zero phases. This result holds for
arbitrary order parameters, i.e., for arbitrary $3\times 3$ matrices $\D$.
In the explicit calculations, we have found that for all transverse
gaps we considered, the suppression factor is given by $e^{-4.5}$. 
However, we have not presented a proof that this result is also 
universal (or a counter-example to show that it is non-universal).  
To our knowledge, this has not been done in the framework
of the QCD gap equation. This problem is also purely technical since 
the conceptual framework has been explained in detail. 

An obvious extension to the discussion of the thermodynamic pressure 
in spin-one color superconductors is the generalization to 
finite temperatures. We have presented the general derivation of the 
effective potential up to a certain point, where we restricted to the 
case $T=0$ in order to proceed analytically. Therefore, a straightforward
numerical study could compare several spin-one phases for nonzero 
temperatures. Besides the temperature, also an external magnetic field 
could be implemented. Obviously, this is of great importance for 
the application to neutron stars. Most likely, for finite 
magnetic fields and finite temperatures one would find results 
deviating from the conclusions in the case of zero temperature and zero 
magnetic field. This conjecture is suggested by the phase diagram 
of $^3$He, where, in the case of an external magnetic field, the ${\rm A}_1$
phase is observed, which is completely absent in systems without 
a magnetic field. 

\bigskip
Finally, let us mention possible future projects that are not straightforward
extensions of the present thesis, but nevertheless suggested by the results.
Remember that in the polar phase, the spins of the Cooper pairs point in one 
fixed direction. Also in other phases, such as the A phase, the spatial 
directions are discriminated by the spin structure. Therefore, it is
a natural question to ask for ferromagnetism in these phases. In 
condensed-matter physics, there has been work related to this topic,
suggesting a possible coexistence of ferromagnetism and 
spin-triplet superconductivity, see for instance Ref.\ \cite{kirkpatrick}.
Also for quark matter, although solely treating color ferromagnetism 
in spin-zero color superconductors, similar studies have been 
done \cite{tatsumi}. The study of ordinary ferromagnetism in spin-one color 
superconductors might be of particular interest for the origin of the
magnetic field in a neutron star, which is, at present, not completely
understood.

A phenomenological implication of the gap nodes is related to the neutrino 
emissivity of a color superconductor. This might be of special interest for
astrophysics since the cooling of a neutron star is dominated by 
neutrino emission \cite{weber,prakash}. It has been argued that in the 
CFL phase, neutrino emission originating from direct Urca processes is 
suppressed
by a factor $\exp(-\phi/T)$ \cite{jaikumar}. The reason is simply that 
a quark Cooper pair has to be broken in order to emit a neutrino, i.e., 
the energy cost of this process is responsible for its suppression.
In a spin-one color superconductor, however, this suppression is likely
to be modified. Because of the nodes in the gap function, present in 
several of the discussed phases, there are gapless excitations (depending
on the direction of the quark momentum). Therefore, neutrino emission from 
Urca processes might become important, and it would be very interesting 
to investigate the consequences for the cooling properties of a neutron
star with a spin-one color-superconducting core.    	 

As mentioned in the introduction as well as in the discussion of the 
results, the structure of superfluid vortices (and magnetic flux tubes)
is crucial in order to understand what is going on in the interior of 
a neutron star. Therefore,
it is a promising project to investigate the vortex structure
in a spin-one color superconductor. For a spin-zero superconductor, 
a similar study, using Ginzburg-Landau theory, has been done in 
Ref.\ \cite{iida3}. In order to point out the basic ideas of the problem, let
us briefly recall some fundamental facts about vortices, based on topological 
arguments \cite{vollhardt}. 

\begin{figure}[ht] 
\begin{center}
\includegraphics[width=8cm]{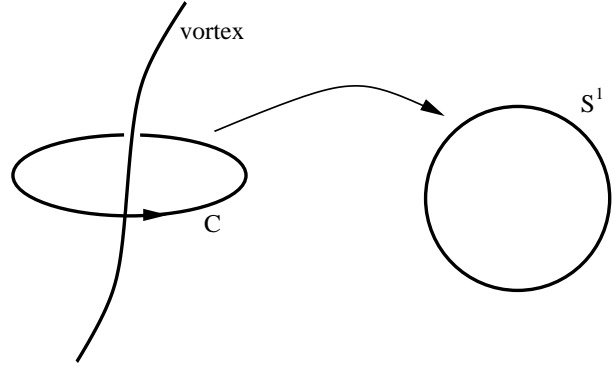}
\vspace{0.5cm}
\caption[Vortex topology]{Mapping of the contour $C$ onto the unit circle 
$S^1$, defined
by the phase of order parameter. The image of the mapping provides
information about the spatial structure of the order parameter and 
therefore determines the topological stability of the vortex. 
}
\label{vortex}
\end{center}
\end{figure}

A vortex in a superconductor is a line along which the system is in the
normal-conducting state. Typically, in a type-II superconductor, there is a 
special pattern of vortices along which a magnetic field can penetrate.
The width of the vortices is characterized by the coherence length.
Under certain conditions, this special configuration of the system 
corresponds to a minimum of the free energy and therefore is preferred over 
the completely superconducting or completely normal-conducting state. 
Without detailed calculations, one can use  
topological arguments for a prediction of the
(topological) stability of the vortex. At this point, it should be mentioned 
that vortices are a special case of ``topological defects'' that may occur in
any system that is characterized by a nonvanishing order parameter (i.e., 
superconductors, superfluids, ferromagnets, etc.). These defects can 
be pointlike \mbox{(= monopoles)}, one-dimensional lines (= vortices) or 
surfaces (= domain walls). Let us briefly discuss the topological stability of 
a line defect in an ordinary superconductor. Employing the notation
introduced in Sec.\ \ref{intro}, in a conventional superconductor 
we have $G=U(1)$, $H=\{{\bf 1}\}$, and \mbox{$G/H = U(1)$}, i.e., the 
electromagnetic
gauge group is spontaneously broken. Now consider a space-dependent order 
parameter 
\be \label{lastequation}
\psi({\bf r}) = \psi_0 \, e^{i\varphi({\bf r})} \,\, .
\ee
Mathematically, $\psi({\bf r})$ is singular along the 
vortex line. This vortex line is called topologically stable if there is no 
continuous deformation of the order parameter field $\psi({\bf r})$ that 
removes the singularity. What does that mean? Consider a closed contour $C$
surrounding the vortex line, cf.\ Fig.\ \ref{vortex}. On each point of this 
line, the order 
parameter (\ref{lastequation}) has a certain value, which is defined by 
its phase $\varphi({\bf r})$, since we assumed the absolute value $\psi_0$
to be constant. Therefore, the order parameter defines a mapping from
$C$ onto the unit circle $S^1$, which is the topological space corresponding 
to the group $G/H=U(1)$.  
The image of this mapping is a closed contour on $S^1$, since the order 
parameter field is continuous along $C$. If this closed contour can be 
contracted to a single point, then there is a continuous deformation 
to remove the singularity. This means that the vortex is 
(topologically) unstable. If it cannot be contracted to a single point, 
the vortex is stable. This is the case if the image goes around the whole
circle (once or more times). The winding number of the image is called the 
topological charge of the vortex. Mathematically speaking, the first homotopy
group $\pi_1$ of the broken part of the group, $G/H$, here 
\be
\pi_1(S^1) = {\cal Z} \,\, ,
\ee
classifies the vortices with respect to their topological charges. 
Therefore, in the 
case of an ordinary superconductor, topological charges of 0 (unstable), 
1, 2, $\ldots$ are possible for vortices. Note that in this 
situation, where $G/H=U(1)$, there cannot be any point defects. 
The reason for that is the trivial second 
homotopy group, $\pi_2(S^1)$, since, for the investigation of 
point defects, one has to consider {\it surfaces} that surround a possible
monopole (in a three-dimensional system). 

The relatively simple situation of an ordinary superconductor 
becomes more involved in the case of a more complicated group $G/H$ that
possibly also allows for point or surface defects.
In the case of a spin-one color superconductor, this group has a rich 
structure. Again, from the theory of $^3$He (where the 
group structure is similar) one could 
borrow the method to tackle the problem of topological defects
in superfluid/superconducting quark matter, especially in 
spin-one color superconductors. Since they are superconductors of type I,
there are no magnetic flux tubes; however, regarding superfluidity, it could be
interesting to study the vortex structure.

    %Open questions%

\begin{appendix}
    \chapter{Projecting onto eigenspaces} \label{proof1}

In this appendix we prove Eq.\ (\ref{projgeneral}) of Sec.\ \ref{exenergies}.

Consider a hermitian matrix $L$ with $n$ different (real) eigenvalues $\l_r$ 
($r=1,\ldots,n$). Then, the projectors $P_r$ onto the corresponding 
eigenspaces can be written as
\be \label{tobeproven}
P_r = \prod_{s\neq r} \frac{L-\l_s}{\l_r-\l_s} \,\, .
\ee
This is Eq.\ (\ref{projgeneral}) in a slightly simplified notation.

Proof: $L$ can be written as
\be
L=\sum_{m=1}^n \l_m P_m \,\, .
\ee 
Since $L$ is hermitian, the eigenspaces are orthogonal to each other,
\be \label{forall}
\forall r\neq s: \qquad P_r P_s = 0 \,\, ,
\ee
and the set of projectors is complete,
\be
\sum_{m=1}^n P_m = {\bf 1} \,\, .
\ee
With these relations, one obtains for a fixed number $s\le n$,
\be
L-\l_s = \sum_{m=1}^n (\l_m - \l_s)\,P_m \,\, .
\ee
Consequently, for a fixed number $r\le n$,
\bea
\prod_{s\neq r} (L-\l_s) &=& \prod_{s\neq r} \sum_{m=1}^n (\l_m - \l_s)\,P_m  \\ 
 &=& \prod_{s\neq r} (\l_r - \l_s)\,P_r \,\, .
\eea
In this product, the only projector that survives is $P_r$, since
this is the only one that occurs in each factor. For every other
projector $P_m$, $m\neq r$, there is one factor that does not contain 
$P_m$, and, consequently, via Eq.\ (\ref{forall}), all these 
projectors disappear. Now, since $\l_r \neq \l_s$ for $r\neq s$, we immediately
conclude for all $r\le n$
\be
P_r = \prod_{s\neq r}\frac{L-\l_s}{\l_r - \l_s} \,\, ,
\ee
which is Eq.\ (\ref{tobeproven}) and thus proves Eq.\ (\ref{projgeneral}).  
   %Proof of projector relation   
    \chapter{Integrating over gluon momentum} \label{AppB}

In this appendix, we compute the integrals over
the gluon 3-momentum $p$ to subleading order in the gap equation, leading to 
Eq.\ (\ref{a1-1}) in Sec.\ \ref{gapsolve}.
We shall see that to this order it is consistent to put
$k = q = \m$.

After replacing $\uk \cdot \uq = (k^2+q^2 -p^2)/(2kq)$,
the coefficients $\eta_{2m}^{\ell, t}(ee',k,q)$
can be read off from the results of the traces 
${\cal T}_{\m\n}^{ee',s}({\bf k},{\bf q})$, defined
in Eq.\ (\ref{T2SC}). One first observes that for all cases considered
here, $\eta_{2m}^{\ell, t}(ee',k,q) = 0$ for $m \geq 3$.
Next, one also realizes that $\eta_{-2}^\ell = 0$, since
there is no term in ${\cal T}_{00}^{ee', i}$ proportional
to $1/p^2$. Consequently, we have to compute the integrals
\begin{subequations}
\be \label{statelec}
{\cal I}_{2m}^\ell = \int_{|k-q|}^{k+q} dp \, p \,
\frac{2}{p^2+3m_g^2} \, \left( \frac{p^2}{kq} \right)^m \,\, , \;\;\;
m = 0,1,2\,\, ,
\ee
for the contribution of static electric gluons to the gap equation,
\be \label{nonstatmagnet}
{\cal I}_{2m}^{t,1} = \int_{M}^{k+q} dp \, p \,
\frac{2}{p^2} \, \left( \frac{p^2}{kq} \right)^m \,\, ,
\;\;\; m = -1,0,1,2\,\, ,
\ee
for the contribution of non-static magnetic gluons, and 
\be \label{statmagn}
{\cal I}_{2m}^{t,2} = \int_{|k-q|}^{M} dp \, p \,
\frac{p^4}{p^6 + M^4 \omega_\pm^2} \, \left( \frac{p^2}{kq} \right)^m \,\, ,
\;\;\; m = -1,0,1,2\,\, ,
\ee
\end{subequations}
with $\omega_\pm \equiv \epsilon_{q,s}^{e'} \pm \epsilon_{k,r}^e$,
for the contribution of almost static magnetic gluons.
The result for the integrals (\ref{statelec}) and (\ref{nonstatmagnet}) is
\begin{subequations} \label{results1}
\bea
{\cal I}_0^\ell & = & 
\ln \left[ \frac{(k+q)^2 + 3 m_g^2}{(k-q)^2 + 3 m_g^2} \right]
\simeq \ln \left( \frac{4\m^2}{3m_g^2} \right) \,\, , \\
{\cal I}_2^\ell & = & 4 - \frac{3 m_g^2}{kq}\,
\ln \left[ \frac{(k+q)^2 + 3 m_g^2}{(k-q)^2 + 3 m_g^2} \right] 
\simeq 4\,\, , \\
{\cal I}_4^\ell & = & 4 \, \frac{k^2 + q^2 - 3m_g^2}{kq}
+ \left( \frac{3 m_g^2}{kq} \right)^2\,
\ln \left[ \frac{(k+q)^2 + 3 m_g^2}{(k-q)^2 + 3 m_g^2} \right] 
\simeq 8\,\, , \\
{\cal I}_{-2}^{t,1} & = & \frac{kq}{M^2} - \frac{kq}{(k+q)^2} 
\simeq \frac{\m^2}{M^2} - \frac{1}{4}\,\, , \label{I-2t1}\\
{\cal I}_0^{t,1} & = & \ln \left[ \frac{(k+q)^2}{M^2} \right] 
\simeq \ln \left( \frac{4 \m^2}{M^2} \right)\,\, , \\
{\cal I}_2^{t,1} & = & \frac{(k+q)^2-M^2}{kq}
\simeq 4\,\, , \\
{\cal I}_4^{t,1} & = &  \frac{(k + q)^4 - M^4}{2(kq)^2} 
\simeq 8\,\, . 
\eea
\end{subequations}
The approximate equalities on the right-hand sides hold
to subleading order in the gap equation. One obtains them
employing two approximations. First, terms proportional to
at least one power of $m_g^2$ or $M^2$ carry at least two
additional powers of $g$, 
which renders them sub-subleading and thus negligible
to the order we are computing. Second, one
utilizes the fact that the $q$ integration in the gap equation
is over a region of size $2 \delta$ around the Fermi surface,
where $\delta \sim m_g$. To subleading order it is thus
accurate to put $k = q = \mu$ (see discussion in Sec.\
\ref{gapsolve}). This then yields the right-hand sides
of Eqs.\ (\ref{results1}).

Note that there is a term $\sim \m^2/M^2 \sim 1/g^2$ in Eq.\
(\ref{I-2t1}). This term is parametrically the
largest and could in principle give the dominant contribution to the
gap equation. However, in all cases considered here, it turns out that
the coefficient $\eta_{-2}^t$ is proportional to at least one power
of $(k-q)^2$. Performing also the $q$ integration in the
gap equation, one then has terms of the form
\be \label{estimate}
g^2 \int_0^\delta \frac{d(q-\m)}{\epsilon_q}\, \frac{(k-q)^2}{M^2} \, 
\phi (\epsilon_q,q)
\sim g^2\, \frac{\phi_0}{M^2} 
\int_0^\d \frac{d \xi}{ \sqrt{\xi^2 + \phi_0^2}}\, \xi^2
\sim g^2 \, \phi_0 \, \frac{\d^2}{M^2} + O\left(\frac{\phi_0^3}{\m^2}
\right)\,\, ,
\ee
where for the purpose of power counting we have neglected
the $q$ dependence of the gap function, $\phi(\e_q,q) \sim \phi_0$,
and we have evaluated the integral on the left-hand side for
$k = \m$.
As long as $\d \sim m_g \sim M$, the leading term
in Eq.\ (\ref{estimate}) is $\sim g^2 \phi_0$,
and thus it is only of sub-subleading order in the gap equation. 
It is obvious that the constant term $- 1/4$ in Eq.\ (\ref{I-2t1})
is parametrically even smaller. The contribution to the term
$\sim \eta_{-2}^t$ from non-static magnetic gluons is therefore
negligible to subleading order. 

Finally, also the integrals ${\cal I}_{2m}^{t,2}$ can be
computed analytically \cite{gradstein}. Defining
$\a \equiv (M^4 \omega_\pm^2)^{1/3}$, the result is
\begin{subequations}
\bea
{\cal I}_{-2}^{t,2} & = & - \frac{kq}{12\a} \,  \left\{ 
\ln \left[\frac{(x+\a)^2}{x^2- \a x + \a^2}\right] - 
2 \sqrt{3}\, {\rm arctg}\, \left( \frac{2x-\a}{\sqrt{3}\,\a} \right) 
\right\}^{M^2}_{(k-q)^2}  \,\, , \label{B1} \\
{\cal I}_0^{t,2} & = & \frac{1}{6} \, 
\ln \left[ \frac{M^6 + \a^3}{(k-q)^6 + \a^3}
\right] \simeq \frac{1}{6} \, \ln \left(
\frac{M^2}{\omega_\pm^2}\right)\,\, , \label{B2} \\
{\cal I}_2^{t,2} & = & \frac{M^2 - (k-q)^2}{2kq} - \frac{\a}{12kq}\, 
\left\{ 
\ln \left[\frac{(x+\a)^2}{x^2- \a x + \a^2}\right] + 2 \sqrt{3}\, {\rm arctg}\,
\left( \frac{2x-\a}{\sqrt{3}\,\a} \right) \right\}^{M^2}_{(k-q)^2} 
\simeq 0 \,\, ,  \label{B3} \\
{\cal I}_4^{t,2} & = & \frac{M^4-(k-q)^4}{4(kq)^2} -
\frac{\a^3}{(kq)^3} 
\, {\cal I}_{-2}^{t,2}
\simeq 0 \,\, . 
\label{B4}
\eea
\end{subequations}
Here, we used the short notation $\{ f(x) \}^{a}_{b} \equiv
f(a) - f(b)$. In order to obtain the approximate equalities on 
the right-hand sides of Eqs.\ (\ref{B2}), (\ref{B3}), and (\ref{B4}),
one employs the fact that typically $(k-q)^2 \sim \omega_\pm^2 \ll M^2$, 
such that parametrically $(k-q)^2 \ll \a \ll M^2$. 
This immediately yields the right-hand side of Eq.\ (\ref{B2}).
For Eqs.\ (\ref{B3}) and (\ref{B4}), we use this estimate in order to
expand the logarithm occurring in Eqs.\ (\ref{B1}) and
(\ref{B3}). One finds that the leading term is $\sim \a/M^2$.
Similarly, one expands the inverse tangent occurring in these equations,
which leads to terms which are even of order $O(1)$. 
Collecting all prefactors, however, all terms in Eqs.\
(\ref{B3}) and (\ref{B4}) are then suppressed by at least one
power of $g^2$. These sub-subleading corrections are negligible
to the order we are computing.

Somewhat more care is necessary in estimating the terms in Eq.\
(\ref{B1}).
Again, one may expand the logarithm and the inverse tangent.
Together with the prefactor, this leads to a term $\sim 1/M^2$ for
the logarithm, and a term $\sim 1/\a$ for the inverse tangent.
The first term is harmless: together with the factor $(k-q)^2$ from
$\eta_{-2}^t$ it leads to an integral of the form (\ref{estimate}),
which was already shown to give a sub-subleading contribution to
the gap equation. The other term leads to the integral
\be
g^2 \int_0^\d \frac{d (q-\m)}{\e_q} \, 
\frac{ (k-q)^2}{\a} \, \phi(\epsilon_q,q)
\sim g^2 \, \frac{\phi_0 }{M^{4/3}} \int_0^\d 
\frac{d\x\, \x^2}{(\x^2 + \phi_0^2)^{5/6}}\,\, ,
\ee
where we used similar power-counting arguments as in Eq.\
(\ref{estimate}).
The last integral is finite even for $\phi_0=0$, so
that we can estimate it to be $\sim \delta^{4/3}$.
For $\d \sim m_g$ this contribution is then again $\sim g^2 \phi_0$
and thus of sub-subleading order in the gap equation.

In conclusion, also the contribution of
almost static magnetic gluons to the term $\sim \eta_{-2}^t$
is of sub-subleading order and can be neglected.
To subleading order, it is therefore consistent to put $\eta_{-2}^t =
0$ from the beginning, provided one chooses $\d \sim m_g$.
   %Gluon momentum integration
    \chapter{Computing eigenvalues} \label{AppA}

In this appendix, we compute the eigenvalues of the matrix $L_{\bf k}$,
defined in Eq.\ (\ref{Ldef}), for several color-superconducting
phases. 
 
The eigenvalues $\lambda_r$ of $L_{\bf k}$ follow from the roots of 
\be
{\rm det} \left( \lambda\, {\bf 1} - L_{\bf k}  \right) = 0 \,\, .
\ee
The left-hand side of this equation can be rewritten in the form
\be
{\rm det} \left( \lambda\, {\bf 1}  - L_{\bf k} \right) \equiv \exp
\left\{ {\rm Tr} \left[ \ln \left( \lambda\, {\bf 1} - L_{\bf k}  \right) 
\right]\right\} \,\, .
\ee
The logarithm of the matrix $\lambda\, {\bf 1} - L_{\bf k}$ is formally
defined in terms of a power series,
\be
{\rm Tr} \left[ 
\ln \left( \lambda\, {\bf 1} - L_{\bf k}  \right) \right] = \ln \lambda 
\, {\rm Tr}\, {\bf 1}
+ {\rm Tr} \left[ \ln \left( 1 - \frac{L_{\bf k}}{\lambda} \right) \right]
= \ln \lambda \, {\rm Tr}\, {\bf 1} 
- \sum_{n=1}^{\infty} \frac{1}{n}\, \lambda^{-n} \,
{\rm Tr}\, L_{\bf k}^n \,\, .
\ee
In order to proceed, one needs to know the trace of the $n$th power of the
matrix $L_{\bf k}$.
In the cases where $L_{\bf k}$ is a projector, we have
$L_{\bf k}^n \equiv L_{\bf k}$. This is the case for instance in the 
2SC phase, cf.\ Eq.\ (\ref{L2SC}). In this case, counting color and flavor 
degrees of freedom, the trace of $L_{\bf k}$ is 4.
Therefore, we obtain for the 2SC phase
\be 
{\rm det} \left( \lambda\, {\bf 1} - L_{\bf k} \right) = 
\lambda^2 \, (\lambda - 1)^4 =0\,\, .
\ee
This yields the eigenvalues given in Eq.\ (\ref{EV2SC}). 

Next, let us discuss some less trivial cases, where $L_{\bf k}$ is 
no projector. First, consider the CFL phase.
In this case, the calculation of $L_{\bf k}^n$ is slightly
more involved. The first step is to notice that $L_{\bf k}^2 = 5\,
L_{\bf k} - 4 \, {\bf 1}$.
Repeated application of this relation allows to reduce an arbitrary
number of powers of $L_{\bf k}$ to a single power, plus a term 
proportional to the unit matrix,
\be
L_{\bf k}^n = a_n\, L_{\bf k} + b_n \,{\bf 1} \,\, .
\ee
Multiplying both sides of this equation by $L_{\bf k}$,
one derives the recursion relation
\be \label{recurs}
a_{n+1} = 5\, a_n - 4\, a_{n-1}
\ee
for the coefficients $a_n$, and the identity
\be
b_{n+1} = - 4\, a_n
\ee
for the coefficients $b_n$. The recursion relation (\ref{recurs})
can be solved with the Ansatz $a_n = p^n$, which yields a quadratic
equation for $p$ with the solutions $p_1 = 4$ and $p_2 = 1$. 
The general solution of the recursion relation is
then $a_n = \eta_1 \, p_1^n + \eta_2\, p_2^n = \eta_1\, 4^n + \eta_2$.
The coefficients $\eta_1$ and $\eta_2$ can be determined from $a_1=1$ 
and $a_2=5$, such that
\be 
a_n = \frac{4^n - 1}{3} \quad, \qquad
b_n = - \frac{4^n - 4}{3} \,\, .
\ee
In the CFL phase, ${\rm Tr} \, L_{\bf k} = 12$ and
${\rm Tr}\, {\bf 1} = 9$. Consequently, 
\be 
{\rm det} \left( \lambda\, {\bf 1} - L_{\bf k} \right) = 
(\lambda-4) \, (\lambda - 1)^8 =0\,\, ,
\ee
which leads to Eq.\ (\ref{EVCFL}).

Next, we compute the eigenvalues of $L_{\bf k}$ in the
planar phase (the calculation for the polar phase is trivial and analogous
to the 2SC phase). 
This can be done for arbitrary coefficients $\a$, $\b$, i.e., 
for general linear combinations of longitudinal and transverse gaps.
To this end, we first write Eq.\ (\ref{Lplanar}) as
\be 
L_{\bf k} = J_1^2 A_1 + J_2^2 A_2 + \{J_1,J_2\}\,B + [J_1,J_2]\,Z \,\, ,
\ee
where we used the abbreviations
\begin{subequations} \label{defs}
\bea
A_{1/2} &\equiv& (\a^2-\b^2)\,\hat{k}_{1/2}^2 + \b^2 \,\, , \label{A12}\\
B&\equiv&(\a^2-\b^2)\,\hat{k}_1\hat{k}_2 \,\, ,\\
Z&\equiv&\b\left\{\a\left[\hat{k}_2\,\g_\perp^1(\uk)-  
\hat{k}_1\,\g_\perp^2(\uk)\right] - \b\left[\g_\perp^1(\uk)\,\g_\perp^2(\uk)-
\hat{k}_1\hat{k}_2\right]\right\} \,\, .
\eea
\end{subequations}
The quantities $A_{1/2}$, $B$, and $Z$ are diagonal in color space. But
while $A_{1/2}$ and $B$ are scalars, $Z$ is a nontrivial $4\times 4$ matrix 
in Dirac space. However, one easily verifies the following relation
\be \label{Zsquared}
Z^2=B^2 -A_1A_2 \,\, .
\ee
With the help of this relation and the (anti-)commutation properties of
the color matrices \mbox{$(J_i)_{jk}=-i\e_{ijk}$} one obtains 
$L_{\bf k}^2 = (A_1 + A_2)\,L_{\bf k}$ and therefore
\be \label{npower}
L_{\bf k}^n = (A_1 + A_2)^{n-1}\,L_{\bf k} \,\, .
\ee
Then, with $\Tr L_{\bf k} = 8(A_1 + A_2)$, we have
\be
{\rm det}(\l - L_{\bf k}) = \l^4[\l-(A_1 + A_2)]^8 \,\, ,
\ee
which, using the definition in Eq.\ (\ref{A12}),  proves 
Eq.\ (\ref{eigenplanar}).

Finally, we discuss the A phase. This case is special in the sense that 
$L_{\bf k}^+ \neq L_{\bf k}^-$. (Remember that in Sec.\ \ref{gapeqsolution}
we introduced the simplified notation $L_{\bf k}\equiv L_{\bf k}^+$ since
$L_{\bf k}^-$ was irrelevant in this section.) For the definition of 
$L_{\bf k}^-$ see Eq.\ (\ref{defL}).
With the abbreviations of Eqs.\ (\ref{defs}) we can write 
\be
L_{\bf k}^\pm = J_3^2[(A_1 + A_2)\pm 2i\,Z] \,\, .
\ee
In order to derive an expression for the $n$-th power of $L_{\bf k}^\pm$,
we notice that
\be \label{Lsquared}
(L_{\bf k}^\pm)^2 = 2(A_1 + A_2)L_{\bf k}^\pm - [(A_1-A_2)^2+4B^2] J_3^2\,\, ,
\ee
where Eq.\ (\ref{Zsquared}) has been used. Since $A_{1/2}$ and $B$ are scalars
and $J_3^2 L_{\bf k}^\pm = L_{\bf k}^\pm$, we have
\be
(L_{\bf k}^\pm)^n = a_n L_{\bf k}^\pm +  b_n J_3^2 \,\, ,
\ee
with real coefficients $a_n$, $b_n$. Applying Eq.\ (\ref{Lsquared}),
one derives the following recursion relations for these coefficients,
\be
a_{n+1} = 2(A_1 + A_2)\,a_n + b_n \,\, , \qquad 
b_{n+1} = -[(A_1-A_2)^2 + 4B^2]\,a_n \,\, .
\ee
With the ansatz of a power series, $a_n = p^n$, one obtains a quadratic 
equation for $p$, which has the two solutions 
\be
p_{1/2} = A_1 + A_2 \pm 2\sqrt{A_1A_2 -B^2} \,\, .
\ee
Therefore, $a_n$ is a 
linear combination of the $n$-th powers of these solutions, 
$a_n=\eta_1\, p_1^n + \eta_2 \,p_2^n$. The coefficients 
$\eta_1$, $\eta_2$ can be
determined from $a_1=1$, $a_2=2(A_1+A_2)$. One finds
\be 
a_n = \frac{1}{4}\frac{1}{\sqrt{A_1A_2-B^2}}(p_1^n - p_2^n) \,\, ,\qquad
b_n = -\frac{1}{4}\frac{(A_1 - A_2)^2 + 4B^2}{\sqrt{A_1A_2-B^2}}
(p_1^{n-1} - p_2^{n-1}) \,\, .
\ee
With $\Tr L_{\bf k}^\pm = 8(A_1 + A_2)$ this yields    
\be
{\rm det}(\l - L_{\bf k}^\pm) = \l^4(\l-p_1)^4(\l-p_2)^4
\,\, .
\ee
Consequently, both $L_{\bf k}^+$ and $L_{\bf k}^-$ have the eigenvalues 0,
$p_1$, and $p_2$, each with degeneracy 4, which proves Eqs.\ (\ref{eigenA}).

%Moreover, we have to show that Eq.\ (\ref{traces}) holds in the A phase.

Finally,  consider the projectors ${\cal P}_{{\bf k},r}^\pm$ corresponding
to the eigenvalues $\l$ (in Sec.\ \ref{gapeqsolution},
${\cal P}_{\bf k}^r\equiv {\cal P}_{{\bf k},r}^+$). With 
Eq.\ (\ref{projgeneral}), 
\be
{\cal P}_{{\bf k},1/2}^\pm=\frac{L_{\bf k}^\pm(L_{\bf k}^\pm-\l_{2/1})}
{\l_{1/2}(\l_{1/2} - \l_{2/1})} \,\, , \qquad 
{\cal P}_{{\bf k},3}^\pm = {\bf 1} - {\cal P}_{{\bf k},1}^\pm - 
{\cal P}_{{\bf k},2}^\pm \,\, .
\ee
This leads to
\begin{subequations} \label{pA}
\bea
{\cal P}_{{\bf k},1}^\pm &=& \frac{1}{2} J_3^2 \left(1\pm\frac{i}
{\sqrt{A_1A_2-B^2}}\,Z\right) \,\, , \label{pA1} \\
{\cal P}_{{\bf k},2}^\pm &=& \frac{1}{2} J_3^2 \left(1\mp\frac{i}
{\sqrt{A_1A_2-B^2}}\,Z\right) \,\, , \label{pA2} \\
{\cal P}_{{\bf k},3}^\pm &=& 1 -J_3^2 \,\, .
\eea
\end{subequations}
Since $\Tr Z = \Tr (\L_{\bf k}^e Z) = 0$, one immediately proves
Eq.\ (\ref{traces}). Note that, making use of Eq.\ (\ref{Zsquared}), 
Eqs.\ (\ref{pA1}) and (\ref{pA2}) can be written as
\begin{subequations} 
\bea
{\cal P}_{{\bf k},1}^\pm &=& \frac{1}{2} J_3^2 \left(1\pm\frac{Z}
{\sqrt{Z^2}} \right) \,\, , \\
{\cal P}_{{\bf k},2}^\pm &=& \frac{1}{2} J_3^2 \left(1\mp\frac{Z}
{\sqrt{Z^2}}\right) \,\, .
\eea
\end{subequations}
 
   %Computing eigenvalues
    \chapter{Proving $d=6$ for longitudinal gaps}
\label{applonggap}

In this appendix, we prove that $d=6$ for any order parameter 
(= for any $3\times 3$ matrix) $\Delta$ in the case of a longitudinal gap.  
The constant $d$ occurs in the value of the gap at the Fermi surface
for $T=0$, cf.\ Eq.\ (\ref{phi0}), and is defined in Eq.\ (\ref{d}).
The longitudinal case, i.e., $(\a,\b)=(1,0)$ in Eq.\ (\ref{Mk}) is 
particularly simple, since in this case the Dirac structure of
the matrix ${\cal M}_{\bf k}$ is trivial. 

In order to compute the constant $d$, we use the method presented in 
Sec.\ \ref{gapsolve}. We have to determine the eigenvalues of $L_{\bf k}$
in order to compute the quantities 
${\cal T}_{00}^{ee',s}({\bf k},{\bf q})$ and
${\cal T}_t^{ee',s}({\bf k},{\bf q})$. 

Let us start with the matrix 
\be
{\cal M}_{\bf k} = \vv_{\bf k}\cdot \vJ \,\, ,
\ee
where we abbreviated $\vv_{\bf k}=(v_{{\bf k},1},v_{{\bf k},2},v_{{\bf k},3})$
with
\be
v_{{\bf k},i} \equiv \sum_{j=1}^3 \Delta_{ij}\hat{k}_j \,\, ,
\qquad i = 1,2,3 \,\, .
\ee
Then,
\be
(L_{\bf k})_{ij} = v_{\bf k}^2\,\d_{ij} - v^*_{{\bf k},j}
\,v_{{\bf k},i} \,\, ,
\ee
where $v_{\bf k}^2\equiv\vv_{\bf k}^*\cdot\vv_{\bf k}$. Now, with 
$L_{\bf k}^2=v_{\bf k}^2\,L_{\bf k}$  
and $\Tr \,L_{\bf k} = 8\,v^2_{\bf k}$ the eigenvalues of $L_{\bf k}$ are 
easily found making use of the method presented in Appendix \ref{AppA}. 
One obtains 
\be
\l_1 = v_{\bf k}^2 \qquad (\mbox{8-fold}) \,\, , 
\qquad \l_2 = 0 \qquad (\mbox{4-fold}) \,\, . 
\ee 
Consequently, the fact that there is one gapped branch with degeneracy 8
and one ungapped branch with degeneracy 4 is completely general, i.e., it 
is true for any order parameter $\D$ in the longitudinal case.  
The corresponding projectors are given by
\be
({\cal P}_{\bf k}^1)_{ij}  = \d_{ij} - 
\frac{v_{{\bf k},i}^* v_{{\bf k},j}}{v^2} \,\, , \qquad
({\cal P}_{\bf k}^2)_{ij}  = \frac{v_{{\bf k},i}^* v_{{\bf k},j}}{v^2} 
\,\, .
\ee
Using Eqs.\ (\ref{T2SC}) and (\ref{Ttransv}), this leads to 
\begin{subequations}
\bea
{\cal T}_{00}^{ee',1}({\bf k},{\bf q})&=&\frac{1}{3}\,
\frac{\vv_{\bf q}\cdot\vv_{\bf k}^*}{v_{\bf k}^2}\,(1+ee'\uq\cdot\uk)\,\, ,
\\
{\cal T}_t^{ee',1}({\bf k},{\bf q})&=&\frac{2}{3}\,
\frac{\vv_{\bf q}\cdot\vv_{\bf k}^*}{v_{\bf k}^2}\,(1-ee'\up\cdot\uq\,
\up\cdot\uk) \,\, ,
\eea
\end{subequations}
and ${\cal T}_{\m\n}^{ee',2}({\bf k},{\bf q})=0$.
In order to perform the angular integration, we choose a 
coordinate system $(q,\theta',\varphi')$,
such that $\uq\cdot\uk=\cos\theta'$. 
With 
${\bf k}=k\,(\sin\theta\cos\varphi,\sin\theta\sin\varphi,\cos\theta)$, 
this is achieved by a rotation 
\be
R(\theta,\varphi)=\left(\begin{array}{ccc} \cos\theta\,\cos\varphi & 
\cos\theta\,\sin\varphi & -\sin\theta \\ -\sin\varphi & \cos\varphi & 0 \\
\sin\theta\,\cos\varphi & \sin\theta\,\sin\varphi & \cos\theta \end{array}
\right) \,\, ,
\ee
that rotates the original 
coordinate system into a new one whose $z'$-axis is parallel to $\uk$. 
Consequently, ${\bf q}$ has to be written as 
${\bf q}=R^{-1}(\theta,\varphi)\,{\bf q}'$ with  
${\bf q}'=q\,(\sin\theta'\cos\varphi',\sin\theta'\sin\varphi',\cos\theta')$.
After this transformation and the integration over $\varphi'$
one obtains (note that there is neither a $\theta$-
nor a $\varphi$-dependence left)
\begin{subequations}
\bea
\frac{1}{2\pi}\int_0^{2\pi}d\varphi'\,{\cal T}_{00}^{ee',s}({\bf k},{\bf q})
&=& 
\frac{1}{3}\,\uq\cdot\uk\,(1+ee'\,\uq\cdot\uk) \,\, , \\ 
\frac{1}{2\pi}\int_0^{2\pi}d\varphi'\,{\cal T}_t^{ee',s}({\bf k},{\bf q})&=& 
\frac{2}{3}\,\uq\cdot\uk\,(1-ee'\,\up\cdot\uq\,\up\cdot\uk) \,\, .
\eea
\end{subequations}
The $\theta'$-integration is performed in terms of an integral over
$p$, cf.\ Sec.\ \ref{gapsolve} and Appendix \ref{AppB}. The result of this 
integration is taken care of by Eq.\ (\ref{d}).
We may set $e=e'$ which is equivalent to neglecting the
antiparticle gap. Finally, via ${\bf p}={\bf k}-{\bf q}$ we find
\be 
\eta_0^\ell=\frac{2}{3} \quad , \qquad 
\eta_2^\ell=-\frac{1}{2} \quad , \qquad 
\eta_4^\ell=\frac{1}{12} \quad , \qquad 
\eta_0^t=\frac{2}{3} \quad , \qquad 
\eta_2^t=-\frac{1}{6} \quad , \qquad 
\eta_4^t=-\frac{1}{12} \,\, .
\ee
With Eq.\ (\ref{d}), this proves the result $d=6$.
   %Longitudinal gaps
    \chapter{Integrating over quark momentum} \label{Appquark}

In this appendix, we prove the results shown in Table \ref{tablevw}, i.e.,
we calculate the integrals over quark momentum defined in 
Eqs.\ (\ref{kintegrals}). 
In this calculation we neglect the antiparticle gap, 
$\phi^-\simeq 0$. 

\section*{Contributions from normal propagators at $T=0$}

We start from Eqs.\ (\ref{defv}) which,
for $p_0=0$ reduce to (cf.\ Eq.\ (\ref{spurious}))
\be \label{v}
v_{e_1e_2}^{rs}=-\frac{\e_{1,r}\e_{2,s}-\x_1\x_2}{2\e_{1,r}\e_{2,s}
(\e_{1,r}+\e_{2,s})}
(1-N_{1,r}-N_{2,s})+\frac{\e_{1,r}\e_{2,s}+\x_1\x_2}{2\e_{1,r}\e_{2,s}
(\e_{1,r}-\e_{2,s})}
(N_{1,r}-N_{2,s}) \,\, .
\ee
Here, we inserted the definition of $n_{i,r}$ given in 
Eq.\ (\ref{abbreviations}) and defined
\be
\x_i\equiv e_ik_i-\m \,\, .
\ee
In the limit $p\to 0$, or, equivalently, $k_1\to k_2$, we obtain for 
$r=1,2$ 
\be
v^{rr}\simeq -\frac{1}{\m^2}\int_0^\infty dk\,k^2\left[
\frac{\l_r\phi^2}{4\e_r^3}(1-2N_r)-\frac{\e_r^2+\x^2}{2\e_r^2}
\frac{dN_r}{d\e_r}   \right] \,\, ,
\ee
where we used $(N_{1,r}-N_{2,r})/(\e_{1,r}-\e_{2,r})\simeq dN_r/d\e_r$
and $\xi\equiv k-\mu$.
All quantities now depend on $k_1=k$ and correspond to positive energies, 
$e_1=e_2=1$, since, due to $\phi^-\simeq 0$, the terms corresponding 
to negative energies vanish. Therefore, we omit the index $i=1,2$. 
Note that, up to subleading order,  the gap function does not depend 
on the index $r$, $\phi_1\simeq\phi_2\equiv\phi$ \cite{schmitt1}. 
We have to distinguish between the cases where $\l_r\neq 0$ and 
where $\l_r=0$. When $\l_r\neq 0$, we obtain for zero temperature $T=0$,
where $N_r=0$, 
\be
v^{rr}=-\frac{1}{\m^2}\int_0^\infty dk\, k^2\frac{\l_r\phi^2}{4\e_r^3}
\simeq -\frac{1}{2}\int_0^\m d\x \l_r\phi^2(\x^2+\l_r\phi^2)^{-3/2}
\simeq -\frac{1}{2} \,\, .
\ee
Here, we restricted the $k$-integration to the region $0\le k\le 2\m$ since
the gap function is strongly peaked around the Fermi surface, $k=\m$.
In this region, we assume that the gap function does not depend on the
momentum $k$. 
Since the eigenvalue $\l_r$ cancels out, this result does not depend on 
$r$. Therefore, it gives rise to the value of $v^{11}$ for all four
considered phases and for $v^{22}$ in the cases of the CFL and
CSL phases.
When $\l_r=0$, we obtain (for arbitrary temperature $T$)
\be
v^{rr}=\frac{1}{\m^2}\int_0^\infty dk \, k^2\frac{dN_F}{dk} \,\, ,
\ee 
where $N_F\equiv 1/[\exp(\x/T)+1]$ is the Fermi distribution. Using
$\m\gg T$, and the substitution $\z=\x/2T$, this can be transformed to
\be
v^{rr}\simeq -\int_0^\infty d\z \frac{1}{\cosh^2\z} = -1 \,\, .
\ee
Let us now compute $v^{rs}$ for $r\neq s$. Using Eq.\ (\ref{v}),
we obtain for two nonvanishing eigenvalues $\l_1,\l_2$ and at zero 
temperature
\be \label{hatv12}
v^{12}=v^{21}=-\frac{1}{\m^2} \int_0^\infty dk\, k^2
\frac{\e_1\e_2-\x^2}{2\e_1\e_2(\e_1+\e_2)}\simeq 
-\int_0^\m d\x \frac{1}{(\l_2-\l_1)\phi^2}\left(\frac{\e_2^2+\x^2}{\e_2}
-\frac{\e_1^2+\x^2}{\e_1}\right)\simeq -\frac{1}{2} \,\, .
\ee
Again, we neglected the integral over the region $k>2\m$.
In the last integral, the leading order contributions cancel. The 
subleading terms give rise to $(\l_2-\l_1)\phi^2/2$. Therefore, 
the eigenvalues cancel and the result does not depend on $\l_1,\l_2\neq 0$.
Eq.\ (\ref{hatv12}) holds for the CFL and CSL phases.

When one of the excitation branches is ungapped (2SC and polar phase), 
for instance, $\l_2=0$, we have
\be
v^{12}=v^{21}=-\frac{1}{\m^2}\int_0^\infty dk\,k^2 
\left[ \frac{n_1}{\e_1+\xi}
(1-N_1-N_F)-\frac{1-n_1}{\e_1-\x}(N_1-N_F)\right] \,\, .
\ee
For $T=0$, this becomes with the standard approximations
\bea
v^{12}&\simeq&-\frac{1}{\m^2}\int_{-\m}^\m d\x \frac{(\x+\m)^2}{2\e_1}
\left[\frac{\e_1-\x}{\e_1+\x}\Theta(\x)+\frac{\e_1+\x}{\e_1-\x}\Theta(-\x)
\right] \nonumber \\ 
&=&-\int_0^\m \frac{d\x}{\e_1}\left(1+\frac{\x^2}{\m^2}\right)
\frac{\e_1-\x}{\e_1+\x}\simeq -\frac{1}{2} \,\, ,
\eea
where the last integral has already been computed in Ref.\ \cite{meissner2}. 

Next, we compute $\bar{v}^{rs}$ for the various cases. Setting the
thermal distribution function for antiparticles to zero, 
we obtain with the definition in Eq.\ (\ref{kintegrals})
\be  \label{tildevrs}
\bar{v}^{rs}=-\frac{1}{\m^2}\int_0^\infty dk\, k^2\left[
\frac{(1-n_r)(1-N_r)}{\e_r^++\e_s^-}-\frac{N_rn_r}{\e_r^+-\e_s^-}
+\frac{(1-n_s)(1-N_s)}{\e_r^-+\e_s^+}-\frac{N_sn_s}{\e_s^+-\e_r^-}\right]
\ee
with $\e_r^\pm\equiv\e_{k,r}^\pm$. Again, we first consider the situation where
$r=s$. In this case, for $T=0$ and $\l_r\neq 0$, (defining $\e\equiv\e^+$)
\bea \label{tildevfinal}
\bar{v}^{rr}&=&-\frac{2}{\m^2}\int_0^\infty dk\, k^2\left[\frac{1-n_r}{\e_r
+k+\m}-\frac{1}{2k}\right] \nonumber \\
&&=\frac{1}{\m^2}\int_0^\infty dk\, k \, \frac{\m(\e_r+\m-k)+\l_r\phi^2}
{\e_r(\e_r+k+\m)}\simeq \frac{1}{2} \,\, .
\eea
Here, $-1/(2k)$ is a vacuum subtraction. The last integral has been computed
in Ref.\ \cite{meissner2}. Note that the result does not depend on the value
of $\l_r\neq 0$. Thus, it is valid for $\bar{v}^{11}$ in the 2SC and 
polar phases as well as for both $\bar{v}^{11}$ and $\bar{v}^{22}$ in
the CFL and CSL phases.

For $\l_r= 0$, we obtain from Eq.\ (\ref{tildevrs}) at zero temperature
and with the vacuum subtraction $1/(2k)$
\be \label{tildevfinal2}
\bar{v}^{rr}=-\frac{2}{\m^2}\int_0^\infty dk\, k^2\left[\frac{\Theta(k-\m)}
{2k}-\frac{1}{2k}\right]=\frac{1}{2} \,\, .
\ee
This result holds for $\bar{v}^{22}$ in the 2SC and polar phases.

Next, we discuss the case $r\neq s$.  From Eqs.\ (\ref{tildevrs}) and 
(\ref{tildevfinal}) it is obvious that, since the result for $\bar{v}^{rr}$ did
not depend on $\l_r$, we get the same result for $\bar{v}^{rs}$ 
with two nonvanishing eigenvalues $\l_1,\l_2 \neq 0$. Thus, in this
case,
\be \label{tildev12}
\bar{v}^{12}=\bar{v}^{21}=\frac{1}{2} \,\, .
\ee
For $\l_2=0$, we obtain at zero temperature
\be
\bar{v}^{12}=\bar{v}^{21}=-\frac{1}{\m^2}\int_0^\infty dk\, k^2\left[
\frac{1-n_1}{\e_1+k+\m}+\frac{\Theta(k-\m)}{2k}-\frac{3}{2k}\right]
\simeq \frac{1}{2} \,\, ,
\ee
where identical integrals as in Eqs.\ (\ref{tildevfinal}) and
(\ref{tildevfinal2}) were performed. Consequently, also for the 2SC and 
polar phases, Eq.\ (\ref{tildev12}) holds.

\section*{Contributions from anomalous propagators at $T=0$}

We start from Eq.\ (\ref{defw}) which, for $p_0=0$, is 
\be \label{wrs}
w_{e_1e_2}^{rs}=\frac{\phi^{e_1}\phi^{e_2}}{2\e_{1,r}\e_{2,s}(\e_{1,r}
+\e_{2,s})}(1-N_{1,r}-N_{2,s})+\frac{\phi^{e_1}\phi^{e_2}}
{2\e_{1,r}\e_{2,s}(\e_{1,r}-\e_{2,s})}(N_{1,r}-N_{2,s}) \,\, .
\ee
Obviously, since $\phi^-\simeq 0$, for all $r$, $s$ and all phases we have 
$\bar{w}^{rs}=0$ (for the definition of $\bar{w}^{rs}$ 
cf.\ Eq.\ (\ref{kintegrals})).
First, we calculate $w^{rs}$ for $r=s$. Neglecting the antiparticle 
gap and taking the limit $k_1\to k_2$, we obtain, 
using the same notation as above, 
$\e_r\equiv \e^+_{k,r}$, 
\be 
w^{rr}=\frac{1}{\m^2}\int_0^\infty dk \, k^2\left[\frac{\phi^2}{4\e_r^3}
(1-2N_r)+\frac{\phi^2}{2\e_r^2}\frac{dN_r}{d\e_r}\right] \,\, .
\ee
For $\l_r\neq 0$ and at zero temperature, where $N_r=0$, this expression 
reduces to
\be
w^{rr}=\frac{1}{\m^2}\int_0^\infty dk \, k^2 \frac{\phi^2}{4\e_r^3}
\simeq\frac{1}{2}\int_0^\m\d\x\,\phi^2(\x^2+\l_r\phi^2)^{-3/2}\simeq
\frac{1}{2\l_r} \,\, .
\ee
Unlike in $v^{rr}$, the eigenvalue $\l_r$ does not cancel out and we
obtain different results for $\l_r=4$ and $\l_r=1$. (Remember the different 
normalization compared to Sec.\ \ref{gapeqsolution}, cf.\ Eqs.\ 
(\ref{VWrescale}) and comments above these equations.) In the cases of 
the 2SC and polar phases we obtain $w^{11}=1/2$ whereas for
the CFL and CSL phases, $w^{11}=1/8$ and $w^{22}=1/2$.
In the former two cases, the quantity $w^{22}$ does not occur
in our calculation. 
Thus we can turn to the case where $r\neq s$. Here, we conclude from 
Eq.\ (\ref{wrs}) 
\be
w^{12}=w^{21}=\frac{1}{\m^2}\int_0^\infty dk \, k^2 
\left[\frac{\phi^2}{2\e_1\e_2(\e_1+\e_2)}(1-N_1-N_2)+ 
\frac{\phi^2}{2\e_1\e_2(\e_1-\e_2)}(N_1-N_2)\right] \,\, .
\ee
For two nonvanishing eigenvalues $\l_1$, $\l_2$, this reads at $T=0$
\bea
w^{12}&=&\frac{1}{\m^2}\int_0^\infty dk \, k^2 \frac{\phi^2}
{2\e_1\e_2(\e_1+\e_2)}\simeq\frac{1}{\l_1-\l_2}
\int_0^\m d\x \, \left(\frac{1}{\e_2}-\frac{1}{\e_1}\right) \nonumber \\
&\simeq&\frac{1}{2}\frac{1}{\l_1-\l_2}\ln\frac{\l_1}{\l_2} \,\, .
\eea
For $\l_1=4$ and $\l_2=1$ (CFL and CSL phases) we obtain 
\be
w^{12}=w^{21}=\frac{1}{3}\ln 2 \,\, .
\ee
The corresponding expression for the case where the second eigenvalue 
vanishes, $\l_2=0$, does not occur in our calculation.

\section*{Integrals in the normal phase, $T\ge T_c$}

For temperatures larger than the transition temperature $T_c$ but still
much smaller than the chemical potential, $T\ll \m$, all integrals defined
in Eqs.\ (\ref{kintegrals}) are easily computed. Since there is no 
gap in this case, $\phi = 0$, all contributions from the anomalous 
propagators vanish trivially, $w^{rs}=\bar{w}^{rs}=0$. From 
Eq.\ (\ref{v}), which holds for all temperatures, we find with
$\phi = 0$
\be
v^{11}=v^{22}=v^{12}=\frac{1}{\m^2}\int_0^\infty dk\, k^2
\frac{dN_F}{d\x}\simeq -1 
\ee
and (with the vacuum subtraction $1/k$)
\be
\bar{v}^{11}=\bar{v}^{22}=\bar{v}^{12}=-\frac{1}{\m^2}
\int_0^\infty dk\, k^2 \left[\frac{1-N_F}{k}-\frac{1}{k}\right]
\simeq \frac{1}{2} \,\, .
\ee
   %Quark momentum integration
    \chapter{Summing over Matsubara frequencies} \label{matsubara}

In this appendix, we prove Eqs.\ (\ref{matsu1}) and (\ref{matsu2}). 
For further details concerning Matsubara frequencies see for instance 
Refs.\ \cite{lebellac,kapusta}. 

In order
to prove Eq.\ (\ref{matsu1}), we compute $\sum_{k_0}(\e_k^2-k_0^2)/T^2$
via performing the sum over fermionic Matsubara frequencies, 
$k_0 = -i\omega_n = -i(2n+1)\pi T$. This calculation can be
divided into four steps:

\begin{enumerate}
\item Rewrite the logarithm as an integral:

\be \label{xint}
\ln\frac{\e_k^2-k_0^2}{T^2} =
\int_{1}^{(\e_k/T)^2}dx^2\, \frac{1}{x^2+(2n+1)^2\pi^2} + \ln [1 + 
(2n+1)^2\pi^2 ] \,\, .
\ee

\item Write the sum over $k_0$ as a contour integral: 

Let us define the integral kernel in Eq.\ (\ref{xint}) as a function 
of $k_0$,
\be
f(k_0)\equiv \frac{1}{x^2+(2n+1)^2\pi^2} = \frac{T^2}{T^2x^2 -k_0^2} \,\, .
\ee
Then, the sum over this function can be written as
\be \label{contour}
T\sum_n \left. f(k_0) \right|_{k_0=-i\omega_n} = \frac{1}{2\pi i}
\oint\limits_C dk_0\,f(k_0)\frac{1}{2}\,\tanh\frac{k_0}{2T} \,\, ,
\ee
where $C$ is a contour that encloses a region in the complex plane 
in which all poles of $\tanh\frac{k_0}{2T}$ are located and where $f(k_0)$ 
is analytical.

Proof of Eq.\ (\ref{contour}): Remember the residue theorem
\be
\frac{1}{2\pi i}\oint\limits_C dz \, f(z) = \sum_n {\rm Res}\,\left. 
f(z) \right|_{z=z_n} \,\, ,
\ee
where $z_n$ are the singularities of $f$ in the area surrounded by $C$.
If $f$ can be written in the form 
\be
f(z) = \frac{\varphi(z)}{\psi(z)} \,\, ,
\ee
then
\be
{\rm Res}\,\left. f(z) \right|_{z=z_n} = \frac{\varphi(z_n)}{\psi'(z_n)} \,\, ,
\ee
where $z_n$ are the roots of $\psi(z)$ and $\psi'\equiv d\psi/dz$. 
We apply this form of the
residue theorem to our calculation. Since
\be
\tanh\frac{k_0}{2T} = \frac{e^{k_0/(2T)} -e^{-k_0/(2T)}}{e^{k_0/(2T)} + 
e^{-k_0/(2T)}} \,\, ,
\ee
the poles of $\frac{1}{2}f(k_0)\tanh\frac{k_0}{2T}$ are given by
\be
k_0 = -i(2n + 1)\,\pi\, T = -i\omega_n \,\, , \qquad n \in {\cal Z} \,\, .
\ee
With 
\be
\frac{d}{dk_0}\left[e^{k_0/(2T)} + e^{-k_0/(2T)}\right]_{k_0=-i\omega_n} 
= \frac{i}{T}(-1)^{n+1} \,\, ,
\ee
and
\be
\left[e^{k_0/(2T)} -e^{-k_0/(2T)}\right]_{k_0=-i\omega_n} = 2i\,(-1)^{n+1} \,\, ,
\ee
we obtain for all $n$, 
\be
{\rm Res}\left[f(k_0)\frac{1}{2}\tanh\frac{k_0}{2T}\right]_{k_0=-i\omega_n}
=T\,f(-i\omega_n) \,\, ,
\ee  
which proves Eq.\ (\ref{contour}).

\item Compute the contour integral:

Since the poles of $\frac{1}{2}f(k_0)\tanh\frac{k_0}{2T}$ are on the imaginary
axis, we choose the contour for the integral on the right-hand side 
of Eq.\ (\ref{contour}) in the following way:
\bea \label{contour2}
\frac{1}{2\pi i}
\oint\limits_C dk_0\,f(k_0)\frac{1}{2}\,\tanh\frac{k_0}{2T} 
&=&\frac{1}{2\pi i}
\int\limits_{-i\infty + \eta}^{i\infty + \eta} dk_0\,f(k_0)\frac{1}{2}\,
\tanh\frac{k_0}{2T} 
+
\frac{1}{2\pi i}
\int\limits_{i\infty - \eta}^{-i\infty - \eta} dk_0\,f(k_0)\frac{1}{2}\,
\tanh\frac{k_0}{2T}  \non
&=& \frac{1}{2\pi i}
\int\limits_{-i\infty + \eta}^{i\infty + \eta} dk_0\,\frac{1}{2}[f(k_0) + 
f(-k_0)]\,
\tanh\frac{k_0}{2T} \non
&=&
\frac{1}{2\pi i}
\int\limits_{-i\infty + \eta}^{i\infty + \eta} dk_0\,\frac{T^2}{T^2x^2-k_0^2}\,
\left(1 - \frac{2}{e^{k_0/T}+1}\right) \non
&=&
\frac{T}{x}\left(\frac{1}{2} - \frac{1}{e^x+1}\right) \,\, ,
\eea
where, in the last step, we closed the integration path in clockwise 
direction (minus sign!) such that one of the poles, namely $k_0=Tx$ is 
enclosed, and used once more the residue theorem.

\item Collect the results:  

With Eqs.\ (\ref{xint}), (\ref{contour}), and (\ref{contour2}), we now
compute the Matsubara sum. Omitting the terms which do not depend on $k$, we 
obtain
\bea 
\sum_{k_0}\ln\frac{\e_k^2-k_0^2}{T^2} &=& 
\int_{1}^{(\e_k/T)^2}dx^2\,\frac{1}{x}\left(\frac{1}{2} - \frac{1}{e^x+1}
\right) \non
&=&
\frac{\e_k}{T} + 2\ln\left[1+\exp\left(-\frac{\e_k}{T}\right)\right] \,\, .
\eea
This is Eq.\ (\ref{matsu1}).

\end{enumerate}

Applying the same method, it is easy to prove also Eq.\ (\ref{matsu2}). 
With the definition 
\be
f(k_0)\equiv \frac{\varphi(k_0)}{k_0^2-\e_k^2} 
\ee
we can proceed along the same lines as described in 2.\ and 3.\ above. Using 
$\varphi(k_0)=\varphi(-k_0)$, we obtain
\bea
\sum_{k_0} f(k_0) &=& \frac{1}{2\pi i}
\int\limits_{-i\infty + \eta}^{i\infty + \eta} dk_0\,f(k_0)\,
\tanh\frac{k_0}{2T} \non
&=&-\frac{\varphi(\e_k)}{2\e_k}\,\tanh\frac{\e_k}{2T} \,\, .
\eea
This is Eq.\ (\ref{matsu2}).

\end{appendix}

%\germanTeX
%\include{zusf} %deutsche Zusammenfassung

%\originalTeX

 %Bibliography

%\germanTeX
%\include {cv}  %Lebenslauf

\end{document}